\documentclass[]{dukedissertation}

 \usepackage{notoccite}
\usepackage[utf8]{inputenc}
\usepackage{amsmath, amssymb, amsfonts, amsthm}
\usepackage{graphicx}
\usepackage{color}
\usepackage{bm}
\usepackage{subfigure}
\usepackage{mathabx}
\usepackage{multirow}
\usepackage{setspace}
\usepackage{float}							
\usepackage{fullpage}
\usepackage{verbatim}
\usepackage{slashed}
\usepackage{stmaryrd}
\usepackage{pstricks}
\usepackage[T1]{fontenc}
\usepackage[utf8]{inputenc}
\usepackage{mathtools} 
\usepackage{xspace}
\usepackage{strobel}
\usepackage[utf8]{inputenc}
\usepackage{multido}
\usepackage{mathrsfs}
\usepackage{xspace}
\usepackage{graphicx}		
\usepackage{graphics}
\usepackage{units}
\usepackage{ulem}
\usepackage{physics}
\usepackage[square,numbers]{natbib}
\setcitestyle{numbers}

\newcommand{\EFT}{$\mathrm{EFT}_{\slashed{\pi}}$\xspace}

\newcommand\StrobelDistance{0.8em}

\newcommand{\vect}[1]{\vec{\boldsymbol{#1}}}

\newcommand{\SixJ}[6]{\left\{\begin{array}{ccc} #1 & #2 & #3\\ #4 & #5 & #6 \end{array}\right\}}
\newcommand{\CG}[6]{C_{#1,#3,#5}^{#2,#4,#6}}

\newcommand{\wave}[3]{\ensuremath{{}^{#1}\mathrm{#2}_{#3}}\xspace}

\newcommand{\oneP}{\wave{1}{P}{1}}
\newcommand{\threePzero}{\wave{3}{P}{0}}
\newcommand{\threePone}{\wave{3}{P}{1}}
\newcommand{\threePtwo}{\wave{3}{P}{2}}

\newcommand{\Pc}{\hat{\mathcal{P}}}

\renewcommand{\emph}[1]{\textit{#1}}

\newcommand{\nnlo}{\mathrm{N}^{2}\mathrm{LO}}
\newcommand{\ntlo}{\mathrm{N}^{3}\mathrm{LO}}

\newcommand{\SJ}[6]{\left\{\begin{array}{ccc} #1 & #2 & #3\\ #4 & #5 & #6 \end{array}\right\}}
\newcommand{\NJ}[9]{\left\{\begin{array}{ccc} #1 & #2 & #3\\ #4 & #5 & #6 \\ #7 & #8 & #9 \end{array}\right\}}

\graphicspath{{./Pictures/}}

\author{Arman Margaryan}
\title{Transverse Asymmetry in Nucleon Deuteron Scattering in Pionless Effective Field Theory}
\supervisor{Roxanne Springer}
\department{Physics} 

\date{2011} 

\copyrighttext{ All rights reserved except the rights granted by the\\
   \href{http://creativecommons.org/licenses/by-nc/3.0/us/}
        {Creative Commons Attribution-Noncommercial Licence}
}

\usepackage[pdftex, pdfusetitle, plainpages=false, 
				letterpaper, bookmarks, bookmarksnumbered,
				colorlinks, linkcolor=black, citecolor=black,
	         filecolor=black, urlcolor=black]
				{hyperref}
\begin{document}

\maketitle

\abstract

In this dissertation we apply the pionless effective field theory (\EFT) to low energy neutron deuteron elastic scattering process. We calculated some of the polarization observables in neutron deuteron scattering to next-to-next-to-next-to-leading-order, in particular the transverse asymmetry $A_y$. 
All of the previous calculations have the same characteristic feature of under-predicting this observable, this is known as the $A_y$-puzzle. 
At this order of the \EFT new two-body $P$-wave interaction terms enter into the Lagrangian. This interaction terms give crucial contributions 
to the $A_y$ observable. By varying the interaction coefficients within the allowed error estimates of the \EFT we find results that at this order are 
consistent with the experimental data.
Our conclusion is that the $A_y$-puzzle is likely to be solved within the next few orders of the \EFT. Other observables 
in neutron-deuteron scattering process are also calculated and are in good agreement with the experimental data. 

\dedication{To my brother David Margaryan.}

\tableofcontents 
\listoffigures	
\abbreviations


\section*{Symbols}


\section*{Abbreviations}

\begin{symbollist}
	\item[EFT] Effective Field Theory
	\item[\EFT] Pionless Effective Field Theory
	\item[QFT] Quantum Field Theory
	\item[QCD] Quantum Chromodynamics
	\item[QED] Quantum Electrodynamics 
	 \item[$\chi$PT] Chiral perturbation theory
	  \item[HQET] Heavy Quark Effective Theory
	  \item[HH$\chi$PT] Heavy Hadron Chiral Perturbation Theory
	   \item[X-EFT] Effective Field Theory for the $X(3872)$
	   \item[SCET] Soft-Collinear Effective Theory
	   \item[PDS] Power Divergence Subtraction
	   \item[ERE] Effective Range Expansion 
	   \item[LO] Leading Order
	   \item[NLO] Next to Leading Order
	    \item[NDA] Naive Dimensional Analysis
	   \item[CM] Center of Mass
	    \item[RG] Renormalization Group 
\end{symbollist}


\acknowledgements

I would like to thank Roxanne Springer for being a great advisor.
I would also like to thank Jared Vanasse for all the discussions during the course of this work. 

This material is based upon work supported by the U.S. Department of Energy, Office of Science, Office
of Nuclear Physics, under Award Number DE-FG02-05ER41368 and Award Number DE-FG02-93ER40756 (JV).

%
%
%
\chapter{Introduction}
\section{Strong interactions at low energies}
In quantum field theories (QFTs) in general the coupling constants are renormalized at some renormalization energy scale $\mu$. This scale is arbitrary and the coupling 
constants change their values depending on $\mu$. This is called the renormalization group (RG) flow or the running of the coupling constants. The equation 
that describes the running of the coupling constant depending on the renormalization scale $\mu$ is called the renormalization group equation. It 
relates the derivative of the coupling constant to a function of the coupling constant Eq.~\eqref{CSeq}.
\begin{equation} \label{CSeq}
\mu \frac{d}{d\mu} g_{\mu}=\beta(g_{\mu}),
\end{equation}
where $g_{\mu}$ is the renormalized coupling at the renormalization scale $\mu$. The right hand side of this equation is called the beta-function 
and is calculated perturbatively. Usually by requiring that the Lagrangian is real and the action of the theory has a minimum it is possible 
to restrict the values of the coupling $g_{\mu}$ to be non-negative. From Eq.~\eqref{CSeq} we see that if the beta-function is positive 
then the coupling constant increases with increasing $\mu$. This is the case for example for a scalar field theory for the field $\phi$ with a $\phi^4$ interaction and quantum electrodynamics (QED). But if the beta-function is negative then the coupling constant decreases with increasing $\mu$. This is called 
the asymptotic freedom. An example theory that has this feature is quantum chromodynamics (QCD). In the regions of $\mu$ 
where the coupling constants are small we can use perturbation theory to calculate matrix 
elements and scattering amplitudes as sums of Feynman diagrams. But when $\mu$ is such 
that the coupling constants are large perturbation theory is impossible. This is called the non-perturbative regime of the theory. The non-perturbative regime of QCD is at low energies where 
the nuclear reactions happen. 

\begin{center}\begin{figure}[ht]
  \begin{center}
  \includegraphics[scale=0.7]{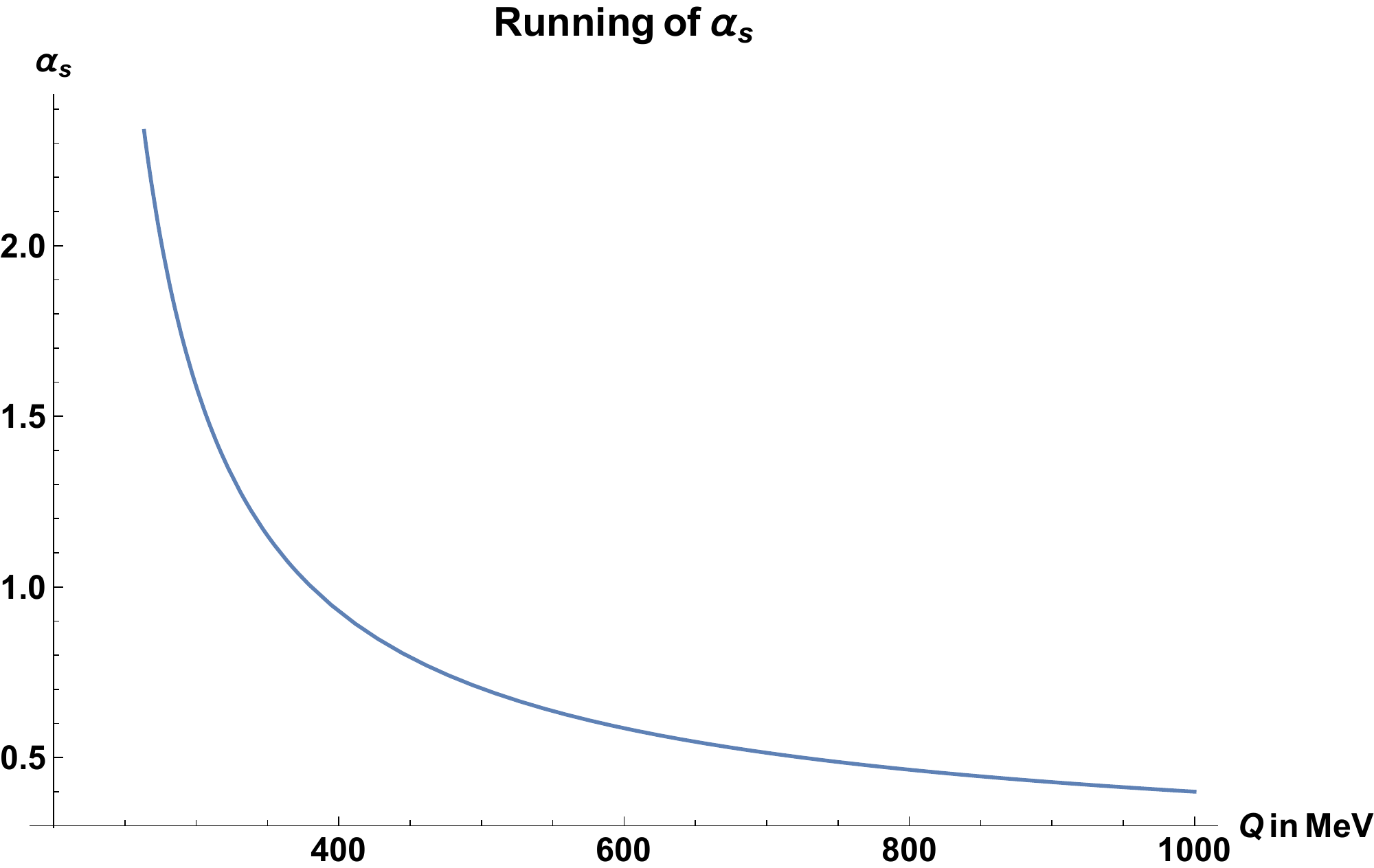}
  \end{center}
   \vspace{0.3cm} \caption{\label{runningalpha}The RG running of the QCD coupling constant.}
\end{figure}\end{center}

The atomic nucleus is composed of neutrons and protons, also called nucleons. Nucleons themselves are composite particles, composed of quarks, which carry color and electromagnetic charge. Quarks interact by strong interactions carried by gluons and by electromagnetic forces carried by photons. The fundamental theory that describes strong interactions is QCD, which is asymptotically free at high energies. This means that 
 the beta-function for the QCD coupling $\alpha_s$ is negative, making the coupling decrease with increasing renormalization scale (see Fig.~\ref{runningalpha}). Solving the RG flow 
 equation Eq.~\eqref{CSeq} for $\alpha_s$ at the lowest order in perturbation theory at which the beta-function gets a non-zero contribution, gives the asymptotic behavior of $\alpha_s(Q)\sim\frac{1}{\log{Q}}$ \cite{Peskin:257493,Weinberg:1996kr}, which is the behavior that we see in Fig.~\ref{runningalpha}. It follows that at high energies $\alpha_s$ approaches to zero and QCD is perturbative, but at low energies 
 the interaction coefficient is large. The behavior of $\alpha_s$ shown in Fig.~\ref{runningalpha} is derived based on the assumption that the coupling $\alpha_s$ is small, so the part 
 of the Fig.~\ref{runningalpha} where $\alpha_s$ is large cannot be trusted. But the fact that $\alpha_s$ decreases with increasing energy scale means that it increases when the 
 energy scale is small. So before reaching the region where $\alpha_s$ is large in Fig.~\ref{runningalpha} the perturbation theory breaks down precisely because $\alpha_s$ becomes large.  
 
 To describe processes involving strongly interacting particles at high energies we can use perturbation theory and calculate Feynman diagrams contributing to the scattering amplitudes. But at lower energies $\alpha_s$ is larger and perturbation theory is not applicable anymore. The low energy few-nucleon systems fall into this energy regime where $\alpha_s$ is large, hence are 
 governed by non-perturbative QCD. Since non-perturbative QCD is extremely difficult and is not yet solved we have to look for other ways to overcome this difficulty and to be able to give quantitative predictions about the low energy few-nucleon systems. I'll describe some of them briefly.

  One of the approaches is the lattice QCD \cite{Wilson:1977nj}. The idea is to discretize the space-time dimensions making it into a lattice. This makes the path integral for the observables into summations over many particle configurations on the lattice. These sums are then calculated numerically and predictions are extracted. Then the lattice spacing is taken to zero and the lattice volume is taken to infinity. The predictions become independent of lattice spacing and this is the continuum limit of lattice QCD. The advantage of this approach is that it gives a non-perturbative result for the observables, but the disadvantage is that it becomes extremely complicated when the number of nucleons increases; a certain class of complications arise called the sign problems. Also it is hard to deal with bound states in lattice QCD.

  Another class of approaches is called potential models \cite{Yukawa:1935xg,Stoks:1994wp,Reid:1968sq,Nagels:1978me,PhysRevC.51.38AV18}. The idea is to construct two-nucleon and three-nucleon interaction potentials based on the known symmetries, then to write the Schr\"{o}dinger equation and determine the energy levels and wave functions by solving the Schr\"{o}dinger equation numerically. The potentials will have some adjustable parameters which are fit to the few nucleon experimental data. The different potential models usually differ on the short range part of the interaction and the long range part is usually taken to be a one-pion exchange Yukawa potential: $\sim \frac{e^{-m_{\pi}r}}{r}$ \cite{Yukawa:1935xg}. The disadvantage of potential models is that they do not give a natural way of estimating the theoretical error associated with a particular model and they are not systematically improvable. Also they don't have a way of making a connection to QCD 
  as opposed to effective field theories (EFTs), which are constructed with the symmetries of the underlying fundamental theory. 
    
  Our approach is the non-relativistic pionless effective field theory (\EFT) \cite{Beane:2000fx,Kaplan:1998we,Bedaque:1999vb,Griesshammer:2004pe,Phillips:1999hh,Vanasse:2016jtc,Bedaque:1999ve,Ji:2012nj,Griesshammer:2005ga,Vanasse:2013sda}. The \EFT is constructed to describe nucleons and their interactions at low energies, such that pions are not explicit degrees of freedom and the appropriate degrees of freedom are 
  the nucleon fields coupled with external electromagnetic fields and weak currents. 
  
  \section{Effective Field Theories}

      EFTs are quantum field theories that approximate an underlying ``true'' field theory such as QED, QCD, the standard model, etc.. \cite{Weinberg:1978kz}.
To describe a given physical process there might be some physical argument why some of the information contained in the underlying 
theory is irrelevant. For example those processes might be at a certain energy regime, which defines an energy scale in the problem,
 and in the underlying theory there might exist particles that are much heavier than this scale and hence the dynamics of those particles are irrelevant for the description of those processes to some order. 
 Another example of how an EFT can approximate an underlying theory might be the following: say again we are trying to describe a process at a certain energy and the underlying theory contains particles that are, this time, much lighter than this scale, then, by neglecting the mass of those particles one might obtain more symmetries in the theory which can be useful in the description of the process.  
   
    There are two ways to construct an EFT: top to bottom and bottom to top. In a certain sense the top to bottom approach can serve as a justification for the bottom to top approach, so I'll describe both of these. 
 
    Here are the steps in the top to bottom approach. Start with the underlying ``true'' theory and write the path integral for the scattering amplitude of a given process. Then figure out the irrelevant particles (degrees of freedom). Start integrating them out of the path-integral. The last step is an order-by-order process, which gives an expansion in powers of a parameter called the power counting parameter. This parameter can be, for example the ratio of the energy scale existing in the problem over the mass of the particles that are being integrated out. At each order the derived theory is equivalent to a theory given by a particular Lagrangian, which is called the ``effective Lagrangian.'' At every order the effective Lagrangian is the same as at the previous order, along with more terms of higher order in the power counting parameter. Those extra terms are constrained only by the symmetries of the original theory. All the parameters appearing at each order, such as masses or coupling constants in the interaction terms, are fixed by the original parameters of the underlying theory and the relevant energy scales. All of these new parameters encode the information contained in the original theory. As a result one gets a list of effective Lagrangians which can describe the process under consideration in simpler terms than the original theory, because these Lagrangians only contain the relevant degrees of freedom. One hopes that at each order the description gets better, because at each subsequent order the new contributions to the Lagrangian are of higher power in power counting parameter. 
  
    In the bottom to top approach the steps are as follows. First figure out and postulate what the symmetries of the underlying ``true'' theory are. Then take all the degrees of freedom which are the fields of the particles relevant to the process under consideration. Figure out what the power counting parameter is. Then write down a completely general Lagrangian in which the terms go up to a certain order in the power counting parameter, respecting all the symmetries of the original theory. Again at each order we may get several new terms in the Lagrangian. For practical purposes the parameters in the Lagrangian are not fixed by matching to the underlying theory. To fix the parameters in the effective Lagrangian at each order experimental data may be used (position of the poles corresponding to bound states, residues around the poles, scattering lengths, scattering cross-sections, etc...). Again the expectation is that at each order the description of a process under consideration gets more accurate. 
    
   QFTs which are constructed to be fundamental theories valid for all energy scales, are required to be renormalizable. This means that all the divergencies at any order can be absorbed into a finite set of interaction parameters. 
    EFTs do not have to satisfy this requirement, because at each order we get new terms in the Lagrangian which can absorb the divergencies. This is called order by order renormalizability. 
    
    Here are some examples of EFTs:
    
    1) Chiral perturbation theory ($\chi$PT) \cite{Bernard:1995dp,Weinberg:1990rz,vanKolck:1994yi,Machleidt:2011zz,Nogga:2005hy,Choudhury:2007bh,Shukla:2008zc,Griesshammer:2015ahu,Pascalutsa:2002pi,Beane:1999uq,Hildebrandt:2003fm,BEANE2005311}. In processes where the masses of the light quarks are negligible we can neglect the mass terms for those quarks in the QCD Lagrangian. Then the four component Dirac spinors representing those quarks can be broken into two two-component spinors, which are called left-handed and right-handed Weyl spinors. After doing so it becomes apparent that there are two $SU(2)$ symmetry groups, one acting on left- and the other on right-handed spinors, leaving the Lagrangian invariant. So the total symmetry group is an SU(2)$_L \times$ SU(2)$_R$ which is called the chiral symmetry group. In constructing the $\chi$PT Lagrangian the terms have to respect all the QCD symmetries and an additional chiral symmetry at leading order (LO). At next-to-leading order (NLO) we should have explicit symmetry breaking for the chiral symmetry as it is broken in QCD by the mass terms of the light quarks. The inclusion of the mass terms, although being NLO in the power counting of the $\chi$PT, may give the LO (first non-zero) contribution for certain observables. We will see the same kind of effect in pionless EFT too, where certain spin-polarization observables get their LO contribution from the N\textsuperscript{2}LO interaction terms in the EFT power counting. The power counting parameters in $\chi$PT are the ratio of the pion masse $m_{\pi}$ over $\Lambda_{QCD}$, and the momentum transfer in a process over $\Lambda_{QCD}$. There is a simultaneous expansion in powers of both of these parameters. 
        
    2) Heavy quark effective theory (HQET) \cite{ISGUR1989113,GEORGI1990447}. This seems to be the opposite situation of the $\chi$PT; here the masses of the heavy quarks are taken to infinity. The heavy quarks are treated as classical, stationary sources. At LO no interaction will rotate their spin, hence no interaction can depend on their spin and we get an additional spin-symmetry, which is the heavy quark spin symmetry. Here the power counting parameter is the ratio of $\Lambda_{QCD}$ over the heavy quark masses. 
    
    3) Heavy hadron chiral perturbation theory (HH$\chi$PT) \cite{PhysRevD.45.R2188,Wise:1993wa,Falk:1991nq,DeFazio:2008xq}. HH$\chi$PT is a combination of $\chi$PT and HQET. Here the heavy hadrons are considered as nonrelativistic particles, so the expansion corresponding to heavy quark spin symmetry is in the powers of the derivatives acting on the rotated heavy hadron fields, because each derivative brings a heavy quark residual momentum down from the exponent. 
    
    4) X-EFT \cite{Fleming:2007rp}. This is a particular type of HH$\chi$PT concerning the physics of the $X(3872)$ particle  \cite{Choi:2003ue,Acosta:2003zx,Abazov:2004kp,Aubert:2004ns,Aaij:2011sn}. It describes 
    D-mesons and pions near the mass of the $X(3872)$. The power counting in X-EFT is the same as for HH$\chi$PT; the operators are 
    arranged in powers of the derivatives acting on them. X-EFT can be used to calculate different decay rates and production mechanisms 
    of the $X(3872)$ \cite{Margaryan:2013tta,Mehen:2011ds}.

    5) Soft-Collinear effective theory (SCET) \cite{Bauer:2000ew}. SCET is constructed to describe processes that involve energetic quark-gluon jets and not-so-energetic (soft) gluons. The power counting parameter is the ratio of the two lightcone-momentum components of the collinear quarks $\lambda=\frac{p_{\perp}}{p^{-}}$. The less $\lambda$ is the more collinear is the quark \cite{Leibovich:2003jd}. 
    
    6) Non-relativistic pionless EFT (\EFT) \cite{Bedaque:2002mn}. \EFT is about nonrelativistic nucleons and their interactions. In the processes analyzed by \EFT the typical momentum exchange in the scattering must be much smaller than the mass of the pion. The power counting 
parameter for \EFT is the ratio $\frac{Q}{\Lambda_{\slashed{\pi}}}$, where $Q$ is the typical momentum exchange  in the scattering and $\Lambda_{\slashed{\pi}}$ is 
the \EFT breakdown scale, $\Lambda_{\slashed{\pi}} \lesssim m_{\pi}$.
  
\section{The tale of $A_y$}

In the past several decades there has been much effort put into understanding the nuclear forces at the level of few nucleons \cite{Stoks:1993tb,Beane:2000fx,Stoks:1994wp}. 
There have been many potentials constructed to describe the interaction between nucleons \cite{Yukawa:1935xg,Stoks:1994wp,Reid:1968sq,Nagels:1978me,PhysRevC.51.38AV18}. To put these potentials to test observables needed to be 
calculated and measured. So the easiest starting point from the point of view of both theory and experiment, was to look at the processes
involving only two nucleons. The quantum mechanical two-body problem is fairly easy to solve; in all the cases of interest it can always be reduced to just a one-body 
problem in a potential. Several potentials can be constructed and adjusted to reproduce the two-nucleon data. 
So the next filter to use to rule out the ``wrong'' potentials or unimportant terms in the potentials is the three-nucleon processes. 
The three-body problem in quantum mechanics is much harder than the two-body one. The first theoretical calculations involving three nucleons 
became possible after Faddeev introduced in 1961 an integral equation that the scattering amplitude must satisfy in a non-relativistic scattering process \cite{Faddeev:1960su}. 
At the three nucleon-sector some of the possible processes are the nucleon deuteron ($Nd$) elastic scattering. 
So we can have either proton deuteron ($pd$) or 
neutron deuteron ($nd$) scatterings. From the point of view of the theory the $nd$ scattering is easier to analyze, 
because there is only one charged particle in this process, the proton inside the deuteron. Thus, the calculation will avoid all the complications coming from 
the Coulomb interactions. But from the point of view of experiment the $pd$ scattering is both easier to perform and collect more precise data. 
This is because the proton beam is easier to control, direct and detect due to the electrical charge. 
One of the observables in $Nd$ scattering is the total unpolarized cross-section. As this observable is averaged over all incoming and outgoing spins 
of the scattering particles, it is not very sensitive to the interaction terms in the potentials that have spin-structure. To be able to gather more information about these terms, more 
sensitive observables are needed. One of those sensitive observables is the vector analyzing power, $A_y$. This observable measures the 
asymmetry of the scattering of polarized nucleons on unpolarized deuterons, the nucleon polarization being perpendicular to the beam axis. 
The asymmetry is between nucleon polarization in opposite directions. 

Until the end of 1970s the experimentalists were gathering data on $pd$ scattering and the theorists were doing calculations on $nd$ scattering. 
Because of the lack of data and calculations physicists were only able to compare the $pd$ data to $nd$ calculations. 
The argument was that the $pd$ and $nd$ elastic scattering processes are related through isospin symmetry, which is an approximate symmetry. 
Therefore, if we neglect the Coulomb interactions these two processes should qualitatively look the same. Real comparison of theoretical calculations to data 
became possible after W. Tornow, \textit{et. al.} published series of experimental papers in 1978, 1982 and 1983 reporting $A_y$ data for $nd$ elastic scattering 
for center of mass energies at $12$ MeV, $10$ MeV and $14.1$ MeV \cite{Tornow:1978ijm,Tornow:1982zz,Tornow:1983zz}. 
In 1987 C. R. Howell, \textit{et. al.} published a paper with new data for $A_y$ in $nd$ scattering, comparing the existing data from $10$ to $14$ MeV with 
theoretical Faddeev calculations \cite{Howell1987}. These calculations used many different potentials (Paris, Bonn, etc.), and the data showed preference for some models
over the others. However, all of these models had the same problem; they failed to predict the correct magnitude of $A_y$ near its maximum. Since then  
and until now this problem persists throughout all of the potential models: they predict the correct shape of $A_y$, the correct position of its maximum, 
but under-predict its magnitude \cite{ Entem:2001tj,Kievsky:1996ca} (see Fig.~\ref{Aysampleillustrate} \footnote{Reprinted from \cite{Kievsky:1996ca}, Nuclear Physics A, Volume 607, Issue 4, A. Kievsky, S. Rosati, W. Tornow, M. Viviani, ``Critical comparison of experimental data and theoretical predictions for N-d scattering below the breakup threshold,'' Pages 402-424, Copyright (1996), with permission from Elsevier.}). 
 \begin{center}\label{bubbleNN}\begin{figure}[ht]
  \begin{center}
  \includegraphics[scale=0.7]{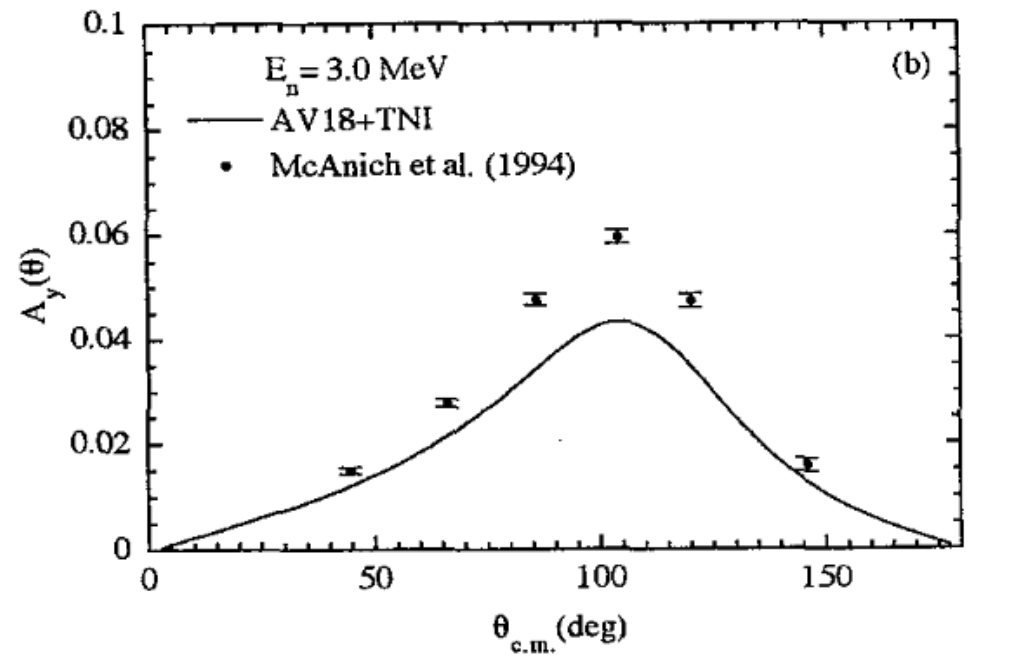}
  \end{center}
   \vspace{0.3cm} \caption{\label{Aysampleillustrate}(Caption and figure reprinted from \cite{Kievsky:1996ca}). Comparison of the neutron analyzing power $A_y$ data of McAninch, \textit{et. al.} \cite{McAninch:1994zz} with the theoretical predictions obtained in the work of A. Kievsky, \textit{et. al.} \cite{Kievsky:1996ca}.}
\end{figure}\end{center}
This problem is known as the ``$A_y$ puzzle'' \cite{WITALA1989157}, and, because it has been measured for $nd$ scattering, there have been 
many attempts to solve it. C. R. Howell, \textit{et. al.} \cite{Howell1987} mentioned in their paper that all of the models see that the $A_y$ is very sensitive 
to the two-nucleon $^3P_J$ phase shifts. This has motivated several authors to seek for the solution of the $A_y$ puzzle by adjusting the $^3P_J$ phase shifts 
to get the correct magnitude of the $A_y$. However changing the $^3P_J$ phase shifts to get the $A_y$ 
right led to inconsistencies with other two or three-body observables. In 1998 D. Huber, \textit{et. al}. \cite{Huber:1998hu} used the AV18 potential to calculate $A_y$. In this potential 
there are eighteen different operator structures, from which D. Huber, \textit{et. al.} determined the five that have non-zero contributions to $A_y$. They 
concluded that the $A_y$ puzzle cannot be solved by just adjusting the two-body forces, so likely it will be solved by including some three-body forces that had not been considered before. 
In 2001 E. Epelbaum et. al. \cite{PhysRevLett.86.4787}
claimed 
that according to their calculations they are able to find the correct magnitude of $A_y$ by including a new two-body force, which is derived from the chiral EFT, in the potential. But in their calculations for two-nucleon processes 
they compared their results not to the observables but to the phase-shifts only. Later in 2001 D. R. Entem and R. Machleidt \cite{Entem:2001tj} noted that the calculations 
need to be compared to the two-nucleon data directly, instead of solely to the phase shifts. So they did the comparison and reported that the problem is 
still not resolved. They also analyzed the possible ways of solving the $A_y$ puzzle and came to the conclusion that, as far as only two-body interactions 
are used, it is impossible to solve the puzzle without ruining the low energy two-nucleon predictions. There also have been attempts to resolve the puzzle by including 
some three-body forces into the potentials \cite{PhysRevC.52.R15,Kievsky1996}. However, for all the three-body forces tested, the predictions for the $A_y$ were either not changing or getting worse. 

The disadvantage of potential models is that there is no systematic theoretical approach to understand 
which potential is better than the other and by how much. Hence, to rule out models, physicists have to entirely rely on comparison of the calculations to the 
experiment. The EFT approach, on the other hand, is constructed with the tool of comparing things in its core. 
This tool is the power counting parameter, which needs to be determined before an EFT is constructed. The power counting parameter is a small 
parameter characteristic to the processes that the EFT is constructed for, and all the terms in an EFT Lagrangian, all the calculated amplitudes and observables 
are expanded in powers of this parameter. As a consequence we get a natural way of determining which term is more significant and to what extent. 

The EFT, which is best fit to describe the non-relativistic few-nucleon systems is \EFT \cite{Kaplan:1998we}. 
Up to and including next-to-next-to-leading-order (N$^2$LO) in powers of the power counting parameter, the \EFT gives very small contributions 
to the $A_y$ observable. Nevertheless, at the next order ($\ntlo$) the two-nucleon $^3P_J$ interactions enter into the \EFT Lagrangian, and these interactions give 
the first substantial contributions to the $A_y$. In this dissertation I am going to present the first $\ntlo$ calculation of the $nd$ scattering amplitude and show what the 
\EFT predictions for $A_y$ look like at this order. The rest of this dissertation is organized as follows: in Chapter \ref{chap:2} \EFT at two- and three-nucleon sectors up to and 
including the second order is described, in Chapter \ref{chap:3} \EFT at two- and three-nucleon sectors at the third order and higher is described. In this chapter the third order calculation of the 
$A_y$ observable is given. In Chapter \ref{Result} the results for the $A_y$ are given along with conclusions and future directions.

\chapter{Non-Relativistic Pionless EFT: Up to Order-2}
\label{chap:2}
\section{Two-nucleon sector at LO}
\label{Two-nucleonsectoratLO}

  The Lagrangian density for the two-nucleon sector is constructed such that the $NN$ scattering amplitude matches onto the effective range expansion (ERE). 
  The ERE exploits the unitarity of the S-matrix to derive an expression for the scattering amplitude. 
  As the S-matrix must be unitary, the diagonal elements (diagonal in partial wave expansion) must be complex numbers with absolute value 1. So we have:  
\begin{equation}
S=e^{2i\delta} ,
\end{equation}
where $\delta$ is a real number and it is called the phase shift. 
  
  From the connection of the S-matrix to the scattering amplitude A: $S=1+i\frac{kM}{2\pi}A$, we can derive an expression for $A$ in terms of the phase shift $\delta$. Here $k$ is the absolute value of the nucleon three-momentum in the center of mass (CM) frame and $M$ is the nucleon mass. 
  
\begin{eqnarray} \label{ERE}
  &&A=\frac{4\pi}{M}\frac{e^{2i\delta} - 1}{2ik} = \frac{4\pi}{M}\frac1{ke^{-i\delta}}\frac{e^{i\delta} - e^{-i\delta}}{2i} = \frac{4\pi}{M}\frac{\sin\delta}{k(\cos\delta - i\sin\delta)} = \frac{4\pi}{M}\frac{1}{k\cot\delta - ik}
\end{eqnarray}
Eq.~\eqref{ERE} is true for any given partial wave and it is shown in \cite{BLANKENBECLER196062,PhysRev.76.18} that $k\cot\delta$ in the $S$-wave is an analytical function of $k$ and that around $k=0$ the expansion is the following:
  \begin{equation}
k\cot(\delta)=-\frac{1}{a}+\frac{1}{2}r_0 k^2+s k^4+...
\end{equation}
Here $a$ is the scattering length, $r_0$ is the effective range, and $s$ is the shape parameter. The derivation of the last formula in  \cite{BLANKENBECLER196062} relies on non-relativistic quantum mechanics and investigation of the Schr\"{o}dinger equation, which is applicable for non-relativistic $NN$ scattering. In the $^3S_1$ channel the deuteron pole exists in the scattering amplitude, so $k\cot\delta$ can be expanded around the deuteron pole.

For the two-nucleon sector at leading order the Lagrange density is:
 \begin{equation}
\mathcal{L}=N^{\dagger}\left(i\partial_0+\frac{\vec{\nabla}^2}{2M_N}\right)N-C_0(N^{T}PN)^{\dagger}(N^{T}PN)
\end{equation}
 where $N$ is the nucleon field, which technically has two indices, one for spin and one for isospin, each of which has two components. 
 Those indices are suppressed because of the common notation. 
 The operator $N$ only destroys a nucleon, it does not create an antinucleon, because this is a non-relativistic field theory. 
 And P is a spin-isospin projector, not containing any derivatives, because this is at leading order and any derivatives would make the operator higher order. To reproduce the deuteron bound state in $NN$ scattering amplitude all the graphs in Fig~\ref{bubbleNN} need to be added up.

  \begin{center}\begin{figure}[ht]
  \begin{center}
  \includegraphics[scale=0.7]{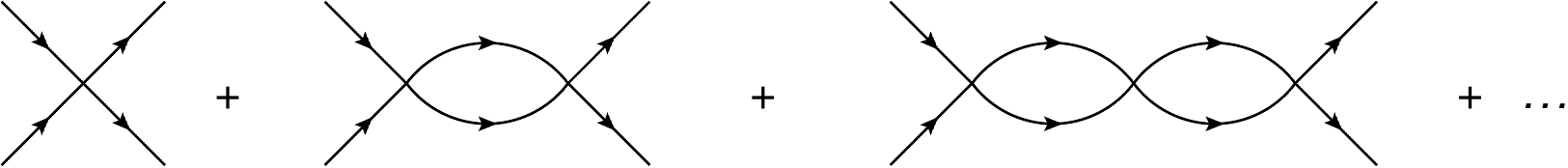}
  \end{center}
   \vspace{0.3cm} \caption{\label{bubbleNN}$NN$ scattering: bubble sum.}
\end{figure}\end{center}

  Luckily this sum is just a geometric series and can be calculated in closed form. In the center of mass frame we have:

   \begin{equation}
A=-C_0-C_0I_0C_0-C_0I_0C_0I_0C_0-...=\frac{-1}{\frac{1}{C_0}-I_0}
\end{equation}
  where $I_0$ is the one-loop integral:

  \begin{equation}
I_0=-i \int \frac{d^4q}{(2\pi)^4} \frac{i}{E+q_0-\frac{q^2}{2M}+i\epsilon} \frac{i}{-q_0-\frac{q^2}{2M}+i\epsilon}.
\end{equation}
Here $E$ is the total incoming momentum in the center of mass frame: $E=\frac{k^2}{M}$.
The $q_0$ integration can be done by closing the contour either in the upper complex plane or the lower, picking up the residue over the pole we find:

  \begin{equation}
I_0=\int \frac{d^3q}{(2\pi)^3} \frac{1}{E-\frac{q^2}{M}+i\epsilon}=-\frac{M}{4\pi}(\frac{2}{\pi}\Lambda+ik)
\end{equation}
  where $\Lambda$ is a sharp cutoff on $|\vec{q}|$. The same result is obtained using the so-called Power Divergence Subtraction (PDS) \cite{Kaplan:1998tg} subtraction scheme. 
  From here we can find the scattering amplitude:
  
   \begin{equation}
A=-\frac{4\pi}{M}\frac{1}{\frac{4\pi}{MC_0}+\frac{2}{\pi}\Lambda+ik}.
\end{equation}
This matches the ERE expression for the scattering amplitude at LO. The pole in this amplitude corresponds to the deuteron bound state:
  
    \begin{equation}
\frac{4\pi}{MC_0}+\frac{2}{\pi}\Lambda=\gamma_t
\end{equation}
where $\gamma_t$ is the binding momentum and is defined by: $\gamma_t=\sqrt{MB}=45.70$ MeV, where $B=2.2$ MeV is the deuteron binding energy. 
Note that the interaction coefficient $C_0$ depends on the regulator, the cutoff in our case, so that the observable $\gamma_t$ matches to its experimental, cutoff-independent value. 
  
  \section{Two-nucleon sector at higher orders}
  \label{Twonucleonsector}
  Here I describe the \EFT in the two-nucleon sector at higher orders, but I only give some of the two-nucleon interactions. Some of the two-nucleon forces such as $SD$ mixing and the $P$-wave interactions will 
  be discussed in the three-nucleon sector. 
  
\begin{center}\begin{figure}[ht]
  \begin{center}
  \includegraphics[scale=0.15]{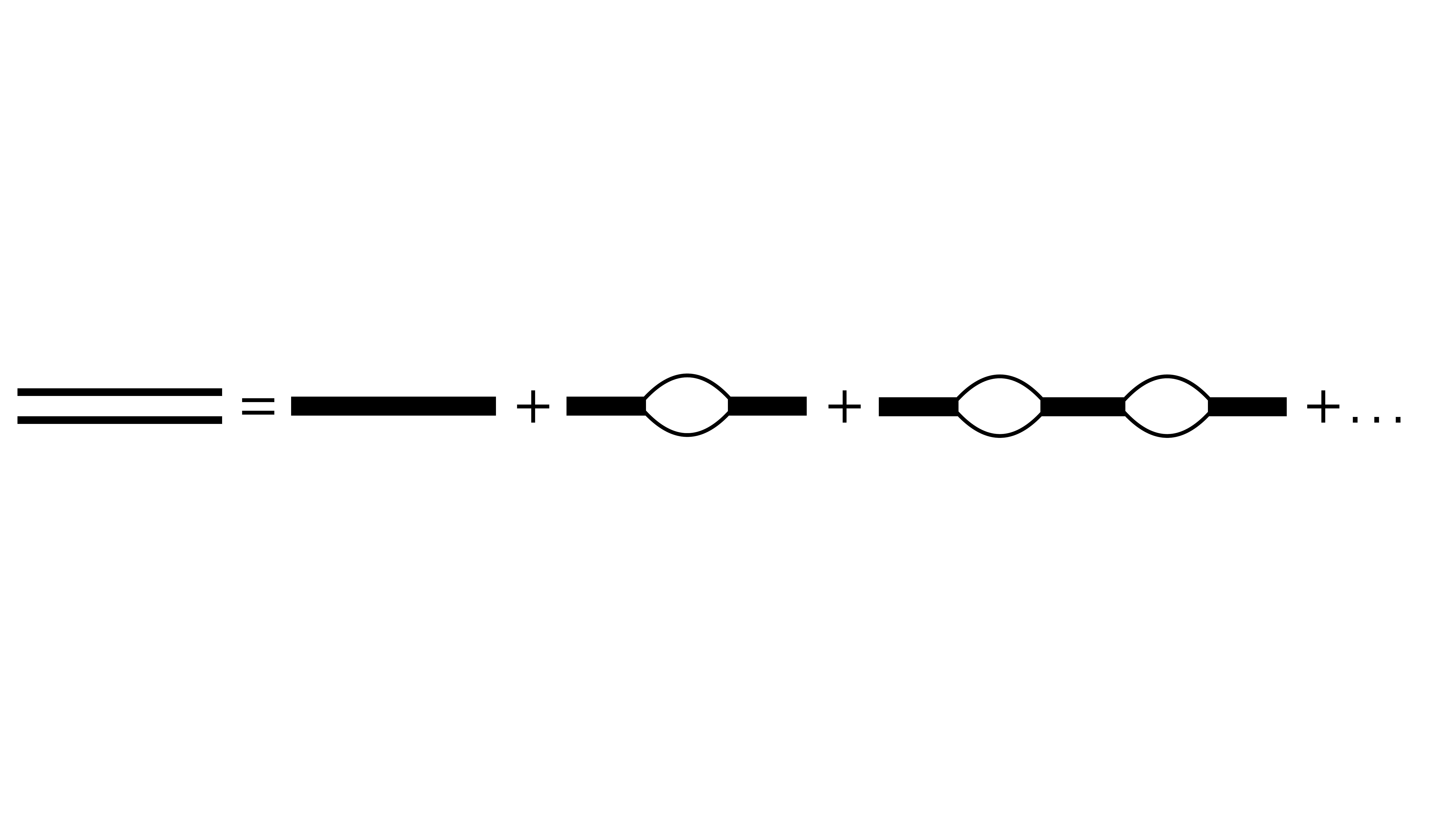}
  \end{center}
  \caption{\label{deuteronpropagator}Double line is the LO full deuteron propagator, thin solid line is the nucleon propagator, thick solid line is the bare deuteron propagator $\frac{i}{\Delta_{t}}$.}
\end{figure}\end{center}

  To be able to consider three-body scattering problems we first need to construct the two-body Lagrangian to higher orders. It turns out to be very useful to introduce auxiliary dibaryon fields $t_i$ for the deuteron, which is a spin-triplet and isospin-singlet and $s_a$, which is a spin-singlet and isospin-triplet. The Lagrangian density in the $S$-wave channel for the two-body sector to all orders is given by:
  
  \begin{eqnarray}
&&\mathcal{L}= N^{\dagger}\left(i\partial_0+\frac{\vec{\nabla}^2}{2M_N}\right)N \nonumber \\
&+&t_{i}^{\dagger}\left(\Delta_{t}-\sum\limits_{n=0}^{\infty}c_{nt}\left(i\partial_0+\frac{\vec{\nabla}^2}{4M_N}+\frac{\gamma_{t}^{2}}{M_N}\right)^{n+1}\right)t_{i} \nonumber \\
&+&s_{a}^{\dagger}\left(\Delta_{s}+\Delta_{s}^{N\textsuperscript{2}LO}\delta^{a}_{-1}-\sum\limits_{n=0}^{\infty}c_{ns}\left(i\partial_0+\frac{\vec{\nabla}^2}{4M_N}+\frac{\gamma_{s}^{2}}{M_N}\right)^{n+1}\right)s_{a} \nonumber \\
&+&y_t(t_{i}^{\dagger}(N^{T}P_{i}N)+\mathrm{h.c.})+y_s(s_{a}^{\dagger}(N^{T}P_{a}N)+\mathrm{h.c.}), \nonumber \\
\end{eqnarray}
where the projectors are defined by: $P_i=\frac{1}{\sqrt{8}}\sigma_2\sigma_i\tau_2$ and $P_a=\frac{1}{\sqrt{8}}\tau_2\tau_a\sigma_2$.
By Gaussian path integration it can be shown that this Lagrangian density is equivalent to a completely general Lagrangian density involving only nucleon fields \cite{Bedaque:1999vb}. The $NN$ elastic scattering amplitude is given by the ERE and the same amplitude can be calculated using this Lagrangian, hence the two answers should match and the $c_{nt}$ ($c_{ns}$) coefficients should be related to the ERE parameters. We will see this matching in the $^3S_1$ channel where the deuteron lives. The deuteron propagator at LO is given by the infinite bubble sum, as shown in Fig.~\ref{deuteronpropagator}.

Without loss of generality the coupling $y_t$ can be chosen to be: $y^{2}_{t}=\frac{4\pi}{M_N}$. This sum is just a geometric series so it can be calculated in a closed form giving the LO deuteron propagator:

 \begin{equation}
iD^{LO}_{t}(p_0,\vec{p})=\frac{i}{\Delta_t+\Lambda-\sqrt{\frac{\vec{p}^2}{4}-M_Np_0-i\epsilon}},
\end{equation}
where $\Lambda$ is the momentum cutoff (it is actually off by the factor $\frac{2}{\pi}$, which is absorbed by just redefining the cutoff). Using this we can find the exact deuteron propagator to all orders, which will be given by another geometric series shown in the figure:

  \begin{center}\begin{figure}[ht]
  \begin{center}
  \includegraphics[scale=0.7]{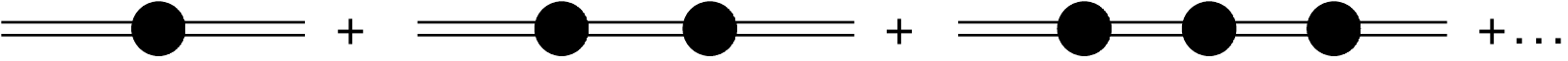}
  \end{center}
  \caption{Deuteron propagator to all orders.}
\end{figure}\end{center}

In this figure the solid circle denotes the single insertion of all the $c_{nt}$ interactions summed over all $n$. This means that the sum given in this figure is not order-by-order, meaning that each of the summands itself has terms at all orders. This sum also can be calculated in a closed form giving:

\begin{equation} 
iD^{\text{all-orders}}_{t}(p_0,\vec{p})=\frac{i}{\Delta_t+\Lambda-\sqrt{\frac{\vec{p}^2}{4}-M_Np_0-i\epsilon}} \sum_{m=0}^{\infty}\left( \frac{\sum\limits_{n=0}^{\infty}c_{nt}(p_0-\frac{\vec{p}^2}{4M_N}+\frac{\gamma_{t}^{2}}{M_N})^{n+1}}{\Delta_t+\Lambda-\sqrt{\frac{\vec{p}^2}{4}-M_Np_0-i\epsilon}}\right)^{m+1}
\end{equation}
Where $p_0$ and $\vec{p}$ is the energy and the three-momentum of the deuteron. Doing the summation over $m$ gives:

\begin{equation}\label{deuterontoallorders}
iD^{\text{all-orders}}_{t}(p_0,\vec{p})=\frac{i}{\Delta_t+\Lambda-\sqrt{\frac{\vec{p}^2}{4}-M_Np_0-i\epsilon}-\sum\limits_{n=0}^{\infty}c_{nt}\left(p_0-\frac{\vec{p}^2}{4M_N}+\frac{\gamma_{t}^{2}}{M_N}\right)^{n+1}}         
\end{equation}
To find the $NN$ scattering amplitude in the center of mass frame we need to attach the nucleon-legs to the deuteron propagator to all-orders and plug in $p_0=\frac{\vec{k}^2}{M_N}, \vec{p}=0$, where $\vec{k}$ is the incoming nucleon momentum. This will give for the amplitude:

\begin{equation}
A=-\frac{4\pi}{M_N}\frac{1}{\Delta_t+\Lambda+ik-\sum\limits_{n=0}^{\infty}c_{nt}(\frac{\vec{k}^2}{M_N}+\frac{\gamma_{t}^{2}}{M_N})^{n+1}}.         
\end{equation}
As was expected this amplitude is exactly the same as the one given by the ERE \eqref{ERE}. Equating both equations at each order in the $k-$expansion we can match the $c_{nt}$ coefficients to ERE parameters. The ERE parameters are matched to experiment so those are 
cutoff independent. This forces the parameter $\Delta_t$ in the Lagrangian to depend on the cutoff, but all the other parameters $c_{nt}$ are cutoff independent \cite{Griesshammer:2004pe}. 

From the matching the \EFT and ERE we know that we can do the substitution $\Delta_t+\Lambda=\gamma_t$ in Eq.~\eqref{deuterontoallorders}. Separating the first term in the sum in the denominator 
Eq.~\eqref{deuterontoallorders} can be rewritten:

\begin{eqnarray}\label{deuteronmidway}
&&D^{\text{all-orders}}_{t}(p_0,\vec{p})\nonumber \\ 
&=&\frac{i}{\gamma_t-\sqrt{\frac{\vec{p}^2}{4}-M_Np_0-i\epsilon}-\frac{c_{0t}}{M_N}\left(M_Np_0-\frac{\vec{p}^2}{4}+\gamma_{t}^{2}\right)-\sum\limits_{n=1}^{\infty}c_{nt}\left(p_0-\frac{\vec{p}^2}{4M_N}+\frac{\gamma_{t}^{2}}{M_N}\right)^{n+1}}   \nonumber \\ 
&=&\frac{1}{\gamma_t-\sqrt{\frac{\vec{p}^2}{4}-M_Np_0-i\epsilon}} \nonumber \\ 
&\times&\frac{1}{1-\frac{c_{0t}}{M_N}\left( \gamma_t+\sqrt{\frac{\vec{p}^2}{4}-M_Np_0} \right)-\sum\limits_{n=1}^{\infty}c_{nt}\left(p_0-\frac{\vec{p}^2}{4M_N}+\frac{\gamma_{t}^{2}}{M_N}\right)^{n}\left(\gamma_t+\sqrt{\frac{\vec{p}^2}{4}-M_Np_0}\right)}.   \nonumber \\ 
\end{eqnarray}
From here we can see that around the deuteron pole where $\gamma_t=\sqrt{\frac{\vec{p}^2}{4}-M_Np_0}$ all the terms in the sum in the second
denominator do not contribute to the residue and we get for the residue the following expression:

\begin{equation}
Z_t=\frac{1}{1-\frac{c_{0t}}{M_N}2\gamma_t}.
\end{equation}
The coefficient $c_{0t}$ is matched onto ERE and this matching gives the numerical value for the residue $Z_t=1.690$. Expressing 
the coefficient $c_{0t}$ in terms of $Z_t$ we can substitute it into Eq.~\eqref{deuteronmidway} we can re-expand the deuteron 
propagator in powers of $\frac{Z_t-1}{2}\approx0.3$ finding:

\begin{eqnarray}
&&D^{\text{all-orders}}_{t}(p_0,\vec{p})\nonumber \\ 
&=&\frac{1}{\gamma_t-\sqrt{\frac{\vec{p}^2}{4}-M_Np_0-i\epsilon}} \nonumber \\ 
&&\left[1+\frac{Z_t-1}{2\gamma_t} \left( \gamma_t+\sqrt{\frac{\vec{p}^2}{4}-M_Np_0} \right) + \left(\frac{Z_t-1}{2\gamma_t}\right)^2
\left( \frac{p^2}{4}-M_Np_0-\gamma_t^2 \right)+... \right].  \nonumber \\ 
\end{eqnarray}
Putting the deuteron propagator into this form is called the Z-parametrization \cite{Griesshammer:2004pe}\cite{Phillips:1999hh}. The same sort of re-expansion can be done for the spin-singlet dibaryon propagator, and this is what we use in our calculations. This way of expanding the dibaryon propagators has its advantages, one of them being that the dibaryon wave-function renormalization is recovered fully at the NLO of 
the perturbation, instead of getting contributions at every order. The other advantages and disadvantages are discussed in the references given above in this paragraph.

\section{Three-nucleon sector at LO, quartet channel}

\begin{center}\begin{figure}[ht]
  \begin{center}
  \includegraphics[scale=0.2]{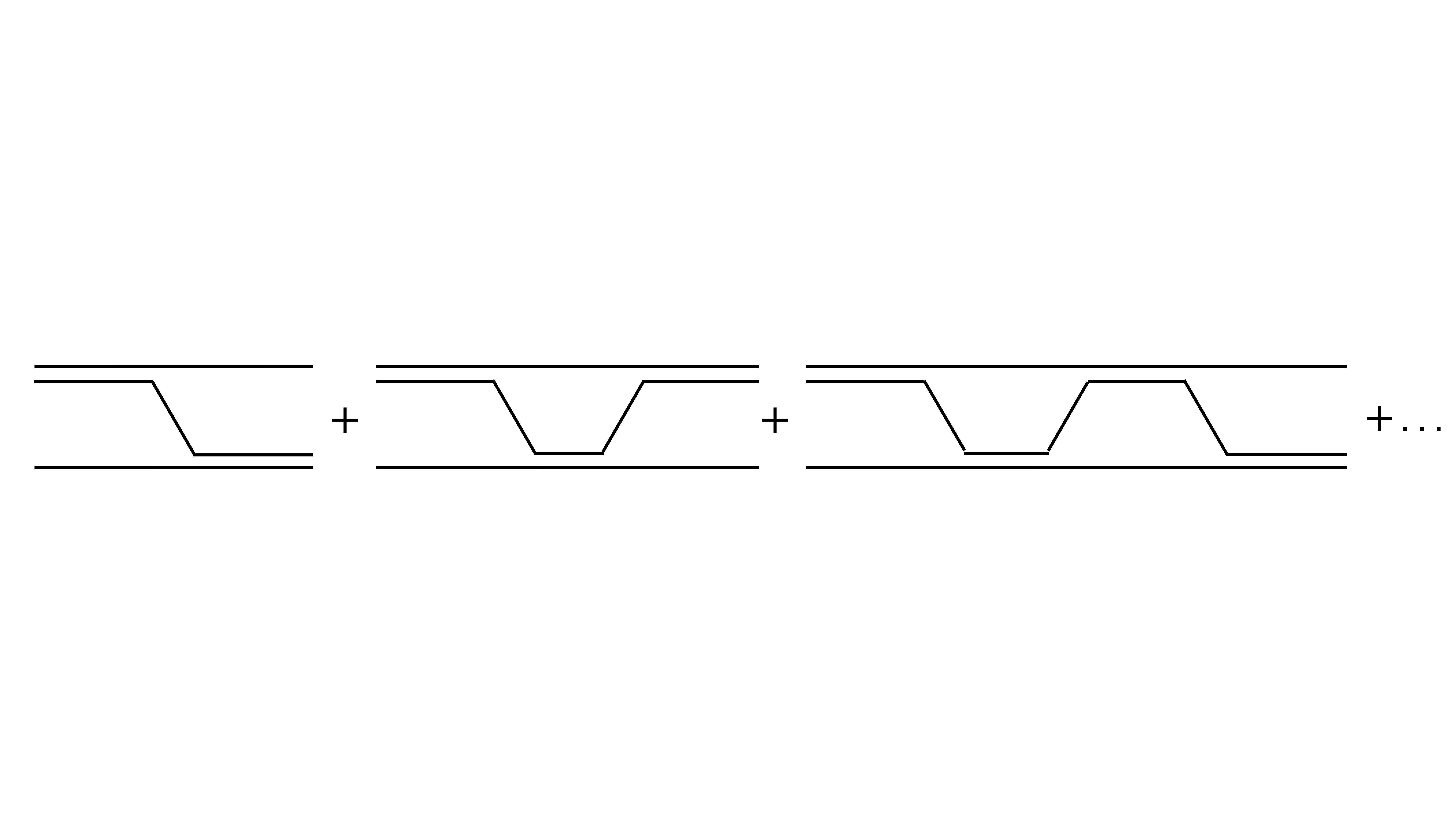}
  \end{center}
  \vspace{-2cm}
  \caption{\label{LO} Pinball diagrams for LO $nd$ scattering.}
\end{figure}\end{center}

\begin{center}\begin{figure}[ht]
  \begin{center}
  \includegraphics[scale=0.4]{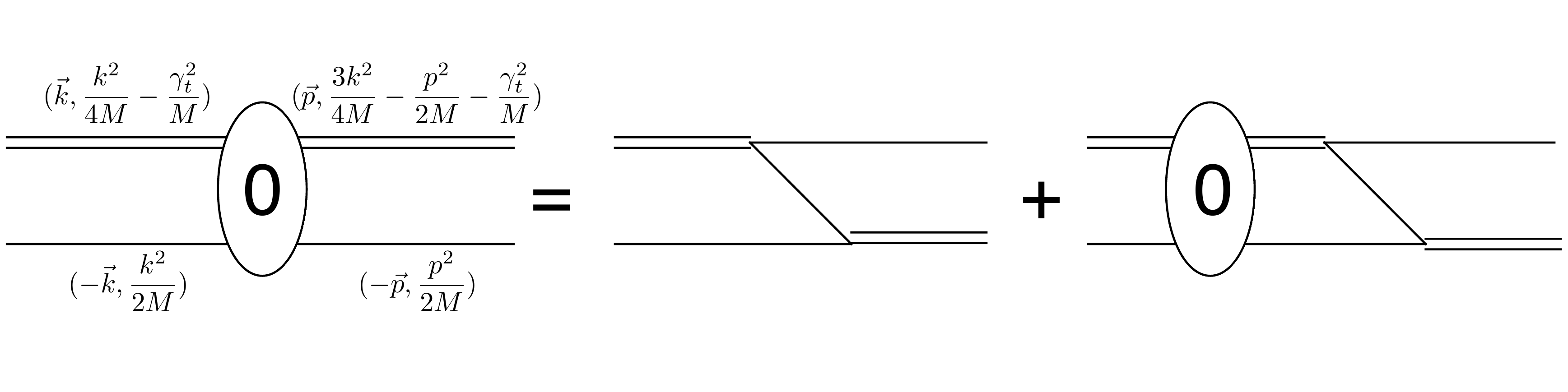}
  \end{center}
  \caption{\label{QuartetLoequationw4momenta}Quartet channel LO $nd$ equation.}
\end{figure}\end{center}

Having the Lagrangian at the two-body sector we are able to consider three-body scattering problems such as neutron-deuteron ($nd$) or proton-deuteron ($pd$) elastic scattering \cite{Vanasse:2016jtc}. 
We will consider $nd$ scattering, because there is no Coulomb interaction and we only need the strong interactions which are encoded in the constructed effective Lagrangian. For calculations in \EFT about the $pd$ scattering process see for example \cite{Vanasse:2014kxa,Vanasse:2015fph}.
As the neutron is a spin $\frac{1}{2}$ particle and the deuteron is a spin $1$ particle the scattering can happen in two different channels: 
total spin-$\frac{1}{2}$, the doublet channel, and total spin-$\frac{3}{2}$, the quartet channel.  We will first take a look at the quartet channel scattering. 
At leading order we have an infinite set of diagrams (see Fig.~\ref{LO}) that contributes to $nd$ scattering in the quartet channel much like the two-body case. 
The difference between this sum and the sum that we find in the two-body case is that this sum is not a simple geometric series, so we don't find a closed form expression for the amplitude as in the two-body case. 
Instead we can see that the scattering amplitude satisfies an integral equation represented diagrammatically in the Fig.~\ref{QuartetLoequationw4momenta} \cite{Vanasse:2016jtc}
The oval with a ``0'' in the Fig.~\ref{QuartetLoequationw4momenta} is the leading order $nd$ scattering amplitude, which is the sum of all the terms 
in the Fig.~\ref{LO}. 
The explicit form of the integral equation is given by:
\begin{eqnarray}
&&(it^{ji}) ^ {\beta b}_{\alpha a} (\vec{k},\vec{p},\frac{p^2}{2M_N})= \frac{y_t^2}{2} (\sigma^i \sigma^j)_{\alpha}^{\beta} \delta_{a}^{b} \frac{i}{\frac{k^2}{4M_N}-\frac{p^2}{2M_N}-\frac{\gamma_t^2}{M_N}-\frac{(\vec{k}+\vec{p})^2}{2M_N}+i\epsilon}\nonumber \\ 
&+&\frac{y_t^2}{2} (\sigma^i \sigma^k)_{\gamma}^{\beta} \delta_{c}^{b} \int \frac{d^4q}{(2\pi)^4} (it^{j k}) ^ {\gamma c}_{\alpha a} (\vec{k},\vec{q},q_0) \nonumber \\ 
&\times& iD_t^{(0)}(E-q_0,\vec{q}) \frac{i}{q_0-\frac{q^2}{2M_N}+i\epsilon} \frac{i}{E-\frac{p^2}{2M_N}-q_0-\frac{(\vec{p}+\vec{q})^2}{2M_N}+i\epsilon} \nonumber \\ 
\end{eqnarray}
where $i$ and $j$ are the initial and final deuteron polarizations, $\alpha$ and $\beta$ are the initial and final nucleon spin, and $a$ and $b$ are the initial and final nucleon isospin. The incoming momentum in the CM frame is $\vec{k}$ and the outgoing momentum in the CM frame is $\vec{p}$, the total energy of the system is $E=\frac{3k^2}{4M_N}-\frac{\gamma_t^2}{M_N}$. The leading order dressed deuteron propagator is $D_t^{(0)}(p_0,\vec{p})$, where the 
superscript $0$ stands for LO.

Here we have put the incoming nucleon, deuteron and the outgoing nucleon on-shell, while the outgoing deuteron is off-shell.
The energy and momentum of the outgoing deuteron is determined by the conservation of energy-momentum. 
The amplitude $t$ in this equation depends on three entries which are respectively the incoming momentum $\vec{k}$, the outgoing momentum $\vec{p}$ and 
the energy of the outgoing nucleon, which is $\frac{\vec{p}^2}{2M_N}$ on the left hand side and $q_0$ on the right hand side. 
On the right hand side inside the loop the  nucleon that is attached to the scattering amplitude has energy-momentum $(q_0,\vec{q})$ 
and this is the loop energy-momentum that we are integrating over.

The first step towards solving this integral equation is to simplify it by doing the $q_0$ integration on the right hand side. Each of the three propagators in the loop has a pole in 
the complex $q_0$ plane and it is easy to see that the two poles coming from the deuteron propagator and the exchanged nucleon propagator are on one side of the real axis and the second 
nucleon propagator is on the other side of the real axis. So we can close the contour of integration so that we pick up only one pole coming from the second nucleon propagator hence putting it on shell:
$q_0=\frac{q^2}{2M_N}$. After doing this we can drop the third variable from the amplitude $t$ and interpret the function $t(\vec{k},\vec{p})$ as the amplitude that has incoming momentum $\vec{k}$,
 outgoing momentum $\vec{p}$, the incoming particles on-shell and the outgoing nucleon on-shell. 
 
 \begin{eqnarray}
&&(it^{ji}) ^ {\beta b}_{\alpha a} (\vec{k},\vec{p})= \frac{y_t^2}{2} (\sigma^i \sigma^j)_{\alpha}^{\beta} \delta_{a}^{b} \frac{i}{E-\frac{k^2}{2M_N}-\frac{p^2}{2M_N}-\frac{(\vec{k}+\vec{p})^2}{2M_N}}\nonumber \\ 
&+&\frac{y_t^2}{2} (\sigma^i \sigma^k)_{\gamma}^{\beta} \delta_{c}^{b} \int \frac{d^3q}{(2\pi)^3} (it^{j k}) ^ {\gamma c}_{\alpha a} (\vec{k},\vec{q}) \nonumber \\ 
&\times& iD_t^{(0)}(E-\frac{q^2}{2M_N},\vec{q}) \frac{i}{E-\frac{p^2}{2M_N}-\frac{q^2}{2M_N}-\frac{(\vec{p}+\vec{q})^2}{2M_N}} \nonumber \\ 
\end{eqnarray}

Now we can project this equation onto the spin quartet channel and do the partial wave decomposition projecting on the $l$-th partial wave. To do this we put the $z$ axis along $\vec{p}$, 
we multiply both sides of the equation by $Y_l^m(\hat{k})$ and integrate by the angles of $\hat{k}$: $d\Omega_{\hat{k}}$. 
From the previous equation we see that the amplitude depends only on the angle $\theta$ between the vectors $\vec{k}$ and $\vec{p}$ and does not depend on the angle $\phi$, which is the azimuthal coordinate of the vector $\vec{k}$.
Doing the integration on the left hand side we will find the projected amplitude for $m=0$ which 
is denoted by $t(k,p)$, and we will find $0$ for all the other values of $m$. The same way for the first first term on the right hand side we will find something proportional to
$\int d\Omega_{\hat{k}} Y_l^m(\hat{k}) \frac{1}{E-\frac{k^2}{2M_N}-\frac{p^2}{2M_N}-\frac{(\vec{k}+\vec{p})^2}{2M_N}}$, which is zero for non-zero values of $m$ and for $m=0$ equals to
$\int dx\ P_l(x) \frac{1}{E-\frac{k^2}{M_N}-\frac{p^2}{M_N}-\frac{kpx}{M_N}}$, where $x=\cos(\theta)$. This will give Legendre polynomials of the second kind. The tricky part of this calculation is the second term of the right 
hand side. Here again we find $m=0$, so from $Y_l^m(\hat{k})$ we have $P_l(\hat{k} \cdot \hat{p})$, which can be expanded using the addition theorem:

\begin{equation}
P_l(\hat{k} \cdot \hat{p})=  \frac{4\pi}{2l+1} \sum_{m=-l}^{l}   Y_l^m(\hat{k}) Y_l^{*m}(\hat{p})       
\end{equation}
In the last equation the $z$ axis is arbitrary so we can chose it to be along $\vec{q}$. No matter whether the $z$ axis is along $\vec{p}$ or $\vec{q}$ the weight $d\Omega_{\hat{k}}$ does not change. 
So we can integrate the amplitude $(it^{j k}) ^ {\gamma c}_{\alpha a} (\vec{k},\vec{q})$ with $Y_l^m(\hat{k})$ over the angles of $\vec{k}$: $d\Omega_{\hat{k}}$ 
to find that the only contribution comes from $m=0$ and is proportional to the projected amplitude $t(k,q)$. Then using the second spherical harmonic with the nucleon propagator,
we can do the integration over the angles of $\vec{q}$ to again find something in terms of Legendre polynomials of the second kind. The final expression after doing all of these manipulations is:
 \begin{eqnarray}\label{ndQLO}
&&t ^ {l}_{0} (k,p)= -\frac{y_t^2M_N}{pk} Q_l\left(\frac{p^2+k^2-M_NE}{pk}\right)\nonumber \\ 
&-& \frac{2}{\pi} \int dq q^2 t ^ {l}_{0} (k,q) \frac{1}{\sqrt{\frac{3q^2}{4}-M_NE-i\epsilon}-\gamma_t} \frac{1}{pq} Q_l \left(\frac{p^2+q^2-M_NE-i\epsilon}{pq}\right) \nonumber \\ 
\end{eqnarray}
Here the Legendre polynomials of the second kind are (this differs from the conventional definition by a phase factor $(-1)^l$):
\begin{equation}
Q_l(a)=  \frac{1}{2} \int_{-1}^1 dx\ \frac{P_l(x)}{x+a}   
\end{equation}
This is the equation that the leading order quartet channel $l$-th partial wave amplitude satisfies. It was first derived by Skornyakov and Ter-Martirosian \cite{G.V.Skornyakov} and 
it can be solved numerically using the Hetherington-Schick method \cite{PhysRev.137.B935}. It is easy to see that discretizing 
the integral turns the equation into a system of algebraic equations, which is then solved numerically. 
I will give more details on the numerical methods used and on the general theory of integral equations in Appendix~\ref{ch:integral-equations}.

Originally we had a three-dimensional integral equation, but after projecting it on different partial waves we get an infinite set of one-dimensional integral equations, one for each partial wave. 
At low energies however, the contributions from higher partial waves gets smaller, so we keep only up to and including G-waves.

We can write the projected integral equation in the following form:
\begin{equation}
t ^ {l}_{0} (k,p)=B_0^l(k,p)+K_0^l(q,p,E) \otimes t ^ {l}_{0} (k,q),
\end{equation}
here the $\otimes$ operation is defined by the following equation:
\begin{equation}\label{otimessymbol}
A(q)\otimes B(q)=\frac{2}{\pi} \int dq\ q^2 A(q)B(q),
\end{equation}
also $B_0^l(k,p)$ and $K_0^l(q,p,E)$ are given by:
\begin{equation}
B_0^l(k,p)=-\frac{y_t^2M_N}{pk} Q_l\left(\frac{p^2+k^2-M_NE}{pk}\right),
\end{equation}
and 
\begin{equation}
K_0^l(q,p,E)=-\frac{1}{\sqrt{\frac{3q^2}{4}-M_NE-i\epsilon}-\gamma_t} \frac{1}{pq} Q_l \left(\frac{p^2+q^2-M_NE}{pq}\right).
\end{equation}

\section{Three-nucleon sector at NLO, quartet channel}

At NLO the scattering amplitude gets a correction which itself satisfies an integral equation given in the following Fig.~\ref{ndnlo}  \cite{Vanasse:2013sda}:
 \begin{center}\begin{figure}[ht]
  \begin{center}
  \includegraphics[scale=0.3]{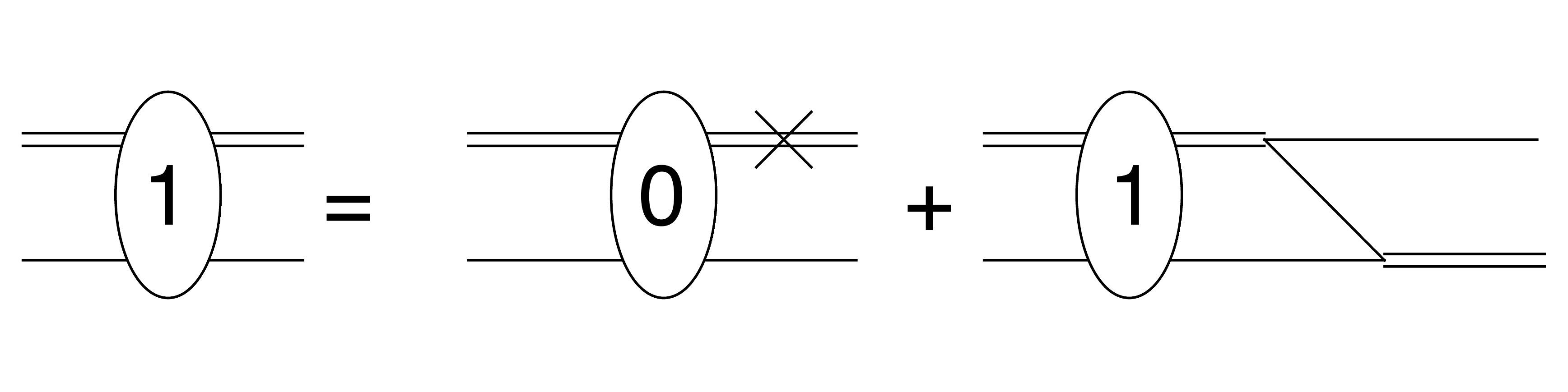}
  \end{center}
  \caption{\label{ndnlo}Quartet channel NLO $nd$ equation.}
\end{figure}\end{center}
The oval with ``1'' inside is the NLO correction to the $nd$ scattering amplitude and the cross on the deuteron represents the effective range correction to the deuteron propagator.
Explicitly, this integral equation can be written as:
\begin{equation}
t ^ {l}_{1} (k,p)=t ^ {l}_{0} (k,p)R_1(p,E)+K_0^l(q,p,E) \otimes t ^ {l}_{1} (k,q),
\end{equation}
where $R_1(p,E)$ is the effective range insertion which is the first order correction to the deuteron propagator, and is given by:
\begin{equation}
R_1(p,E)=\frac{Z_t-1}{2\gamma_t}\left(\gamma_t+\sqrt{\frac{3}{4}p^2-M_NE}   \right).
\end{equation}

\section{Three-nucleon sector at N$^2$LO, quartet channel}

The integral equation satisfied by the N$^2$LO correction to the scattering amplitude is given in the Fig.~\ref{ndn2lo}:

\begin{center}\begin{figure}[ht]
  \begin{center}
  \includegraphics[scale=0.5]{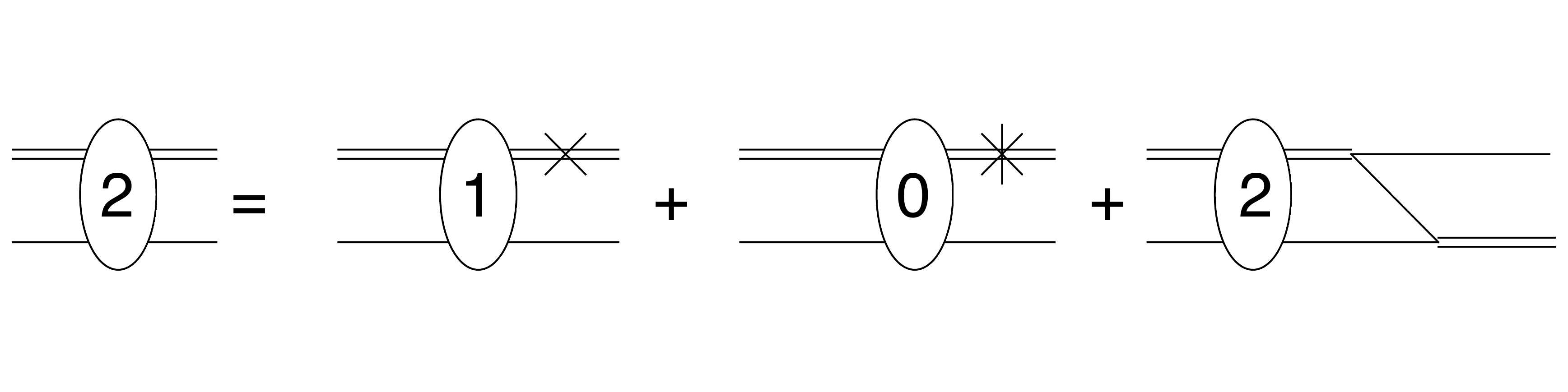}
  \end{center}
  \caption{\label{ndn2lo}Quartet channel $\nnlo$ $nd$ equation.}
\end{figure}\end{center}

Here the oval with the ``2'' in it is the second order correction to the amplitude, also the star on the deuteron is a higher order effective range correction to 
the deuteron propagator. Explicitly:
\begin{equation}
t ^ {l}_{2} (k,p)=[t ^ {l}_{1} (k,p) -(Z_t-1)t ^ {l}_{0} (k,p)] R_1(p,E)+K_0^l(q,p,E) \otimes t ^ {l}_{2} (k,q).
\end{equation}
Note that the kernels of the LO, NLO and N$^2$LO integral equations are all the same, which makes the numerical solution of those equations easier. This point is also discussed in the beginning of chapter \ref{chap:3}.

The reason why we look at the quartet channel first is two-fold. Firstly in the quartet channel the diagrams involving the spin-singlet auxiliary field $s_a$ do not contribute, because this field has spin zero.
In the doublet channel on the other hand, diagrams that involve $s_a$ fields do contribute. As a result in the quartet channel we have one integral equation for each partial wave, 
but in the doublet channel we have a system of two coupled integral equations for each partial wave (more on this in chapter \ref{chap:3}). Secondly the asymptotic analysis of the solutions for the quartet channel integral
equations shows that these solutions become independent of the cutoff as the cutoff goes to infinity. The same result is found by numerically looking at the cutoff dependence 
of the solution. This is not always the case for the doublet channel. The doublet $S$-wave amplitude shows a very strong dependence on the cutoff which does not go away 
for big cutoffs, which makes it necessary to include a three-body counterterm into the Lagrangian. This modifies the doublet $S$-wave equation so that the solution of the new 
equation becomes cutoff independent for large cutoffs. The same type of problem arises when one considers the three-body scattering problem in an EFT that is the equivalent of the \EFT 
when instead of spin-$\frac{1}{2}$ nucleons one considers spinless bosons. So now I will describe the three-boson problem and show how the asymptotic analysis is done in this simpler setting.
Then we can go on to the discussion of the doublet channel $nd$ scattering.

\section{Three-boson system in \EFT and asymptotic analysis}
Following \cite {Bedaque:1998km},
the Lagrangian describing non-relativistic bosons with contact interactions is given by:
\begin{equation}\label{3bosonlagrangian}
\mathcal{L}=\psi^{\dagger}\left(i\partial_0+\frac{\vec{\nabla}^2}{2M}\right)\psi+\Delta T^{\dagger}T -\frac{g}{\sqrt2}(T^{\dagger}\psi\psi+h.c.)+hT^{\dagger}T\psi^{\dagger}\psi.
\end{equation}
Here $\psi$ is the non-relativistic boson field and $T$ is the dimeron field which represents the bound state of the two bosons. The field $T$ is the analog to the $t_i$ and $s_a$ fields in \EFT for nucleons.
Here too one can recover the Lagrangian involving only the $\psi$ field by integrating the $T$ field out of the theory. In this Lagrangian the parameter $\Delta$ is a constant unphysical parameter, 
the dimeron field has a bare propagator equal to $i/\Delta$. The parameter $g$ determines the two body interaction and we have inserted the three-body interaction term determined by the parameter $h$. 
For now we will consider $h$ to be zero and show that the scattering amplitudes have strong cutoff dependence, this will prove the need to insert the three-body interaction. 

The process of calculating the three-body scattering amplitude is the same as for nucleons, but without having to do the spin-isospin projections. We first calculate the dressed dimeron propagator, then write the integral equation that the three-body scattering 
amplitude must satisfy, then solve it numerically. So let's start with the dressed dimeron propagator. 
 
 The propagator of the bosonic field $\psi$ with four momentum $p$ is just the usual non-relativistic propagator:
 
\begin{equation}
iS(p)=\frac{i}{p_0-\frac{p^2}{2m}+i\epsilon}.
\end{equation}

From here the dressed dimeron propagator is represented by the same figure as the dressed deuteron propagator (see Fig.~\ref{deuteronpropagator}) and it is also given by a geometric series which sums up to:

\begin{equation}
i\Delta(p)=\frac{-i}{-\Delta+\frac{mg^2}{4\pi} \sqrt{-mp_0+\frac{p^2}{4}-i\epsilon} }.
\end{equation}

Now we can look at the boson-dimeron scattering amplitude. This amplitude satisfies an analogous integral equation as before and we can again perform partial wave decomposition
for this equation. The complications arise only in the S-wave channel, so we will only look at the integral equation in this channel.

\begin{eqnarray}
&&t(k,p)= \frac{mg^2}{pk} \log{ \left(  \frac{p^2+pk+k^2-mE}{p^2-pk+k^2-mE}  \right) }\nonumber \\ 
&+& \frac{2}{\pi} \int dq q^2 t(k,q) \frac{1}{\sqrt{\frac{3q^2}{4}-M_NE-i\epsilon}-\frac{1}{a_2}} \frac{1}{pq} \log{ \left(  \frac{p^2+pq+q^2-mE-i\epsilon}{p^2-pq+q^2-mE-i\epsilon}  \right) } \nonumber \\ 
\end{eqnarray}

Here $a_2$ is the two body scattering length and $t(k,p)$ is the boson-dimeron scattering amplitude in the $S$-wave channel.
On the right hand side inside the loop integration we have the dimeron propagator which has a pole, so to make this pole structure simpler and to get rid of the square root structure from the kernel
 it is customary to do the following substitution:
 
 \begin{equation}
\frac{a(k,p)}{p^2-k^2}=\frac{1}{mg^2} \frac{t(k,p)}{\sqrt{\frac{3p^2}{4}-M_NE}-\frac{1}{a_2}}.
\end{equation}

Substituting this back into the integral equation gives an equation for the new function $a(k,p)$. To derive the equation for this new function we use the facts that the total energy $E$ is given by 
$E=\frac{3k^2}{4m}-B_2$, where $B_2$ is the two body binding energy and it is related to the two body scattering length by: $B_2=\frac{1}{ma_2^2}$. After all the simplifications the equation we get is:
\begin{equation}\label{aamplitude}
a(k,p)=M(k,p,k)+\frac{2}{\pi} \int dq M(q,p,k) \frac{q^2}{q^2-k^2-i\epsilon} a(k,q)
\end{equation}
where $M(q,p,k)$ is the new kernel and is defined by:
\begin{equation}
M(q,p,k)=\frac{4}{3} \left (\sqrt{\frac{3p^2}{4}-M_NE}+\frac{1}{a_2} \right ) \frac{1}{pq} \log{ \left(  \frac{p^2+pq+q^2-mE-i\epsilon}{p^2-pq+q^2-mE-i\epsilon}  \right) }
\end{equation}

To understand the behavior of the function $a(k,p)$ for large $p$ we can derive an approximate equation from Eq.~\eqref{aamplitude}. We can see that the inhomogeneous part of this equation can be neglected as $p$ gets large,
also the main contribution in the integral comes from the region where the argument of the $\log$ is the biggest, which is the region where $q \sim p$. After doing all of these approximations the integral equation
simplifies to:

\begin{equation}\label{aamplitudeasymp}
a(k,p)=\frac{4}{\sqrt{3}\pi} \int \frac{dq}{q} \log{\left(\frac{q^2+pq+p^2}{q^2-pq+p^2}\right)} a(k,q)
\end{equation}

The solution to Eq.~\eqref{aamplitude} is unique for any given cutoff $\Lambda$, but Eq.~\eqref{aamplitudeasymp} does not have to have a unique solution. In fact it is obvious that from any given solution to Eq.~\eqref{aamplitudeasymp} one can construct infinitely many 
solutions by just multiplying it with any number $C$. So the most general solution to Eq.~\eqref{aamplitudeasymp} will be parametrized by some family of parameters. The solution to Eq.~\eqref{aamplitude} then has to be matched to the solution of
Eq.~\eqref{aamplitudeasymp} at some intermediate region of the variable $p$ and this matching will give one condition that this family of parameters has to satisfy.


At this point the solution for this integral equation is just guessed to be of the form $a(k,p)=p^s$ and to find $s$ we just substitute it into the equation:

\begin{equation}
p^s=\frac{4}{\sqrt{3}\pi} \int_0^\infty dq\ q^{s-1} \log{\left(\frac{q^2+pq+p^2}{q^2-pq+p^2}\right)} 
\end{equation}

The integration limits in this equation are from $0$ to $\infty$, so we can change the integration variable $q \rightarrow  xp$ without changing the integration limits. After doing this, $p^s$ cancels out 
and what we are left with is an equation for $s$:

\begin{equation} \label{findings}
1=\frac{4}{\sqrt{3}\pi} \int_{0}^{\infty} dx\ x^{s-1} \log{\left(\frac{x^2+x+1}{x^2-x+1}\right)}.
\end{equation}
Denote the integral on the right hand side $I=\int_{0}^{\infty} dx\ x^{s-1} \log{\left(\frac{x^2+x+1}{x^2-x+1}\right)}$. Doing change of integration variable $x \rightarrow-x$ we find 
$\int_{-\infty}^{0} dx\ x^{s-1} \log{\left(\frac{x^2+x+1}{x^2-x+1}\right)=(-1)^{s}I}$. 
Note that it may seem like doing the same change of integration variable twice we find $(-1)^{2s}=1$ which is obviously not the case if $s$ 
is anything but an integer. In case $s$ is not an integer we have the formula $(-x)^{s-1}=(-1)^{s-1}x^{s-1}$ only if $x>0$. 

From the previous paragraph we find:

\begin{equation}
(1+(-1)^{s})I=\int_{- \infty }^{\infty} dx\ x^{s-1} \log{\left(\frac{x^2+x+1}{x^2-x+1}\right)}
\end{equation}

Substitute Eq.~ \eqref{findings} into the previous equation to find:

\begin{equation}
(1+(-1)^{s})=\frac{4}{\sqrt{3}\pi} \int_{-\infty}^{\infty} dx\ x^{s-1} \log{\left(\frac{x^2+x+1}{x^2-x+1}\right)} 
\end{equation}

To calculate the integral on the right hand side we can do integration by parts taking $x^{s-1}$ under the differential:

\begin{equation}
(1+(-1)^{s})=\frac{4}{\sqrt{3}s\pi} \left ( x^{s} \log{\left(\frac{x^2+x+1}{x^2-x+1}\right)}  \bigg|_{-\infty}^{\infty}   -   \int_{-\infty}^\infty dx\ x^{s} \frac{f'(x)}{f(x)} \right )
\end{equation}
 where $f(x)=\left( \frac{x^2+x+1}{x^2-x+1} \right)$ is the argument of the $\log$.
 
For Eq.~\eqref{findings} to be valid $s$ has to be so that the first term on the right hand side of the last equation goes to zero in the limits $-\infty$ and $\infty$. This puts a restriction on possible values for $s$: $Re(s)<1$. So we are left with:

\begin{equation} \label{findings2}
(1+(-1)^{s})=-\frac{4}{\sqrt{3}s\pi}  \int_{-\infty}^{\infty} dx\ x^{s} \frac{f'(x)}{f(x)} 
\end{equation}

To calculate the integral on the right hand side we can close the integration contour in the upper half complex plane and find a complex integral over a contour.

In general if we have an integral of the form $\int dx\ g(x) \frac{f'(x)}{f(x)}$ over a closed contour where $g(x)$ is a regular function inside the contour, $f(x)$ has zeros $a_i$ of order $\alpha_i$ and poles $b_j$
of order $\beta_j$ then this integral can be calculated with the following formula:

\begin{equation}
\frac{1}{2\pi i}\int dx\ g(x) \frac{f'(x)}{f(x)}=\sum_{i} \alpha_{i} g(a_i)-\sum_{j} \beta_{j} g(b_j)
\end{equation}

This formula can be easily verified by expanding the functions $f$ and $g$ around the zeros and the poles of $f$ and using the residue theorem.

We use this formula to calculate the integral in Eq.~\eqref{findings2}. In the upper half complex plane the function $f(x)$ has one zero $a=e^{i\frac{2}{3}\pi}$ of order one and one pole $b=e^{i\frac{1}{3}\pi}$ of order one.
Using this Eq.~\eqref{findings2} becomes:

\begin{equation}
(1+e^{is\pi})=-\frac{8i}{\sqrt{3}s} (e^{i\frac{2}{3}s\pi}-e^{i\frac{1}{3}s\pi}),
\end{equation}
which simplifies to:

\begin{equation}
s=\frac{8}{\sqrt{3}}  \frac{  \sin{\frac{\pi s}{6} }    }{  \cos{ \frac{\pi s}{2} }  }.
\end{equation}
To see another derivation of this equation see \cite{JiC.:2012diss}.
This equation has two complex solutions $s=\pm is_0$, with $s_0$ approximately equal to $s_0\approx1.00624$. Those two solutions give two different linearly independent solutions to Eq.~\eqref{aamplitudeasymp} from which one can 
construct a general solution that is real and has two arbitrary undetermined parameters:

\begin{equation}\label{solutiona}
a(k,p)=C \cos \left( s_0 \log{\frac{p}{p_*}} \right)
\end{equation}
with undetermined constants $C$ and $p_*$.

If we take the solution to Eq.~\eqref{aamplitude} with very large cutoff $\Lambda$, and look at its form when $\Lambda>p\gg$ then we see that Eq.~\eqref{aamplitude} reduces to Eq.~\eqref{aamplitudeasymp}, where there are no $k$-related 
scales left, and this solution approaches to Eq.~\eqref{solutiona}, with parameters $C$ and $p_*$ that can't depend on $k$. All that they can depend on is the cutoff $\Lambda$ used in the original Eq.~\eqref{aamplitude}.
So this solution does not converge to a unique function when $\Lambda \rightarrow \infty$. 

To regulate the theory and get rid of this cutoff dependence we have to have a cutoff dependent counterterm added to the Lagrangian, which will modify the integral equation that the amplitude must satisfy, 
and will render it cutoff independent. This counterterm is given in the Lagrangian in Eq.~\eqref{3bosonlagrangian} by the interaction term proportional to $h$. It will modify Eq.~\eqref{aamplitude} to an equation that looks the same but
has a different function $M(q,p,k)$:

\begin{equation}
M(q,p,k)=\frac{4}{3} \left (\sqrt{\frac{3p^2}{4}-M_NE}+\frac{1}{a_2} \right ) \left[ \frac{1}{pq} \log{ \left(  \frac{p^2+pq+q^2-mE}{p^2-pq+q^2-mE}  \right) } +\frac{h}{mg^2}\right]
\end{equation}
with cutoff dependent $h=h(\Lambda)$. To determine $h$ the modified Eq.~\eqref{aamplitude} is solved numerically, then from the solution an observable is constructed, such as the three-body scattering length or three-body
binding energy and it is held fixed but the cutoff is varied. Doing this the dependence of $h(\Lambda)$ on the cutoff is determined numerically. The approximate form of the function $h(\Lambda)$ can be derived analytically too when the cutoff approaches to infinity by choosing the function
$h=h(\Lambda)$ so that the cutoff dependence of the solution of Eq.~\eqref{aamplitude} goes away \cite{Bedaque:1998kg,Hammer:2000nf}. Here I will only mention that the function $h(\Lambda)$ shows a non-trivial, oscillatory cutoff dependence and that 
this is a purely non-perturbative effect and it cannot be seen at any order of perturbation theory~\cite{Bedaque:1998km}. In the next section, when I discuss the doublet channel $nd$ scattering, I'll show why a three-body 
interaction counterterm is needed there for the same reason as it is needed for the three-boson case. I will also show the numerically determined function $h(\Lambda)$.

\section{Three-nucleon sector at LO, doublet channel: S-wave}
\label{Three-nucleon sector at LO, doublet channel: S-wave}

The purpose of this section is to show the equivalence of the cutoff dependence problems of the two situations: one with spin-${\frac{1}{2}}$ nucleons, the other with spin-$0$ bosons \cite{Bedaque:1999ve}.
Diagrammatically the equations for $nd$ scattering in the $S$-wave doublet channel without any three-body forces can be expressed as in the following Fig.~\ref{nddoubletchannel}:

\begin{center}\begin{figure}[ht]
  \begin{center}
  \includegraphics[scale=0.5]{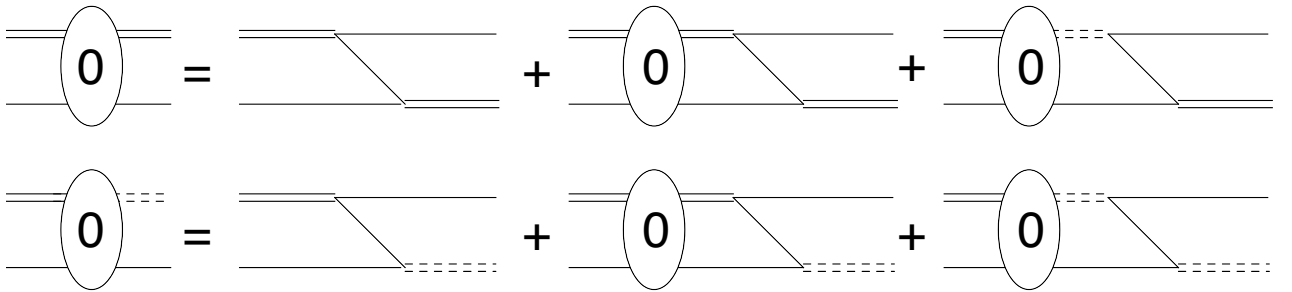}
  \end{center}
  \caption{\label{nddoubletchannel}Doublet channel $S$-wave LO $nd$ equation without three-body forces.}
\end{figure}\end{center}

Here the oval with a ``0'' is the leading order amplitude, the double lines stand for the deuteron and the double dashed lines stand for the spin-singlet isospin-triplet dibaryon $s_a$. We can see that 
here we have two coupled integral equations as opposed to the quartet channel $nd$ scattering. After writing the equations explicitly and projecting on the appropriate channel the 
following explicit form is obtained:

\begin{equation}
\frac{3}{2} \left(\frac{1}{a_2^t} + \sqrt{\frac{3p^2}{4}-M_NE} \right)^{-1} a(p,k)=K(p,k)+\frac{2}{\pi} \int_{0}^{\Lambda} \frac{q^2dq}{q^2-k^2-i\epsilon} K(p,q)(a(q,k)+3b(q,k))
\end{equation}

\begin{equation}
2 \frac{\left(-\frac{1}{a_2^s} + \sqrt{\frac{3p^2}{4}-M_NE} \right)}{p^2-k^2} b(p,k)=3K(p,k)+\frac{2}{\pi} \int_{0}^{\Lambda} \frac{q^2dq}{q^2-k^2-i\epsilon} K(p,q)(3a(q,k)+b(q,k))
\end{equation}
where 
\begin{equation}
K(p,q)=\frac{1}{2pq} \log{ \left(  \frac{p^2+pq+q^2-mE-i\epsilon}{p^2-pq+q^2-mE-i\epsilon}  \right) },
\end{equation}
also $a(p,k)$ and $b(p,k)$ are the two scattering amplitudes corresponding to the left hand sides of the equations in the Fig.~\ref{nddoubletchannel}. $a_2^t$ and $a_2^s$ are the two-body scattering lengths in the $^3S_1$ and $^1S_0$ channels respectively.

As we are only interested in the cutoff dependence of the amplitudes $a(p,k)$ and $b(p,k)$, we want to apply the same kind of asymptotic analysis as in the 3-boson case. In the limit where the 
cutoff $\Lambda$ is large and $p$ is large with $p<\Lambda$ the difference between $a_2^t$ and $a_2^s$ does not play an important role hence to figure out the cutoff dependence of the 
solutions we can take $a_2^t=a_2^s=a_2$. In this case we can define new amplitudes formed from $a(p,k)$ and $b(p,k)$ so that the equations for the new amplitudes decouple. The new amplitudes 
are $a_+=a(p,k)+b(p,k)$ and $a_-=a(p,k)-b(p,k)$ and the decoupled equations that they need to satisfy are:

\begin{equation}
\frac{3}{4} \left(\frac{1}{a_2} + \sqrt{\frac{3p^2}{4}-M_NE} \right)^{-1} a_+(p,k)=2K(p,k)+\frac{2}{\pi} \int_{0}^{\Lambda} \frac{q^2dq}{q^2-k^2-i\epsilon} 2K(p,q)a_+(q,k)
\end{equation}

\begin{equation}
\frac{3}{4} \left(\frac{1}{a_2} + \sqrt{\frac{3p^2}{4}-M_NE} \right)^{-1} a_-(p,k)=-K(p,k)-\frac{2}{\pi} \int_{0}^{\Lambda} \frac{q^2dq}{q^2-k^2-i\epsilon} 2K(p,q)a_-(q,k)
\end{equation}
Comparing these two equations to Eq.~\eqref{aamplitude} and Eq.~\eqref{ndQLO} we see that $a_-$ satisfies that same equation as the quartet channel $nd$ scattering amplitude, hence it has no cutoff 
dependence as $\Lambda\rightarrow\infty$, and $a_+$ satisfies the same equation as the three boson scattering amplitude, so it has cutoff dependence as $\Lambda\rightarrow\infty$. This indicates that a three-body force is necessary to 
regulate this cutoff dependence and ultimately render the $nd$ scattering amplitude and other observables cutoff independent. The interaction Lagrangian that describes this 
three-body force is given by:

\begin{equation}
\mathcal{L}=\frac{M_NH_0(\Lambda)}{3\Lambda^2}\left(  y_t N^{\dagger} (\vec{t} \cdot \vec{\sigma})^{\dagger}-y_s N^{\dagger} (\vec{s} \cdot \vec{\tau})^{\dagger} \right)
\left(  y_t N (\vec{t} \cdot \vec{\sigma})-y_s N (\vec{s} \cdot \vec{\tau})\right).
\end{equation}

\begin{center}\begin{figure}[H]
  \begin{center}
  \includegraphics[scale=0.6]{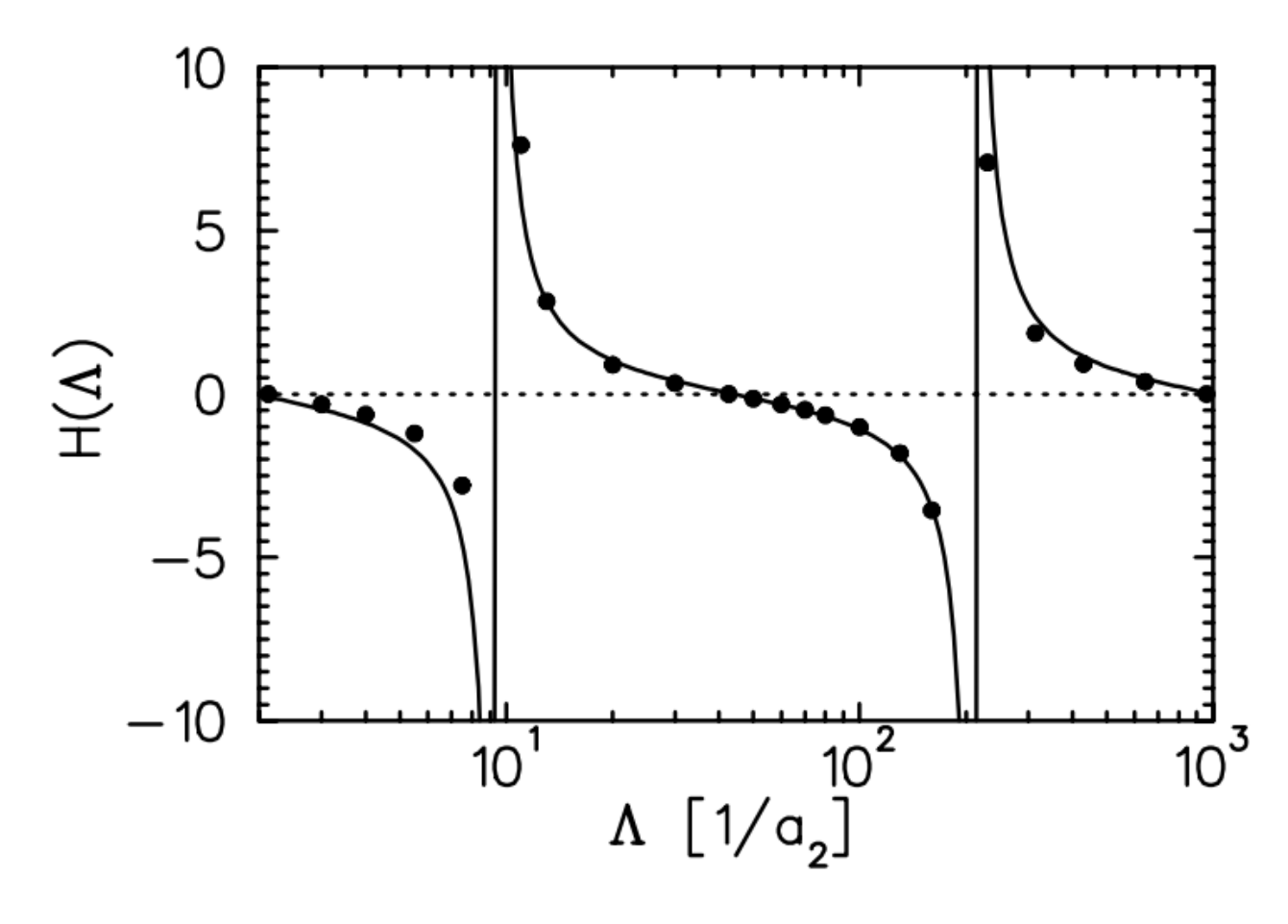}
  \end{center}
  \caption{\label{3bf}(Caption and figure reprinted from \cite{Bedaque:1998kg}). Three-body force as a function of the cutoff $\Lambda$: numerical solution (dots) and the analytical approximate solution (solid line).}
\end{figure}\end{center}

This will modify the integral equations satisfied by the doublet channel $nd$ scattering amplitude and the modified equations will be given diagrammatically by the following Fig.~\ref{nddoubletchannel3bf}.
\begin{center}\begin{figure}[ht]
  \begin{center}
  \includegraphics[scale=0.5]{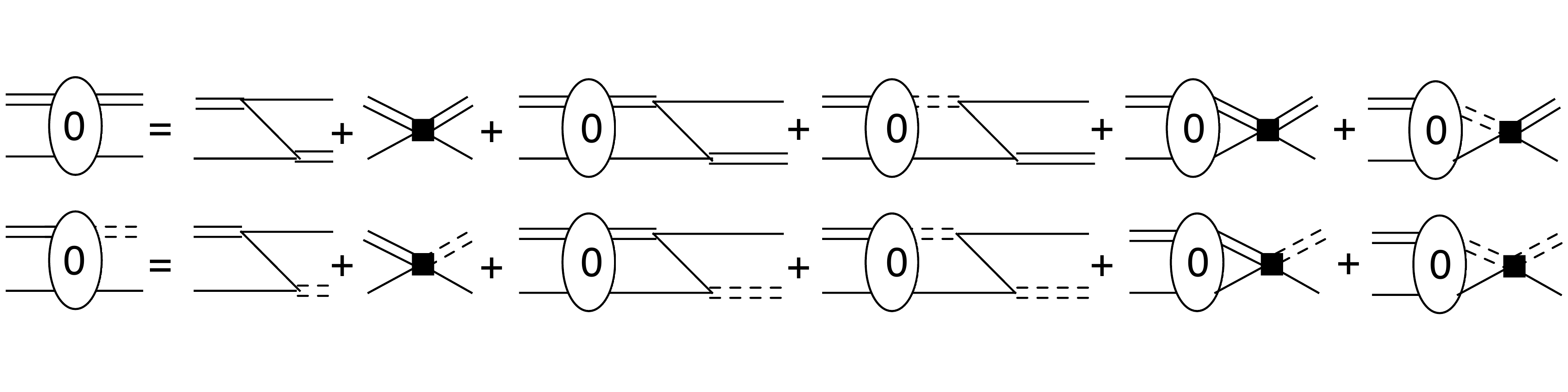}
  \end{center}
  \caption{\label{nddoubletchannel3bf}Doublet channel $S$-wave LO $nd$ equation with three-body forces.}
\end{figure}\end{center}
and have the form:
\begin{eqnarray}
&&t^0_{0,Nt \rightarrow Nt}(k,p)=2\pi \left( \frac{1}{pk}\log{ \left(  \frac{p^2+pk+k^2-mE}{p^2-pk+k^2-mE}  \right) }+\frac{2H_0(\Lambda)}{\Lambda^2} \right) \nonumber \\
&+&  \int \left[ \frac{1}{2pq} \log{ \left(  \frac{p^2+pq+q^2-mE-i\epsilon}{p^2-pq+q^2-mE-i\epsilon}  \right) }+\frac{H_0(\Lambda)}{\Lambda^2}  \right]\nonumber \\
&\times& \left(- D_t^{(0)} (E,q)t^0_{0,Nt \rightarrow Nt}(k,q)  +3D_s^{(0)} (E,q)t^0_{0,Nt \rightarrow Ns}(k,q) \right) \nonumber \\
&& \nonumber \\
&&t^0_{0,Nt \rightarrow Ns}(k,p)=-6\pi \left( \frac{1}{pk}\log{ \left(  \frac{p^2+pk+k^2-mE}{p^2-pk+k^2-mE}  \right) }+\frac{2H_0(\Lambda)}{\Lambda^2} \right) \nonumber \\
&+&  \int \left[ \frac{1}{2pq} \log{ \left(  \frac{p^2+pq+q^2-mE-i\epsilon}{p^2-pq+q^2-mE-i\epsilon}  \right) }+\frac{H_0(\Lambda)}{\Lambda^2}  \right] \nonumber \\
&\times&\left(3 D_t^{(0)} (E,q)t^0_{0,Nt \rightarrow Nt}(k,q)  -D_s^{(0)} (E,q)t^0_{0,Nt \rightarrow Ns}(k,q)\right). \nonumber \\
\end{eqnarray} 
These equations are solved numerically and the three-body force $H_0(\Lambda)$ is fixed by an observable such as the three-body scattering length or the three-body 
binding energy. The dependence of the three-body force on the cutoff is oscillatory, this RG behavior is called a limit-cycle \cite{Bedaque:1999ve,Ji:2012nj,Griesshammer:2005ga,Bedaque:1998km} (see Fig.~\ref{3bf}\footnote{Reprinted from \cite{Bedaque:1998kg}, Phys. Rev. Lett., Volume 82,  Bedaque, Paulo F. and Hammer, H. W. and van Kolck, U., ``Renormalization of the three-body system with short range interactions,'' Pages 463-467, Copyright (1999), by the American Physical Society.} \cite{Bedaque:1998kg}). This three body force gets corrections at all higher orders of perturbation theory, at each order being fixed to the same observable.

\section{Three-nucleon sector at LO, doublet channel: Higher partial waves}

\begin{center}\begin{figure}[ht]
  \begin{center}
  \includegraphics[scale=0.5]{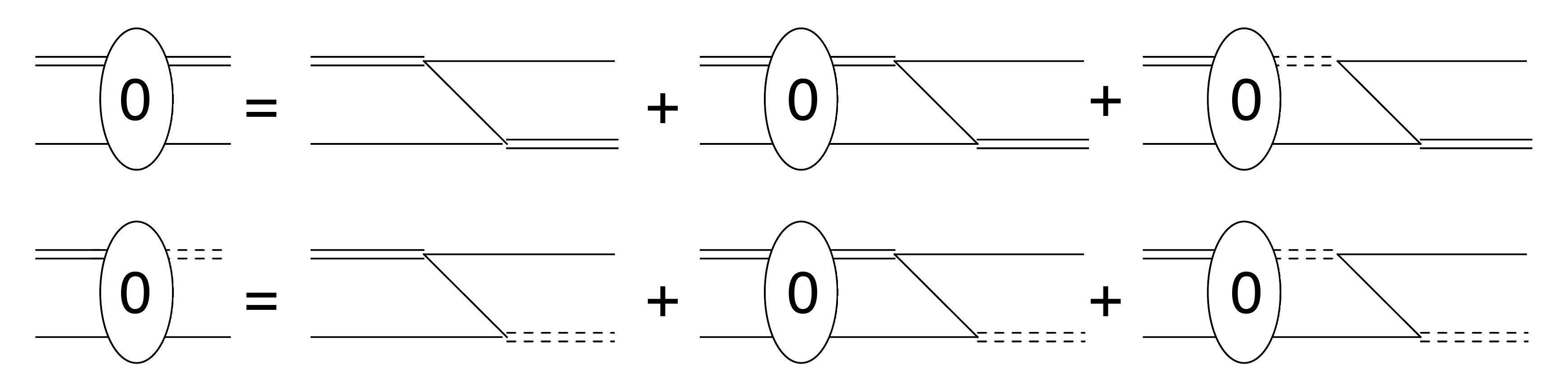}
  \end{center}
  \caption{Doublet channel LO $nd$ equation higher partial waves.}
  \label{fig:dbhw}
\end{figure}\end{center}

For higher partial waves the LO equations are expressed by the same diagrams as the $S$-wave equation without the three-body force, as shown if Fig.~\ref{fig:dbhw}.
Explicitly these equation are Eq.~\eqref{nddoubletchanneleq}  \cite{Vanasse:2016jtc}.
\begin{equation} \label{nddoubletchanneleq}
t^l_{0,d}(k,p)=B_0^l(k,p,E)+K_0^l(q,p,E)\otimes t^l_{0,d}(k,q).
\end{equation}
where $t^l_{0,d}(k,p)$ is the doublet channel scattering amplitude in so-called cluster configuration space, with the first component being the
$Nt \rightarrow Nt$ scattering amplitude and the second component being the $Nt \rightarrow Ns$ scattering amplitude:

\begin{equation}
t^l_{n,d}(k,p)=\left(\begin{matrix}
t^l_{n,Nt \rightarrow Nt}(k,p) \\
t^l_{n,Nt \rightarrow Ns}(k,p)
\end{matrix} \right),
\end{equation}
where the superscript $l$ stands for the $l$-th partial wave. The subscript $n$ stands for $n$-th order, and $d$ for doublet channel. 

 $B_0^l(k,p,E)$ is defined by:

\begin{equation}
B_0^l(k,p,E)=\left(\begin{matrix}
 \frac{2\pi}{pk}Q_l(\frac{p^2+k^2-ME}{pk})\\
-\frac{6\pi}{pk}Q_l(\frac{p^2+k^2-ME}{pk})
\end{matrix} \right)
\end{equation}
and the kernel matrix is given by:
\begin{eqnarray}
&&K_0^l(q,p,E)=\nonumber \\
&&  \frac{1}{2qp}Q_l\left (\frac{p^2+q^2-M_NE-i\epsilon}{pq}\right ) 
\left(\begin{matrix}  \frac{1}{\sqrt{\frac{3q^2}{4}-M_NE-i\epsilon}-\gamma_t}&\frac{-3}{\sqrt{\frac{3q^2}{4}-M_NE-i\epsilon}-\gamma_s}  \\
\frac{-3}{\sqrt{\frac{3q^2}{4}-M_NE-i\epsilon}-\gamma_t} & \frac{1}{\sqrt{\frac{3q^2}{4}-M_NE-i\epsilon}-\gamma_s} 
\end{matrix} \right).
\nonumber \\
\end{eqnarray} 
The symbol $\otimes$ is defined in Eq.~\eqref{otimessymbol}, but it is also accompanied by a matrix multiplication:
\begin{equation}
\left( \begin{matrix} A(q) & B(q)\\
C(q) & D(q)
\end{matrix} \right)
\otimes \left( \begin{matrix} U(q)\\
V(q) 
\end{matrix} \right) =
\left( \begin{matrix} A(q)\otimes U(q) +B(q)\otimes V(q)\\
C(q)\otimes U(q) +D(q)\otimes V(q)
\end{matrix} \right).
\end{equation}

 \section{Three-nucleon sector at NLO, doublet channel}
 
 The NLO correction to the amplitude is determined from an integral equation diagrammatically expressed in the following Fig.~\ref{nlodnd}:
 
 \begin{center}\begin{figure}[ht]
  \begin{center}
  \includegraphics[scale=0.5]{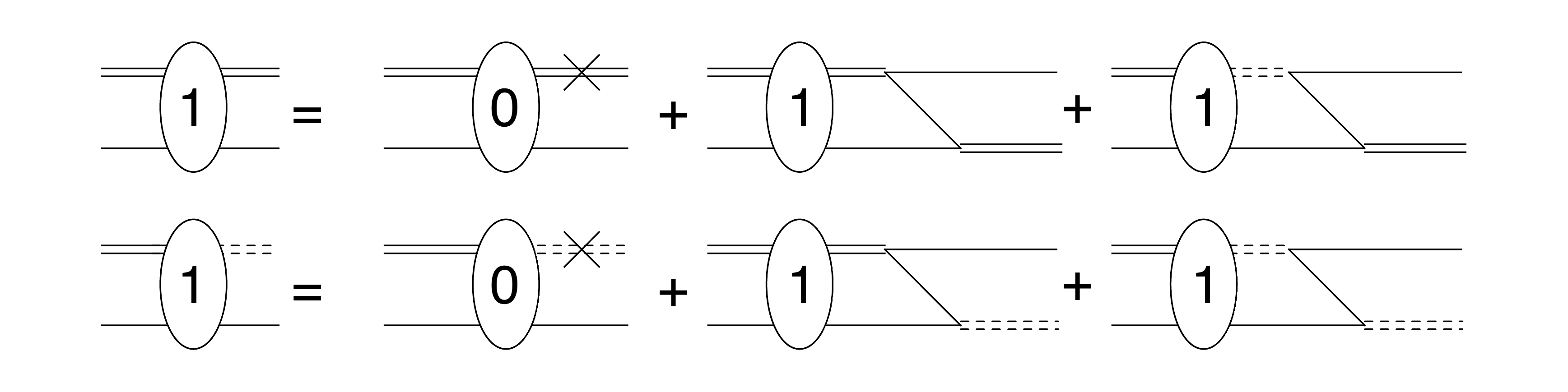}
  \end{center}
  \caption{\label{nlodnd}Doublet channel NLO $nd$ equation.}
\end{figure}\end{center}
and explicitly by the equation Eq.~\eqref{ndnlodoubletchanneleq}:
\begin{equation} \label{ndnlodoubletchanneleq}
t^l_{1,d}(k,p)=t^l_{0,d}(k,p)  \circ R_1(p,E)+K_0^l(q,p,E)\otimes t^l_{1,d}(k,q),
\end{equation}
where:
\begin{equation}
R_1(p,E)=\left(\begin{matrix}
\frac{z_t-1}{2\gamma_t}(\gamma_t+\sqrt{\frac{3p^2}{4}}-M_NE) \\
\frac{z_s-1}{2\gamma_s}(\gamma_s+\sqrt{\frac{3p^2}{4}}-M_NE) 
\end{matrix} \right),
\end{equation}
and the $\circ$ is the Schur product:
\begin{equation}
\left( \begin{matrix} A\\
B
\end{matrix} \right)
\circ
\left ( \begin{matrix} C\\
D
\end{matrix} \right) =
\left(  \begin{matrix} 
AC\\
BD
\end{matrix} \right).
\end{equation}

 \section{Three-nucleon sector at N$^2$LO, doublet channel}
 
The N$^2$LO correction also satisfies similar diagrammatic equation Fig.~\ref{n2lodnd}:
 \begin{center}\begin{figure}[ht]
  \begin{center}
  \includegraphics[scale=0.5]{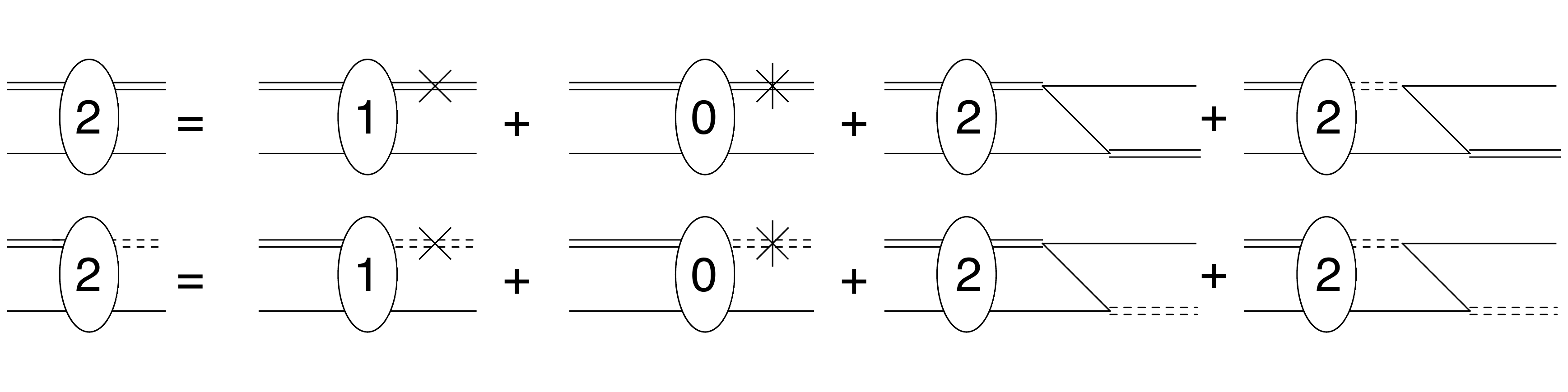}
  \end{center}
  \caption{\label{n2lodnd}Doublet channel $\nnlo$ $nd$ equation.}
\end{figure}\end{center}
 which explicitly becomes Eq.~\eqref{ndn2lodoubletchanneleq}:
 \begin{equation}\label{ndn2lodoubletchanneleq}
t^l_{2,d}(k,p)=(t^l_{1,d}(k,p)-c_1 \circ t^l_{0,d}(k,p)) \circ R_1(p,E)+K_0^l(q,p,E)\otimes t^l_{2,d}(k,q),
\end{equation}
where $c_1$ is given by:
\begin{equation}
c_1=
\left(  \begin{matrix} 
Z_t-1\\
Z_s-1
\end{matrix} \right).
\end{equation}

Note that the kernels of the LO,
 NLO and N$^2$LO equations are all the same, which makes the numerical part of the calculations easier; once the $(1-$Kernel$)$ matrix is inverted to solve the LO equation, 
 that inverse can be used to calculate higher order amplitudes. This statement generalizes to higher orders as we will see later.
  The LO amplitude contributes only to the 
 inhomogeneous parts of the NLO and the N$^2$LO equations, the same way the NLO amplitude only contributes to the inhomogeneous part of the 
 N$^2$LO equation.
 
 At N$^2$LO there is a new interaction that enters in the two-body sector called the $SD$ mixing. It gives a vertex that is an incoming deuteron with two outgoing nucleons attached to it or vise versa. 
In the deuteron the nucleons are in relative $S$-wave and the other two nucleons that are attached to this vertex are always in a relative $D$-wave, otherwise this interaction vertex is just zero, and hence the name.
There are two derivatives acting on the nucleon fields, which makes this term enter at N$^2$LO, also making the two nucleons outside of the deuteron to be in relative a $D$-wave. 
This interaction is given by the Lagrangian:
\begin{equation}
\mathcal{L}^{SD}_{2}=y_{SD}\hat{t}_{i}^{\dagger}\left[\hat{N}^{T}\left((\stackrel{\rightarrow}{\partial}-\stackrel{\leftarrow}{\partial})^{i}(\stackrel{\rightarrow}{\partial}-\stackrel{\leftarrow}{\partial})^{j}-\frac{1}{3}\delta^{ij}(\stackrel{\rightarrow}{\partial}-\stackrel{\leftarrow}{\partial})^{2}\right)P_{j}\hat{N}\right]+\mathrm{H.c.},
\end{equation}
where, just like before, the projector is defined by $P_j=\frac{1}{\sqrt{8}}\sigma_2\sigma_j\tau_2$.
The coupling constant $y_{SD}$ is fixed by the two-body scattering data. A detailed discussion of how to fix $y_{SD}$ and the calculation of the contributions of this term to the three-body scattering amplitude 
can be found in \cite{Vanasse:2013sda,Fleming:1999bs}. Here I will just note that this is the first interaction term that gives a non-zero contribution to the observable $A_{y}$. At $\ntlo$ new interaction 
terms enter (see section \ref{tbPw}) into the Lagrangian giving more substantial contributions to this observable.

\section{Two-body $P$-wave interactions at two-nucleon sector}
\label{tbPw}
Two-body $P$-wave contact interactions first occur at $\ntlo$.  The ${}^{3}\!P_{J}$ terms are given by the ${}^{3}\!P_{J}$ Lagrangian~\cite{Chen:1999bg},  
\begin{align}
\label{eq:PwaveLagRupak}
&\mathcal{L}_{2}^{{}^{3}P_{J}}=\left(C_{2}^{({}^3\!P_{0})}\delta_{xy}\delta_{wz}+C_{2}^{({}^3\!P_{1})}[\delta_{xw}\delta_{yz}-\delta_{xz}\delta_{yw}]+C_{2}^{({}^3\!P_{2})}\left[2\delta_{xw}\delta_{yz}+2\delta_{xz}\delta_{yw}-\frac{4}{3}\delta_{xy}\delta_{wz}\right]\right)\\\nonumber
&\hspace{3cm}\times\frac{1}{4}(\hat{N}^{T}\mathcal{O}_{xyA}^{(1,P)}\hat{N})^{\dagger}(\hat{N}^{T}\mathcal{O}_{wzA}^{(1,P)}\hat{N})
\end{align}
where
\begin{equation}
\mathcal{O}_{ijA}^{(1,P)}=\stackrel{\leftarrow}{\nabla}_{i}P_{jA}^{P}-P_{jA}^{P}\stackrel{\rightarrow}{\nabla}_{i}
\end{equation}
and the projector is defined as $P_{iA}^{P}=\frac{1}{\sqrt{8}}\sigma_{2}\sigma_{i}\tau_{2}\tau_{A}$.  
This interaction Lagrangian is also considered in the paper Ref.~\cite{Chen:1999bg}, in which the authors consider the process $np\rightarrow d \gamma$. To calculate this particular
scattering amplitude they need the projector $P_{iA}^{P}$ to be something particular so that it projects to the isospin quantum numbers of the initial state $np$. In our case 
we are interested in $nd$ scattering, hence we can have $nn$ and $np$ scattering through this interaction. For this reason we need the projector $P_{iA}^{P}$ to be a more
general one than given in Ref.~\cite{Chen:1999bg}.

At this order also the two-body ${}^{1}\!P_{1}$  contact interaction appears:
\begin{equation}
\label{eq:Pwave1P1}
 \mathcal{L}_{2}^{{}^{1}\!P_{1}}=C_{2}^{({}^{1}\!P_{1})}\frac{1}{4}(\hat{N}^{T}\mathcal{O}_{x}^{(0,P)}\hat{N})^{\dagger}(\hat{N}^{T}\mathcal{O}_{x}^{(0,P)}\hat{N}),
\end{equation}
The operator $\mathcal{O}_{i}^{(0,P)}$ is defined by
\begin{equation}
\mathcal{O}_{i}^{(0,P)}=\stackrel{\leftarrow}{\nabla}_{i}P^{P}-P^{P}\stackrel{\rightarrow}{\nabla}_{i},
\end{equation}
where the projector is $P^{P}=\frac{1}{\sqrt{8}}\sigma_{2}\tau_{2}$.  

To understand the index structure of the projectors $P_{iA}^{P}=\frac{1}{\sqrt{8}}\sigma_{2}\sigma_{i}\tau_{2}\tau_{A}$ and $P^{P}=\frac{1}{\sqrt{8}}\sigma_{2}\tau_{2}$ 
consider the following. The nucleons interacting through the ${}^{3}\!P_{J}$ interaction terms have to be in relative a $P-$wave, which is antisymmetric, with total spin $S=1$, 
which is symmetric, so to have in total an antisymmetric configuration the total isospin should be also symmetric hence $I=1$. 
This is why the projector $P_{iA}^{P}=\frac{1}{\sqrt{8}}\sigma_{2}\sigma_{i}\tau_{2}\tau_{A}$ has a spin index $i$ and an isospin index $A$. 
The same way in ${}^{1}\!P_{1}$ channel we have antisymmetric $P-$wave, antisymmetric $S=0$ state, so we must have antisymmetric $I=0$ state hence there are no 
spin and isospin indices on the $P^{P}=\frac{1}{\sqrt{8}}\sigma_{2}\tau_{2}$ projector.

The interaction coefficients are $C_2^{^3P_J}$ with $J=1,2,3$ and $C_2^{^1P_1}$.
 The subscript $2$ refers to two-body, in the superscripts $^3P_J$ and $^1P_1$ the standard spectroscopic notation is used: $^{(2S+1)}L_J$. If $S=1$ and $L=1$ then $J=S+L$ can be only $1, 2,$ and $3$, so we have three terms here.
  The same with $^1P_1$: $S=0$, $L=1$ hence $J=1$ and we have one interaction term here. These interaction are constructed so that each of these interaction terms only contributes to the $NN$ scattering amplitude in the corresponding channel.

To determine the interaction coefficients we use these interactions to calculate the $np$ scattering amplitude in the CM frame. This calculation is fairly straightforward, because we already 
know the channels in which we are going to get non-zero contributions. As an example let's do the calculation of the contribution that comes from $C^{3P_0}$ term. Denoting the incoming momentum 
by $\vec{k}$ and the outgoing momentum by $\vec{p}$ we get for this amplitude:
\begin{eqnarray}\label{fixing}
&&T^J_{LS,L'S'}=4\frac{1}{4}\frac{1}{8} C_2^{3P_0} \sum_{i,j,m_1,m_2,m'_1,m'_2} \CG{L}{m_L}{S}{m_S}{J}{M} \CG{L'}{m'_L}{S'}{m'_S}{J'}{M'}  \CG{\frac{1}{2}}{m_1}{\frac{1}{2}}{m_2}{S}{m_S} \CG{\frac{1}{2}}{m'_1}{\frac{1}{2}}{m'_2}{S}{m'_S}  \nonumber \\
&& \bra{L'_,m'_L}2k_i2p_j\ket{L,m_L} \bra{\frac{1}{2} m_2}\sigma_2 \sigma_i \ket{\frac{1}{2} m_1} \bra{\frac{1}{2} m'_2} \sigma_j \sigma_2  \ket{\frac{1}{2} m'_1} \nonumber \\
\end{eqnarray} 
in this equation the first factor $4$ is a symmetry factor, the ratio $\frac{1}{4}$ comes from the operator, the ratio $\frac{1}{8}$ comes from the projectors. The first Clebsch-Gordan coefficient combines the spin and the orbital angular momentum $S$ and $L$ into the total angular momentum $J$ in the initial state, the second does the same for the final state. The third and fourth Clebsch-Gordan coefficients combine 
the spins of the nucleons into the total spins $S$ and $S'$. $m_1$ and $m_2$ stand for initial nucleons' spin magnetic quantum numbers, and correspondingly $m'_1$ and $m'_2$ for final. 

The sum in Eq.~\eqref{fixing} is easily calculated using the methods described in Appendix \ref{Projections}. The answer  
turns out to be non-zero only if $J=J'=0$, $L=L'=1$, $S=S'=1$ as was expected, because the operators by construction should give 
contributions only in the corresponding channels. After we have the amplitude $T^J_{LS,L'S'}$, the phase shift is calculated by the Eq.~\eqref{psh}.
\begin{equation}
\label{psh}
\delta=\frac{Mk}{4\pi} T^J_{LS,L'S'},
\end{equation}
where we have assumed $\delta$ to be small and we have expanded $e^{2i\delta}$, keeping only the linear term in $\delta$.
Having a numerical value for $\delta$ at a given scattering energy we find the coefficient $C_2^{3P_0}$. And the same way we can determine all the other interaction coefficients in the Lagrangians in Eqs.\eqref{eq:PwaveLagRupak} and \eqref{eq:Pwave1P1}. 
The numerical values for the phase shifts are taken from the Nijmegen phase shifts~\cite{Stoks:1994wp}. The coefficients we get are: 
\begin{equation}
\label{eq:CPJnpfit}
C^{{}^{3}\!P_{0}}=6.27~\mathrm{fm}^{4}, \ \ C^{{}^{3}\!P_{1}}=-5.75~\mathrm{fm}^{4}, \ \ C^{{}^{3}\!P_{2}}=.522~\mathrm{fm}^{4},\,\,\mathrm{and}\,\, C^{{}^{1}\!P_{1}}=-19.8~\mathrm{fm}^{4}.  
\end{equation}
%

\chapter{Non-Relativistic Pionless EFT: Order-3 and Higher}
\label{chap:3}

\section{Three-nucleon sector at N$^n$LO, both channels}

\begin{figure}[ht]
\begin{center}
\includegraphics[width=90mm]{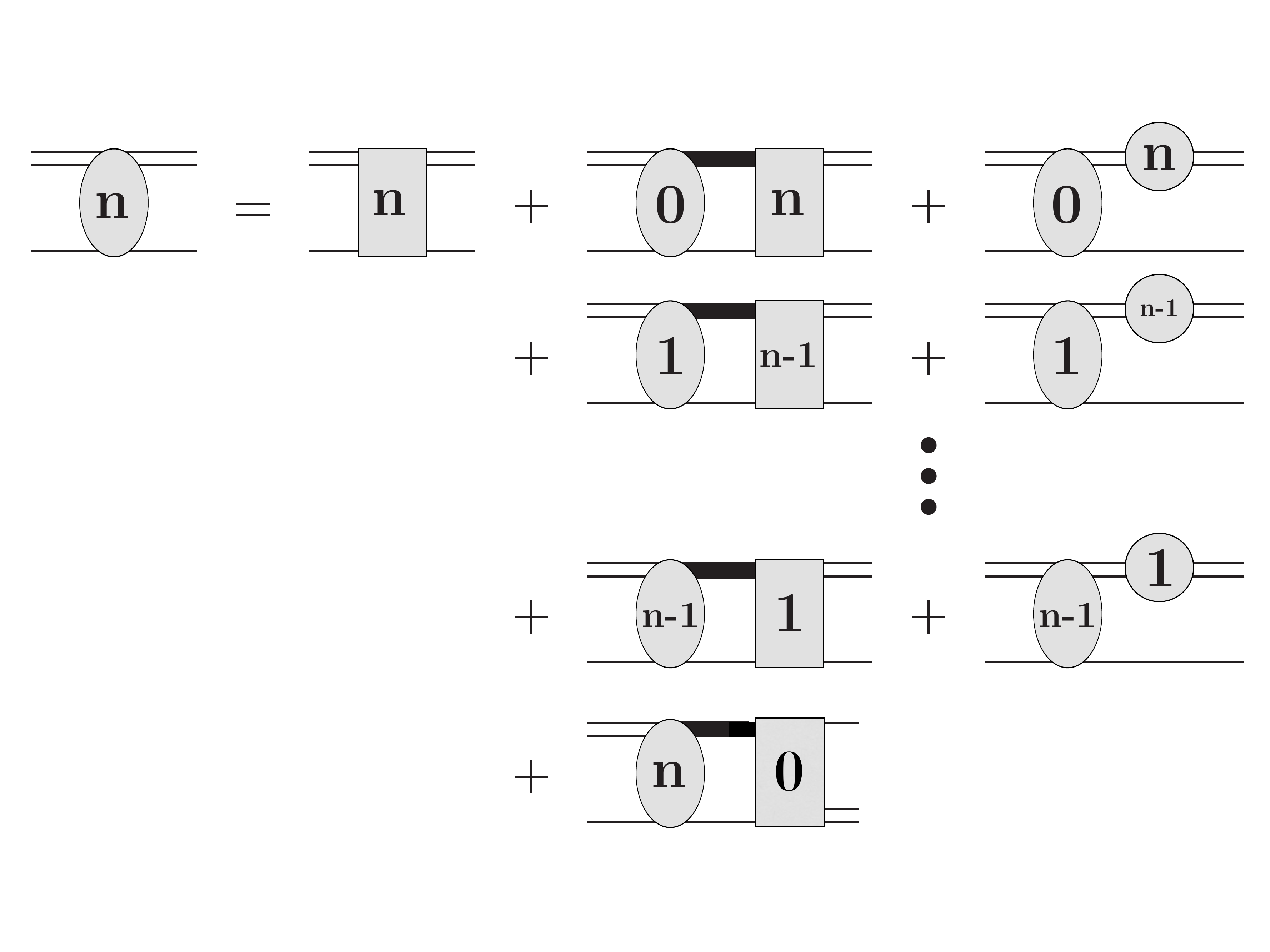}
\includegraphics[width=90mm]{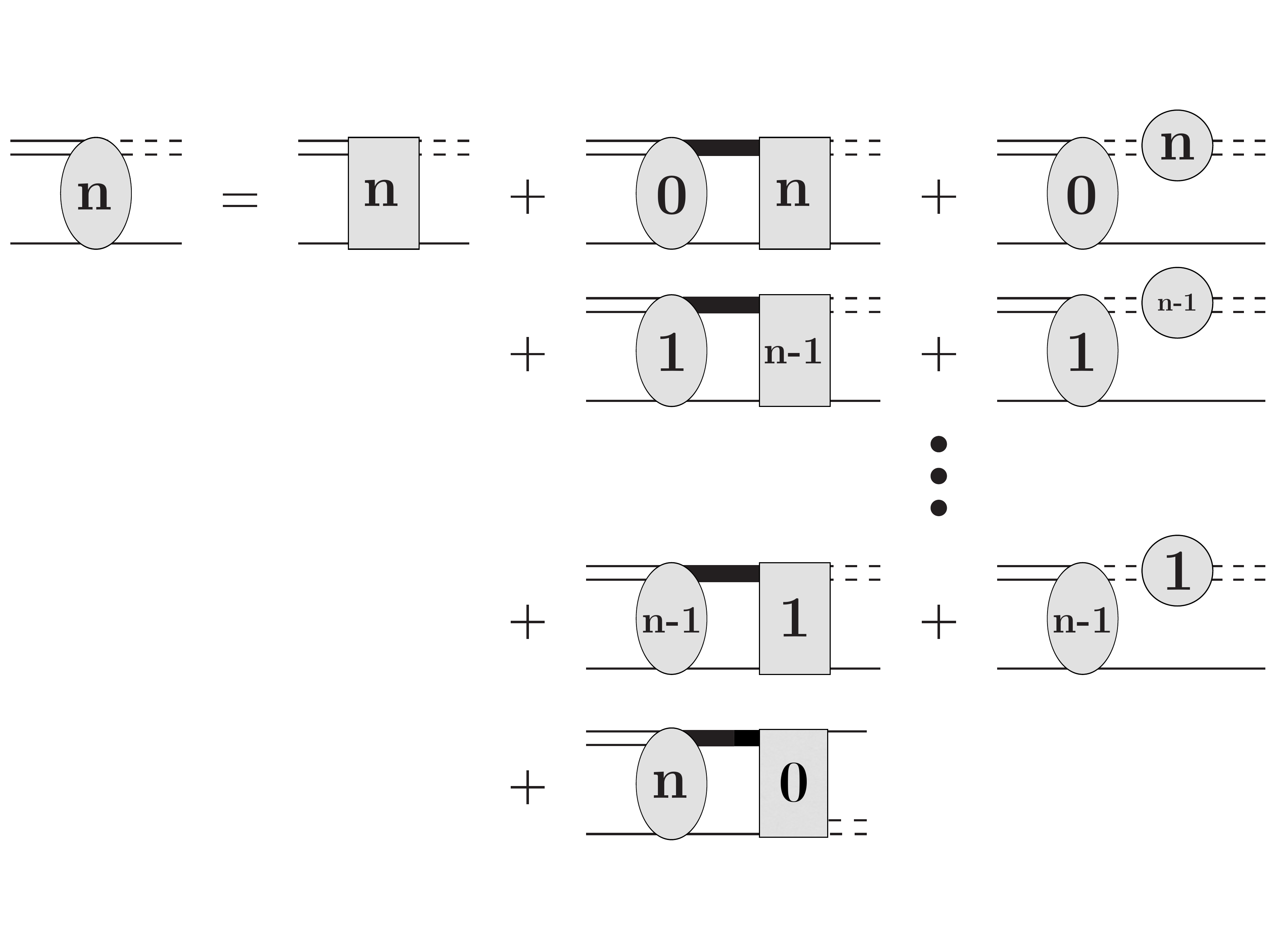}
\end{center}
\caption{\label{fig:PertCorrectionDiagrams}(Caption and figures are from \cite{Margaryan:2015rzg}). Single lines represent nucleons, double lines spin-triplet dibaryons, and double dashed lines spin-singlet dibaryons.  Thick solid lines denote a sum over both spin-triplet and spin-singlet dibaryons.  The LO $nd$ scattering amplitude is the oval with a ``0'' inside and the oval with the $n$ inside is the $\mathrm{N}^{n}\mathrm{LO}$ correction to the $nd$ scattering amplitude.  The circle with the $n$ inside is the $\mathrm{N}^{n}\mathrm{LO}$ correction to the dibaryon propagators (see Fig.~\ref{fig:TwoBodyContributions}), and the rectangle with the $n$ inside is the $n$th order ``three-body'' correction (see Fig.~\ref{fig:N3LOndScatteringDiagrams}).}
\end{figure}

The process of going to higher orders is the same as before. We solve the LO equation, getting the LO amplitude numerically.
Then this amplitude contributes to the inhomogeneous part of the $\mathrm{N}\mathrm{LO}$ equation. Then the LO and $\mathrm{N}\mathrm{LO}$
amplitudes contribute to the inhomogeneous part of the $\mathrm{N}^{2}\mathrm{LO}$ equation, etc. At each subsequent order all of the lower order
amplitudes, which are known numerically, contribute to the inhomogeneous parts of the equations at that order and not to the kernel. 
The kernel of the equations is always determined by just one nucleon exchange, which contributes in all channels, 
and the LO three-body force, which only contributes in the doublet $S$-wave channel. Using the technology developed earlier to represent the LO, $\mathrm{N}\mathrm{LO}$ and $\mathrm{N}^{2}\mathrm{LO}$ equations 
diagrammatically we can generalize it to represent the $\mathrm{N}^{n}\mathrm{LO}$ equations diagrammatically as in Fig.~\ref{fig:PertCorrectionDiagrams} (caption and figures are from \cite{Margaryan:2015rzg}).
\pagebreak

In the Fig.~\ref{fig:PertCorrectionDiagrams} we use the same notation as before, except a few things need to be generalized:

1) Note that this system of equations is for the $n$-th order amplitude, which is the oval with an ``$n$.'' 

\begin{figure}[ht]
\begin{center}
\includegraphics[width=105mm]{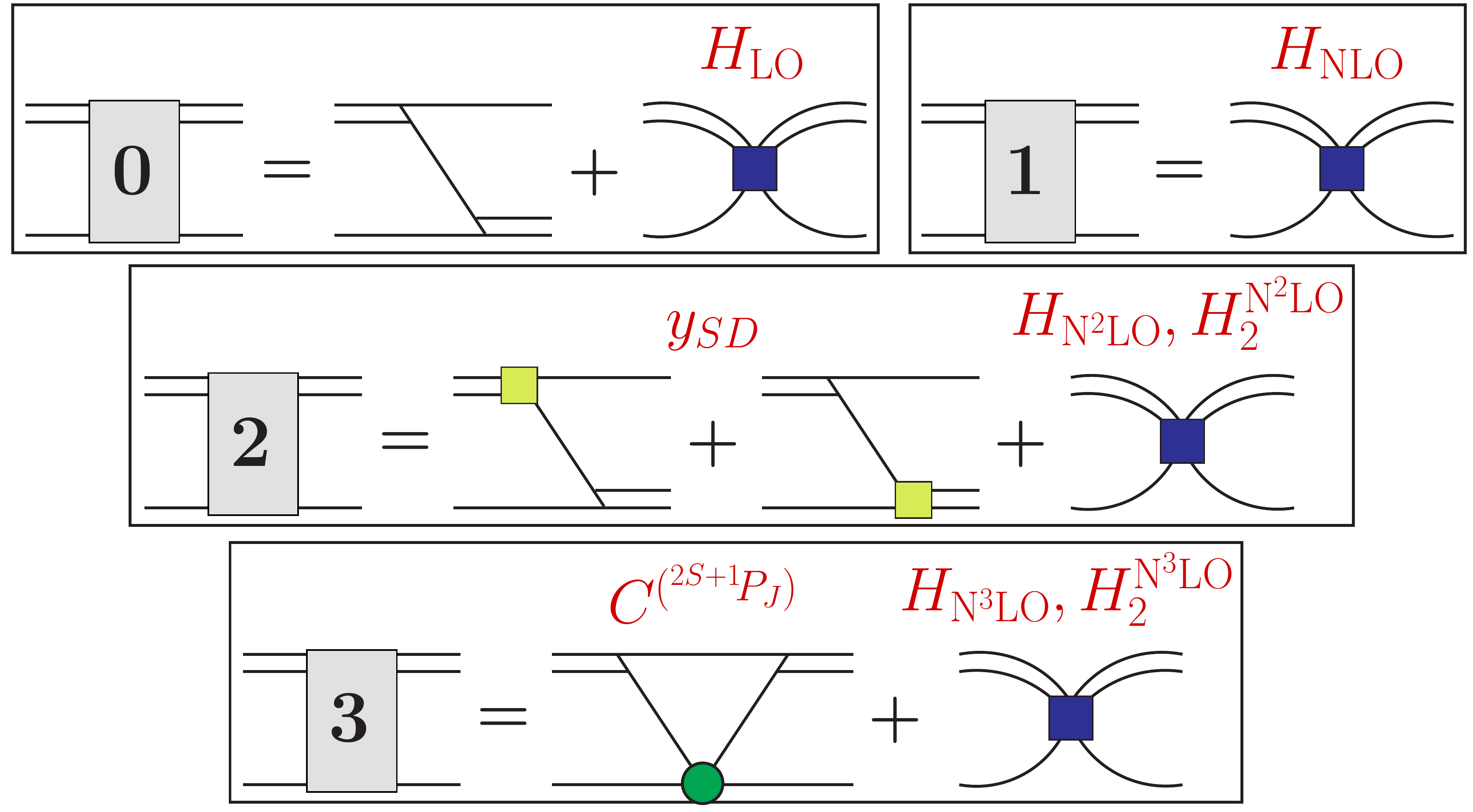}
\caption{\label{fig:N3LOndScatteringDiagrams}(Caption and figures are from \cite{Margaryan:2015rzg}). ``three-body'' contributions to integral equations (used in the diagrams of Fig.~\ref{fig:PertCorrectionDiagrams}).  The LO terms are nucleon exchange plus in the doublet $S$-wave channel the LO three-body force (dark square) .  The NLO term is a NLO correction to the LO three-body force.  At $\nnlo$ there are contributions from the two-body $SD$-mixing term (coupling indicated by pale square), the $\nnlo$ correction to the LO three-body force, $H_{\nnlo}$, and a new energy dependent three-body force, $H_{2}^{\nnlo}$.  The $\ntlo$ contributions are from the two-body $P$-wave contact interactions (green circle), the $\ntlo$ correction to the LO three-body force, and the $\ntlo$ correction to the $\nnlo$ energy dependent three-body force.}
\end{center}
\end{figure}

2) The $n$-th order, not-resummed , and using only the leading order dibaryon propagators, contributions to the $n$-th order amplitude are collected in the rectangle with an ``$n$'' inside, see Fig.~\ref{fig:PertCorrectionDiagrams} and Fig.~\ref{fig:N3LOndScatteringDiagrams} (caption and figures are from \cite{Margaryan:2015rzg}); this only makes 
contributions to the inhomogeneous parts of the equations except for the leading order. 

At LO this box is the one-nucleon exchange contributing in all channels plus the LO three-body force contributing only in the doublet $S$-wave channel. This is the only order when the box has a contribution to the kernel: it is the kernel.

At $\mathrm{N}\mathrm{LO}$ it is only the first order correction to the LO three-body force and nothing else.

At $\mathrm{N}^{2}\mathrm{LO}$ it has contributions from $SD$ mixing, the $\nnlo$ correction to the LO three-body force and the new energy-dependent three-body force $H_{2}^{(\nnlo)})$.

At $\mathrm{N}^{3}\mathrm{LO}$ it gets the expected contributions, which are the third order correction to the LO three-body force and the first order correction to the energy-dependent three-body force.
Most importantly it gets a contribution from the two-body $P$-wave interactions (see Fig.~\ref{fig:N3LOndScatteringDiagrams}, in the box on the third line, first contribution to the rectangle with a ``$3$'' in it), 
which determine the leading magnitude and behavior of the $A_{y}$ observable. 

\begin{figure}[ht]
\begin{center}
\includegraphics[width=110mm]{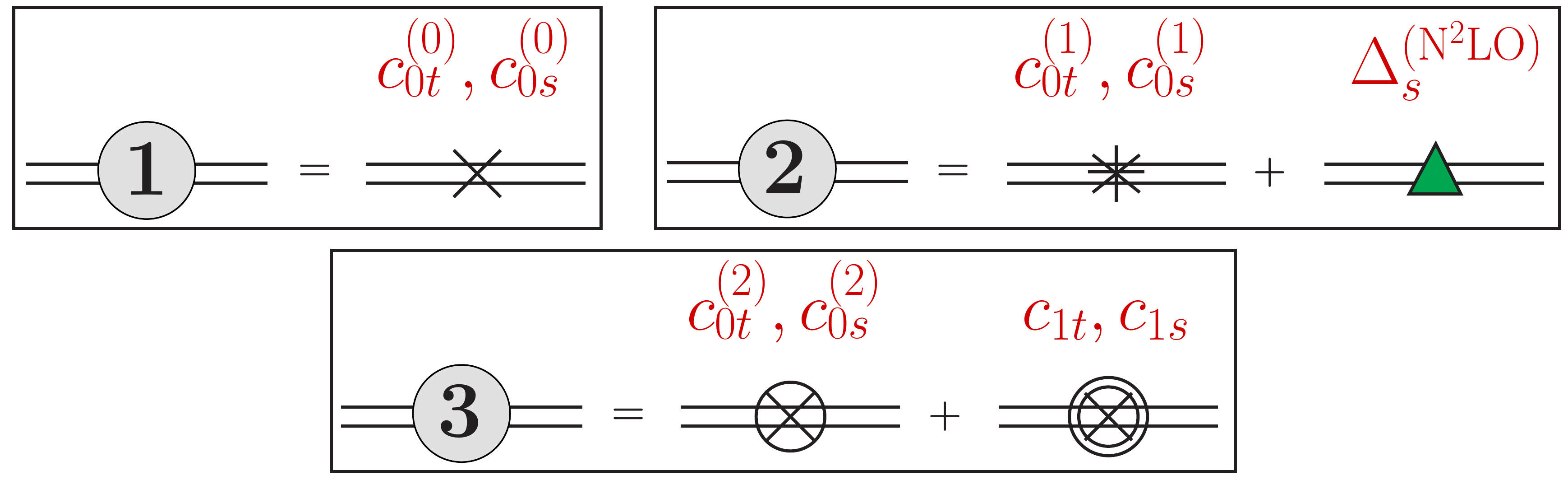}
\caption{\label{fig:TwoBodyContributions}(Caption and figures are from \cite{Margaryan:2015rzg}). Higher order corrections to dibaryon propagators (used in the diagrams of Fig.~\ref{fig:PertCorrectionDiagrams}). 
The NLO (n=1) corrections are range corrections from $c_{0t}^{(0)}$ and $c_{0s}^{(0)}$. 
At $\nnlo$ (n=2) the dibaryons receive further range corrections $c_{0t}^{(1)}$ and $c_{0s}^{(1)}$ in the $Z$-parametrization,
as well as the $\Delta^{(\nnlo)}$ correction from splitting between the $nn$ and $np$ spin-singlet scattering lengths. 
The $\ntlo$ (n=3) corrections are from higher order range corrections $c_{0t}^{(2)}$ and $c_{0s}^{(2)}$ in the $Z$-parametrization, 
and shape parameter corrections $c_{1t}$ and $c_{1s}$.}
\end{center}
\end{figure}

3) We also generalized the ERE corrections to the 
dibaryon propagators ~\cite{Griesshammer:2004pe}. We collect all the $n-$th order corrections to the dibaryon propagator into a circle with an ``$n$'' 
on the dibaryon double line Fig.~\ref{fig:TwoBodyContributions} (caption and figures are from \cite{Margaryan:2015rzg}). 

4) The last piece of new notation used here is the thick line, which denotes a sum over spin-triplet and 
spin-singlet leading order dibaryon propagators. This notation is used to make the picture more concise.

This coupled system of integral equations is not yet projected onto spin-angular momentum channels. 
The $nd$ scattering can happen in either spin-quartet or spin-doublet channel. After projecting this system of equations onto the spin-quartet channel we get only one 
integral equation instead of two coupled ones. The reason for this is that the spin-singlet dibaryon can combine with a spin-doublet nucleon only to form a total spin-doublet 
state, so the amplitudes involving the double dashed lines do not exist in the quartet channel and we are left with only one integral equation. After doing the spin-angular momentum projections
these equations can be put in the following general form:
\begin{align}\label{nthordereq}
&\mathbf{t}_{n;L'S',LS}^{J}(k,p,E)=\mathbf{K}_{n;L'S',LS}^{J}(k,p,E)\mathbf{v}_{p}+\sum_{i=1}^{n}\mathbf{t}_{n-i;L'S',LS}^{J}(k,p,E)\circ\mathbf{R}_{i}(p,E)\\\nonumber
&+\sum_{L'',S''}\sum_{i=0}^{n-1}\mathbf{K}_{n-i;L'S',L''S''}^{J}(q,p,E)\  \mathbf{D} \!\left(\! E-\frac{q^{2}}{2M_{N}},\vect{q}\right)\otimes\mathbf{t}_{i;L''S'',LS}^{J}(k,q,E)\\\nonumber
&+\sum_{L'',S''}\mathbf{K}_{0;L'S',L''S''}^{J}(q,p,E) \ \mathbf{D} \! \left(\! E-\frac{q^{2}}{2M_{N}},\vect{q}\right)\otimes\mathbf{t}_{n;L''S'',LS}^{J}(k,q,E),
\end{align}
where the $\otimes$ symbol is defined by:
\begin{equation}
A(q)\otimes B(q)=\frac{1}{2\pi^2} \int dq\ q^2 A(q)B(q).
\end{equation}
Note that this definition is different from Eq.~\eqref{otimessymbol}. This equation is written in cluster configuration space as were our equations for the doublet channel amplitudes given before in Eq.~\eqref{nddoubletchanneleq}, Eq.~\eqref{ndnlodoubletchanneleq}, Eq.~\eqref{ndn2lodoubletchanneleq}. Everything is collected into two dimensional vectors and matrices in cluster configuration space. 

$\mathbf{D}(E,\vect{q})$ is given by:
\begin{equation}
\label{eq:DibMatrix}
\mathbf{D}(E,\vect{q})=
\left(
\begin{array}{cc}
D_{t}(E,\vect{q}) & 0 \\
0&D_{s}(E,\vect{q})  
\end{array}\right),
\end{equation}
where $D_{t}$ and $D_{s}$ are the LO dibaryon propagators. 

\begin{figure}[hbt]
\begin{center}
\includegraphics[width=110mm]{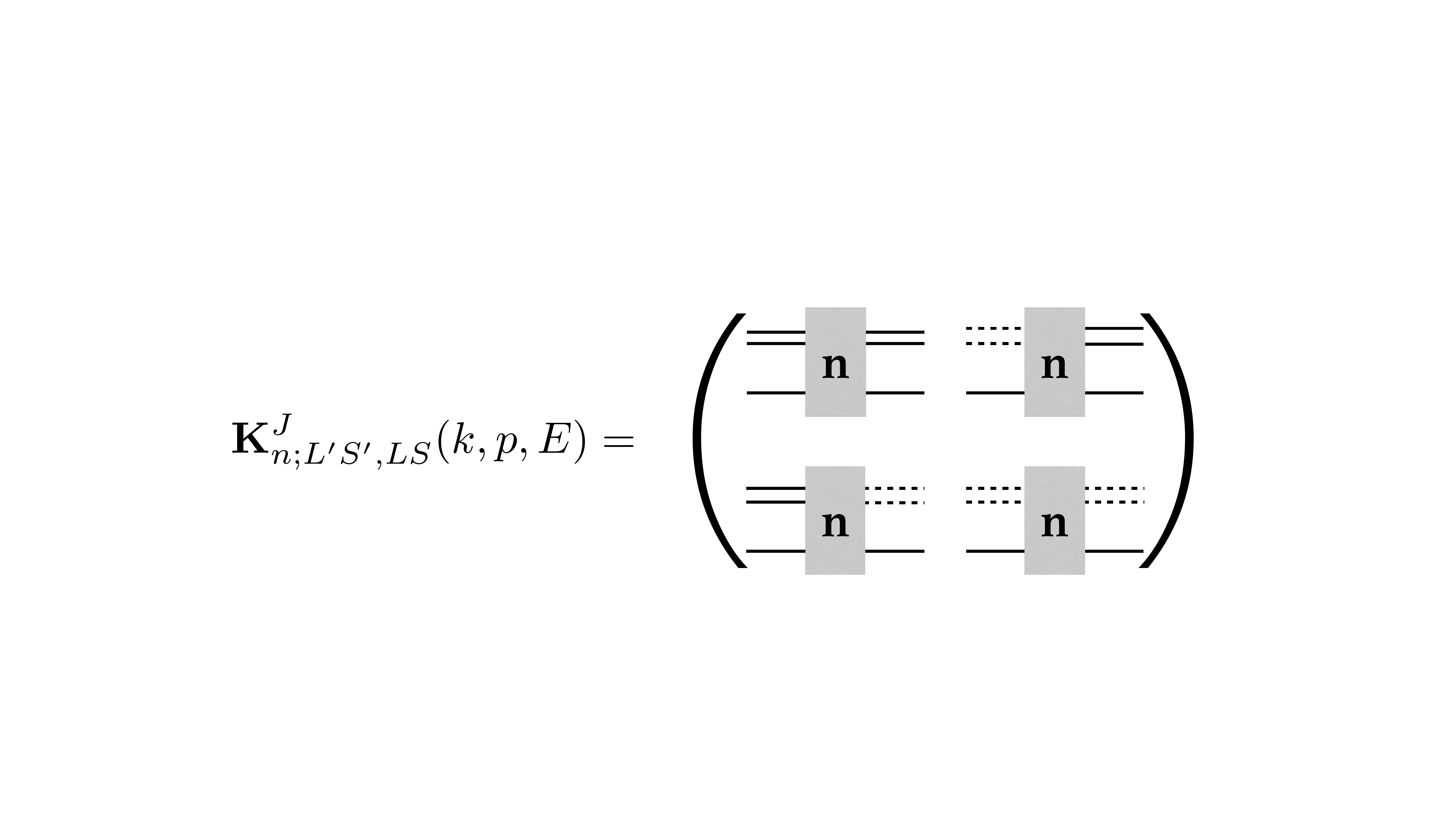}
\caption{\label{Kmatrix}Diagrammatic definition of $K$-matrices.}
\end{center}
\end{figure}

$\mathbf{t}^{J}_{n,L'S',LS}(k,p,E)$ is the $n$-th order correction to the scattering amplitude in cluster configuration space, with the first component being the
$Nt \rightarrow Nt$ scattering amplitude and the second component being the $Nt \rightarrow Ns$ scattering amplitude:
\begin{equation}
\hspace{-.85cm}\label{eq:tDef}
\mathbf{t}^{J}_{n,L'S',LS}(k,p,E)=\left(
\begin{array}{c}
t^{J;Nt\to Nt}_{n;L'S',LS}(k,p,E)\\[2mm]
t^{J;Nt\to Ns}_{n;L'S',LS}(k,p,E)
\end{array}\right).
\end{equation}
The $K$-matrices are defined by the boxes introduced in Fig.~\ref{fig:PertCorrectionDiagrams} as in Fig~\ref{Kmatrix}.

\pagebreak

\begin{figure}[hbt]
\begin{center}
\includegraphics[width=60mm]{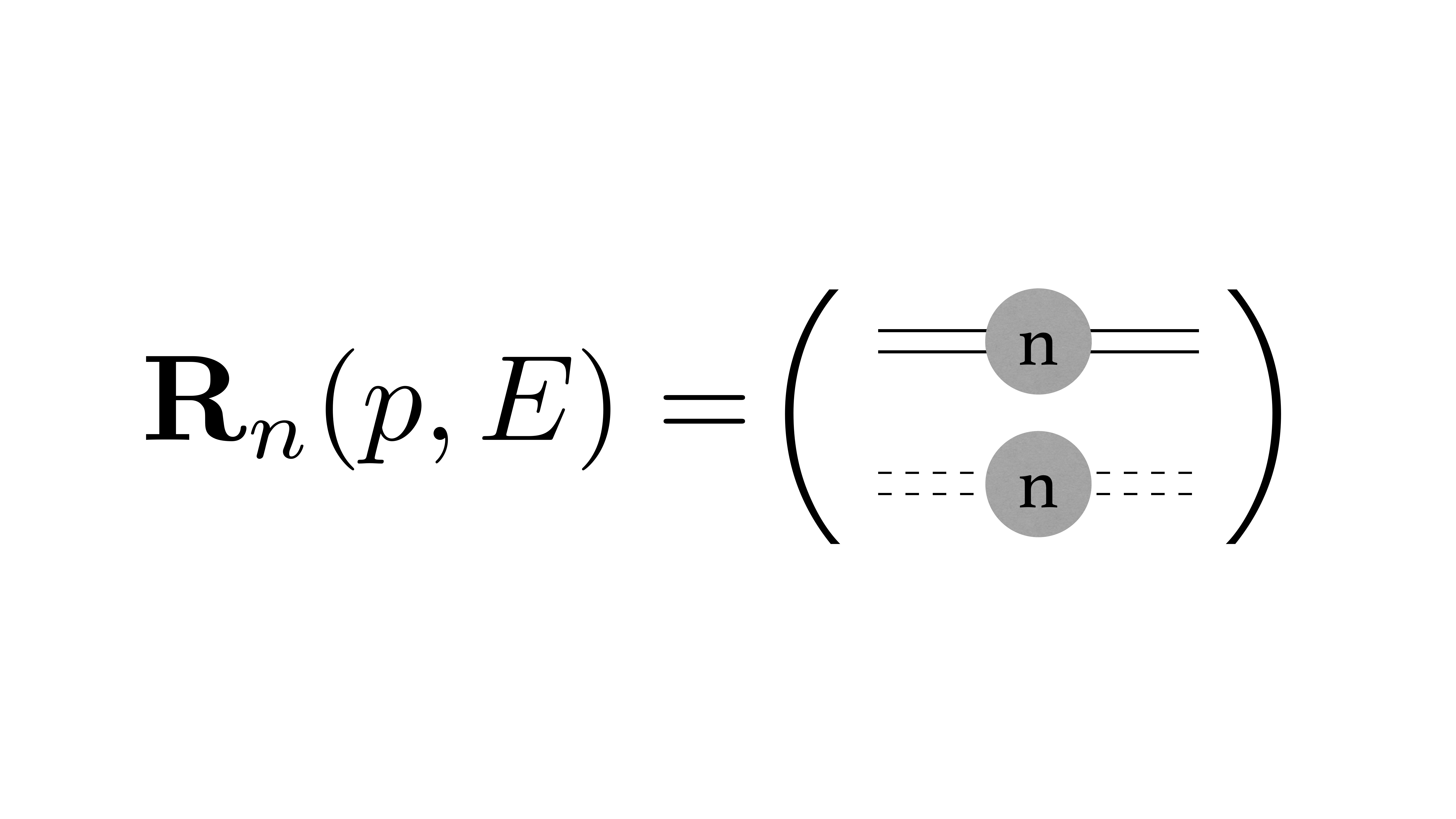}
\caption{\label{Rfunctions}Diagrammatic definition of $R$-functions.}
\end{center}
\end{figure}

Note that the actual kernel of these equations is only the matrix $\mathbf{K}^{J}_{0;L'S',LS}(q,p,E)$, which is the matrix of the rectangles with a ``$0$'' in them. 
The functions $\mathbf{R}_{n}(p,E)$ (in what follows I call those $R$-functions) are defined as two dimensional vectors in cluster configuration space with entries equal to the $n$-th order corrections to the dibaryon propagators ($n=1,2,3...$) Fig~\ref{Rfunctions}.

I have already described the explicit forms of these equations up to and including N$^2$LO. Here I will summarize all of that, 
talk about two-body $P$-wave interactions that enter at N$^3$LO and give the explicit forms of the equations satisfied by scattering amplitudes up to and including N$^3$LO. 

At LO we only have the LO $K$-matrix, which is the kernel for the integral equations at all orders:
\begin{align}
&\mathbf{K}^{J}_{0;L'S',LS}(q,p,E)=\\\nonumber
&\hspace{1cm}\delta_{LL'}\delta_{SS'}\left\{\begin{array}{cc}
-\frac{2\pi}{qp}Q_{L}\left(\frac{q^{2}+p^{2}-M_{N}E-i\epsilon}{qp}\right)\left(
\begin{array}{rr}
1 & -3 \\
-3 & 1
\end{array}\right)-\pi H_{\mathrm{LO}}\delta_{L0}
\left(\!\!\begin{array}{rr}
1 & -1 \\
-1 & 1 
\end{array}\!\right) & ,\ \ S=\nicefrac{1}{2}\\[8mm]
-\frac{4\pi}{qp}Q_{L}\left(\frac{q^{2}+p^{2}-M_{N}E-i\epsilon}{qp}\right)\left(
\begin{array}{cc}
1 & 0\\
0 & 0
\end{array}\right) & .\ \ S=\nicefrac{3}{2}\\
\end{array}\right.
\end{align}
At this order we don't have any $R$-functions.

At NLO the $K$-matrix is:
\begin{equation}
\hspace{-.5cm}\mathbf{K}^{J}_{1;L'S',LS}(q,p,E)=-\pi H_{\mathrm{NLO}}\delta_{L0}\delta_{LL'}\delta_{SS'}\delta_{S\nicefrac{1}{2}}
\left(\!\!\begin{array}{rr}
1 & -1 \\
-1 & 1 
\end{array}\!\right).
\end{equation}
$H_{\mathrm{NLO}}$ is the first order correction to the LO three-body force, and the $R$-function at this order is:
\begin{equation}
\mathbf{R}_{1}(p,E)=
\left(\begin{array}{c}
\frac{(Z_{t}-1)}{2\gamma_{t}}\left(\gamma_{t}+\sqrt{\frac{3}{4}p^{2}-M_{N}E-i\epsilon}\right)\\[3mm]
\frac{(Z_{s}-1)}{2\gamma_{s}}\left(\gamma_{s}+\sqrt{\frac{3}{4}p^{2}-M_{N}E-i\epsilon}\right)
\end{array}\right).
\end{equation}

At N$^2$LO the $K$-matrix is:
\begin{align}
&\hspace{-.5cm}\left[\mathbf{K}^{J}_{2;L'S',LS}(q,p,E)\right]_{zx}=\frac{y \, y_{SD} \, M_{N}}{2}\left(Z_{SD}^{(1)}(J,L',S',L,S,x,z)\frac{1}{kp}\left[4p^{2}Q_{L}(a)+k^{2}Q_{L'}(a)\right]\right.\\\nonumber
&+Z_{SD}^{(1)}(J,L,S,L',S',z,x)\frac{1}{kp}\left[p^{2}Q_{L}(a)+4k^{2}Q_{L'}(a)\right]\\\nonumber
&\left.+\sum_{L''}\left[Z_{SD}^{(2)}(J,L',S',L,S,x,z,L'')+Z_{SD}^{(2)}(J,L,S,L',S',z,x,L'')\right]Q_{L''}(a)\right)\\\nonumber
&-\pi (H_{\nnlo}+\frac{4}{3}(M_{N}E+\gamma_{t}^{2})H_{2}^{(\nnlo)})\delta_{L0}\delta_{LL'}\delta_{SS'}\delta_{S\nicefrac{1}{2}}(-1)^{x+z},
\end{align}
where the subscripts $x$ and $z$ determine the matrix element of the $K$-matrix in cluster configuration space. They run over the values zero and one, zero standing for spin-singlet dibaryon and one standing for the deuteron.

 Functions $Z_{SD}^{(1)}(\cdots)$ and $Z_{SD}^{(2)}(\cdots)$ are calculated using the Pauli-matrix structure of the $SD$ mixing diagrams. These functions calculate the spin-angular momentum projections of 
 the diagrams onto particular scattering channels. I will show an example calculation of similar projection functions for the $P$-wave contribution diagrams in Appendix \ref{Projections}. 
The functions $Z_{SD}^{(1)}(\cdots)$ and $Z_{SD}^{(2)}(\cdots)$ are defined by the equations:
\begin{align}
&Z_{SD}^{(1)}(J,L',S',L,S,x,z)=2\,\sqrt{\widehat{x}\widehat{z}\widehat{(1-z)}\widehat{S}\widehat{S'}\widehat{L}}\,\sqrt{\frac{10}{3}}\,(-1)^{\nicefrac{1}{2}+x+z+L+S+S'-J}\SJ{z}{\frac{1}{2}}{\frac{1}{2}}{1}{S'}{\frac{1}{2}}\\\nonumber
&\hspace{5cm}\times\SJ{2}{1}{x}{\frac{1}{2}}{S}{S'}\SJ{S'}{2}{S}{L}{J}{L'}\CG{L}{0}{2}{0}{L'}{0},
\end{align}
and
\begin{align}
&Z_{SD}^{(2)}(J,L',S',L,S,x,z,L'')=\sum_{L''}8\,\sqrt{\widehat{x}\widehat{z}\widehat{(1-z)}\widehat{S}\widehat{S'}\widehat{L}\widehat{L''}}\,(-1)^{z+L''+L}\SJ{1}{\frac{1}{2}}{\frac{1}{2}}{z}{S'}{\frac{1}{2}}\\\nonumber
&\times\left(\NJ{\frac{1}{2}}{x}{S}{1}{L''}{L}{S'}{L'}{J}+\SJ{\frac{1}{2}}{1}{S'}{L'}{J}{L''}\SJ{L}{x}{L''}{\frac{1}{2}}{J}{S}+\frac{1}{3}(-1)^{1+L''+L}\frac{1}{\widehat{S}\widehat{L}}\delta_{LL'}\delta_{SS'}\right)\CG{L}{0}{1}{0}{L''}{0}\CG{L''}{0}{1}{0}{L'}{0},
\end{align}
where the hat is defined by $\hat{x}=2x+1$.
The $R$-function at this order is:
\begin{equation}
\mathbf{R}_{2}(p,E)=-
\left(\begin{array}{cc}
\frac{(Z_{t}-1)^{2}}{2\gamma_{t}}\left(\gamma_{t}+\sqrt{\frac{3}{4}p^{2}-M_{N}E-i\epsilon}\right) &\\[3mm]
\frac{(Z_{s}-1)^{2}}{2\gamma_{s}}\left(\gamma_{s}+\sqrt{\frac{3}{4}p^{2}-M_{N}E-i\epsilon}\right) &-\,\, \frac{2}{3}\Delta_{s}^{(\nnlo)}D_{s} \! \left(E-\frac{p^{2}}{2M_{N}},p\right)
\end{array}\right) .
\end{equation}

At $\ntlo$ the $K$-matrix gets three different contributions: the $\ntlo$ correction to the LO three-body force, the NLO correction to the energy dependent three-body force and the $P$-wave diagram. 

\begin{equation}
\hspace{-.5cm}\mathbf{K}^{J}_{3;L'S',LS}(q,p,E)=\mathbf{K}_{3-bforces}+\mathbf{K}_{P-wave}
\end{equation}

The first term is given by:

\begin{equation}
\hspace{-.5cm}\mathbf{K}_{3-bforces}=-\pi\left(H_{\ntlo}+\frac{4}{3}(M_{N}E+\gamma_{t}^{2})H_{2}^{\ntlo}\right)\delta_{L0}\delta_{LL'}\delta_{SS'}\delta_{S\nicefrac{1}{2}}
\left(\!\!\begin{array}{rr}
1 & -1 \\
-1 & 1 
\end{array}\!\right),
\end{equation}

and the $\ntlo$ $R$-function is given by:
\begin{equation}
\mathbf{R}_{3}(p,E)=
\left(\begin{array}{c}
\left(\gamma_{t}+\sqrt{\frac{3}{4}p^{2}-M_{N}E-i\epsilon}\right)\left[\frac{(Z_{t}-1)^{3}}{2\gamma_{t}}+\rho_{t1}\left(\frac{3}{4}p^{2}-M_{N}E-\gamma_{t}^{2}\right)\right]\\[3mm]
\left(\gamma_{s}+\sqrt{\frac{3}{4}p^{2}-M_{N}E-i\epsilon}\right)\left[\frac{(Z_{s}-1)^{3}}{2\gamma_{s}}+\rho_{s1}\left(\frac{3}{4}p^{2}-M_{N}E-\gamma_{s}^{2}\right)\right]
\end{array}\right).
\end{equation}
The only thing that I have not talked about yet is the $\mathbf{K}_{P-wave}$ contribution to the $K$-matrix. I want to mention again that this is the most important piece that contributes to the $A_{y}$ observable up to $\ntlo$. 
The calculation of $\mathbf{K}_{P-wave}$ can be done in two different ways. First we can calculate directly the one-loop Feynman diagram given in Fig.~\ref{fig:N3LOndScatteringDiagrams}, 
in the box on the third line, first contribution to the rectangle with a ``$3$'' in it. The second way, suggested by Jared Vanasse, is to introduce a new set of auxiliary fields, which have quantum numbers corresponding 
to a dibaryon in which the two nucleons are in a relative $P$-wave state. This second way is completely equivalent to the direct calculation, but more preferable because it makes the coding easier. I will elaborate on this at the end of section \ref{secondway}.
Practically all it does is to change the order of some of the steps in the calculation. 

In what follows I describe both of the ways of calculating $\mathbf{K}_{P-wave}$. I note that we have done the third order calculation in both ways and gotten the same results for the amplitude.





















\section{First way of computing $P$-wave contribution: direct calculation}


The first way of calculating the $P$-wave contributions to the $nd$ scattering amplitude is to calculate directly the one-loop Feynman diagram given in Fig.~\ref{fig:N3LOndScatteringDiagrams}, 
in the box on the third line, first contribution to the rectangle with a ``$3$'' in it. Then we can do the spin-angular momentum projections and use that in the third order equations. 
Here I will only show the calculation of this diagram. The results of the spin projections are given in the appendices. 

 Denoting this diagram by $t$ we have:
\begin{eqnarray} \label{eqn:pwavecontributionamplitude}
&&64t= \int  \frac{d^3q}{(2\pi)^3}  \frac{1}{\frac{k^2}{4M}-\frac{\gamma^2}{M}-\frac{q^2}{2M}-\frac{(\vec{k}-\vec{q})^2}{2M}} \frac{1}{\frac{k^2}{4M}-\frac{\gamma^2}{M}+\epsilon-\frac{q^2}{2M}-\frac{(\vec{p}-\vec{q})^2}{2M}}\frac{1}{4}(2\vec{k}-\vec{q})_w (2\vec{p}-\vec{q})_x \nonumber \\ 
&\times& \bigg[ \sigma_y \sigma_i \sigma_j \sigma_z \bigg(C_2^{^3P_0}\delta^{xy}\delta^{wz}+C_2^{^3P_1}[\delta^{xw}\delta^{yz}-\delta^{xz}\delta^{yw}]+C_2^{^3P_2}[2\delta^{xw}\delta^{yz}+2\delta^{xz}\delta^{yw}-\frac{4}{3}\delta^{xy}\delta^{wz}]\bigg)  \nonumber \\
&+&\delta^{xw}C_2^{^1P_1}\sigma_i \sigma_j \bigg],        \nonumber \\
\end{eqnarray}
where $\epsilon=\frac{k^2-p^2}{2M_N}$.

After some notations and simplifications the integration part of this expression is reduced to three different types of integrals given by the Eqs.~\eqref{eqn:tensor}, \eqref{eqn:vector} and \eqref{eqn:scalar} respectively.
Introducing more notations given in Eqs.~\eqref{not1}, \eqref{not2} we get for $t$ Eq.~\eqref{t3}.

\begin{align} \label{t3}
&64t= Z_{wx}\Big{[}\sigma_y \sigma_i \sigma_j \sigma_z \Big(C_2^{^3P_0}\delta^{xy}\delta^{wz}+C_2^{^3P_1}[\delta^{xw}\delta{yz}-\delta^{xz}\delta^{yw}]  \nonumber \\    
&+C_2^{^3P_2}[2\delta^{xw}\delta^{yz}+2\delta^{xz}\delta^{yw}-\frac{4}{3}\delta^{xy}\delta^{wz}]\Big)+\delta^{xw}C_2^{^1P_1}\sigma_i \sigma_j\Big{]},             \nonumber \\                                       
\end{align}
where,
\begin{equation}\label{not1}
a^2=\gamma^2,
\end{equation}
\begin{equation}\label{not2}
b^2=\gamma^2+\frac{\vec{p}^2}{4}-\frac{\vec{k}^2}{4}-\epsilon M,
\end{equation}
which after simplifications becomes:

\begin{align}
&64t=   (C_2^{^3P_1}+2C_2^{^3P_2})Z^2\sigma_k \sigma_i \sigma_j \sigma_k+(2C_2^{^3P_2}-C_2^{^3P_1})Z_{kl}\sigma_k \sigma_i \sigma_j \sigma_l \nonumber \\   
&+(C_2^{^3P_0}-\frac{4}{3}C_2^{^3P_2})Z_{lk}\sigma_k \sigma_i \sigma_j \sigma_l+Z^2C_2^{^1P_1}\sigma_i \sigma_j,            \nonumber \\                                 
\end{align}
where $Z^2=Z_{kk}$ and:

\begin{equation} 
Z_{kl}=M^2(\frac{I_{kl}(\vec{k},\vec{p},a^2,b^2)}{4}-k_k \frac{I_{l}(\vec{k},\vec{p},a,b))}{2}-p_l \frac{I_{k}(\vec{k},\vec{p},a,b)}{2}+k_kp_lI(\vec{k},\vec{p},a,b))
\end{equation}
 Eqs.~\eqref{eqn:tensor}, \eqref{eqn:vector} and \eqref{eqn:scalar} are substituted into the previous equation giving the result:

\begin{eqnarray}
&&Z_{kl}/M^2= \frac{\pi}{3(2\pi)^3}\Lambda\delta_{kl}\nonumber \\
&-&\frac{\pi^2}{16(2\pi)^3}\delta_{kl} \left\{ a+b-\frac{4(a^2-b^2)^2}{(a+b)(\vec{k}-\vec{p})^2}+8\frac{\tan^{-1}(\alpha)}{|\vec{k}-\vec{p}|}\frac{(a^2-b^2)^2+(a^2+b^2)\frac{(\vec{k}-\vec{p})^2}{2}+\frac{(\vec{k}-\vec{p})^4}{16}}{(\vec{k}-\vec{p})^2} \right\} \nonumber \\
&-&\frac{\pi^2}{16(2\pi)^3}k_kk_l\Bigg\{ \frac{12(a^2-b^2)^2}{(a+b)(\vec{k}-\vec{p})^4}+\frac{13b-11a}{(\vec{k}-\vec{p})^2} \nonumber \\
&-& 8\frac{\tan^{-1}(\alpha)}{|\vec{k}-\vec{p}|}\frac{3(a^2-b^2)^2+(-5a^2+7b^2)\frac{(\vec{k}-\vec{p})^2}{2}+\frac{13(\vec{k}-\vec{p})^4}{16}}{(\vec{k}-\vec{p})^4} \Bigg\} \nonumber \\
&-&\frac{\pi^2}{16(2\pi)^3}p_kp_l\Bigg\{ \frac{12(b^2-a^2)^2}{(a+b)(\vec{k}-\vec{p})^4}+\frac{13a-11b}{(\vec{k}-\vec{p})^2} \nonumber \\
&-& 8\frac{\tan^{-1}(\alpha)}{|\vec{k}-\vec{p}|}\frac{3(a^2-b^2)^2+(-5b^2+7a^2)\frac{(\vec{k}-\vec{p})^2}{2}+\frac{13(\vec{k}-\vec{p})^4}{16}}{(\vec{k}-\vec{p})^4} \Bigg\} \nonumber \\ 
&+&\frac{\pi^2}{16(2\pi)^3}k_lp_k\left\{  \frac{12(a^2-b^2)^2}{(a+b)(\vec{k}-\vec{p})^4}+\frac{a+b}{(\vec{k}-\vec{p})^2}-8\frac{\tan^{-1}(\alpha)}{|\vec{k}-\vec{p}|}\frac{3(a^2-b^2)^2+(a^2+b^2)\frac{(\vec{k}-\vec{p})^2}{2}-\frac{(\vec{k}-\vec{p})^4}{16}}{(\vec{k}-\vec{p})^4} \right \}  \nonumber \\
&+&\frac{\pi^2}{16(2\pi)^3}k_kp_l\left\{  \frac{12(a^2-b^2)^2}{(a+b)(\vec{k}-\vec{p})^4}+\frac{a+b}{(\vec{k}-\vec{p})^2}-8\frac{\tan^{-1}(\alpha)}{|\vec{k}-\vec{p}|}\frac{3(a^2-b^2)^2+(a^2+b^2)\frac{(\vec{k}-\vec{p})^2}{2}-\frac{97(\vec{k}-\vec{p})^4}{16}}{(\vec{k}-\vec{p})^4} \right \}.  \nonumber \\
\end{eqnarray}
Defining:
\begin{equation}
Z_{kl}=A\delta_{kl}+Bk_kk_l+Cp_kp_l+Dk_lp_k+Ek_kp_l,
\end{equation}
and this gives:
\begin{equation}
Z^2=3A+Bk^2+Cp^2+(D+E)\vec{k} \cdot \vec{p}.
\end{equation}
Substituting into the expression for $t$ we get:
\begin{eqnarray} \label{oneloopdiagramamplitude}
&&64t=   (C_2^{^3P_1}+2C_2^{^3P_2})Z^2\sigma_k \sigma_i \sigma_j \sigma_k+(2C_2^{^3P_2}-C_2^{^3P_1})Z_{kl}\sigma_k \sigma_i \sigma_j \sigma_l\nonumber \\ 
&+&(C_2^{^3P_0}-\frac{4}{3}C_2^{^3P_2})Z_{lk}\sigma_k \sigma_i \sigma_j \sigma_l+Z^2C_2^{^1P_1}\sigma_i \sigma_j \nonumber \\                                            
&=&\left\{Z^2(C_2^{^3P_1}+2C_2^{^3P_2})+A(C_2^{^3P_0}-C_2^{^3P_1}+\frac{2}{3}C_2^{^3P_2})\right \}\sigma_k \sigma_i \sigma_j \sigma_k \nonumber \\       
&+&B(C_2^{^3P_0}-C_2^{^3P_1}+\frac{2}{3}C_2^{^3P_2})k_kk_l\sigma_k \sigma_i \sigma_j \sigma_l \nonumber \\ 
&+&C(C_2^{^3P_0}-C_2^{^3P_1}+\frac{2}{3}C_2^{^3P_2})p_kp_l\sigma_k \sigma_i \sigma_j \sigma_l \nonumber \\ 
&+&\left\{D(2C_2^{^3P_2}-C_2^{^3P_1})+E(C_2^{^3P_0}-\frac{4}{3}C_2^{^3P_2})\right \}k_lp_k\sigma_k \sigma_i \sigma_j \sigma_l \nonumber \\ 
&+&\left\{E(2C_2^{^3P_2}-C_2^{^3P_1})+D(C_2^{^3P_0}-\frac{4}{3}C_2^{^3P_2})\right \}k_kp_l\sigma_k \sigma_i \sigma_j \sigma_l \nonumber \\ 
&+& Z^2C_2^{^1P_1}\sigma_i \sigma_j.
\end{eqnarray}
In this expression the index $i$ is contracted with final deuteron polarization and the index $j$ is contracted with initial deuteron polarization. 
Appendix \ref{Projections} gives the results of spin-angular momentum and isospin projections of all the pieces of this expression.

\section{Second way of computing $P$-wave contribution: introducing P-wave auxiliary fields} \label{secondway}

Before we turn to the introduction of $P$-wave auxiliary fields let's separate the $P$-wave contribution to the full third order scattering amplitude from the equation that this amplitude satisfies. 
To do that consider the integral equation:

\begin{equation}
t(p)=f(p)+g(p)+\int K(p,q)t(q)dq,
\end{equation}
where $K(p,q)$ is the kernel of this equation and $f(p)$ and $g(p)$ are known functions. 
The sum $f(p)+g(p)$ is the inhomogeneous part of this integral equation. 

If we have a solution $t_f(p)$ to an integral equation with the same kernel, but with inhomogeneous part equal to $f(p)$ only,
and we have a solution $t_g(p)$ to an integral equation with the same kernel, but with inhomogeneous part equal to $g(p)$ only, 
then the sum $t_f(p)+t_g(p)$ is a solution for the original equation: indeed summing the two equations:

\begin{equation}
t_f(p)=f(p)+\int K(p,q)t_f(q)dq,
\end{equation}
\begin{equation}
t_g(p)=g(p)+\int K(p,q)t_g(q)dq,
\end{equation}
we see that the function $t(p)=t_f(p)+t_g(p)$ satisfies the original equation. Now returning to our third order equation (Eq.~\eqref{nthordereq} with $n=3$), we see that 
the three-body diagram with the two-body $P$-wave interactions contributes to only the inhomogeneous part of the equation and not the kernel. So we can take two 
integral equations with the same kernel but one with inhomogeneous part equal to everything else except the two-body $P$-wave contributions and another with only 
the two-body $P$-wave contributions. Here we only consider the second equation, the solution to which we call the ``$3_P$ amplitude'': 
$\mathbf{t}_{3_{\mathrm{P}};L'S',LS}^{J}(k,p,E)$.

Let's take a look at the 2-body $P$-wave interaction Lagrangian again:
\begin{align}
\label{eq:PwaveLagRupakreminder}
&\mathcal{L}_{2}^{{}^{3}P_{J}}=\left(C_{2}^{({}^3\!P_{0})}\delta_{xy}\delta_{wz}+C_{2}^{({}^3\!P_{1})}[\delta_{xw}\delta_{yz}-\delta_{xz}\delta_{yw}]+C_{2}^{({}^3\!P_{2})}\left[2\delta_{xw}\delta_{yz}+2\delta_{xz}\delta_{yw}-\frac{4}{3}\delta_{xy}\delta_{wz}\right]\right)\\\nonumber
&\hspace{3cm}\times\frac{1}{4}(\hat{N}^{T}\mathcal{O}_{xyA}^{(1,P)}\hat{N})^{\dagger}(\hat{N}^{T}\mathcal{O}_{wzA}^{(1,P)}\hat{N})
\end{align}
where
\begin{equation}
\mathcal{O}_{ijA}^{(1,P)}=\stackrel{\leftarrow}{\nabla}_{i}P_{jA}^{P}-P_{jA}^{P}\stackrel{\rightarrow}{\nabla}_{i}
\end{equation}
and the projector is defined as $P_{iA}^{P}=\frac{1}{\sqrt{8}}\sigma_{2}\sigma_{i}\tau_{2}\tau_{A}$.  
\begin{equation}
\label{eq:Pwave1P1reminder}
 \mathcal{L}_{2}^{{}^{1}\!P_{1}}=C_{2}^{({}^{1}\!P_{1})}\frac{1}{4}(\hat{N}^{T}\mathcal{O}_{x}^{(0,P)}\hat{N})^{\dagger}(\hat{N}^{T}\mathcal{O}_{x}^{(0,P)}\hat{N}),
\end{equation}
The operator $\mathcal{O}_{i}^{(0,P)}$ is defined by
\begin{equation}
\mathcal{O}_{i}^{(0,P)}=\stackrel{\leftarrow}{\nabla}_{i}P^{P}-P^{P}\stackrel{\rightarrow}{\nabla}_{i},
\end{equation}
where the projector is $P^{P}=\frac{1}{\sqrt{8}}\sigma_{2}\tau_{2}$.  

We can introduce a new set of auxiliary fields $\left(\Pc_{kA}^{\wave{3}{P}{J}}\right)$ through a new interaction Lagrangian given by:
\begin{align}
\label{eq:PwaveLag}
&\mathcal{L}_{2}^{P}=-{\Pc_{0A}^{\threePzero}}{}^{\dagger}\Delta^{(\threePzero)}\Pc_{0A}^{\threePzero}-{\Pc_{iA}^{\threePone}}{}^{\dagger}\Delta^{(\threePone)}\Pc_{iA}^{\threePone}-{\Pc_{iA}^{\threePtwo}}{}^{\dagger}\Delta^{(\threePtwo)}\Pc_{iA}^{\threePtwo}
-{\Pc_{i}^{\oneP}}{}^{\dagger}\Delta^{(\oneP)}\Pc_{i}^{\oneP}\\\nonumber
&\hspace{2cm}+\frac{1}{2}\sum_{J=0}^{2}y^{\wave{3}{P}{J}}\left[\CG{1}{i}{1}{j}{J}{k}\left(\Pc_{kA}^{\wave{3}{P}{J}}\right)^{\dagger}\hat{N}^{T}i\mathcal{O}_{jiA}^{(1,P)}\hat{N}+\mathrm{H.c.}\right]\\\nonumber
&\hspace{2cm}+\frac{1}{2}y^{\oneP}\left[\left(\Pc_{i}^{\oneP}\right)^{\dagger}\hat{N}^{T}i\mathcal{O}_{i}^{(0,P)}\hat{N}+\mathrm{H.c.}\right],
\end{align}
so that if we integrate out the $\left(\Pc_{kA}^{\wave{3}{P}{J}}\right)$ fields we will recover the original form of the $P$-wave interaction Lagrangian Eq.~\eqref{eq:PwaveLagRupakreminder} and Eq.~\eqref{eq:Pwave1P1reminder}. 
In Eq.~\eqref{eq:PwaveLag} all of the $\Delta^{(\wave{3}{P}{J})}$ coefficients are unphysical constants that are just chosen so that this Lagrangian matches onto the original $P$-wave interaction Lagrangian Eq.~\eqref{eq:PwaveLagRupakreminder} and Eq.~\eqref{eq:Pwave1P1reminder}. The indices $i,j,k$ are spherical indices, they take values $-1,0,1$. 

\begin{figure}[hbt]
\begin{center}
\includegraphics[width=60mm]{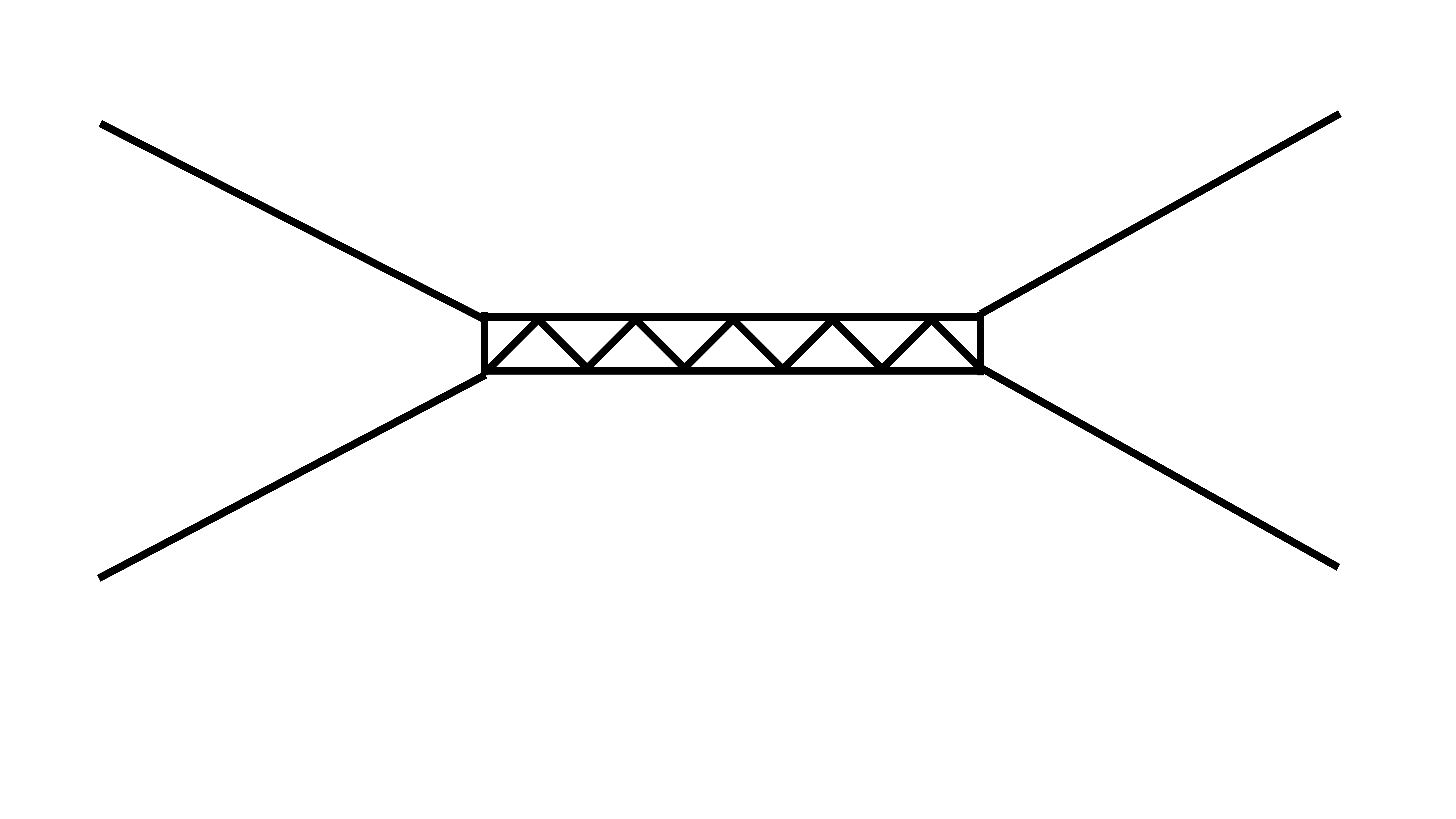}
\caption{\label{tbes}Two-body $P$-wave elastic scattering. The $\Pc_{kA}^{\wave{3}{P}{J}}$ field propagator is the double line with a zig-zag in it and is given by the constant $\frac{i}{\Delta^{\wave{3}{P}{J}}}$.} 
\end{center}
\end{figure}

To see the matching between the two Lagrangians at the two-body sector we can calculate the two-body elastic scattering with the new Lagrangian. The tree diagram 
that contributes to this scattering is given by the Fig.~\ref{tbes}.

Given two vectors $\vec{v}$ and $\vec{w}$ the dot product in terms of space indices is given by $\sum_i v_i w_i$, but when these vectors are given in terms of spherical indices
the same dot product is given by $\sum_i (-1)^i v_iw_{-i}$. Using this the diagram in Fig.~\ref{tbes} can be calculated to give (here I only do the matching for the $\wave{3}{P}{J}$ channel, 
the \wave{1}{P}{1} is analogous and easier):

\begin{equation}
\label{2ba}
it=\frac{(y^{\wave{3}{P}{J}})^2}{4\Delta^{\wave{3}{P}{J}}} \sum_{i,j.m.l.k;A}(-1)^k \CG{1}{i}{1}{j}{J}{k} \CG{1}{m}{1}{l}{J}{-k} (i\mathcal{O}_{jiA}^{(1,P)}) (i\mathcal{O}_{lmA}^{(1,P)})^\dagger
\end{equation}

In this expression the operators $\mathcal{O}$ are supposed to be calculated between nucleon spin states, but that part matches automatically, so I just skip it. 
It can be shown that going between spherical coordinates and cartesian coordinates the following substitutions can be made:

\begin{equation}
\CG{1}{i}{1}{j}{0}{0}\CG{1}{m}{1}{l}{0}{0}\to\frac{1}{3}\delta_{ij}\delta_{ml},
\end{equation}
\begin{equation}
\sum_{k}\CG{1}{i}{1}{j}{1}{k}\CG{1}{m}{1}{l}{1}{-k}(-1)^k\to\frac{1}{2}(\delta_{il}\delta_{jm}-\delta_{im}\delta_{jl}),
\end{equation}
\begin{equation}
\sum_{k}\CG{1}{i}{1}{j}{2}{k}\CG{1}{m}{1}{l}{2}{-k}(-1)^k\to\frac{1}{2}(\delta_{il}\delta_{jm}+\delta_{im}\delta_{jl}-\frac{2}{3}\delta_{ij}\delta_{ml})
\end{equation}
In here the coordinates on the left hand side are spherical and the ones on the right hand side are cartesian. Using these relations it becomes clear how Eq.~\eqref{2ba} matches 
onto the $\wave{3}{P}{J}$ interaction terms in the original Lagrangian Eq.~~\eqref{eq:PwaveLagRupakreminder} giving the following relations between the different parameters in 
the two Lagrangians:
\begin{equation}
C^{{}^{3}\!P_{0}}=\frac{1}{3}\frac{(y^{{}^{3}\!P_{0}})^{2}}{\Delta^{({}^{3}\!P_{0})}}, \ \
C^{{}^{3}\!P_{1}}=-\frac{1}{2}\frac{(y^{{}^{3}\!P_{1}})^{2}}{\Delta^{({}^{3}\!P_{1})}}, \ \
C^{{}^{3}\!P_{2}}=\frac{1}{4}\frac{(y^{{}^{3}\!P_{2}})^{2}}{\Delta^{({}^{3}\!P_{2})}}, \ \
C^{{}^{1}\!P_{1}}=\frac{(y^{{}^{1}\!P_{1}})^{2}}{\Delta^{({}^{1}\!P_{1})}}.
\end{equation}

\begin{figure}[hbt]
\begin{center}
\includegraphics[width=80mm]{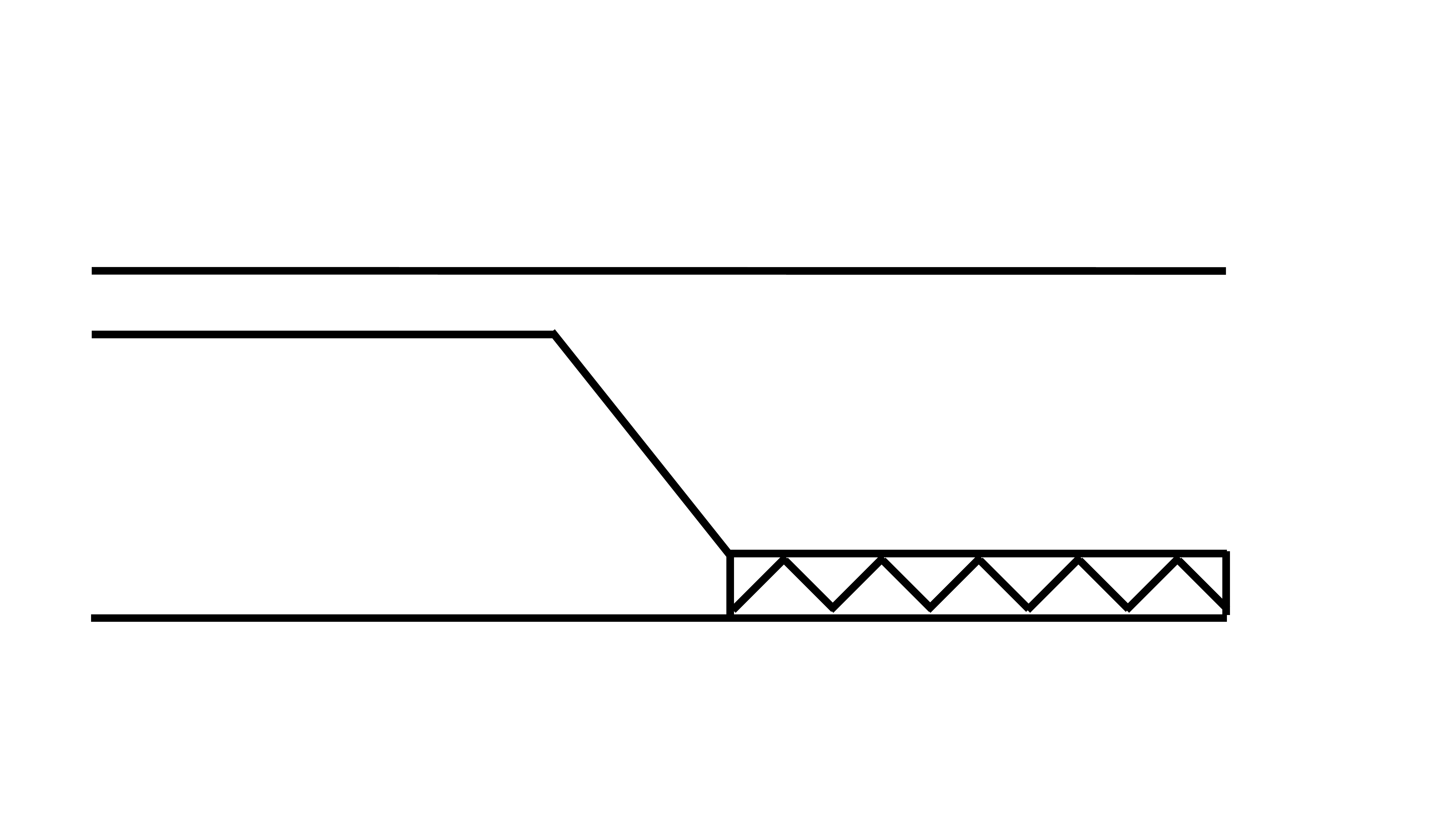}
\caption{\label{fig:treepwave}Tree-level contribution to $nd\rightarrow nP$-wave dibaryon.}
\end{center}
\end{figure}

At this point we can describe the second way of calculating the $3_P$ amplitude. Instead of calculating the one loop diagram as in the previous section we can just 
calculate the tree diagram given in the Fig~\ref{fig:treepwave} and do the spin-angular momentum projections for this diagram (In the figure I only give the diagram with the 
initial dibaryon being the deuteron, but the second diagram with initial dibaryon being the spin-singlet also has to be calculated). 

\begin{figure}[hbt]
\begin{center}
\includegraphics[width=100mm]{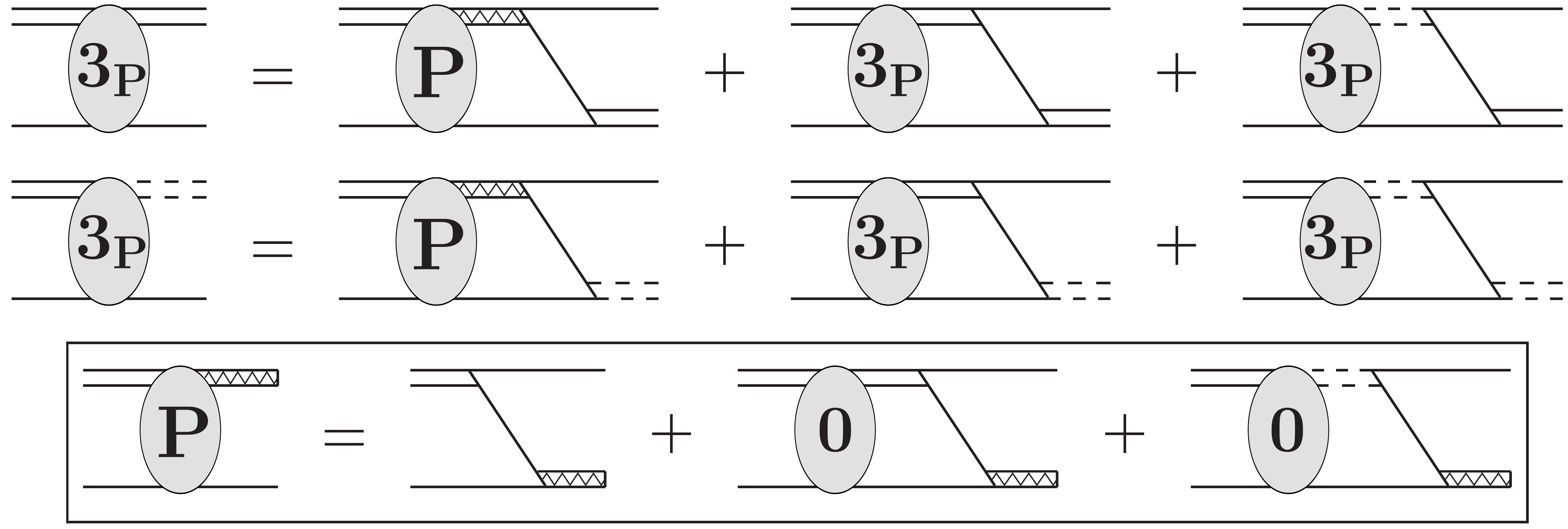}
\caption{\label{fig:TwoBodyPwave}(Caption and figures are from \cite{Margaryan:2015rzg}). Unboxed diagrams are the integral equations for the $\ntlo$ contribution to the $nd$ scattering amplitude from the two-body $P$-wave contact interactions. The double lines with a zig-zag in the middle are the $P$-wave dibaryon propagator, given by $i/\Delta^{\wave{2R+1}{P}{J}}$. The boxed diagrams represent the equation for the ``P'' amplitude used in the unboxed integral equations above.  The notation ``3$_{\rm P} $'' in the oval indicates that this is a $\ntlo$ correction but one that only involves the two-body $P$-wave contributions.}
\end{center}
\end{figure}

After this we can construct the $P$-wave amplitude as in the boxed area in Fig~\ref{fig:TwoBodyPwave}, which attached to the same diagram will give the 
inhomogeneous part of the equation that the  $3_P$ amplitude satisfies (the first terms of the right hand side of the first two lines of Fig~\ref{fig:TwoBodyPwave}).

Explicitly the $P$-wave amplitude, which is the boxed area in Fig~\ref{fig:TwoBodyPwave} is given by:
\begin{equation}
\label{eq:Pamp}
t^{J({}^{2R+1}\!P_{z})}_{L'S',LS}(k,p,E)=\left[\mathbf{K}^{J({}^{2R+1}\!P_{z})}_{L'S',LS}(k,p,E)\right]_{1}+\mathbf{K}^{J({}^{2R+1}\!P_{z})}_{L'S',LS}(q,p,E)\otimes \mathbf{t}_{0;LS,LS}^{J}(k,q,E),
\end{equation}
where the $K$-matrices are determined by the diagram in Fig~\ref{fig:treepwave} and are given by:
\begin{equation}
\left[\mathbf{K}_{L'S',LS}^{J({}^{3}\!P_{z})}(k,p,E)\right]_{x}=-\frac{M_{N} \, y \, y^{\wave{3}{P}{z}}}{4kp}\mathcal{Z}^{\wave{3}{P}{z}}\left(J,L',S',L,S,x,z\right)\left(2k \, Q_{L'}(a)+p \, Q_{L}(a)\right)
\end{equation}
\begin{equation}
\left[\mathbf{K}_{L'S',LS}^{J({}^{1}\!P_{1})}(k,p,E)\right]_{x}=-\frac{M_{N} \, y \, y^{\oneP}}{4kp}\mathcal{Z}^{\oneP}\left(J,L',S',L,S,x\right)\left(2k \, Q_{L'}(a)+p \, Q_{L}(a)\right)
\end{equation}
and the $Z$-functions are the spin-angular momentum projections of the diagram in Fig~\ref{fig:treepwave}, given by:
\begin{align}
&\mathcal{Z}^{\wave{3}{P}{z}}\left(J,L',S',L,S,x,z\right)=12(-1)^{\nicefrac{3}{2}+S+S'+L-J}\sqrt{\widehat{x}\widehat{(1-x)}\widehat{z}\widehat{S}\widehat{S'}\widehat{L}}\,\SixJ{x}{\nicefrac{1}{2}}{\nicefrac{1}{2}}{1}{S}{\nicefrac{1}{2}}\\\nonumber
&\hspace{1cm}\times\SixJ{\nicefrac{1}{2}}{1}{S}{1}{S'}{z}\SixJ{S}{1}{S'}{L'}{J}{L}\SixJ{1-x}{\nicefrac{1}{2}}{\nicefrac{1}{2}}{1}{\nicefrac{1}{2}}{\nicefrac{1}{2}}\CG{L}{0}{1}{0}{L'}{0},
\end{align}
and
\begin{align}
\mathcal{Z}^{\oneP}\left(J,L',S',L,S,x\right)=(-1)^{\nicefrac{3}{2}+L-J}\sqrt{\widehat{x}\widehat{(1-x)}\widehat{S'}\widehat{L}}\,\SixJ{S}{1}{S'}{L'}{J}{L}\delta_{S,\nicefrac{1}{2}}\CG{L}{0}{1}{0}{L'}{0}.
\end{align}
In terms of all of these newly defined objects the equation that the $3_P$ amplitude satisfies is:
\begin{align}
&\mathbf{t}_{3_{\mathrm{P}};L'S',LS}^{J}(k,p,E)=\\\nonumber
&\hspace{2cm}\sum_{R=0}^{1} \ \sum_{z=|R-1|}^{R+1} \ \sum_{L'',S''}\frac{(-1)^{z}}{\Delta^{({}^{2R+1}\!P_{z})}}\left[\mathbf{K}^{J({}^{2R+1}\!P_{z})}_{L''S'',L'S'}(p,q,E)\right]^{T}\otimes t_{L''S'',LS}^{J({}^{2R+1}\!P_{z})}(k,q,E)\\\nonumber
&\hspace{5cm}+\mathbf{K}^{J}_{0;L'S',L'S'}(q,p,E)\otimes \mathbf{t}_{3_{\mathrm{P}};L'S',LS}^{J}(k,q,E).
\end{align}
%
From how the new Lagrangian with auxiliary $P$ fields matches onto the original Lagrangian the equivalence of the two methods, described in the 
previous sections, for calculating the $P$-wave contributions to the three-body scattering amplitude becomes obvious. The only essential difference is that in the method of direct calculation we 
do the loop-integration analytically first, then we do the spin-angular momentum projections, but in the second method the order is changed. 
In the second method we do the spin-angular momentum projections first, then the loop-integration, but the loop-integration is done numerically. 
The second method proves to be preferable because in calculating the angular momentum projections we only see functions like the 
Legendre polynomials of the second kind, the analytical properties of which are well understood and which are already coded, 
because the same functions arise in the angular momentum projections of 
the lower order equations. As opposed to this, in the method of direct calculation, the loop-integration is done analytically, but in the angular 
momentum projections we get much more complicated functions then the Legendre polynomials of the second kind. With these new functions 
new numerical issues arise associated with the analytical behavior of these functions near the collinear limit where the vectors $\vec{k}$ and $\vec{p}$ become parallel. 
\section{Power Counting}
\label{powercountingsection}
\begin{center}\begin{figure}[ht]
  \begin{center}
  \includegraphics[scale=0.7]{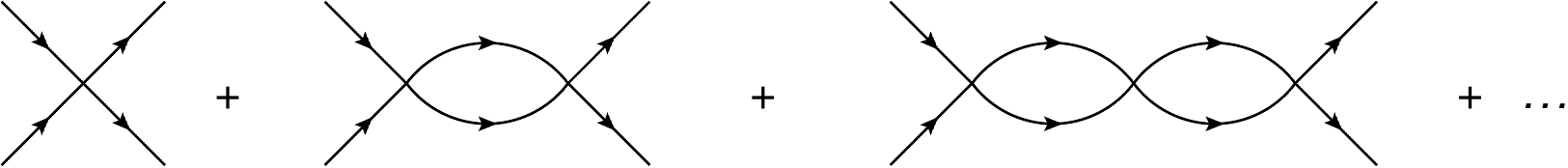}
  \end{center}
   \vspace{0.3cm} \caption{\label{bubbleNNpowercounting}$NN$ scattering: bubble sum.}
\end{figure}\end{center}

The power of an EFT is that when we calculate an observable using it, we are also able to make a prediction on the accuracy of the calculation. 
To construct an EFT in the first place we need to identify the relevant scales of the theory. Having done that we can form a small parameter in powers
of which we can organize the infinite number of interaction terms in the Lagrangian. Also all the observables are calculated order by order, 
as a power series expansion in powers of this small parameter. The \EFT is constructed to describe non-relativistic nucleons which interact 
through contact interactions, so pions are integrated out. From here we see that the small energy scale in this theory is the typical momentum
exchange in Feynman diagrams denoted by $Q$, and the large energy scales are the \EFT breaking scale $\Lambda_{\slashed{\pi}}$, which is of the order of the pion mass $\Lambda_{\slashed{\pi}}\sim m_{\pi}$ and the nucleon mass $M_N$. 
The binding momentum of the deuteron $\gamma_t=45.7025$ MeV is much smaller than the pion mass $m_{\pi} \approx 140$ MeV, so it is also considered to be a small energy scale $\sim Q$. 
All of the interactions in the Lagrangian and the observables are ordered in the powers of $Q$. For instance the more derivatives 
in an interaction term the higher order that term is in the Lagrangian. The scaling of an interaction in \EFT is not always easy to figure out, 
and it can depend on the regularization scheme used to render the loop-integrals finite \cite{Bedaque:1999vb,Griesshammer:2004pe}. Here I will talk about the scaling of some of the interactions and diagrams. 
Let's start with the only-nucleon-fields formulation of \EFT at the LO two-body sector:
\begin{equation}\label{two-nucleonsectoratleadingordertheLagrangedensity}
\mathcal{L}=N^{\dagger}\left(i\partial_0+\frac{\vec{\nabla}^2}{2M_N}\right)N-C_0(N^{T}PN)^{\dagger}(N^{T}PN).
\end{equation}
From this Lagrangian we see that in loops with nucleon propagators, if we denote the loop four momentum by $(l^0,\vec{l})$, the $l^0$ integration picks up a pole 
at a nucleon energy, so we have that $l^0$ integration scales like $\frac{Q^2}{M_N}$ and each component of the $\vec{l}$ integration scales like $Q$. Also the non-relativistic nucleon propagator scales like $\frac{M_N}{Q^2}$. 
Using naive dimensional analysis (NDA) the scaling of the interaction coefficient $C_0$ in Eq.~\eqref{two-nucleonsectoratleadingordertheLagrangedensity} can be estimated to be $C_0\sim \frac{1}{M_N\Lambda_{\slashed{\pi}}}$.
As it was shown in Section \ref{Two-nucleonsectoratLO} with this Lagrangian we can calculate the sum of all the diagrams 
contributing to the $NN$ scattering amplitude given in Fig.~\ref{bubbleNNpowercounting}. 
Using the scalings described above and the NDA estimate for $C_0$, we can see that each subsequent term in the series 
given in Fig.~\ref{bubbleNNpowercounting} is suppressed by a factor of $\frac{Q}{\Lambda_{\slashed{\pi}}}$. This makes each 
subsequent term in this series to be of higher order in \EFT power counting. If this was the correct scaling of the terms then, to do a calculation to a 
given order we would have to truncate the series. This would have been 
a valid conclusion if the $NN$ scattering length was of the natural or ``naive'' size $a\sim \frac{1}{\Lambda}$. 
But as we know this is not the case, in fact the scattering length is unnaturally large $a=\frac{1}{\gamma_t}\sim \frac{1}{Q}$. Hence we need to find 
the deuteron pole at the leading order in \EFT.
In order to generate the deuteron bound state at LO,
the appropriate power counting needs to find
 that all the terms in the series in Fig.~\ref{bubbleNNpowercounting} contribute at the same leading
 order.
This is impossible to achieve using NDA to estimate the size of $C_0$. To make the power counting correct PDS was invented \cite{Kaplan:1998tg}. In PDS regularization scheme for the calculation of the loop integrals in Fig.~\ref{bubbleNNpowercounting} the momentum cutoff $\Lambda$ is replaced by an unphysical arbitrary 
parameter $\mu$ and since it is arbitrary we can choose it to be of the order $Q$. With this choice for $\mu$, matching the 
calculated $NN$ scattering amplitude to the ERE result gives a different scaling for the interaction coefficient 
$C_0 \sim \frac{1}{M_NQ}$. Motivated by the need of getting the shallow deuteron pole at the LO the auxiliary field $t$ was 
introduced into the Lagrangian:
\begin{eqnarray}\label{pctwoblag}
&&\mathcal{L}= N^{\dagger}(i\partial_0+\frac{\vec{\partial}^2}{2M_N})N \nonumber \\
&+&t_{i}^{\dagger}\left(\Delta_{t}\right)t_{i} \nonumber \\
&+&y_t(t_{i}^{\dagger}(N^{T}P_{i}N)+h.c.). \nonumber \\
\end{eqnarray}
It was shown in \cite{Bedaque:1999vb} by integrating out the auxiliary field $t$ that this Lagrangian is physically equivalent to the one given in Eq.~\eqref{two-nucleonsectoratleadingordertheLagrangedensity}. 
The field $t$ is constructed such that it has the quantum numbers of a deuteron. From this Lagrangian we can find 
the following list for the scaling of different things.
\vspace{0.3cm}


$\gamma_t \sim Q$

dressed deuteron propagator $\sim \frac{1}{Q}$ (this is proven below)

the non-relativistic nucleon propagator $\sim \frac{M_N}{Q^2}$

as a convention we use $y_t^2 \sim \frac{1}{M_N}$

non-relativistic loop integration $\sim \frac{Q^5}{M_N}$

\vspace{1cm}

\begin{center}\begin{figure}[ht]
  \begin{center}
  \includegraphics[scale=0.15]{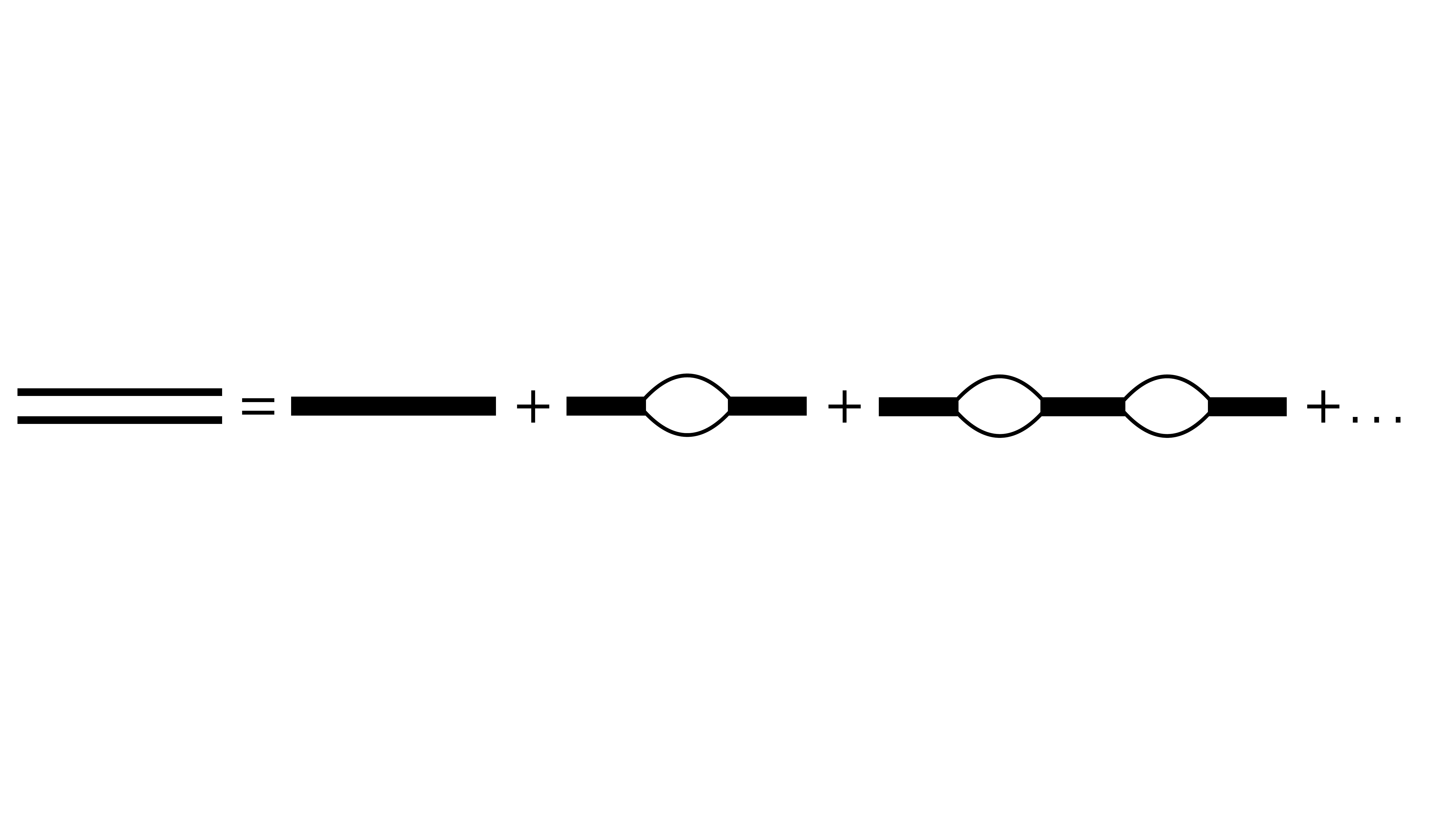}
  \end{center}
  \caption{\label{dp}Double line is the LO full deuteron propagator, thin solid line is the nucleon propagator, thick solid line is the bare deuteron propagator $\frac{i}{\Delta_{t}}$.}
\end{figure}\end{center}

From the Lagrangian in Eq.~\eqref{pctwoblag} the dressed deuteron propagator can be calculated as a geometric series given in 
Fig.~\ref{dp}.
Here again to regulate the loop integrals in Fig.~\ref{dp} PDS is used and the scaling of the parameter $\Delta_t$ is $\Delta_t \sim Q$. Using 
the list given above we can power count all the terms in the series in Fig.~\ref{dp}.
The first term scales as  $\sim \frac{1}{Q}$. The second term-$y_t^2*$(bare deuteron propagator)$^2*$(nucleon propagator)$^2*$ loop integration
$=\sim \frac{1}{M_N} \frac{1}{Q^2} \frac{M_N^2}{Q^4} \frac{Q^5}{M_N}=\frac{1}{Q}$.
From here it becomes clear that all the terms in this series contribute at the LO and scale as $\sim \frac{1}{Q}$ (which was the purpose of PDS) justifying the need to sum all the terms. 
Looking at the dressed deuteron propagator,
\begin{equation}
iD^{LO}_{t}(p_0,\vec{p})=\frac{i}{\gamma_t-\sqrt{\frac{\vec{p}^2}{4}-M_Np_0-i\epsilon}},
\end{equation}
we see that it also scales as $\sim \frac{1}{Q}$. This is the advantage of using PDS instead of cutoff with NDA. With PDS all the terms contributing to the dressed deuteron are of the same order (LO) as the final result. 

Using dimensional analysis, from the requirement that the action is dimensionless and from the nucleon kinetic term in Eq.~\eqref{pctwoblag} 
we see that the mass dimension of a nucleon field is $[N]=M^{\frac{3}{2}}$ 
(here I use the notation that the square brackets denote the mass dimension). From the convention we use for 
the $dNN$ coupling constant: $y_t^2=\frac{4\pi}{M_N}$ and from the third term in Eq.~\eqref{pctwoblag} we see that the 
mass dimension of a dibaryon field is the same as for the nucleon field $[t]=M^{\frac{3}{2}}$. From this and the 
second term in Eq.~\eqref{pctwoblag} we find the mass dimension of $\Delta_{t}$ to be $[\Delta_{t}]=M$.

From the $SD$-mixing Lagrangian:
\begin{equation}
\mathcal{L}^{SD}_{2}=y_{SD}\hat{t}_{i}^{\dagger}\left[\hat{N}^{T}\left((\stackrel{\rightarrow}{\partial}-\stackrel{\leftarrow}{\partial})^{i}(\stackrel{\rightarrow}{\partial}-\stackrel{\leftarrow}{\partial})^{j}-\frac{1}{3}\delta^{ij}(\stackrel{\rightarrow}{\partial}-\stackrel{\leftarrow}{\partial})^{2}\right)P_{j}\hat{N}\right]+\mathrm{H.c.},
\end{equation}
and the dimensions of the nucleon and deuteron fields we find the dimension and the scaling of the $SD$-mixing interaction coefficient $y_{SD}\sim \frac{1}{M_N^{\frac{5}{2}}}$. The same way from the two-body $P$-wave interaction 
Lagrangian:
\begin{align}
\label{rupakpowercounting}
&\mathcal{L}_{2}^{{}^{3}P_{J}}=\left(C_{2}^{({}^3\!P_{0})}\delta_{xy}\delta_{wz}+C_{2}^{({}^3\!P_{1})}[\delta_{xw}\delta_{yz}-\delta_{xz}\delta_{yw}]+C_{2}^{({}^3\!P_{2})}\left[2\delta_{xw}\delta_{yz}+2\delta_{xz}\delta_{yw}-\frac{4}{3}\delta_{xy}\delta_{wz}\right]\right)\\\nonumber
&\hspace{3cm}\times\frac{1}{4}(\hat{N}^{T}\mathcal{O}_{xyA}^{(1,P)}\hat{N})^{\dagger}(\hat{N}^{T}\mathcal{O}_{wzA}^{(1,P)}\hat{N})
\end{align}
we can find the dimensions of the interaction coefficients $C_{2}^{({}^3\!P_{J})}$. Each of the operators $\mathcal{O}$ in Eq.~\eqref{rupakpowercounting} has one derivative, which has mass dimension one. With this and the dimension of 
the nucleon fields we find for the dimension of the interaction coefficients $C_{2}^{({}^3\!P_{J})}\sim \frac{1}{M_N^4}$.

As we just showed NDA fails to give the correct scaling of the interaction coefficient $C_0$. In general NDA can not be trusted 
to give to correct scaling of operators in the Lagrangian or to give the correct scaling of the interaction coefficients. 
In order to find the correct scaling one needs to understand the renormalization conditions and use RG analysis. 

We can also use the Z-parametrization to compute the dressed deuteron propagator as described in section \ref{Twonucleonsector}. 
The deuteron propagator in the Z-parametrization \cite{Griesshammer:2004pe} up to N$^2$LO is given by:
 \begin{eqnarray}
&&iD_{t}(p_0,\vec{p})=\frac{i}{\gamma_t-\sqrt{\frac{\vec{p}^2}{4}-M_Np_0-i\epsilon}} \nonumber \\
&\times& \left[1+\frac{Z_t-1}{2\gamma_t}\left(\gamma_t+\sqrt{\frac{\vec{p}^2}{4}-M_Np_0-i\epsilon}\right)+\left(\frac{Z_t-1}{2\gamma_t}\right)^2\left(-\gamma_t^2+\frac{\vec{p}^2}{4}-M_Np_0\right)\right] \nonumber \\
\end{eqnarray}
In the Z-parametrization, since $\frac{Z_t-1}{2}$ is about $0.3$, $Z_t-1$ is also considered to be of order $Q$. Using that we see that each subsequent term is one order higher in powers of $Q$. 
So we conclude that the circle with an ``$n$'' in it on the dibaryon line (the $R$-functions) scales like $Q^n$.

\begin{center}\begin{figure}[ht]
  \begin{center}
  \includegraphics[scale=0.2]{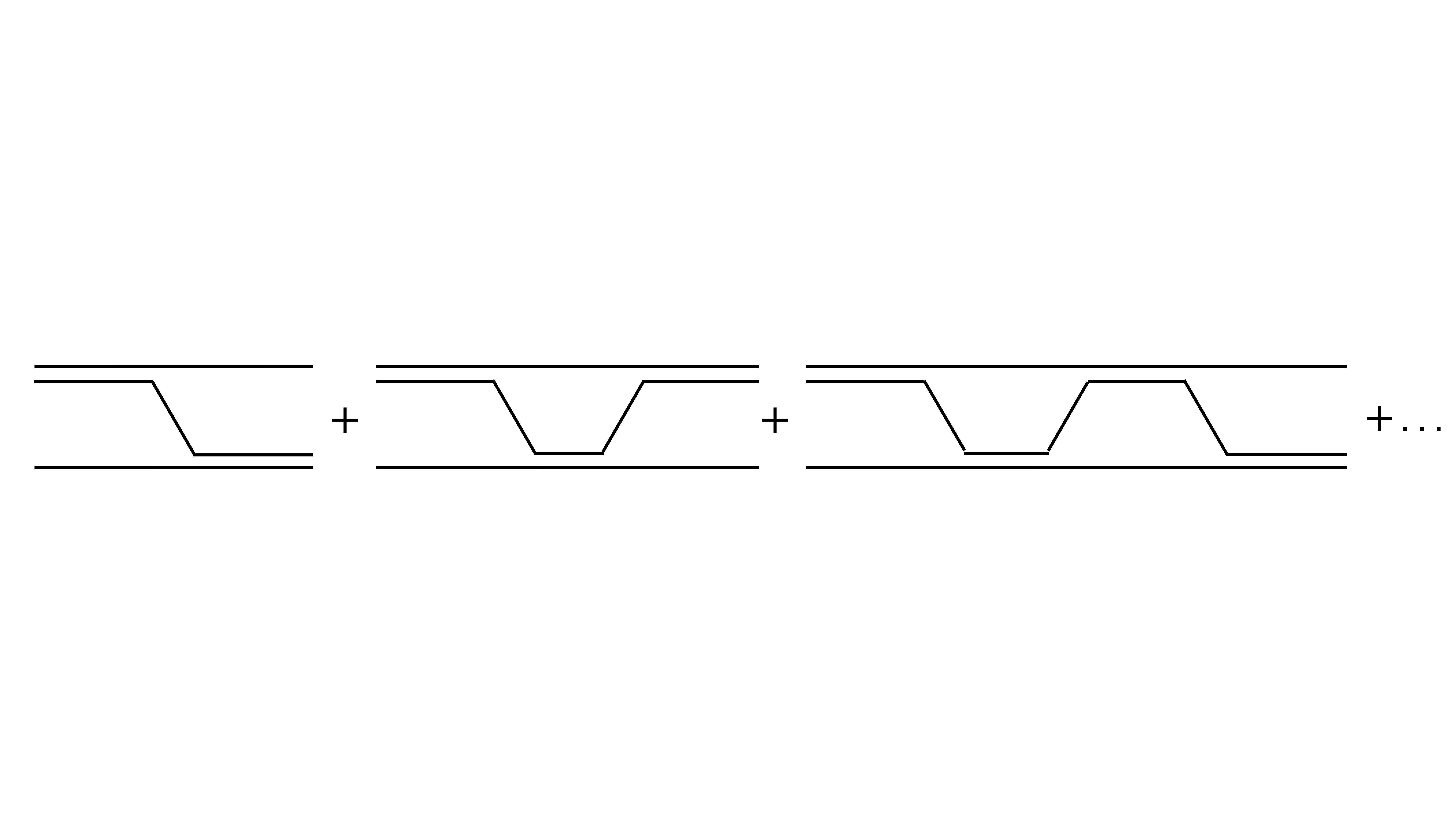}
  \end{center}
  \vspace{-2cm}
  \caption{\label{nd}Pinball diagrams for LO $nd$ scattering.}
\end{figure}\end{center}

To see how things scale at the three-body sector let's look at the diagrams contributing to the $nd$ scattering amplitude at the leading order Fig~\ref{nd}. 
The first diagram is just a tree and using our rules we get that it scales like $\sim\frac{1}{Q^2}$. In the second diagram we have one loop and 
three more propagators, it scales like $\sim (y_t)^4\frac{M_N^3}{Q^6} \frac{1}{Q} \frac{Q^5}{M_N}=\frac{1}{Q^2}$. Again we see that all the diagrams 
in the expansion are going to contribute to the same order, again justifying the need to sum all of them at the LO. From here naively we would assume 
that the LO $nd$ scattering amplitude scales like $\sim\frac{1}{Q^2}$. 

\begin{center}\begin{figure}[ht]
  \begin{center}
  \includegraphics[scale=0.15]{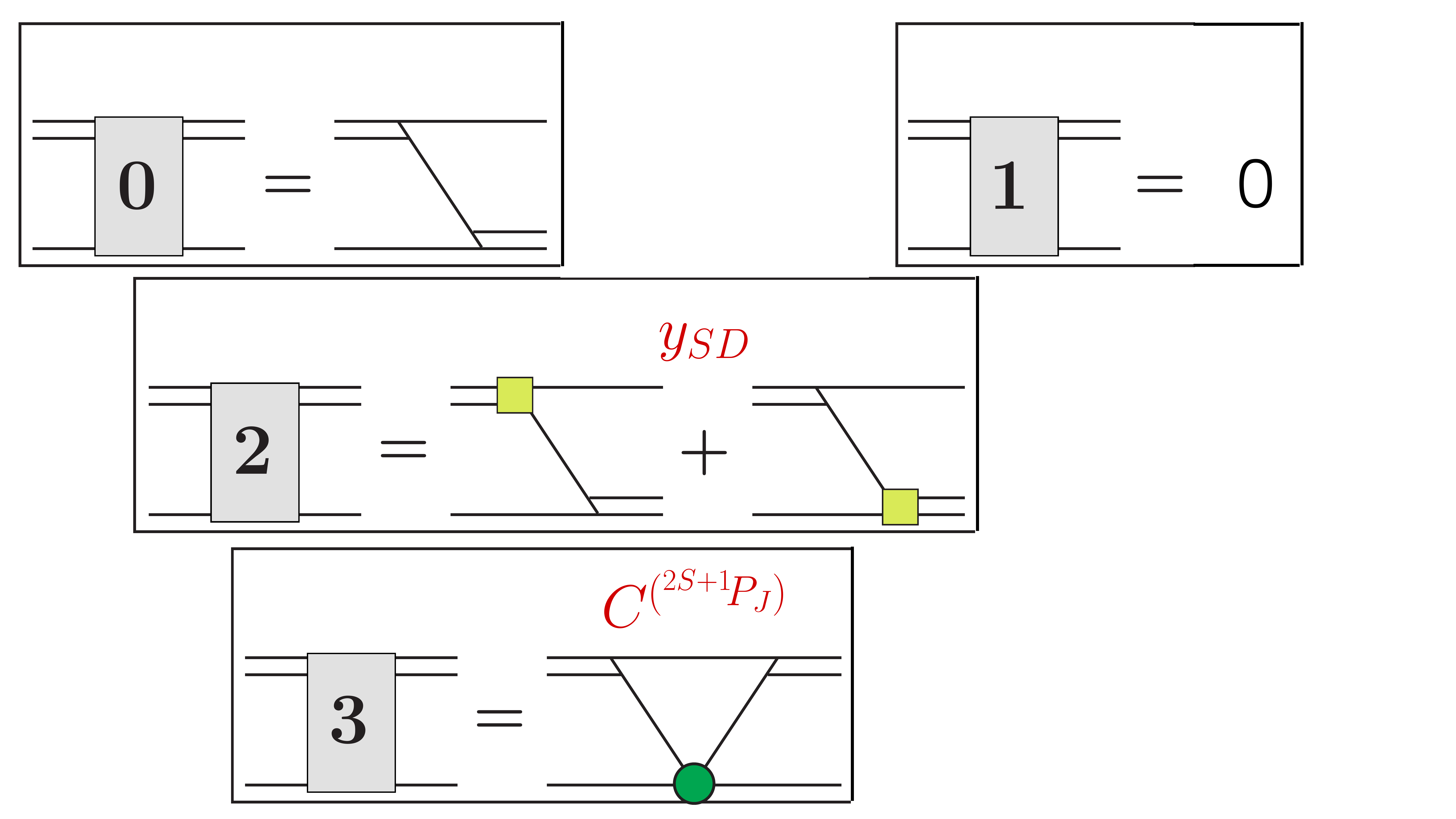}
  \end{center}
  \vspace{1cm}
  \caption{\label{kernelsno3bf}Three-body diagrams without three-body forces.}
\end{figure}\end{center}

Let's also look at the three-body contributions which previously were defined as the rectangles with an ``$n$'' in them, 
except without all of the three-body forces as in the Fig~\ref{kernelsno3bf}.

Here we have already figured out that the rectangle with the number ``0'' scales like $\sim\frac{1}{Q^2}$. 

The rectangle with ``1'' is zero, so we can think of it as scaling like $\sim\frac{1}{Q}$. 

The rectangle with ``2'' has the $SD$ mixing interaction, which has two derivatives, so it scales like $\sim y_{SD} Q^2 y_t\frac{M_N}{Q^2}$,
and as $(y_{SD} y_t)$ scales as $\sim\frac{1}{M_N^3}$, we get that this rectangle scales as $\sim \frac{1}{M_N^2}\sim \frac{1}{Q^2}\frac{Q^2}{M_N^2}$.

The rectangle with ``3'' has the $P$-wave interactions, which have two derivatives, so it scales like $\sim (y_t)^2 C^{{}^{3}\!P_{J}} Q^2 \frac{M_N^3}{Q^6}\frac{Q^5}{M_N}$, and the coefficient $C^{{}^{3}\!P_{J}}$ scales like $\frac{1}{M_N^4}$, giving for the rectangle scaling 
$\sim \frac{Q}{M_N^3}\sim \frac{1}{Q^2}\frac{Q^3}{M_N^3}$.

Generalizing the previous results we conclude that the rectangle with an ``$n$'' in it by construction scales like $\sim Q^{n-2}$ in powers of $Q$.

\begin{center}\begin{figure}[ht]
  \begin{center}
  \includegraphics[scale=0.15]{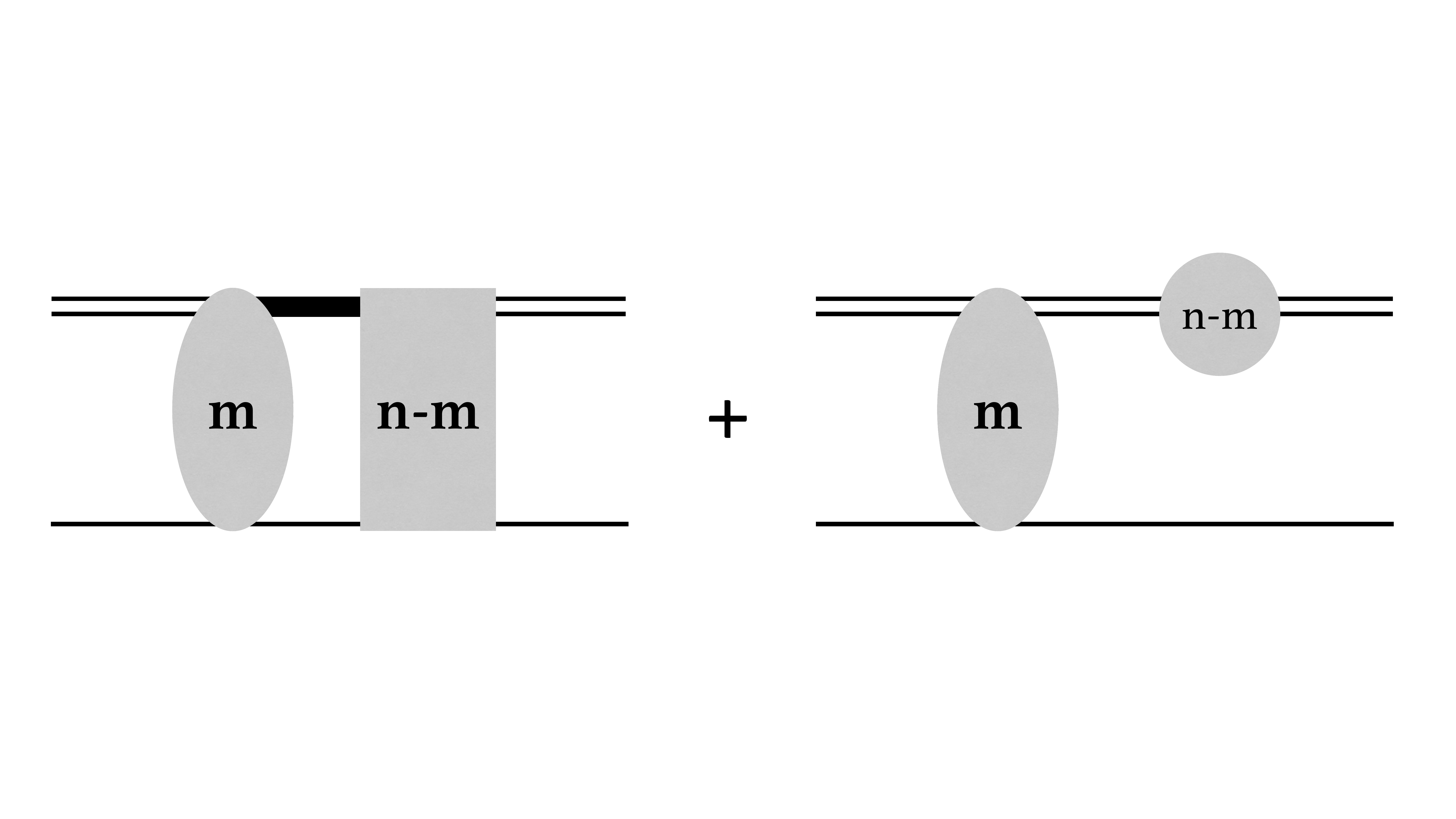}
  \end{center}
  \caption{\label{generalterm}General terms in $n$-th order equation.}
\end{figure}\end{center}

At this point we are ready to power count all the terms appearing on the right hand side of the $n$-th order equations given in the Fig~\ref{fig:PertCorrectionDiagrams}. 
As I already mentioned above with the naive approach to how things scale we conclude that the LO $nd$ scattering amplitude 
scales like each of the terms contributing to it, which is $\frac{1}{Q^2}$. Here I prove that with the same naive approach we will 
come to the conclusion that the N$^n$LO $nd$ scattering amplitude 
(the oval with an ``$n$'' in it, Fig~\ref{fig:PertCorrectionDiagrams}) scales like each of the terms contributing to it: $\sim Q^{n-2}$. To prove this we can induce on $n$;
assuming that everything works at up to and including N$^{n-1}$LO we can look at the two general types of diagrams contributing 
to the $n$-th order amplitude Fig~\ref{generalterm}. 

Using our rules for power counting we see that the first term scales like: $\sim m$-th order amplitude $*$ $(n-m)$-th order rectangle $*$
dibaryon propagator $*$ nucleon propagator $*$ loop integration $=Q^{m-2}Q^{n-m-2}\frac{1}{Q}\frac{M_N}{Q^2} \frac{Q^5}{M_N}=Q^{n-2}$.
And the second term: $\sim Q^{m-2} Q^{n-m}=Q^{n-2}$. Now assuming that this amplitude also scales as all the terms contributing to it 
we find that it scales like $\sim Q^{n-2}$, which is the naive conclusion. 

In reality though things are not so simple. At each order we have infinite number of diagrams contributing to the amplitude. 
Although each one of these contributions scales the same, there is no guarantee that the total amplitude will scale the same. One of the 
most important examples when this analysis fails is the LO doublet channel $S$-wave amplitude. According to the analysis this 
amplitude should be well-behaved and scale like $\frac{1}{Q^2}$, but as it is shown it has strong cutoff dependence. To regulate this 
cutoff dependence we are forced to include a three-body contact interaction at LO, which is naturally expected to be included 
at N$^2$LO \cite{Bedaque:1999ve,Ji:2012nj,Griesshammer:2005ga,Bedaque:1998km}. To see this let's 
look at the interaction Lagrangian for the three-body force again:

\begin{equation}\label{3bfpcsection}
\mathcal{L}=\frac{M_NH_0(\Lambda)}{3\Lambda^2}\left(  y_t N^{\dagger} (\vec{t} \cdot \vec{\sigma})^{\dagger}-y_s N^{\dagger} (\vec{s} \cdot \vec{\tau})^{\dagger} \right)
\left(  y_t N (\vec{t} \cdot \vec{\sigma})-y_s N (\vec{s} \cdot \vec{\tau})\right).
\end{equation}
We can easily find the scaling of the contribution of this term to the three-body scattering amplitude. 
This contribution is just a tree diagram and it scales as $\sim \frac{1}{\Lambda^2}=\frac{1}{Q^2}\frac{Q^2}{\Lambda^2}$ (here $\Lambda$ is the momentum cutoff in three-body loop diagrams). From the analysis of the terms in Fig~\ref{nd} we know that the LO contribution to the three-body scattering amplitude scales as 
$\frac{1}{Q^2}$. From here we find that using NDA we would conclude that the three-body force contribution comes at N$^2$LO, which is the wrong conclusion as we found in Section \ref{Three-nucleon sector at LO, doublet channel: S-wave}.

\chapter{Results and conclusions}
\label{Result}
\section{Observables}
\begin{center}\begin{figure}[ht]
  \begin{center}
  \includegraphics[scale=0.15]{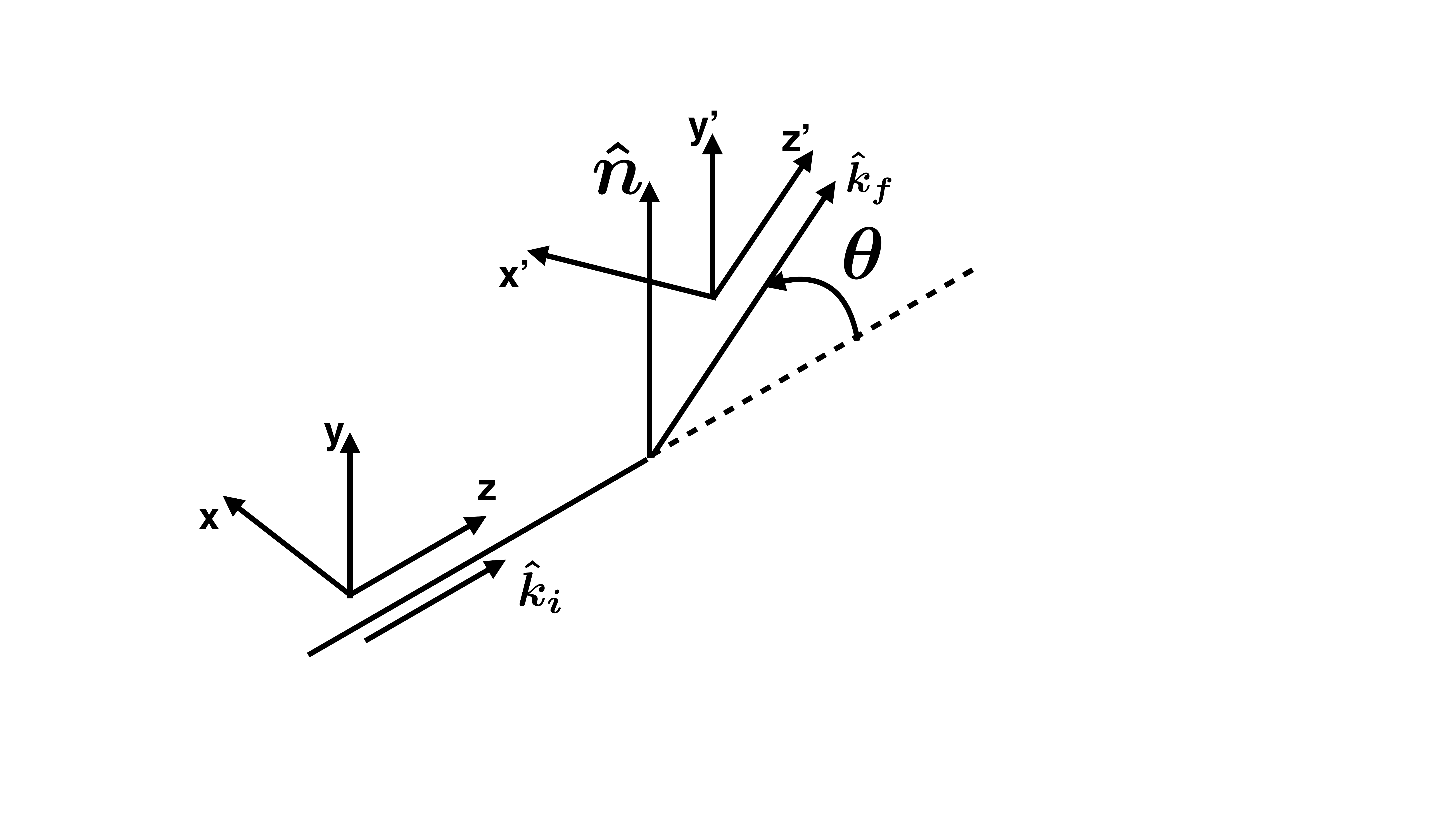}
  \end{center}
  \caption{\label{coordinate}Coordinate systems.}
\end{figure}\end{center}

In $nd$ scattering there are several observables related to initial beam and target polarizations. Let's take 
for example initial polarized neutron beam and unpolarized deuteron target. We need to define a coordinate system 
to describe the observables. We define two coordinate systems (see Fig.~\ref{coordinate}) in the laboratory reference frame; one attached to the initial neutron ($x$, $y$, $z$) and 
the other attached to final (scattered) neutron ($x'$, $y'$, $z'$).
In the laboratory frame denote the initial neutron momentum as $\vec{k}_i$ and the final neutron momentum 
as $\vec{k}_f$. The $z$ axis is defined along $\vec{k}_i$, the $y$ axis is defined along the cross 
product $\vec{k}_i \cross \vec{k}_f$ and the $x$ axis is defined so that ($x$, $y$, $z$) is right handed. The same way 
$z'$ axis is defined along $\vec{k}_f$, the $y'$ axis is defined along the cross product $\vec{k}_i \cross \vec{k}_f$ 
and the $x'$ axis is defined so that \\ ($x'$, $y'$, $z'$) is right handed.

We can denote the spinor that describes the initial nucleon by $\chi_i$ and the final nucleon by $\chi_f$. 
The transition matrices $M$ are related to the amplitudes that we calculate solving the integral equations numerically. These matrices depend on the incoming energy, the scattering angle $\theta$, also they
have indices corresponding to the initial and the final deuteron spin components and the initial and final neutron spin components. As, for now we are only 
considering unpolarized deuteron target the indices that correspond to the deuterons are going to be summed over, so for now we suppress those indices 
to make the notations more concise. The transition matrix for this process is $M_{m_i m_f}$, where $m_i$ and $m_f$ are the initial and final neutron 
spin components. This matrix connects the initial and final neutron spinors so we have:

\begin{align} \label{transitionM}
\chi_f=M\chi_i.
\end{align}
If we have a beam of $N$ neutrons in the initial state then the density matrix that describes it can be written as:
\begin{align} \label{densityi}
\rho_i=\sum_{n=1}^{N}\chi^{(n)}_i \left[\chi^{(n)}_i\right]^{\dagger},
\end{align}
and the same way for the final state we have:
\begin{align} \label{densityf}
\rho_f=\sum_{n=1}^{N}\chi^{(n)}_f \left[\chi^{(n)}_f\right]^{\dagger},
\end{align}
where the superscript $(n)$ denotes the particle number. 
Putting the Eqs.~\eqref{transitionM}, \eqref{densityi}, \eqref{densityf} together it is easy to see:
\begin{align} \label{}
\rho_f=M\rho_iM^{\dagger}.
\end{align}
Denoting the components of a spinor:
\begin{align} \label{}
\chi^{(n)}= \left(  \begin{matrix} 
      a_1^{(n)}  \\
      a_2^{(n)} \\
   \end{matrix} \right)
\end{align}
and using the definition for the density matrix we find:
\begin{align} \label{densitycomponents}
\rho= \left( \begin{matrix} 
      \sum\limits_{n=1}^{N} |a_1^{(n)}|^2 & \sum\limits_{n=1}^{N}a_1^{(n)} a_2^{(n)*} \\
      \sum\limits_{n=1}^{N} a_2^{(n)} a_1^{(n)*} & \sum\limits_{n=1}^{N} |a_2^{(n)}|^2  \\
   \end{matrix}
\right).
\end{align}
The total averaged spin polarization components of the beam can be calculated using the Pauli matrices and the density matrix Eq.~\eqref{densitycomponents} finding the expressions in Eq.~\eqref{polcomp}.
\begin{align} \label{polcomp}
p_x=\Tr (\rho\sigma_x)=\sum\limits_{n=1}^{N}2\Re(a_1^{(n)} a_2^{(n)*}), \nonumber\\
p_y=\Tr (\rho\sigma_y)=\sum\limits_{n=1}^{N}2\Im(a_1^{(n)} a_2^{(n)*}),\nonumber\\
p_z=\Tr (\rho\sigma_z)=\sum\limits_{n=1}^{N} (|a_1^{(n)}|^2-|a_2^{(n)}|^2).\nonumber\\
\end{align}
These spin polarization components components correspond to the initial beam as a whole. To find 
the average spin polarization per particle we need to divide the three lines of the Eq.~\eqref{polcomp} by $N$.
From Eq.~\eqref{polcomp} we see that if the initial beam is not polarized then from the requirements $p_x=p_y=p_z=0$ we find that 
the density matrix describing that beam will be proportional to the $2\cross2$ identity matrix $I$, Eq.~\eqref{densityidentity}.
\begin{align} \label{densityidentity}
\rho_i= \frac{N}{2} I=\frac{N}{2} \left( \begin{matrix} 
      1 & 0\\
      0 & 1  \\
   \end{matrix}
\right).
\end{align}
Here we assumed that the spinors $\chi_i^{(n)}$ are normalized to one: $|a_1^{(n)}|^2+|a_2^{(n)}|^2=1$.

As the density matrix is a hermitian operator and as the identity matrix together with the Pauli matrices span the 
space of the $2\cross2$ hermitian matrices, we can express the density matrix in terms of the identity matrix and the Pauli matrices. Denoting the $2\cross2$ identity matrix as $\sigma_0$ we have  Eq.~\eqref{densitylinear}.
\begin{align} \label{densitylinear}
\rho=\sum_{j=0}^{3}a_j\sigma_j
\end{align}
To find the coefficients $a_j$ in Eq.~\eqref{densitylinear} note that the matrices $\sigma_j$ with $j=0,1,2,3$, satisfy 
an orthogonality relation with a scalar product defined as the trace of the product (see Eq.~\eqref{orth}).
\begin{align} \label{orth}
\Tr(\sigma_i\sigma_j)=2\delta_{ij}. 
\end{align}
From Eq.~\eqref{densitylinear} and Eq.~\eqref{orth} we find:

\begin{align} \label{}
a_j=\frac{1}{2}\Tr(\rho\sigma_j)=\frac{1}{2}p_j, \ j=1,2,3
\end{align}
where $p_j$ is defined for $j=1,2,3$ in Eq.~\eqref{polcomp} and for $j=0$ we have:
 \begin{align} \label{}
a_0=\frac{1}{2}\Tr(\rho\sigma_0)=\frac{1}{2}\Tr(\rho)=\frac{N}{2}.
\end{align}
For the density matrix we find:
\begin{align} \label{densityfinal}
\rho=\frac{1}{2}(N\sigma_0+ \sum_{j=1}^{3}p_j\sigma_j).
\end{align}
Again we see that if the beam is not polarized we recover Eq.~\eqref{densityidentity} for the density matrix. 

In terms of the density matrices the total cross section of the scattering of a polarized beam on an unpolarized target is given by (noting that $\Tr(\rho_i)=N$):
\begin{align} \label{polcross}
\frac{d\sigma}{d\Omega}(\theta,\phi)=\frac{\Tr(\rho_f)}{\Tr(\rho_i)}=\frac{\Tr(M\rho_iM^{\dagger})}{N},
\end{align}
and for the completely unpolarized scattering, for which the initial beam is non-polarized and hence $\rho_i$ is given by Eq.~\eqref{densityidentity}, the cross section Eq.~\eqref{polcross} reduces to:
\begin{align} \label{unpolcross}
\frac{d\bar{\sigma}}{d\Omega}(\theta)=\frac{\Tr(\rho_f)}{\Tr(\rho_i)}=\frac{1}{2}\Tr(MM^{\dagger}),
\end{align}
where the bar over the $\sigma$ indicates that this is the unpolarized cross section.
Using the expansion Eq.~\eqref{densityfinal} for the initial density matrix the polarized cross section 
Eq.~\eqref{polcross} can be put into the following form: 
\begin{align} \label{polcrossexpand}
\frac{d\sigma}{d\Omega}(\theta,\phi)=\frac{d\bar{\sigma}}{d\Omega}(\theta)(1+\sum_{j=1}^{3}\frac{p_j}{N}A_j(\theta)),
\end{align}
where the $A_j(\theta)$ are the analyzing powers and are given by:
 \begin{align} \label{Ajformula}
 A_j(\theta)=\frac{\Tr(M\sigma_jM^{\dagger})}{\Tr(MM^{\dagger})}.
\end{align}
If we have an initial beam that is polarized along the $y$ axis, 
then in Eq.~\eqref{polcrossexpand} we only have the term proportional to $p_y$:
\begin{align} \label{polcrossexpandy}
\frac{d\sigma}{d\Omega}(\theta,\phi)=\frac{d\bar{\sigma}}{d\Omega}(\theta)(1+\frac{p_y}{N}A_y(\theta)),
\end{align}
where the fraction $\frac{p_y}{N}$ is the average polarization per particle. 
The function $A_y(\theta)$ is the $A_y$ observable that is called the transverse asymmetry.
To give it the interpretation of an ``asymmetry'' note that from rotational symmetry the unpolarized 
cross section has to be an even function of $\theta$: $\frac{d\bar{\sigma}}{d\Omega}(-\theta)=\frac{d\bar{\sigma}}{d\Omega}(\theta)$ and $A_y(\theta)$ has to be an odd function of $\theta$: $A_y(-\theta)=-A_y(\theta)$. 
Taking all the particles in the initial beam polarized along the $y$ axis we can substitute $\frac{p_y}{N}=1$ in Eq.~\eqref{polcrossexpandy}. 
Writing this equation for $\pm\theta$, then adding and subtracting from each other we get:

\begin{align} \label{}
\frac{d\sigma}{d\Omega}(\theta,\phi)+\frac{d\sigma}{d\Omega}(-\theta,\phi)=2\frac{d\bar{\sigma}}{d\Omega}(\theta)\end{align}
and
\begin{align} \label{Ayinterpret}
A_y(\theta)=\frac{\frac{d\sigma}{d\Omega}(\theta,\phi)-\frac{d\sigma}{d\Omega}(-\theta,\phi)}{\frac{d\sigma}{d\Omega}(\theta,\phi)+\frac{d\sigma}{d\Omega}(-\theta,\phi)}.
\end{align}
Using Eq.~\eqref{Ayinterpret} we interpret $A_y$ as the asymmetry between the number of the left- and right-scattered particles. We also see that the denominator in Eq.~\eqref{Ayinterpret} is the total unpolarized cross section. For practical calculations of $A_y$ though we use  Eq.~\eqref{Ajformula} with $j=2$. 
The Pauli matrix $\sigma_2$ is given by:
\begin{align}\label{Pauli2}
 \left( \begin{matrix} 
      0 & -i\\
      i & 0  \\
   \end{matrix}
\right).
\end{align}
Substituting this into Eq.~\eqref{Ajformula} we get for $A_y$:
 \begin{align} \label{}
 A_y(\theta)=\frac{\Tr(M\sigma_2M^{\dagger})}{\Tr(MM^{\dagger})}=\frac{\sum\limits_{m_f,m_1,m_2}(M_{m_fm_1}(\sigma_2)_{m_1m_2}M^{\dagger}_{m_2m_f})}{\Tr(MM^{\dagger})}.
\end{align}
Using the definition of $M^{\dagger}$:
\begin{align} \label{}
 A_y(\theta)=\frac{\sum\limits_{m_f,m_1,m_2}(M_{m_fm_1}(\sigma_2)_{m_1m_2}M^{*}_{m_fm_2})}{\Tr(MM^{\dagger})},
\end{align}
and using the explicit from of $\sigma_2$ Eq.~\eqref{Pauli2}:
\begin{align} \label{Aypractical}
 A_y(\theta)=\frac{\sum\limits_{m_f}i(-M_{m_f\frac{1}{2}}M^{*}_{m_f-\frac{1}{2}}+M_{m_f-\frac{1}{2}}M^{*}_{m_f\frac{1}{2}})}{\Tr(MM^{\dagger})}=\frac{\sum\limits_{m_f}2\Im(M_{m_f\frac{1}{2}}M^{*}_{m_f-\frac{1}{2}})}{\Tr(MM^{\dagger})}.
\end{align}
Now redefining magnetic quantum numbers as $m_1$ and $m'_1$ for the initial and final deuteron, and $m_2$ and 
$m'_2$ for initial and final neutron we get the final form of the formula to calculate $A_y$:
 \begin{equation} \label{Aypracticalfinal}
A_{y}(\theta,\phi)=\frac{\displaystyle\sum_{m_{1}}\sum_{m_{1}',m_{2}'}2\,\mathrm{Im}\left[M_{m_{1}',m_{2}';m_{1},\nicefrac{1}{2}} \, M^{*}_{m_{1}',m_{2}';m_{1},-\nicefrac{1}{2}}\right]}{\displaystyle\sum_{m_{1},m_{2}}\sum_{m_{1}',m_{2}'}\left|M_{m_{1}',m_{2}';m_{1},m_{2}}\right|^{2}}.
\end{equation}
The connection between the transition matrix $M$ and the on-shell scattering amplitude calculated solving the integral equations is given by Eq.~\eqref{Mtconnection}.
\begin{equation}\label{Mtconnection}
M_{L'S',LS}^{J}(k)=Z_{\mathrm{LO}}t_{L'S',LS}^{J;Nt\to Nt}(k,k,E),
\end{equation}
where $Z_{LO}$ is the leading order deuteron wavefunction renormalization given by: 
$Z_{LO}=\frac{2\gamma_t}{M_N}$.
If the deuterons in the target are polarized we get more polarization observable, for which formulas analogous to Eq.~\eqref{Aypracticalfinal} can be derived using the density-matrix formulation described earlier \cite{nla.cat-vn1931225,Ohlsen:1972zz,glockle1983quantum,Fukukawa:2010wx}. Here are the formulas we use to compute these observables:
\begin{equation}
iT_{11}(\theta,\phi)=-\sqrt{\frac{3}{2}} \ \frac{\displaystyle\sum_{m_{2}}\sum_{m_{1}',m_{2}'}\,\mathrm{Im}\left[M_{m_{1}',m_{2}';-1,m_{2}}M^{*}_{m_{1}',m_{2}';0,m_{2}}+M_{m_{1}',m_{2}';0,m_{2}}M^{*}_{m_{1}',m_{2}';1,m_{2}}\right]}{\displaystyle\sum_{m_{1},m_{2}}\sum_{m_{1}',m_{2}'}\left|M_{m_{1}',m_{2}';m_{1},m_{2}}\right|^{2}},
\end{equation}
\begin{equation}
T_{20}(\theta)=\frac{1}{\sqrt{2}} \ \frac{\displaystyle\sum_{m_{2}}\sum_{m_{1}',m_{2}'}\left\{\left|M_{m_{1}',m_{2}';1,m_{2}}\right|^{2}-2\left|M_{m_{1}',m_{2}';0,m_{2}}\left|^{2}+\right|M_{m_{1}',m_{2}';-1,m_{2}}\right|^{2}\right\}}{\displaystyle\sum_{m_{1},m_{2}}\sum_{m_{1}',m_{2}'}\left|M_{m_{1}',m_{2}';m_{1},m_{2}}\right|^{2}},
\end{equation}
\begin{equation}
T_{21}(\theta,\phi)=-\sqrt{\frac{3}{2}} \ \frac{\displaystyle\sum_{m_{2}}\sum_{m_{1}',m_{2}'}\,\mathrm{Re}\left[M_{m_{1}',m_{2}';0,m_{2}}\left(M^{*}_{m_{1}',m_{2}';1,m_{2}}-M^{*}_{m_{1}',m_{2}';-1,m_{2}}\right)\right]}{\displaystyle\sum_{m_{1},m_{2}}\sum_{m_{1}',m_{2}'}\left|M_{m_{1}',m_{2}';m_{1},m_{2}}\right|^{2}},
\end{equation}
and
\begin{equation}
T_{22}(\theta,\phi)=\sqrt{3} \ \frac{\displaystyle\sum_{m_{2}}\sum_{m_{1}',m_{2}'}\mathrm{Re}\left[M_{m_{1}',m_{2}';1,m_{2}}M_{m_{1}',m_{2}';-1,m_{2}}^{*}\right]}{\displaystyle\sum_{m_{1},m_{2}}\sum_{m_{1}',m_{2}'}\left|M_{m_{1}',m_{2}';m_{1},m_{2}}\right|^{2}},
\end{equation}

\section{Results}

Here I present the up to $\nnlo$ results of the \EFT calculation for the total unpolarized cross-section and 
the $\ntlo$ results for the observables discussed in the previous section for the elastic $nd$ scattering in comparison 
to the data and some potential model calculations. 

\begin{figure}[hbt]
	\begin{center}
		\hspace{-1cm}\includegraphics[width=90mm]{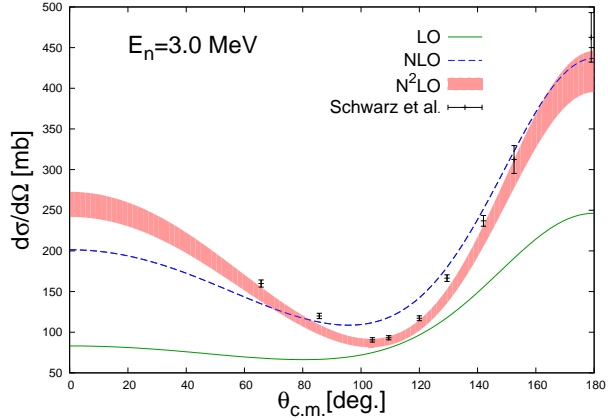} 
	\end{center}
\caption{\label{fig:sigma}(Caption and figures are from \cite{Margaryan:2015rzg}). $nd$ scattering cross section for $E_{n}=3.0$~MeV  with experimental data from Schwarz, \textit{et. al.}~\cite{SCHWARZ19831}.  The LO prediction (without theoretical errors) is the solid green line, the dashed blue line the NLO prediction (without theoretical errors), and the solid red band the $\nnlo$ prediction with a 6\% error estimate.}
\end{figure}

The results for the cross section calculations at $3$ MeV-nucleon energy are given in Fig.~\ref{fig:sigma} (caption and figures are from \cite{Margaryan:2015rzg}). 
In this figure we can clearly see that the \EFT results get better order by order. 
There is an excellent agreement of the $\nnlo$ result with the experiment. The curve that corresponds to the $\nnlo$ 
calculation is given with a band that represents the estimated theoretical error. 

\begin{figure}[H]
	\begin{center}
	\begin{tabular}{cc}
	\hspace{-1cm}\includegraphics[width=90mm]{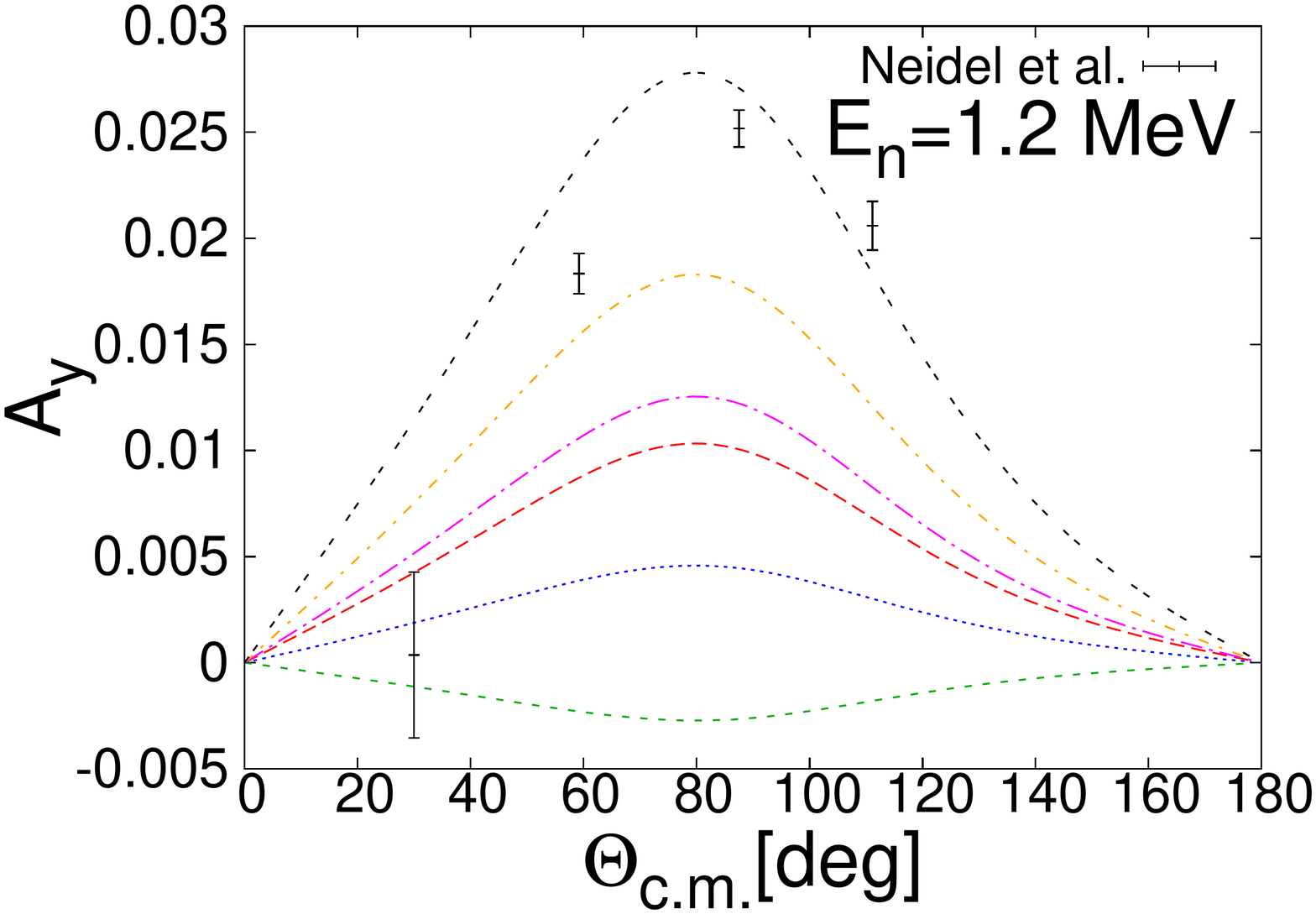} &
	\hspace{-1cm}\includegraphics[width=90mm]{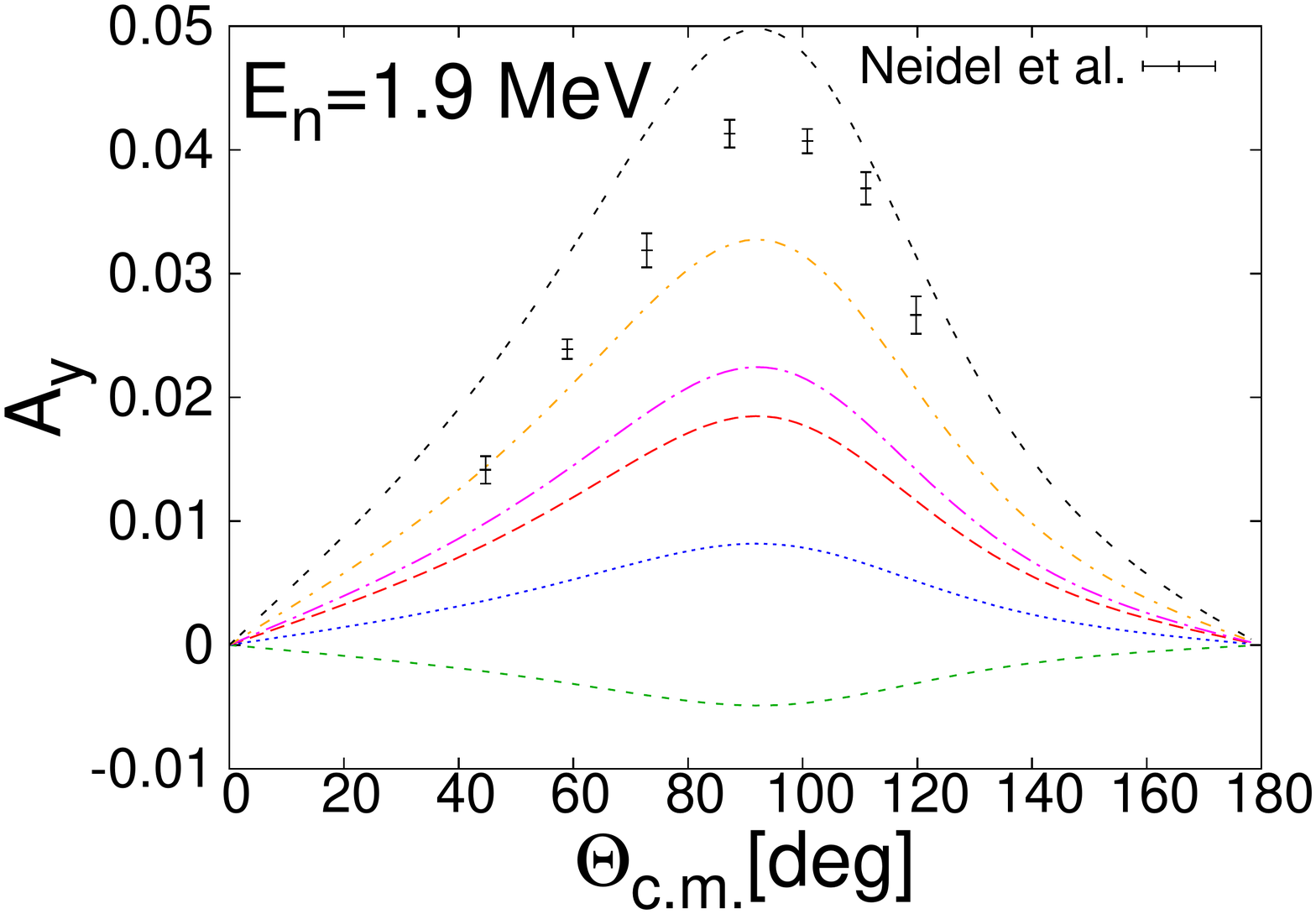} \\
	\end{tabular}

	\vspace{-.5cm}
	
	\includegraphics[width=90mm]{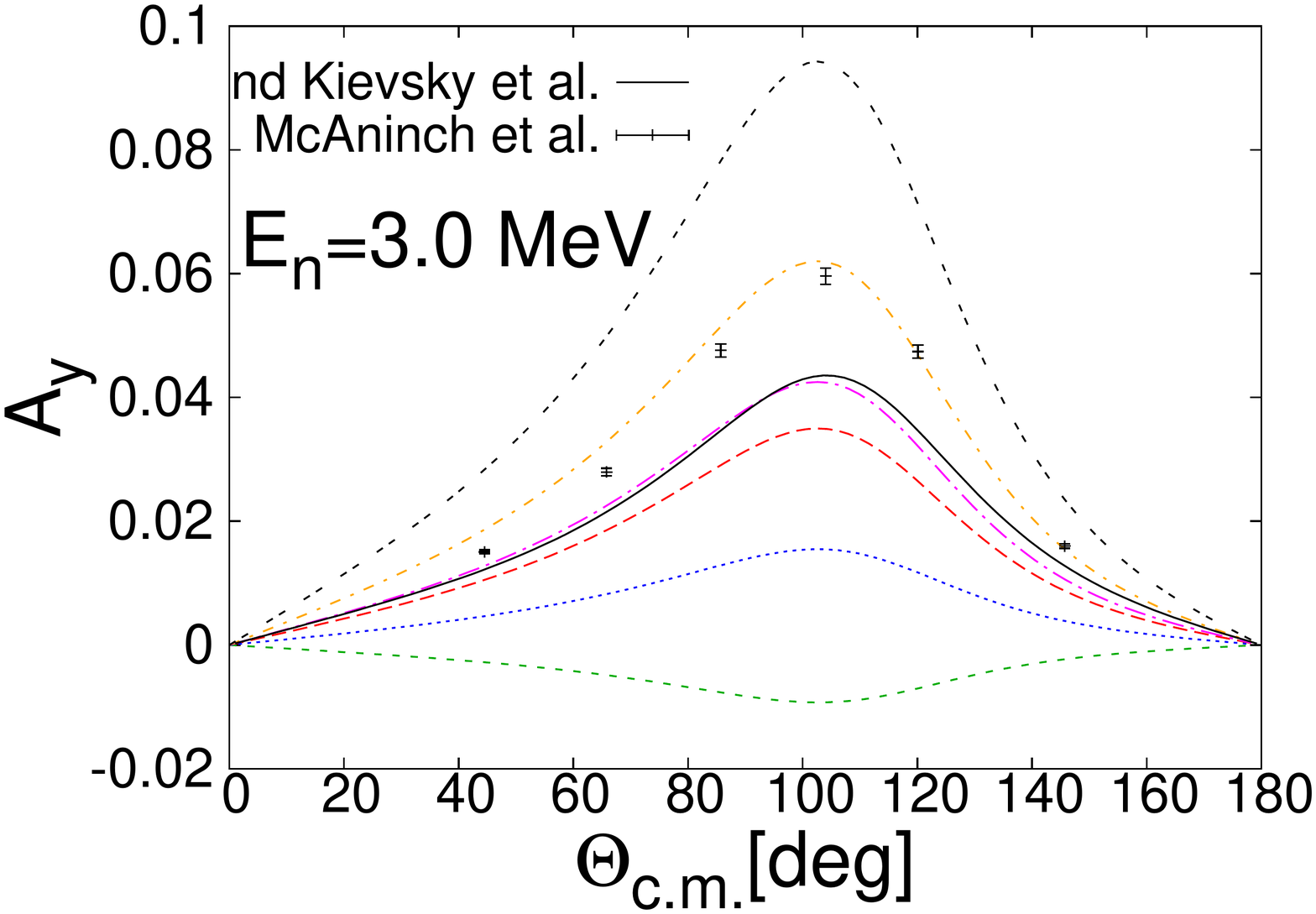}
	\end{center}

\caption{\label{fig:Aybands}(Caption and figures are from \cite{Margaryan:2015rzg}). The dashed lines are \EFT results for $A_{y}$ for several sets of $C^{{}^{3}\!P_{J}}$ coefficients varied by 15\% around their central values.   Top Left: $E_{n}=1.2$~MeV, experimental data from Neidel et al.~\cite{neidel2003new}. Top Right: $E_{n}=1.9$~MeV, experimental data from Neidel, \textit{et. al.}~\cite{neidel2003new}. Bottom: $E_{n}=3.0$~MeV, the solid line is a PMC calculation from Kievsky, \textit{et. al.} using AV-18+UR~\cite{Kievsky:1996ca}, with experimental data from McAninch, \textit{et. al.}~\cite{McAninch:1994zz}. In the following, ``$+$'' stands for 15 percent above central values given in Eq.~(\ref{eq:CPJnpfit}); ``$0$'' is at central value; and ``$-$'' is 15 percent below central value.  The coefficient values ($C^{{}^{3}\!P_{0}}$,$C^{{}^{3}\!P_{1}}$,$C^{{}^{3}\!P_{2}}$) used to produce the curves shown are (from lowest \EFT curve to highest \EFT curve on the plots): big dots (green)=$(+,-,+)$; small dots (blue)=$(+,0,+)$; long dash (red)=$(0,0,0)$; long-dash-dot (purple) = $(0,0,+)$; short-dash-dot (orange) = $(-,0,0)$; double-dot (black) = $(-,+,+)$.}
\end{figure}

A good energy range of processes that we expect \EFT to work well is given by typical momentum exchange $Q$ 
such that the power counting parameter is about $\frac{Q}{\Lambda_{\slashed{\pi}}}\sim0.3$, which is used 
to estimate the theoretical error. As the terms in the Lagrangian and the amplitudes 
are calculated as a power series in powers of $0.3$ we can estimate that each subsequent term is going to be 
of the order of $0.3\times$ previous term. So we estimate that the $\ntlo$ correction to an amplitude
is of the order of $0.3\times0.3\times0.3\times$ the LO result$\approx0.03\times$the LO result, which is the $3\%$ 
of the LO result. Cross sections, being proportional to the square of the amplitudes, get the double of that error estimate. This is why we give a $6\%$ theoretically estimated error band on the $\nnlo$ cross section in Fig.~\ref{fig:sigma} \cite{Margaryan:2015rzg}.
Our $\ntlo$ results for the observable $A_y$ are given in Fig.~\ref{fig:Aybands} (caption and figures are from \cite{Margaryan:2015rzg}) for three different energies. 
Just as we calculated an estimate for the theoretical error for the cross section, we can do the same also for $A_y$. 
The main contributions to this observable come from the ${}^{3}\!P_{J}$ interactions Eq.~\eqref{eq:PwaveLagRupak} 
with interaction coefficients $C_2^{^3P_J}$. At the next order of the \EFT the correction to the observable 
is estimated to be the same order as the change in the observable if we vary the $C_2^{^3P_J}$ interaction coefficients within a $30\%$ ($\pm15\%$) band around their central values given in Eq.~\eqref{eq:CPJnpfit}. 
Fig.~\ref{fig:Aybands} contains various curves corresponding to different values of the $C_2^{^3P_J}$ coefficients \cite{Margaryan:2015rzg}.
We can make a couple different conclusions from the results in Fig.~\ref{fig:Aybands}. First of all we see that 
the $A_y$ observable is indeed very sensitive to the ${}^{3}\!P_{J}$ interactions as was already discovered by various 
potential model calculations previously. Secondly we can see from the $3$ MeV plot that the maximum of $A_y$ is 
around $\theta=105^{\degree}$, which coincides with the minimum of the unpolarized cross-section (see Fig.~\ref{fig:sigma}). This is justified by the fact that the total unpolarized cross section is in the denominator in Eq.~\eqref{Ayinterpret}. 
Finally, we see that the experimental data is well within the reach 
of the \EFT. This means that if the \EFT at higher than $\ntlo$ orders gives corrections according to the given 
error estimates, as is expected, then the $A_y$-puzzle could be solved in the next order. 

\begin{figure}[H]
	\begin{center}
\hspace{-2.5cm} \includegraphics[width=190mm]{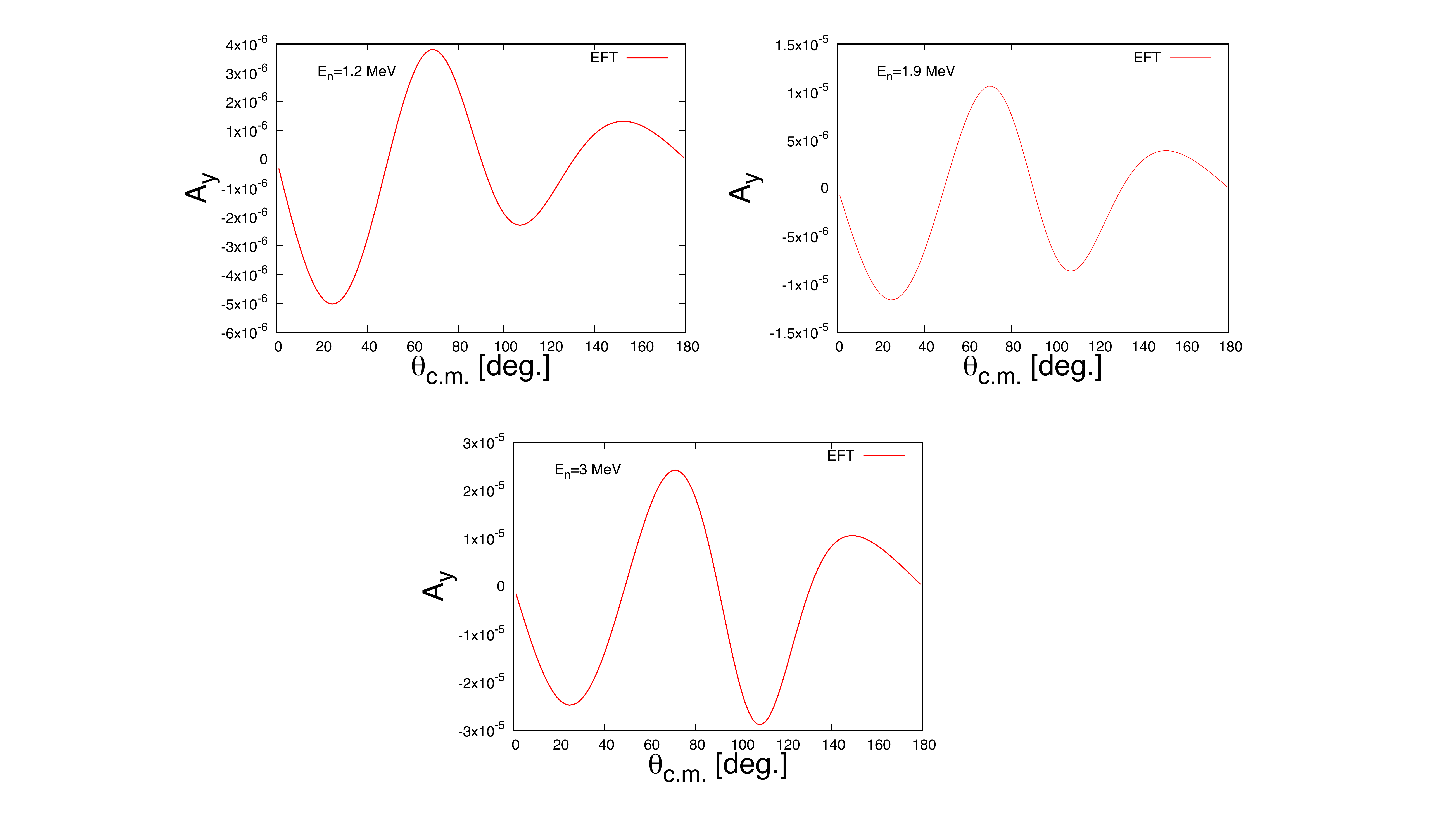} 
	\end{center}
\caption{\label{fig:AySD}\EFT results for $A_y$ only coming from the $SD$-mixing.}
\end{figure}

We have previously claimed that the contribution to $A_y$ coming from the two-body $SD$-mixing term is 
negligible compared to the contributions that come from the ${}^{3}\!P_{J}$ interactions. To see this we plot only the 
$SD$-mixing contribution to the $A_y$ in Fig.~\ref{fig:AySD}. As we can see from the comparison of the two figures
Figs.~\ref{fig:Aybands} and \ref{fig:AySD} the $SD$-contribution is about three orders of magnitude smaller than the ${}^{3}\!P_{J}$ contributions.

\begin{figure}[H]
\begin{center}
	\begin{center}
	\begin{tabular}{cc}
	\hspace{-1cm}\includegraphics[width=90mm]{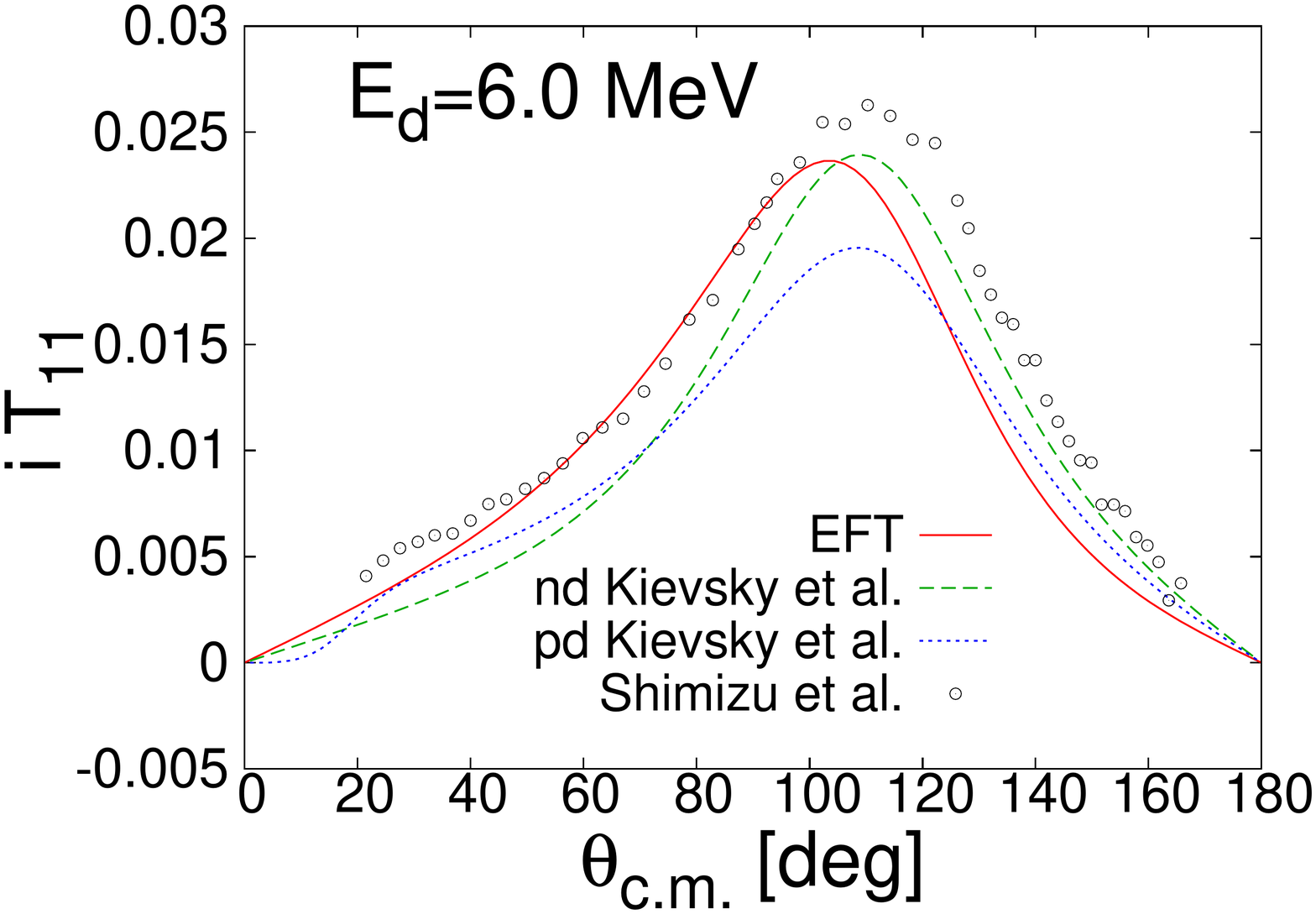} &
	\hspace{-1cm}\includegraphics[width=90mm]{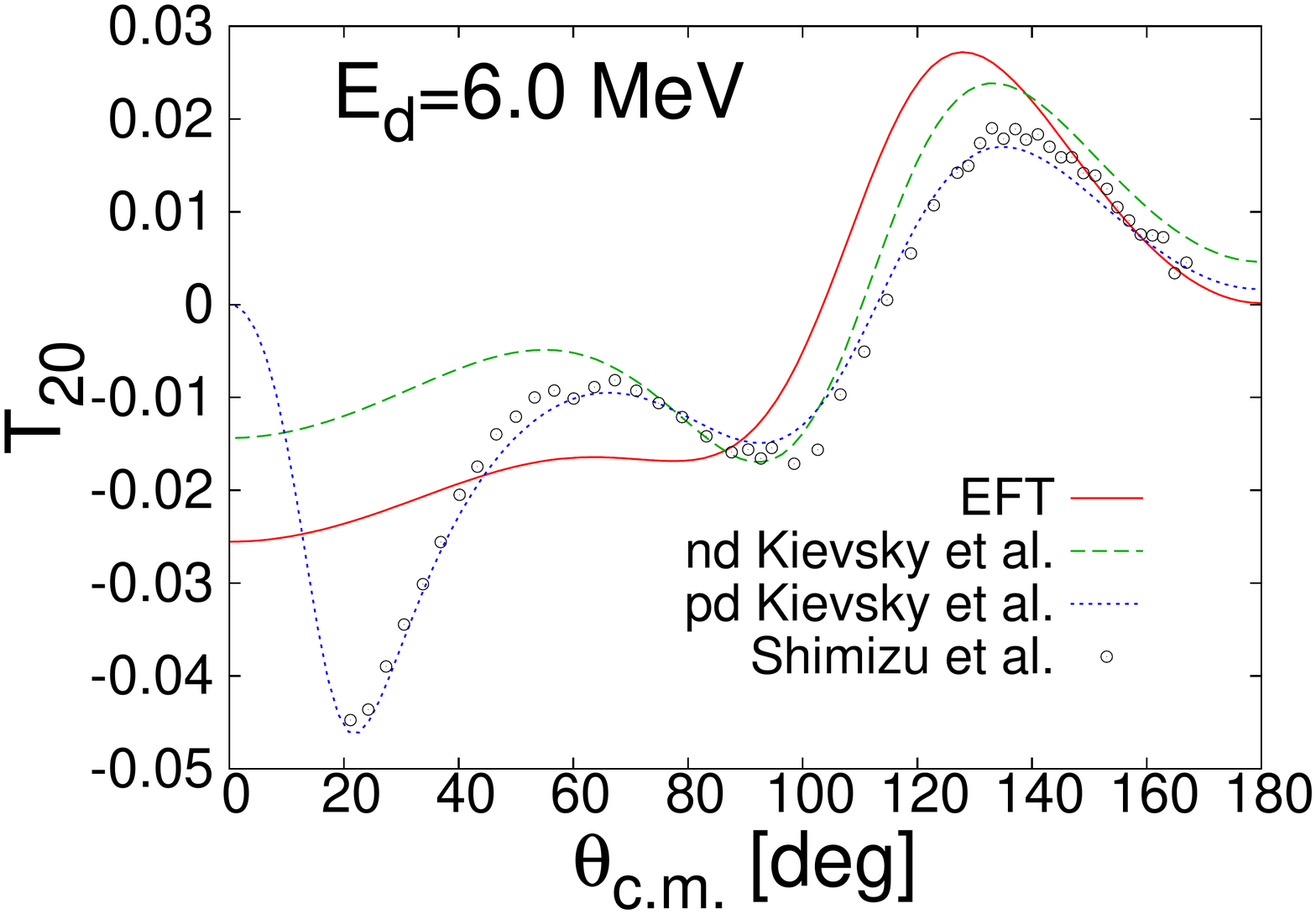} \\[-.5cm]

	\hspace{-1cm}\includegraphics[width=90mm]{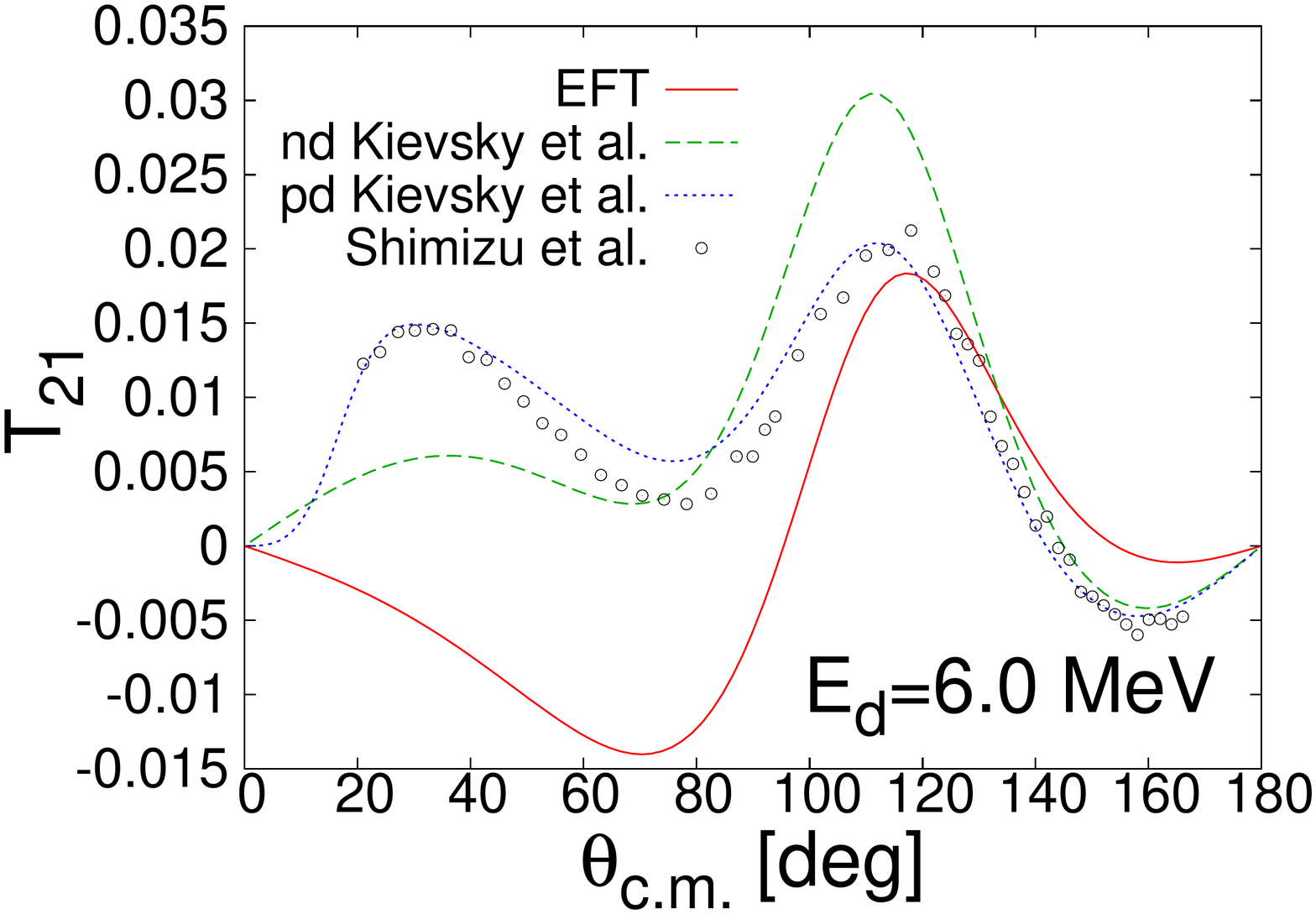} &
	\hspace{-1cm}\includegraphics[width=90mm]{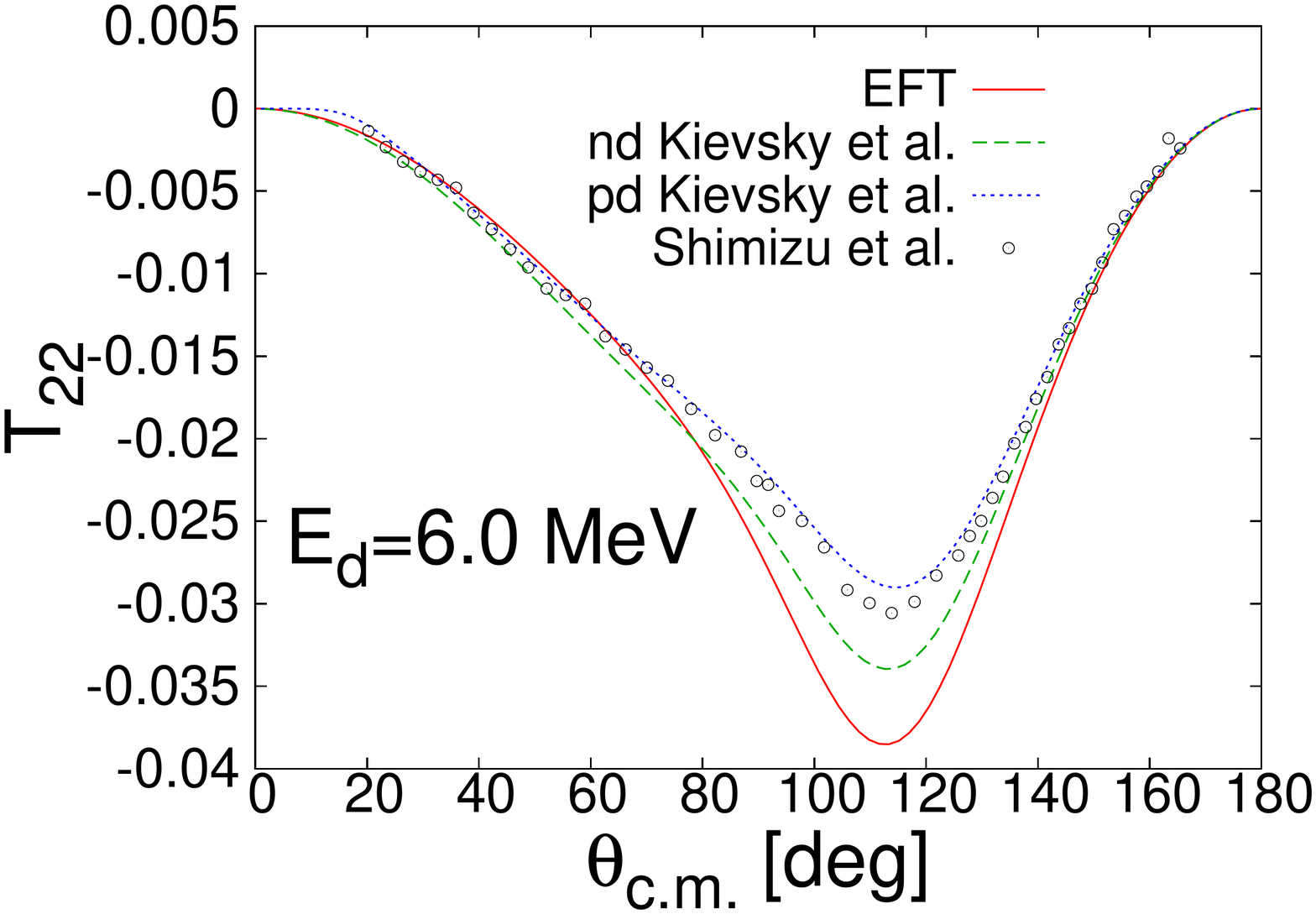}
	\end{tabular}
	\end{center}


\caption{\label{fig:Tnn}(Caption and figures are from \cite{Margaryan:2015rzg}). The solid red line is the \EFT prediction (without theoretical error bars) for deuteron polarization observables in $nd$ scattering, the dashed green line PMC calculations using AV18+UR for $nd$ scattering~\cite{Kievsky:1996ca} , and the dotted blue line PMC calculations using AV18+UR for $pd$ scattering~\cite{Kievsky:1996ca}. All experimental data is for $pd$ scattering from Shimizu, \textit{et. al.}~\cite{Shimizu:1995zz} at a laboratory deuteron energy of $E_{d}=6.0$~MeV.}
\end{center}
\end{figure}

To develop a feeling about how well the \EFT works for few nucleon processes it is also useful to look at
the results that it gives for the other $nd$ polarization observables Fig.~\ref{fig:Tnn} (caption and figures are from \cite{Margaryan:2015rzg}). On this figure the \EFT results
for deuteron polarization observables are plotted along with some potential model calculations and experimental data. The experimental data is not for $nd$ elastic scattering though, it is for $pd$ elastic scattering. Given the 
approximate isospin symmetry and that the Coulomb interactions become less and less important for backward angles 
and higher scattering energies, we expect qualitative agreement between the theory and experiment, which is observed.


\section{Conclusions}

In this dissertation the $\ntlo$ calculation of the $A_y$ observable for $nd$ elastic scattering process in \EFT is described. This observable at $\ntlo$ gets contributions from the two-body $SD$ 
mixing and the two-body $P$-wave interactions. We find that the contributions from $SD$ mixing are about three orders of magnitude smaller than the contributions from the two-body 
$P$-wave interactions. The $A_y$ observable is found to be very sensitive to the two-body $P$-wave interactions. We see that varying the two-body $P$-wave interaction coefficients within the estimated theoretical error we find a very wide range of theoretical curves for the $A_y$ observable. Most importantly 
the experimental data is within this range, which, at this order, eliminates the long-standing disagreement between the theory and the experiment, but 
dictates that a higher order calculation of this observable is necessary to unambiguously test \EFT. As we go to higher orders in an EFT 
we get more interaction terms in the Lagrangian and this increases the number of parameters that can be matched onto the experiment. 
Some of the possible results of a higher order calculation might be as follows. 
One of the possibilities is that at the next order the new three-body $P$-wave interaction coefficient is matched onto the $A_y$ observable at some particular scattering energy and the predictions of the theory for the other energies agree well with the experiment. 
This would solve the $A_y$-puzzle and justify \EFT as a well-constructed low-energy approximation to QCD, which gives a good description of the low energy nuclear physics at two- and three-body sectors. 
Another possibility is that the higher order theoretical prediction either grossly disagrees with the experiment or gets better but is still unsatisfactory. This would possibly be an indication of a beyond the standard model physics, but more likely this will be taken as a first sign that the power counting of \EFT needs to be modified for this observable. In this case a possible solution might be the extension of \EFT in such a way that certain three-body diagrams are resummed in order for the theory to be able to include more of the non-perturbative aspects of QCD. 

The $A_y$ observable has the total unpolarized $nd$ scattering cross section in its denominator. Strictly perturbatively for $\ntlo$ result this denominator is 
supposed to be expended and truncated at NLO, because the first nonzero contribution to the numerator of $A_y$ is at N$^2$LO. Doing so we discovered 
that the position of the peak of $A_y$ does not match with the experimental result. Keeping the denominator at N$^2$LO is equivalent to resumming 
some of the higher order contributions and makes the position of the maximum of $A_y$ to match with the experiment. The total unpolarized $nd$ scattering cross section is in the denominators of all of the polarization observables, so the interesting physics about 
the polarization observables is concentrated in the numerators. This justifies keeping the denominator unexpended when comparing to 
experiments. 

We have not done the calculation of the $\ntlo$ total unpolarized cross-section for the $nd$ elastic scattering process. The reason is that at this order 
there is a new three-body force that we have not included in the calculation. This three body force gives zero contribution to the $A_y$ but is important 
in the calculation of the total cross section, which will be necessary for higher order calculations of $A_y$ (see \cite{Margaryan:2015rzg} for a more detailed discussion).

The calculation of the same observable for the $pd$ elastic scattering vs $nd$ elastic scattering is complicated by Coulomb interactions and by the 
necessity to include more three-body counter-terms that are not necessary in the $nd$ system. Calculations on $pd$ system are of interest because 
there is more experimental data for this system to compare to. The qualitative agreement between the \EFT calculations for the $nd$ 
system and the experimental data for the $pd$ system in Fig.~\ref{fig:Tnn} is encouraging. This figure shows that as expected the 
Coulomb interactions are more important for forward angles, so to get the correct behavior for these observables at forward angles 
it might be necessary to include the Coulomb effects non-perturbatively. 

\section{Future Directions}

From the theory point of view nuclear physics is the simplest when the number of nucleons is fewer and the processes happen in the low-energy 
regime. So this seems to be a good starting point in exploring nuclear forces, and the \EFT is specifically designed to deal with processes like that. 
Given this, the next steps to advance our understanding of nuclear physics can be: $1.$ to increase the accuracy of our predictions in the few-nucleon low-energy 
regime, $2.$ to increase the number of particles involved in a process, $3.$ to consider the higher-energy regimes. 

$1.$ We can increase the accuracy of the predictions by just going into higher orders of the \EFT. In fact our hope and conclusion is that 
the $A_y$-puzzle will be solved in the next few orders of the \EFT, because at N$^4$LO we will have new 3-body $P$-wave interactions 
entering into the Lagrangian, which will probably give the missing contributions to the $A_y$. 

Some of the difficulties with this direction are that in the higher orders the number of diagrams increases excessively and the difficulty of both analytical and 
numerical calculations increases too. Besides in higher orders we always get new interaction terms allowed by the symmetries, which need to 
be fixed by implementing more and more experimental data into the theory, losing the predictive power of the theory.
For example for the $A_y$ in particular we might have to fix one of the new interactions to this observable at least at some scattering energy 
and then the prediction of the theory will be perefect at that scattering energy, but it will not really be a prediction, the theory will still be able to provide predictions for other scattering energies. 

$2.$ Increasing the number of particles we can explore a wider range of processes involving both short-range and long-range interactions and 
it should in principle allow us to predict binding energies of the different nuclei. 
The difficulties in this direction are partly the same as in previous point: it becomes numerically expensive to deal with a larger number of particles and an 
analytical approach is even harder. We have started to look at the LO four-body problem in \EFT and firstly we would like to know whether a four-body contact interaction is needed at LO, as was the case with the three-body sector.
To answer this question our strategy is to disregard any -four-body interactions, calculate the four-body binding energy numerically and 
check to see whether or not this binding energy has cutoff dependence. If it does have a cutoff dependence then it will most definitely prove 
that a four-body contact interaction will be necessary at LO so that the theory is properly renormalized, but if the predictions are cutoff independent then there is no need for the four-body contact interaction \cite{Platter:2004if,Platter:2004he,Platter:2004zs,Nogga:2000uu,Epelbaum:2000mx}. 

There is a special family of nuclei though called the halo-nuclei, which have a structure with a dense nuclear core with one or two nucleons 
on higher orbits. The EFT that is constructed to describe these nuclei is called Halo EFT and has similar features to \EFT
\cite{Hammer:2017tjm,Epelbaum:2008ga,Bertulani:2002sz,Bedaque:2003wa,Vanasse:2016hgn,Fernando:2011ts,Phillips:2013msa}. 

$3.$ As we go higher in energy, we will encounter particles that had been integrated out before and we will have to include them 
into the theory as explicit degrees of freedom. The first particles to come into the theory are the pions. The low-energy effective field theory with pions is the $\chi$PT, which was created before the \EFT. The $\chi$PT allows one to do calculations 
on the same processes that are calculated in the \EFT and many more, and the applicability range of the $\chi$PT is wider. 
One of the difficulties of the $\chi$PT is its power counting though; as opposed to the \EFT its power counting is not un-ambiguous.

And finally one shouldn't lose the hope that QCD will eventually be solved in its non-perturbative regime. That would ultimately give answers to all of our questions about nuclear physics.

\chapter{Lorem ipsum prompta platonem intellegat ut est Tamquam epicurei ne duo Et quidam iuvaret eum}


\appendix
\chapter{Projections}
\label{Projections}
To do the spin-angular momentum projections we use integrals proven in \cite{Jared:2012diss} and diagrammatic approach to Clebsch-Gordan coefficients and $3nJ$-symbols described in \cite{Strobel:1967bxz}.
Here I will give the integrals used, describe the diagrammatic approach and do a sample calculation of one of the diagrams. 

The integrals necessary are given in the following equations. 

\begin{eqnarray} \label{kkint}
&&\frac{k^2}{3} \int d\Omega_k \int d\Omega_p f(\cos{\theta}) (Y_{L'}^{m'_L}(\hat{\bf{p}}))^{*} Y_{L}^{m_L}(\hat{\bf{k}}) Y_{1}^{m_1}(\hat{\bf{k}}) Y_{1}^{m_2}(\hat{\bf{k}})\nonumber \\
&=& k^2 \sum_{L''}\sum_{m''}  \sqrt{\frac{\bar{L}}{\bar{L'}}} \CG{1}{m_1}{1}{m_2}{L''}{m''} \CG{L}{m_L}{L''}{m''}{L'}{m'_L}  \CG{1}{0}{1}{0}{L''}{0} \CG{L}{0}{L''}{0}{L'}{0} \frac{1}{2}\int_{-1}^{1}P_{L'}(x) f(x)dx \nonumber \\
\end{eqnarray}

\begin{eqnarray} \label{ppint}
&&\frac{p^2}{3} \int d\Omega_k \int d\Omega_p f(\cos{\theta}) (Y_{L'}^{m'_L}(\hat{\bf{p}}))^{*} Y_{L}^{m_L}(\hat{\bf{k}}) Y_{1}^{m_1}(\hat{\bf{p}}) Y_{1}^{m_2}(\hat{\bf{p}})\nonumber \\
&=& p^2 \sum_{L''}\sum_{m''}  \sqrt{\frac{\bar{L}}{\bar{L'}}} \CG{1}{m_1}{1}{m_2}{L''}{m''} \CG{L}{m_L}{L''}{m''}{L'}{m'_L}  \CG{1}{0}{1}{0}{L''}{0} \CG{L}{0}{L''}{0}{L'}{0} \frac{1}{2}\int_{-1}^{1}P_{L}(x) f(x)dx \nonumber \\
\end{eqnarray}

\begin{eqnarray} \label{kpint}
&&\frac{kp}{3} \int d\Omega_k \int d\Omega_p f(\cos{\theta}) (Y_{L'}^{m'_L}(\hat{\bf{p}}))^{*} Y_{L}^{m_L}(\hat{\bf{k}}) Y_{1}^{m_1}(\hat{\bf{k}}) Y_{1}^{m_2}(\hat{\bf{p}})\nonumber \\
&=& kp \sum_{L''}\sum_{m''}  \sqrt{\frac{\bar{L}}{\bar{L'}}} \CG{L}{m_L}{1}{m_1}{L''}{m''} \CG{L''}{m''}{1}{m_2}{L'}{m'_L}  \CG{L}{0}{1}{0}{L''}{0} \CG{L''}{0}{1}{0}{L'}{0} \frac{1}{2}\int_{-1}^{1}P_{L''}(x) f(x)dx \nonumber \\
\end{eqnarray}
where $\theta$ is the angle between $\hat{\bf{k}}$ and $\hat{\bf{p}}$. These integrals arise when we have to project expressions like 
$f(\cos{\theta})k_ik_j$, $f(\cos{\theta})p_ip_j$ and $f(\cos{\theta})k_ip_j$ onto an angular-momentum state basis. The vectors $\vec{k}$ and $\vec{p}$ are first 
represented as spherical vectors arriving at expressions like the left hand sides of the previous equations, then the integrals are evaluated using these 
equations. Usually for the function $f(x)$ we have $f(x)=\frac{1}{a+x}$, which on the right hand side gives (up to a sign) 
Legendre polynomials of the second kind. 

In evaluating the spin-angular momentum projections we encounter large expressions which are sums of products of many Clebsch-Gordan coefficients. The sums run over magnetic quantum numbers. 
We simplify these expressions by reducing them into products of $6J$- and $9J$- symbols which are independent of magnetic quantum numbers. In general the $3nJ$-symbols arise when one 
has to combine more than two angular momenta together to form a total angular momentum of a system. The diagrammatic approach we use is described below. I give identities from \cite{Strobel:1967bxz} which are the most relevant to our calculations. 

A Clebsch-Gordan coefficient is drawn as an arrow:

\begin{equation}
\CG{J_1}{m_1}{J_2}{m_2}{J_3}{m_3}=\StrobelLine{m_2}{m_1}{m_3}
\end{equation}

On the arrow we only put the magnetic quantum numbers, keeping in mind that it is actually a function of six variables: 
the magnetic quantum numbers and the angular momentum $J$-values.

With this notation the symmetry properties of Clebsch-Gordan coefficients can be expressed as:

\begin{eqnarray} \label{cgproperties}
&&\StrobelLine{m_1}{m_2}{m_3} \hspace{0.1cm} =\hspace{0.1cm} (-1)^{J_1+m_1} (\frac{\bar{J_3}}{\bar{J_2}})^{\frac{1}{2}} \StrobelLine{m_3}{-m_1}{m_2} \hspace{0.1cm} = \hspace{0.1cm}
(-1)^{J_2-m_2} (\frac{\bar{J_3}}{\bar{J_1}})^{\frac{1}{2}} \StrobelLine{-m_2}{m_3}{m_1} \nonumber \\ 
&=&(-1)^{J_1+J_2-J_3}\StrobelLine{-m_1}{-m_2}{-m_3} \hspace{0.1cm} = \hspace{0.1cm}
(-1)^{J_1+J_2-J_3} \StrobelLine{m_2}{m_1}{m_3} \nonumber \\ 
\end{eqnarray}
 where $\bar{J}=2J+1$.
 
The orthogonality and completeness relations are written as:

\begin{equation}\label{orthogonality}
\sum_{m_1,m_2} \StrobelLine{m_1}{m_2}{m_3} \StrobelLine{m_1}{m_2}{m'_3}= \delta_{J_3,J'_3}\delta_{m_3,m'_3}
\end{equation}
and 

\begin{equation}\label{completeness}
\sum_{J_3,m_3} \StrobelLine{m_1}{m_2}{m_3} \StrobelLine{m'_1}{m'_2}{m_3}= \delta_{m_1,m'_1}\delta_{m_2,m'_2}
\end{equation}
In the last two equations it is to be understood that the first two $J$-values in the two arrows are the same. 

In our expressions with Clebsch-Gordan coefficients we have each magnetic quantum number appearing either once or twice. 
If it appears once then it is a free index and the final expression is going to depend on it. If it appears twice, then it is on two 
different Clebsch-Gordan coefficients, the $J$- values corresponding to the two places where it appears are always the same 
(otherwise the expression would not make sense), and it is understood that there is a sum going over that magnetic quantum number. 
If we have two arrows, which have two coinciding magnetic quantum numbers, meaning there is a sum over it, 
then we can draw these arrows intersecting or touching at that magnetic quantum number. With this new notation the $6J$-symbols 
are defined by:

\begin{equation}
(\bar{J_6}\bar{J_4}\bar{J_5}\bar{J_6})^{\frac{1}{2}}(-1)^{J_1+J_3+J_4+J_5}\begin{Bmatrix}
 J_1&J_2 & J_4 \\
J_3 & J_6 & J_5
\end{Bmatrix}=
\StrobelTriangleA{m_3}{m_4}{m_6}{m_5}{m_1}{m_2}
\end{equation}

Using the properties of Clebsch-Gordan coefficients and the $6J$-symbols the following equations can be derived:

\begin{equation}
\StrobelTriangleC{m_3}{m_4}{m_6}{m_7}{m_5}{m_1}{m_2}=
(\bar{J_4}\bar{J_5})^{\frac{1}{2}}(-1)^{J_1+J_3+J_4+J_5}\begin{Bmatrix}
 J_1&J_2 & J_4 \\
J_3 & J_6 & J_5
\end{Bmatrix}
\delta_{J_6,J_7} \delta_{m_6,m_7}
\end{equation}

and 

\begin{equation} \label{tri3}
\StrobelTriangleB{m_3}{m_4}{m_6}{m_5}{m_1}{m_2}=
(\bar{J_4}\bar{J_5})^{\frac{1}{2}}(-1)^{J_1+J_3+J_4+J_5}\begin{Bmatrix}
 J_1&J_2 & J_4 \\
J_3 & J_6 & J_5
\end{Bmatrix}
\StrobelLine{m_1}{m_5}{m_6}
\end{equation}

Using same notation a $9J$-symbol is defined by the following equation:

\begin{equation}
\begin{Bmatrix}
J_1 & J_2& J_3 \\
J_4& J_5  & J_6\\
J_7 & J_8 & J_9
\end{Bmatrix} 
=(\bar{J_3}\bar{J_6}\bar{J_9}\bar{J_7}\bar{J_8}\bar{J_9})^{-\frac{1}{2}} \StrobelBoxA{m_1}{m_4}{m_7}{m_2}{m_3}{m_5}{m_6}{m_8}{m_9}
\end{equation}

And the following identities can be proven:

\begin{equation} \label{box1}
\text{\StrobelBoxAA{m_1}{m_4}{m_7}{m_2}{m_3}{m_5}{m_6}{m_8}{m_9}{m'_9}}
=(\bar{J_8}\bar{J_6}\bar{J_7}\bar{J_3})^{\frac{1}{2}} 
\begin{Bmatrix}
J_1 & J_2& J_3 \\
J_4& J_5  & J_6\\
J_7 & J_8 & J_9
\end{Bmatrix} 
\delta_{J_9,J'_9}\delta_{m_9,m'_9}
\end{equation}
and

\begin{equation}
\StrobelBoxB{m_1}{m_4}{m_7}{m_2}{m_3}{m_5}{m_6}{m_8}{m_9}
= (\bar{J_8}\bar{J_6}\bar{J_7}\bar{J_3})^{\frac{1}{2}} 
\begin{Bmatrix}
J_1 & J_2& J_3 \\
J_4& J_5  & J_6\\
J_7 & J_8 & J_9
\end{Bmatrix} 
\StrobelLine{m_3}{m_6}{m_9}
\end{equation}

As an example let's calculate the projections of a term in the two-body $P$-wave loop Feynman diagram derived above:

\begin{eqnarray}
&&64t= \left\{Z^2(C_2^{^3P_1}+2C_2^{^3P_2})+A(C_2^{^3P_0}-C_2^{^3P_1}+\frac{2}{3}C_2^{^3P_2})\right \}\sigma_k \sigma_i \sigma_j \sigma_k \nonumber \\       
&+&B(C_2^{^3P_0}-C_2^{^3P_1}+\frac{2}{3}C_2^{^3P_2})k_kk_l\sigma_k \sigma_i \sigma_j \sigma_l \nonumber \\ 
&+&C(C_2^{^3P_0}-C_2^{^3P_1}+\frac{2}{3}C_2^{^3P_2})p_kp_l\sigma_k \sigma_i \sigma_j \sigma_l \nonumber \\ 
&+&\left\{D(2C_2^{^3P_2}-C_2^{^3P_1})+E(C_2^{^3P_0}-\frac{4}{3}C_2^{^3P_2})\right \}k_lp_k\sigma_k \sigma_i \sigma_j \sigma_l \nonumber \\ 
&+&\left\{E(2C_2^{^3P_2}-C_2^{^3P_1})+D(C_2^{^3P_0}-\frac{4}{3}C_2^{^3P_2})\right \}k_kp_l\sigma_k \sigma_i \sigma_j \sigma_l \nonumber \\ 
&+& Z^2C_2^{^1P_1}\sigma_i \sigma_j
\end{eqnarray}

The notation used in what follows is that all the not-primed variables correspond to initial states,
all the primed variables correspond to final states, \\
$J$ stands for total angular momentum and $M$ its projection,\\
$L$ stands for orbital angular momentum and $m_L$ its projection\\
$S$ stands for total spin and $m_S$ its projection\\
$m_N$ stands for nucleon spin projection\\
$m_d$ stands for deuteron spin projection\\

Let's take the term proportional to $k_kp_l$. This needs to be projected onto a total angular momentum state basis, so we have:
\begin{eqnarray}
&&P_{kl}=\bra{J',M'} f(\cos{\theta}) k_kp_l\sigma_k \sigma_i \sigma_j \sigma_l \ket{J,M} \nonumber \\ 
&=& \CG{L}{m_L}{S}{m_S}{J}{M} \CG{L'}{m'_L}{S'}{m'_S}{J'}{M'}  \bra{L'_,m'_L}f(\cos{\theta})k_kp_l\ket{L,m_L}
\bra{S',m'_S} \sigma_k \sigma_i \sigma_j \sigma_l \ket{S,m_S}\nonumber \\ 
&=& \CG{L}{m_L}{S}{m_S}{J}{M} \CG{L'}{m_L}{S'}{m'_S}{J'}{M'} \CG{\frac{1}{2}}{m_N}{1}{m_d}{S}{m_S} \CG{\frac{1}{2}}{m'_N}{1}{m'_d}{S}{m'_S}  \nonumber \\ 
&&\bra{L'_,m'_L}f(\cos{\theta})k_kp_l\ket{L,m_L}
\bra{\frac{1}{2},m'_N} \sigma_k \sigma_i \sigma_j \sigma_l \ket{\frac{1}{2},m_N} \epsilon_d^{*m'_d}  \epsilon_d^{m_d}  \nonumber \\ 
\end{eqnarray}
In here the first equation is just a notation, the second and third equations come from the definition of Clebsch-Gordan coefficients. 
$\epsilon_d^{m_d}$ and $\epsilon_d^{*m'_d}$ are the initial and final deuteron polarizations. The indices $j$ and $i$ are contracted 
with initial and final deuteron polarization projections, so going to spherical indices we can do the substitution  
$\sigma_j \epsilon_d^{m_d}= \sigma^{m_d}$ and $\sigma_i \epsilon_d^{*m'_d}=(-1)^{m'_d} \sigma^{-m'_d}$, 
the minus signs come from the complex conjugation on $\epsilon_d^{*m'_d} $.
\begin{eqnarray}
P_{kl}=(-1)^{m'_d}\CG{L}{m_L}{S}{m_S}{J}{M} \CG{L'}{m'_L}{S'}{m'_S}{J'}{M'} \CG{\frac{1}{2}}{m_N}{1}{m_d}{S}{m_S} \CG{\frac{1}{2}}{m'_N}{1}{m'_d}{S}{m'_S} \nonumber \\ 
 \bra{L'_,m'_L}f(\cos{\theta})k_kp_l\ket{L,m_L}
\bra{\frac{1}{2},m'_N} \sigma_k \sigma^{-m'_d} \sigma^{m_d} \sigma_l \ket{\frac{1}{2},m_N} \nonumber \\ 
\end{eqnarray}
 We can also switch to spherical indices for $k$ and $l$ 
doing the substitutions:
\begin{equation}
k_k  \sigma_k=(-1)^{m_k}\frac{1}{\sqrt{3}} k Y_{1}^{-m_k}(\hat{\bf{k}}) \sigma^{m_k}
\end{equation}
and 
\begin{equation}
p_l  \sigma_l=(-1)^{m_l}\frac{1}{\sqrt{3}} p Y_{1}^{m_l}(\hat{\bf{p}}) \sigma^{-m_l}.
\end{equation}
The placement of the minus sign on the Pauli matrix 
or the spherical harmonic is un-important; the equations are true in both ways. With these substitutions we get:
\begin{eqnarray}
&&P_{kl}=(-1)^{m'_d+m_k+m_l}\CG{L}{m_L}{S}{m_S}{J}{M} \CG{L'}{m'_L}{S'}{m'_S}{J'}{M'} \CG{\frac{1}{2}}{m_N}{1}{m_d}{S}{m_S} \CG{\frac{1}{2}}{m'_N}{1}{m'_d}{S}{m'_S} \nonumber \\ 
&& \frac{1}{3}kp\bra{L'_,m'_L}f(\cos{\theta})Y_{1}^{-m_k}(\hat{\bf{k}}) Y_{1}^{m_l}(\hat{\bf{p}}) \ket{L,m_L}
\bra{\frac{1}{2},m'_N} \sigma^{m_k} \sigma^{-m'_d} \sigma^{m_d} \sigma^{-m_l} \ket{\frac{1}{2},m_N} \nonumber \\ 
\end{eqnarray}
Here we recognize the first matrix element on the second line as the Eq.~\eqref{kpint}. Doing the substitution we find:
\begin{eqnarray}
&&P_{kl}=(-1)^{m'_d+m_k+m_l}\CG{L}{m_L}{S}{m_S}{J}{M} \CG{L'}{m'_L}{S'}{m'_S}{J'}{M'} \CG{\frac{1}{2}}{m_N}{1}{m_d}{S}{m_S} \CG{\frac{1}{2}}{m'_N}{1}{m'_d}{S}{m'_S} \nonumber \\ 
&&kp \sum_{L''}  \sqrt{\frac{\bar{L}}{\bar{L'}}} \CG{L}{m_L}{1}{-m_k}{L''}{m''} \CG{L''}{m''}{1}{m_l}{L'}{m'_L}  \CG{L}{0}{1}{0}{L''}{0} \CG{L''}{0}{1}{0}{L'}{0} \frac{1}{2}\int_{-1}^{1}P_{L''}(x) f(x)dx
\nonumber \\ 
&&\bra{\frac{1}{2},m'_N} \sigma^{m_k} \sigma^{-m'_d} \sigma^{m_d} \sigma^{-m_l} \ket{\frac{1}{2},m_N} \nonumber \\ 
\end{eqnarray}
The summation over $m''$ is understood as are summations over other magnetic quantum numbers. After rearranging the terms in 
previous equation and denoting $\frac{1}{2}\int_{-1}^{1}P_{L''}(x) f(x)dx=\tilde{Q}_{L''}$ we have:
\begin{eqnarray}
&&P_{kl}=kp \sum_{L''} \tilde{Q}_{L''} \sqrt{\frac{\bar{L}}{\bar{L'}}} \CG{L}{0}{1}{0}{L''}{0} \CG{L''}{0}{1}{0}{L'}{0}
\nonumber \\ 
&&(-1)^{m'_d+m_k+m_l}\CG{L}{m_L}{S}{m_S}{J}{M} \CG{L'}{m'_L}{S'}{m'_S}{J'}{M'} \CG{\frac{1}{2}}{m_N}{1}{m_d}{S}{m_S} \CG{\frac{1}{2}}{m'_N}{1}{m'_d}{S}{m'_S} \nonumber \\ 
&&  \CG{L}{m_L}{1}{-m_k}{L''}{m''} \CG{L''}{m''}{1}{m_l}{L'}{m'_L}  \nonumber \\ 
&&\bra{\frac{1}{2},m'_N} \sigma^{m_k} \sigma^{-m'_d} \sigma^{m_d} \sigma^{-m_l} \ket{\frac{1}{2},m_N} \nonumber \\ 
\end{eqnarray}
Now we can insert complete set of states $\ket{\frac{1}{2},m^{(1)}}\bra{\frac{1}{2},m^{(1)}}$, 
$\ket{\frac{1}{2},m^{(2)}}\bra{\frac{1}{2},m^{(2)}}$, $\ket{\frac{1}{2},m^{(3)}}\bra{\frac{1}{2},m^{(3)}}$ between the Pauli matrices and 
using the fact that $\bra{\frac{1}{2},m_1}\sigma^{m}\ket{\frac{1}{2},m_2}=\sqrt{3}\CG{\frac{1}{2}}{m_2}{1}{m}{\frac{1}{2}}{m_1}$, which 
is proven using the Wigner-Eckart Theorem \cite{Sakurai:1167961}, we find:
\begin{eqnarray}
&&P_{kl}=9 kp \sum_{L''} \tilde{Q}_{L''} \sqrt{\frac{\bar{L}}{\bar{L'}}} \CG{L}{0}{1}{0}{L''}{0} \CG{L''}{0}{1}{0}{L'}{0}
\nonumber \\ 
&&(-1)^{m'_d+m_k+m_l}\CG{L}{m_L}{S}{m_S}{J}{M} \CG{L'}{m'_L}{S'}{m'_S}{J'}{M'} \CG{\frac{1}{2}}{m_N}{1}{m_d}{S}{m_S} \CG{\frac{1}{2}}{m'_N}{1}{m'_d}{S}{m'_S} \nonumber \\ 
&&  \CG{L}{m_L}{1}{-m_k}{L''}{m''} \CG{L''}{m''}{1}{m_l}{L'}{m'_L}  \nonumber \\ 
&& \CG{\frac{1}{2}}{m^{(1)}}{1}{m_k}{\frac{1}{2}}{m'_N}
 \CG{\frac{1}{2}}{m^{(2)}}{1}{-m'_d}{\frac{1}{2}}{m^{(1)}} 
  \CG{\frac{1}{2}}{m^{(3)}}{1}{m_d}{\frac{1}{2}}{m^{(2)}}
   \CG{\frac{1}{2}}{m_N}{1}{-m_l}{\frac{1}{2}}{m^{(3)}}  \nonumber \\ 
\end{eqnarray}
As I described above we found an expression which is a sum of products of Clebsch-Gordan coefficients and the sum runs over 
all of the magnetic quantum numbers except $M$ and $M'$. To proceed from here it is convenient to start using the diagrammatic 
notation for Clebsch-Gordan coefficients. 
\begin{eqnarray}
&&P_{kl}=9 kp \sum_{L''} \tilde{Q}_{L''} \sqrt{\frac{\bar{L}}{\bar{L'}}} \CG{L}{0}{1}{0}{L''}{0} \CG{L''}{0}{1}{0}{L'}{0} \nonumber \\ 
&&(-1)^{m'_d+m_k+m_l}
\StrobelLine{m_S}{m_L}{M}\hspace{\StrobelDistance}
 \StrobelLine{m'_S}{m'_L}{M'}\hspace{\StrobelDistance}
 \StrobelLine{m_d}{m_N}{m_S}\hspace{\StrobelDistance}
 \StrobelLine{m'_d}{m'_N}{m'_S}  \hspace{\StrobelDistance}  \StrobelLine{-m_k}{m_L}{m''}\hspace{\StrobelDistance}
  \StrobelLine{m_l}{m''}{m'_L} \hspace{\StrobelDistance} \StrobelLine{m_k}{m^{(1)}}{m'_N}\hspace{\StrobelDistance}
\StrobelLine{-m'_d}{m^{(2)}}{m^{(1)}}\hspace{\StrobelDistance}
 \StrobelLine{m_d} {m^{(3)}} {m^{(2)}}\hspace{\StrobelDistance}
\StrobelLine{-m_l}{m_N}{m^{(3)}} \nonumber \\ 
\end{eqnarray}
Now we can use Eq.~\eqref{cgproperties} to make the signs of all the magnetic quantum numbers to be the same in our expression:
\begin{eqnarray}
&&P_{kl}=9 kp \sum_{L''} \tilde{Q}_{L''} \sqrt{\frac{\bar{L}}{\bar{L'}}} \CG{L}{0}{1}{0}{L''}{0} \CG{L''}{0}{1}{0}{L'}{0} \nonumber \\ 
&&(-1)^{m'_d+m_k+m_l}
\StrobelLine{m_S}{m_L}{M}\hspace{\StrobelDistance}
 \StrobelLine{m'_S}{m'_L}{M'}\hspace{\StrobelDistance}
 \StrobelLine{m_d}{m_N}{m_S}\hspace{\StrobelDistance}
 \StrobelLine{m'_d}{m'_N}{m'_S}    \hspace{\StrobelDistance}
 \StrobelLine{m''}{m_k}{m_L} \hspace{\StrobelDistance}  \nonumber \\ 
 && \left(\frac{\bar{L''}}{\bar{L}}\right)^{\frac{1}{2}} (-1)^{1-m_k}
  \StrobelLine{m_l}{m''}{m'_L}  \hspace{\StrobelDistance}
  \StrobelLine{m_k}{m^{(1)}}{m'_N}\hspace{\StrobelDistance}
\StrobelLine{m^{(1)}}{m'_d}{m^{(2)}}  \nonumber \\  
&&(-1)^{1-m'_d}\hspace{\StrobelDistance}
 \StrobelLine{m_d} {m^{(3)}} {m^{(2)}}\hspace{\StrobelDistance}
\StrobelLine{m^{(3)}}{m_l}{m_N}  (-1)^{1-m_l} \hspace{\StrobelDistance} \nonumber \\ 
&=&-9 kp \sum_{L''} \tilde{Q}_{L''} \sqrt{\frac{\bar{L''}}{\bar{L'}}} \CG{L}{0}{1}{0}{L''}{0} \CG{L''}{0}{1}{0}{L'}{0} \nonumber \\ 
&&
\StrobelLine{m_S}{m_L}{M}\hspace{\StrobelDistance}
 \StrobelLine{m'_S}{m'_L}{M'}\hspace{\StrobelDistance}
 \StrobelLine{m_d}{m_N}{m_S}\hspace{\StrobelDistance}
 \StrobelLine{m'_d}{m'_N}{m'_S}    \hspace{\StrobelDistance}
 \StrobelLine{m''}{m_k}{m_L} \hspace{\StrobelDistance}
  \StrobelLine{m_l}{m''}{m'_L}  \hspace{\StrobelDistance}
  \StrobelLine{m_k}{m^{(1)}}{m'_N}\hspace{\StrobelDistance}
\StrobelLine{m^{(1)}}{m'_d}{m^{(2)}} \hspace{\StrobelDistance}
 \StrobelLine{m_d} {m^{(3)}} {m^{(2)}}\hspace{\StrobelDistance}
\StrobelLine{m^{(3)}}{m_l}{m_N}   \nonumber \\ 
\end{eqnarray}
Now again using Eq.~\eqref{cgproperties} to change the order of the first two entries in two of the arrows and rearranging the terms we find: 
\begin{eqnarray}
&&P_{kl}=-9 kp \sum_{L''} \tilde{Q}_{L''} \sqrt{\frac{\bar{L''}}{\bar{L'}}} \CG{L}{0}{1}{0}{L''}{0} \CG{L''}{0}{1}{0}{L'}{0} \nonumber \\ 
&& \StrobelLine{m_S}{m_L}{M} \hspace{\StrobelDistance}
 \StrobelLine{m'_S}{m'_L}{M'}\hspace{\StrobelDistance}
 \StrobelLine{m''}{m_k}{m_L} \hspace{\StrobelDistance}
  \StrobelLine{m_l}{m''}{m'_L}  \hspace{\StrobelDistance}
  \StrobelLine{m_k}{m^{(1)}}{m'_N}\hspace{\StrobelDistance}
\StrobelLine{m'_d}{m^{(1)}}{m^{(2)}} \hspace{\StrobelDistance}
 \StrobelLine{m'_d}{m'_N}{m'_S}  \hspace{\StrobelDistance}
 \StrobelLine{m_d} {m^{(3)}} {m^{(2)}}\hspace{\StrobelDistance}
\StrobelLine{m_l}{m^{(3)}}{m_N}   \hspace{\StrobelDistance}
 \StrobelLine{m_d}{m_N}{m_S} \nonumber \\ 
\end{eqnarray}
Now we see that the last six arrows fall into patterns as in Eq.~ \eqref{tri3}:
\begin{eqnarray}
&&P_{kl}=-9 kp \sum_{L''} \tilde{Q}_{L''} \sqrt{\frac{\bar{L''}}{\bar{L'}}} \CG{L}{0}{1}{0}{L''}{0} \CG{L''}{0}{1}{0}{L'}{0} \nonumber \\ 
&& \StrobelLine{m_S}{m_L}{M}\hspace{\StrobelDistance}
 \StrobelLine{m'_S}{m'_L}{M'}\hspace{\StrobelDistance}
 \StrobelLine{m''}{m_k}{m_L} \hspace{\StrobelDistance}
  \StrobelLine{m_l}{m''}{m'_L}  \hspace{\StrobelDistance}
  \StrobelTriangleB{m'_d}{m'_N}{m'_S}{m^{(2)}}{m_k}{m^{(1)}}
 \StrobelTriangleB{m_d}{m_N}{m_S}{m^{(2)}}{m_l}{m^{(3)}}
 \nonumber \\ 
 &=& -36 kp \sum_{L''} \tilde{Q}_{L''} \sqrt{\frac{\bar{L''}}{\bar{L'}}} \CG{L}{0}{1}{0}{L''}{0} \CG{L''}{0}{1}{0}{L'}{0} \nonumber \\  
 &&\begin{Bmatrix}
 1& \frac{1}{2} & \frac{1}{2} \\[0.4em]
1 & S & \frac{1}{2}
\end{Bmatrix}
\begin{Bmatrix}
 1& \frac{1}{2} & \frac{1}{2} \\[0.4em]
1 & S' & \frac{1}{2}
\end{Bmatrix}
 \StrobelLine{m_S}{m_L}{M} \hspace{\StrobelDistance}
 \StrobelLine{m'_S}{m'_L}{M'}\hspace{\StrobelDistance}
 \StrobelLine{m''}{m_k}{m_L} \hspace{\StrobelDistance}
  \StrobelLine{m_l}{m''}{m'_L}  \hspace{\StrobelDistance}
   \StrobelLine{m_l}{m^{(2)}}{m_S}\hspace{\StrobelDistance}
    \StrobelLine{m_k}{m^{(2)}}{m'_S}\hspace{\StrobelDistance}
\end{eqnarray}
And on the last step we change the order of the first two entries on the first, second and fourth arrows using Eq.~\eqref{cgproperties},
and see that the remaining six arrows fall into the pattern of the Eq.~\eqref{box1} giving the final expression:
\begin{eqnarray}
&&P_{kl}=-36 kp \sum_{L''} \tilde{Q}_{L''} \sqrt{\frac{\bar{L''}}{\bar{L'}}} \CG{L}{0}{1}{0}{L''}{0} \CG{L''}{0}{1}{0}{L'}{0} \nonumber \\  
 &&\begin{Bmatrix}
 1& \frac{1}{2} & \frac{1}{2} \\
1 & S & \frac{1}{2}
\end{Bmatrix}
\begin{Bmatrix}
 1& \frac{1}{2} & \frac{1}{2} \\
1 & S' & \frac{1}{2}
\end{Bmatrix}
 \StrobelLine{m_S}{m_L}{M} \hspace{\StrobelDistance}
 \StrobelLine{m'_S}{m'_L}{M'}\hspace{\StrobelDistance}
 \StrobelLine{m''}{m_k}{m_L} \hspace{\StrobelDistance}
  \StrobelLine{m_l}{m''}{m'_L}  \hspace{\StrobelDistance}
   \StrobelLine{m_l}{m^{(2)}}{m_S}\hspace{\StrobelDistance}
    \StrobelLine{m_k}{m^{(2)}}{m'_S}\nonumber \\  
    &=&-36 kp \sum_{L''} (-1)^{S+L-J} (-1)^{S'+L'-J'} (-1)^{L''+1-L'} \tilde{Q}_{L''} \sqrt{\frac{\bar{L''}}{\bar{L'}}} \CG{L}{0}{1}{0}{L''}{0} \CG{L''}{0}{1}{0}{L'}{0} \nonumber \\  
 &&\begin{Bmatrix}
 1& \frac{1}{2} & \frac{1}{2} \\
1 & S & \frac{1}{2}
\end{Bmatrix}
\begin{Bmatrix}
 1& \frac{1}{2} & \frac{1}{2} \\
1 & S' & \frac{1}{2}
\end{Bmatrix}
 \StrobelLine{m_L}{m_S}{M}\hspace{\StrobelDistance}
 \StrobelLine{m'_L}{m'_S}{M'}\hspace{\StrobelDistance}
 \StrobelLine{m''}{m_k}{m_L} \hspace{\StrobelDistance}
  \StrobelLine{m''}{m_l}{m'_L}  \hspace{\StrobelDistance}
   \StrobelLine{m_l}{m^{(2)}}{m_S}\hspace{\StrobelDistance}
    \StrobelLine{m_k}{m^{(2)}}{m'_S}\nonumber \\     
 &=&36 kp \sum_{L''} (-1)^{S+S'+L+L''-J-J'} \tilde{Q}_{L''} \sqrt{\frac{\bar{L''}}{\bar{L'}}} \CG{L}{0}{1}{0}{L''}{0} \CG{L''}{0}{1}{0}{L'}{0} \nonumber \\  
 &&\begin{Bmatrix}
 1& \frac{1}{2} & \frac{1}{2} \\
1 & S & \frac{1}{2}
\end{Bmatrix}
\begin{Bmatrix}
 1& \frac{1}{2} & \frac{1}{2} \\
1 & S' & \frac{1}{2}
\end{Bmatrix}
\text{\StrobelBoxAA{m''}{m_l}{m'_L}{m_k}{m_L}{m^{(2)}\hspace{-1cm}}{m_S}{m'_S}{M'}{M}}
 \nonumber \\  
  &=&36 kp \sum_{L''} (-1)^{S+S'+L+L''-2J} \tilde{Q}_{L''} \sqrt{\bar{L''} \bar{L} \bar{S} \bar{S'}} \CG{L}{0}{1}{0}{L''}{0} \CG{L''}{0}{1}{0}{L'}{0} \nonumber \\  
 &&\begin{Bmatrix}
 1& \frac{1}{2} & \frac{1}{2} \\
1 & S & \frac{1}{2}
\end{Bmatrix}
\begin{Bmatrix}
 1& \frac{1}{2} & \frac{1}{2} \\
1 & S' & \frac{1}{2}
\end{Bmatrix}
\begin{Bmatrix}
L'' & 1 & L \\
1& \frac{1}{2}  & S\\
L' & S' & J
\end{Bmatrix} 
\delta_{J,J'}\delta_{M,M'}
 \nonumber \\  
\end{eqnarray}
which coincides with Eq.~\eqref{sampleprojection}, from which some parts are suppressed because they are common in all the terms. 
The Kronecker deltas in the final answer confirm that the operators we had preserve the total angular momentum and its projection. 
We see that in the final expression we have none of the magnetic quantum numbers, those are ``integrated over'' and the answers depends 
only on the angular momenta. 

\section{Two-body $P$-wave contributions to $nd$ scattering, one-loop diagram: Projections}

Here I give the results of projecting all the terms in Eq.~\eqref{oneloopdiagramamplitude}, which are calculated using the techniques described in this Appendix. Once these answers are 
combined with the orbital angular-momentum projections the answer needs to be summed over $L''$. In general in all the answers that I give in this Appendix, all the angular momenta that are not $S$, $S'$, $L$, $L'$ and $J$ are to be summed over. 

\subsection{Deuteron to Deuteron Projections Spin}

$\delta_{kl}$ -term:

\begin{equation}
  =36\delta_{S S'}\delta_{L L'}
  \begin{Bmatrix}
1 & \frac{1}{2} & \frac{1}{2} \\
1 & S' & \frac{1}{2} 
\end{Bmatrix}
\begin{Bmatrix}
1 & \frac{1}{2} & \frac{1}{2} \\
1 & S & \frac{1}{2} 
\end{Bmatrix} 
\end{equation}

$k_k p_l$ -term:

\begin{equation} \label{sampleprojection}
  =36(-1)^{S+S'+L+L''-2J}C^{000}_{L1L''}C^{000}_{L''1L'}(\bar{L} \bar{L}''\bar{S}\bar{S}')^{\frac{1}{2}}
\begin{Bmatrix}
1 & \frac{1}{2} & \frac{1}{2} \\
1 & S' & \frac{1}{2} 
\end{Bmatrix}
\begin{Bmatrix}
1 & \frac{1}{2} & \frac{1}{2} \\
1 & S & \frac{1}{2} 
\end{Bmatrix} 
\begin{Bmatrix}
L'' & 1& L' \\
1 & \frac{1}{2}  & S'\\
L & S& J
\end{Bmatrix} 
\end{equation}

$k_l p_k$ -term:

\begin{equation}
  =36(-1)^{S+S'+L+L''-2J}C^{000}_{L1L''}C^{000}_{L''1L'}(\bar{L} \bar{L}''\bar{S}\bar{S}')^{\frac{1}{2}}
\begin{Bmatrix}
1 & \frac{1}{2} & \frac{1}{2} \\
1 & S' & \frac{1}{2} 
\end{Bmatrix}
\begin{Bmatrix}
1 & \frac{1}{2} & \frac{1}{2} \\
1 & S & \frac{1}{2} 
\end{Bmatrix} 
\begin{Bmatrix}
\frac{1}{2} & 1& S' \\
L' & J& L''
\end{Bmatrix} 
\begin{Bmatrix}
L & 1& L'' \\
\frac{1}{2}  & J& S
\end{Bmatrix} 
\end{equation}

$k_l k_k$ -term:

\begin{equation}
  =36(-1)^{\frac{3}{2}+2S'+L+L''-J}C^{000}_{11L''}C^{000}_{LL''L'}(\bar{L} \bar{L}''\bar{S}\bar{S}')^{\frac{1}{2}}
\begin{Bmatrix}
1 & \frac{1}{2} & \frac{1}{2} \\
1 & S' & \frac{1}{2} 
\end{Bmatrix}
\begin{Bmatrix}
1 & \frac{1}{2} & \frac{1}{2} \\
1 & S & \frac{1}{2} 
\end{Bmatrix} 
\begin{Bmatrix}
L'' & 1& 1 \\
\frac{1}{2} & S& S'
\end{Bmatrix} 
\begin{Bmatrix}
S' & L''& S \\
L  & J& L'
\end{Bmatrix} 
\end{equation}

$^1P_1$-term:
\begin{equation}
=3 \delta_{S \frac{1}{2}}\delta_{S' \frac{1}{2}}\delta_{L L'}
\end{equation}

\subsection{Singlet to Deuteron Projections Spin}

$\delta_{kl}$ -term:

\begin{equation}
  =6\sqrt{3}\delta_{S' \frac{1}{2}}\delta_{LL'}\begin{Bmatrix}
1 & \frac{1}{2} & \frac{1}{2} \\
1 & S' & \frac{1}{2} 
\end{Bmatrix}
\end{equation}

$k_k p_l$ -term:

\begin{equation}
  =6\sqrt{3}(-1)^{\frac{1}{2}+S'+L+L''-2J}C^{000}_{L1L''}C^{000}_{L''1L'}(\bar{L} \bar{L}''\bar{S}\bar{S}')^{\frac{1}{2}}
\begin{Bmatrix}
1 & \frac{1}{2} & \frac{1}{2} \\
1 & S' & \frac{1}{2} 
\end{Bmatrix}
\begin{Bmatrix}
L'' & 1& L' \\
1 & \frac{1}{2}  & S'\\
L &\frac{1}{2} & J
\end{Bmatrix} 
\end{equation}

$k_l p_k$ -term:

\begin{equation}
  =6\sqrt{3}(-1)^{\frac{1}{2}+S'+L+L''-2J}C^{000}_{L1L''}C^{000}_{L''1L'}(\bar{L} \bar{L}''\bar{S}\bar{S}')^{\frac{1}{2}}
\begin{Bmatrix}
1 & \frac{1}{2} & \frac{1}{2} \\
1 & S' & \frac{1}{2} 
\end{Bmatrix}
\begin{Bmatrix}
\frac{1}{2} & 1& S' \\
L' & J& L''
\end{Bmatrix} 
\begin{Bmatrix}
L & 1& L'' \\
\frac{1}{2}  & J& \frac{1}{2} 
\end{Bmatrix} 
\end{equation}

$k_l k_k$ -term:

\begin{equation}
  =6\sqrt{3}(-1)^{\frac{3}{2}+2S'+L+L''-J}C^{000}_{11L''}C^{000}_{LL''L'}(\bar{L} \bar{L}''\bar{S}\bar{S}')^{\frac{1}{2}}
\begin{Bmatrix}
1 & \frac{1}{2} & \frac{1}{2} \\
1 & S' & \frac{1}{2} 
\end{Bmatrix}
\begin{Bmatrix}
L'' & 1& 1 \\
\frac{1}{2} & \frac{1}{2} & S'
\end{Bmatrix} 
\begin{Bmatrix}
S' & L''& \frac{1}{2}  \\
L  & J& L'
\end{Bmatrix} 
\end{equation}

$^1P_1$-term:
\begin{equation}
=\sqrt{3} \delta_{S' \frac{1}{2}}\delta_{L L'}
\end{equation}

\subsection{Deuteron to Singlet Projections Spin}

$\delta_{kl}$ -term:

\begin{equation}
  =6\sqrt{3}\delta_{S \frac{1}{2}}\delta_{LL'}\begin{Bmatrix}
1 & \frac{1}{2} & \frac{1}{2} \\
1 & S & \frac{1}{2} 
\end{Bmatrix}
\end{equation}

$k_k p_l$ -term:

\begin{equation}
  =6\sqrt{3}(-1)^{\frac{1}{2}+S+L+L''-2J}C^{000}_{L1L''}C^{000}_{L''1L'}(\bar{L} \bar{L}''\bar{S}\bar{S}')^{\frac{1}{2}}
\begin{Bmatrix}
1 & \frac{1}{2} & \frac{1}{2} \\
1 & S & \frac{1}{2} 
\end{Bmatrix}
\begin{Bmatrix}
L'' & 1& L' \\
1 & \frac{1}{2}  & \frac{1}{2}\\
L & S & J
\end{Bmatrix} 
\end{equation}

$k_l p_k$ -term:

\begin{equation}
  =6\sqrt{3}(-1)^{\frac{1}{2}+S+L+L''-2J}C^{000}_{L1L''}C^{000}_{L''1L'}(\bar{L} \bar{L}''\bar{S}\bar{S}')^{\frac{1}{2}}
\begin{Bmatrix}
1 & \frac{1}{2} & \frac{1}{2} \\
1 & S & \frac{1}{2} 
\end{Bmatrix}
\begin{Bmatrix}
\frac{1}{2} & 1& \frac{1}{2} \\
L' & J& L''
\end{Bmatrix} 
\begin{Bmatrix}
L & 1& L'' \\
\frac{1}{2}  & J& S 
\end{Bmatrix} 
\end{equation}

$k_l k_k$ -term:

\begin{equation}
  =6\sqrt{3}(-1)^{\frac{1}{2}+L+L''-J}C^{000}_{11L''}C^{000}_{LL''L'}(\bar{L} \bar{L}''\bar{S}\bar{S}')^{\frac{1}{2}}
\begin{Bmatrix}
1 & \frac{1}{2} & \frac{1}{2} \\
1 & S & \frac{1}{2} 
\end{Bmatrix}
\begin{Bmatrix}
L'' & 1& 1 \\
\frac{1}{2} & S & \frac{1}{2}
\end{Bmatrix} 
\begin{Bmatrix}
S' & L''& S \\
L  & J& L'
\end{Bmatrix} 
\end{equation}

$^1P_1$-term:
\begin{equation}
=\sqrt{3} \delta_{S \frac{1}{2}}\delta_{L L'}
\end{equation}

\subsection{Singlet to Singlet Projections Spin}

$\delta_{kl}$ -term:

\begin{equation}
  =3\delta_{LL'}
\end{equation}

$k_k p_l$ -term:

\begin{equation}
  =3(-1)^{1+L+L''-2J}C^{000}_{L1L''}C^{000}_{L''1L'}(\bar{L} \bar{L}''\bar{S}\bar{S}')^{\frac{1}{2}}
\begin{Bmatrix}
L'' & 1& L' \\
1 & \frac{1}{2}  & \frac{1}{2}\\
L & \frac{1}{2} & J
\end{Bmatrix} 
\end{equation}

$k_l p_k$ -term:

\begin{equation}
  =3(-1)^{1+L+L''-2J}C^{000}_{L1L''}C^{000}_{L''1L'}(\bar{L} \bar{L}''\bar{S}\bar{S}')^{\frac{1}{2}}
\begin{Bmatrix}
\frac{1}{2} & 1& \frac{1}{2} \\
L' & J& L''
\end{Bmatrix} 
\begin{Bmatrix}
L & 1& L'' \\
\frac{1}{2}  & J& \frac{1}{2}  
\end{Bmatrix} 
\end{equation}

$k_l k_k$ -term:

\begin{equation}
  =3(-1)^{\frac{1}{2}+L+L''-J}C^{000}_{11L''}C^{000}_{LL''L'}(\bar{L} \bar{L}''\bar{S}\bar{S}')^{\frac{1}{2}}
\begin{Bmatrix}
L'' & 1& 1 \\
\frac{1}{2} & \frac{1}{2} & \frac{1}{2}
\end{Bmatrix} 
\begin{Bmatrix}
S' & L''& \frac{1}{2} \\
L  & J& L'
\end{Bmatrix} 
\end{equation}

$^1P_1$-term:
\begin{equation}
= \delta_{L L'}
\end{equation}

\subsection{Isospin Projections}

Deuteron to Deuteron $3^P_J$-terms: 3
\vspace{0.5cm}

Deuteron to Deuteron $1^P_1$-terms: 1
\vspace{0.5cm}

Singlet to Deuteron $3^P_J$-terms: $-\sqrt{3}$
\vspace{0.5cm}

Singlet to Deuteron $1^P_1$-terms: $\sqrt{3}$
\vspace{0.5cm}

Deuteron to Singlet $3^P_J$-terms: $-\sqrt{3}$
\vspace{0.5cm}

Deuteron to Singlet $1^P_1$-terms: $\sqrt{3}$
\vspace{0.5cm}

Singlet to Singlet $3^P_J$-terms: 1
\vspace{0.5cm}

Singlet to Singlet $1^P_1$-terms: 3

\section{Magnetic photon exchange contributions to $nd$ scattering, one-loop vertex-photon diagram: Projections}

Here I give the results of projecting all the terms in the diagram in Fig.~\ref{photonvertex}. The Lagrangian that describes the photon vertices are given in Appendix \ref{RelevantIntegrals}. 

\begin{center}\begin{figure}[ht]
  \begin{center}
  \includegraphics[scale=0.07]{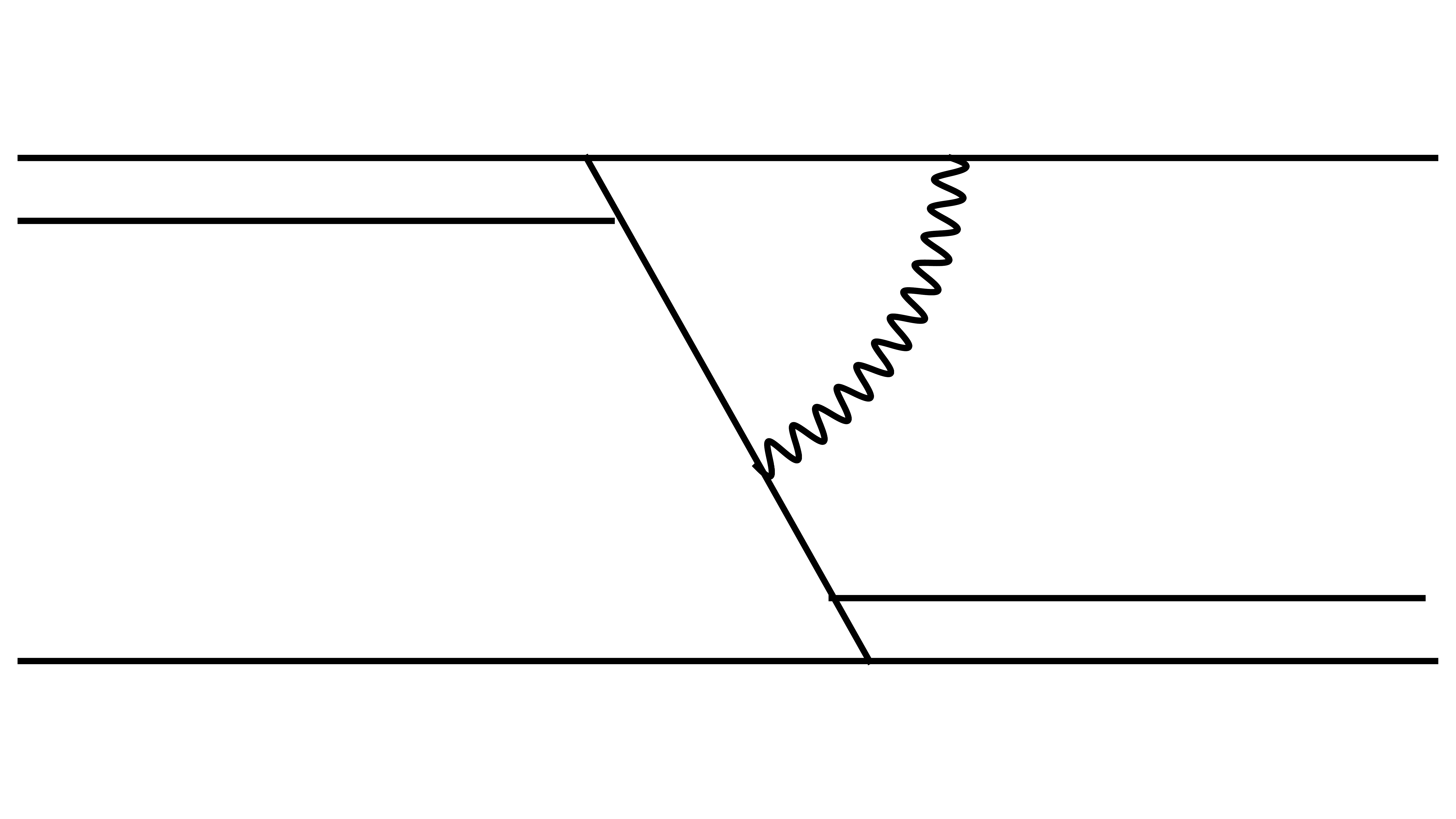}
  \end{center}
  \caption{\label{photonvertex}Magnetic photon exchange on the vertex.}
\end{figure}\end{center}

\subsection{Deuteron to Deuteron Projections Spin}

$\delta_{kl}$ -term:

\begin{equation}
  =36\delta_{S S'}\delta_{L L'}\begin{Bmatrix}
1 & \frac{1}{2} & \frac{1}{2} \\
1 & S & \frac{1}{2} 
\end{Bmatrix}
\begin{Bmatrix}
1 & \frac{1}{2} & \frac{1}{2} \\
1 & \frac{1}{2} & \frac{1}{2} 
\end{Bmatrix} 
\end{equation}

$k_k p_l$ -term:

\begin{eqnarray}
 && =36(-1)^{1+L+S-S'+L''+2J_4-2J}C^{000}_{L1L''}C^{000}_{L''1L'}\bar{J_4}\bar{J_6}(\bar{L} \bar{L}''\bar{S}\bar{S}')^{\frac{1}{2}} \nonumber \\
&\times& \begin{Bmatrix}
1 & \frac{1}{2} & \frac{1}{2} \\
1 & J_4 & \frac{1}{2} 
\end{Bmatrix}
\begin{Bmatrix}
1 & \frac{1}{2} & \frac{1}{2} \\
1 & S & J_4 
\end{Bmatrix} 
\begin{Bmatrix}
1 & \frac{1}{2} & \frac{1}{2} \\
1 & J_6 & S'
\end{Bmatrix} 
\begin{Bmatrix}
1 & \frac{1}{2} & J_4 \\
1 & S & J_6
\end{Bmatrix} 
\begin{Bmatrix}
L & 1& L'' \\
J_6 & J & S
\end{Bmatrix} 
\begin{Bmatrix}
L'' & 1& L' \\
S' & J  & J_6
\end{Bmatrix} 
\end{eqnarray}

$k_l p_k$ -term:

\begin{eqnarray}
 && =36(-1)^{1-L+L''-S+S'+2J_4+2J_6}C^{000}_{L1L''}C^{000}_{L''1L'}\bar{J_4}\bar{J_6}(\bar{L} \bar{L}''\bar{S}\bar{S}')^{\frac{1}{2}}\nonumber \\
&\times&\begin{Bmatrix}
1 & \frac{1}{2} & \frac{1}{2} \\
1 & J_4 & \frac{1}{2} 
\end{Bmatrix}
\begin{Bmatrix}
1 & \frac{1}{2} & \frac{1}{2} \\
1 & S & J_4 
\end{Bmatrix} 
\begin{Bmatrix}
1 & \frac{1}{2} & \frac{1}{2} \\
1 & J_6 & S'
\end{Bmatrix} 
\begin{Bmatrix}
1 & \frac{1}{2} & J_4 \\
1 & S & J_6
\end{Bmatrix} 
\begin{Bmatrix}
J_6 & 1 & S'\\
1 & L'' & L' \\
S & L & J
\end{Bmatrix} 
\end{eqnarray}

$k_l k_k$ -term:

\begin{eqnarray}
 && =36(-1)^{1+2S'+S-J}C^{000}_{11L''}C^{000}_{LL''L'}\bar{J_4}(\bar{L} \bar{L}''\bar{S}\bar{S}')^{\frac{1}{2}}\nonumber \\
&\times&\begin{Bmatrix}
1 & \frac{1}{2} & \frac{1}{2} \\
1 & J_4 & \frac{1}{2} 
\end{Bmatrix}
\begin{Bmatrix}
1 & \frac{1}{2} & \frac{1}{2} \\
1 & S & J_4 
\end{Bmatrix} 
\begin{Bmatrix}
1& 1 & L'' \\
\frac{1}{2} & J_4& \frac{1}{2}
\end{Bmatrix} 
\begin{Bmatrix}
1 & \frac{1}{2}& S' \\
L''  & S& J_4
\end{Bmatrix} 
\begin{Bmatrix}
S' & L''& S \\
L  & J& L'
\end{Bmatrix} 
\end{eqnarray}

\subsection{Singlet to Deuteron Projections Spin}

$\delta_{kl}$ -term:

\begin{equation}
  =3\delta_{S' \frac{1}{2}}\delta_{S \frac{1}{2}}\delta_{LL'}
\end{equation}

$k_k p_l$ -term:

\begin{equation}
  =6\sqrt{6}(-1)^{\frac{3}{2}+3S'+L+L''}C^{000}_{L1L''}C^{000}_{L''1L'}\bar{J_4}(\bar{L} \bar{L}''\bar{S}')^{\frac{1}{2}}
\begin{Bmatrix}
1 & \frac{1}{2} & \frac{1}{2} \\
1 & J_4 & \frac{1}{2} 
\end{Bmatrix}
\begin{Bmatrix}
1 & \frac{1}{2} & \frac{1}{2} \\
1 & J_4 & S' 
\end{Bmatrix}
\begin{Bmatrix}
L'' & 1& L \\
\frac{1}{2}  & J & J_4
\end{Bmatrix} 
\begin{Bmatrix}
L'' & 1& L' \\
S'  & J & J_4
\end{Bmatrix} 
\end{equation}

$k_l p_k$ -term:

\begin{equation}
  =12\sqrt{3}(-1)^{\frac{1}{2}+S'-L+L''}C^{000}_{L1L''}C^{000}_{L''1L'}\bar{J_4}(\bar{L} \bar{L}'')^{\frac{1}{2}}
\begin{Bmatrix}
1 & \frac{1}{2} & \frac{1}{2} \\
1 & J_4 & \frac{1}{2} 
\end{Bmatrix}
\begin{Bmatrix}
1 & \frac{1}{2} & \frac{1}{2} \\
1 & J_4 & S' 
\end{Bmatrix}
\begin{Bmatrix}
J_4 & 1 & S' \\
1 & L'' & L'\\
\frac{1}{2}  & L & J
\end{Bmatrix} 
\end{equation}

$k_l k_k$ -term:

\begin{equation}
  =6\sqrt{6}(-1)^{1+L+J_4-J}C^{000}_{11L''}C^{000}_{LL''L'}\bar{J_4}(\bar{L} \bar{L}''\bar{S}')^{\frac{1}{2}}
\begin{Bmatrix}
1 & \frac{1}{2} & \frac{1}{2} \\
1 & J_4 & \frac{1}{2} 
\end{Bmatrix}
\begin{Bmatrix}
1 & \frac{1}{2} & \frac{1}{2} \\
1 & J_4 & S' 
\end{Bmatrix}
\begin{Bmatrix}
1 & 1 & L''  \\
S' & \frac{1}{2} & J_4
\end{Bmatrix} 
\begin{Bmatrix}
L & L'' & L'  \\
S' & J & \frac{1}{2}
\end{Bmatrix} 
\end{equation}

\subsection{Deuteron to Singlet Projections Spin}

$\delta_{kl}$ -term:

\begin{equation}
  =6\sqrt{3}\delta_{S \frac{1}{2}}\delta_{LL'}\begin{Bmatrix}
1 & \frac{1}{2} & \frac{1}{2} \\
1 & \frac{1}{2}  & \frac{1}{2} 
\end{Bmatrix}
\end{equation}

$k_k p_l$ -term:
\begin{equation}
  =6\sqrt{3}(-1)^{\frac{1}{2}+S+L+L''-2J}C^{000}_{L1L''}C^{000}_{L''1L'}(\bar{L} \bar{L}''\bar{S}\bar{S}')^{\frac{1}{2}}
\begin{Bmatrix}
1 & \frac{1}{2} & \frac{1}{2} \\
1 & S & \frac{1}{2} 
\end{Bmatrix}
\begin{Bmatrix}
\frac{1}{2} & 1& \frac{1}{2} \\
L' & J& L''
\end{Bmatrix} 
\begin{Bmatrix}
L & 1& L'' \\
\frac{1}{2}  & J& S 
\end{Bmatrix} 
\end{equation}

$k_l p_k$ -term:

\begin{equation}
  =6\sqrt{3}(-1)^{\frac{1}{2}+S+L+L''-2J}C^{000}_{L1L''}C^{000}_{L''1L'}(\bar{L} \bar{L}''\bar{S}\bar{S}')^{\frac{1}{2}}
\begin{Bmatrix}
1 & \frac{1}{2} & \frac{1}{2} \\
1 & S & \frac{1}{2} 
\end{Bmatrix}
\begin{Bmatrix}
L'' & 1& L' \\
1 & \frac{1}{2}  & \frac{1}{2}\\
L & S & J
\end{Bmatrix} 
\end{equation}

$k_l k_k$ -term:

\begin{equation}
  =6\sqrt{3}(-1)^{\frac{1}{2}+L+L''-J}C^{000}_{11L''}C^{000}_{LL''L'}(\bar{L} \bar{L}''\bar{S}\bar{S}')^{\frac{1}{2}}
\begin{Bmatrix}
1 & \frac{1}{2} & \frac{1}{2} \\
1 & S & \frac{1}{2} 
\end{Bmatrix}
\begin{Bmatrix}
L'' & 1& 1 \\
\frac{1}{2} & S & \frac{1}{2}
\end{Bmatrix} 
\begin{Bmatrix}
S' & L''& S \\
L  & J& L'
\end{Bmatrix} 
\end{equation}

\subsection{Singlet to Singlet Projections Spin}

$\delta_{kl}$ -term:

\begin{equation}
  =3\delta_{LL'}
\end{equation}

$k_k p_l$ -term:

\begin{equation}
  =3(-1)^{1+L+L''-2J}C^{000}_{L1L''}C^{000}_{L''1L'}(\bar{L} \bar{L}''\bar{S}\bar{S}')^{\frac{1}{2}}
\begin{Bmatrix}
L'' & 1& L' \\
1 & \frac{1}{2}  & \frac{1}{2}\\
L & \frac{1}{2} & J
\end{Bmatrix} 
\end{equation}

$k_l p_k$ -term:

\begin{equation}
  =3(-1)^{1+L+L''-2J}C^{000}_{L1L''}C^{000}_{L''1L'}(\bar{L} \bar{L}''\bar{S}\bar{S}')^{\frac{1}{2}}
\begin{Bmatrix}
\frac{1}{2} & 1& \frac{1}{2} \\
L' & J& L''
\end{Bmatrix} 
\begin{Bmatrix}
L & 1& L'' \\
\frac{1}{2}  & J& \frac{1}{2}  
\end{Bmatrix} 
\end{equation}

$k_l k_k$ -term:

\begin{equation}
  =3(-1)^{\frac{1}{2}+L+L''-J}C^{000}_{11L''}C^{000}_{LL''L'}(\bar{L} \bar{L}''\bar{S}\bar{S}')^{\frac{1}{2}}
\begin{Bmatrix}
L'' & 1& 1 \\
\frac{1}{2} & \frac{1}{2} & \frac{1}{2}
\end{Bmatrix} 
\begin{Bmatrix}
S' & L''& \frac{1}{2} \\
L  & J& L'
\end{Bmatrix} 
\end{equation}

\subsection{Isospin Projections}

\hspace{.6cm}Deuteron to Deuteron: $\kappa_{0}^2-\kappa_{1}^2$
\vspace{0.5cm}

Singlet to Deuteron: $\sqrt{3}\kappa_{0}^2+\frac{1}{\sqrt{3}}\kappa_{1}^2-\frac{4}{\sqrt{3}}\kappa_{0}\kappa_{1}$
\vspace{0.5cm}

Deuteron to Singlet: $\sqrt{3}(\kappa_{0}^2-\kappa_{1}^2)$
\vspace{0.5cm}

Singlet to Singlet: $-\kappa_{0}^2-\frac{1}{3}\kappa_{1}^2+\frac{4}{3}\kappa_{0}\kappa_{1}$

\section{Magnetic photon exchange contributions to $nd$ scattering, one-loop bubble diagram: Projections}

Here I give the results of projecting all the terms in the diagram in Fig.~\ref{photonbubble}. The Lagrangian that describes the photon vertices are given in Appendix \ref{RelevantIntegrals}. 

\begin{center}\begin{figure}[ht]
  \begin{center}
  \includegraphics[scale=0.07]{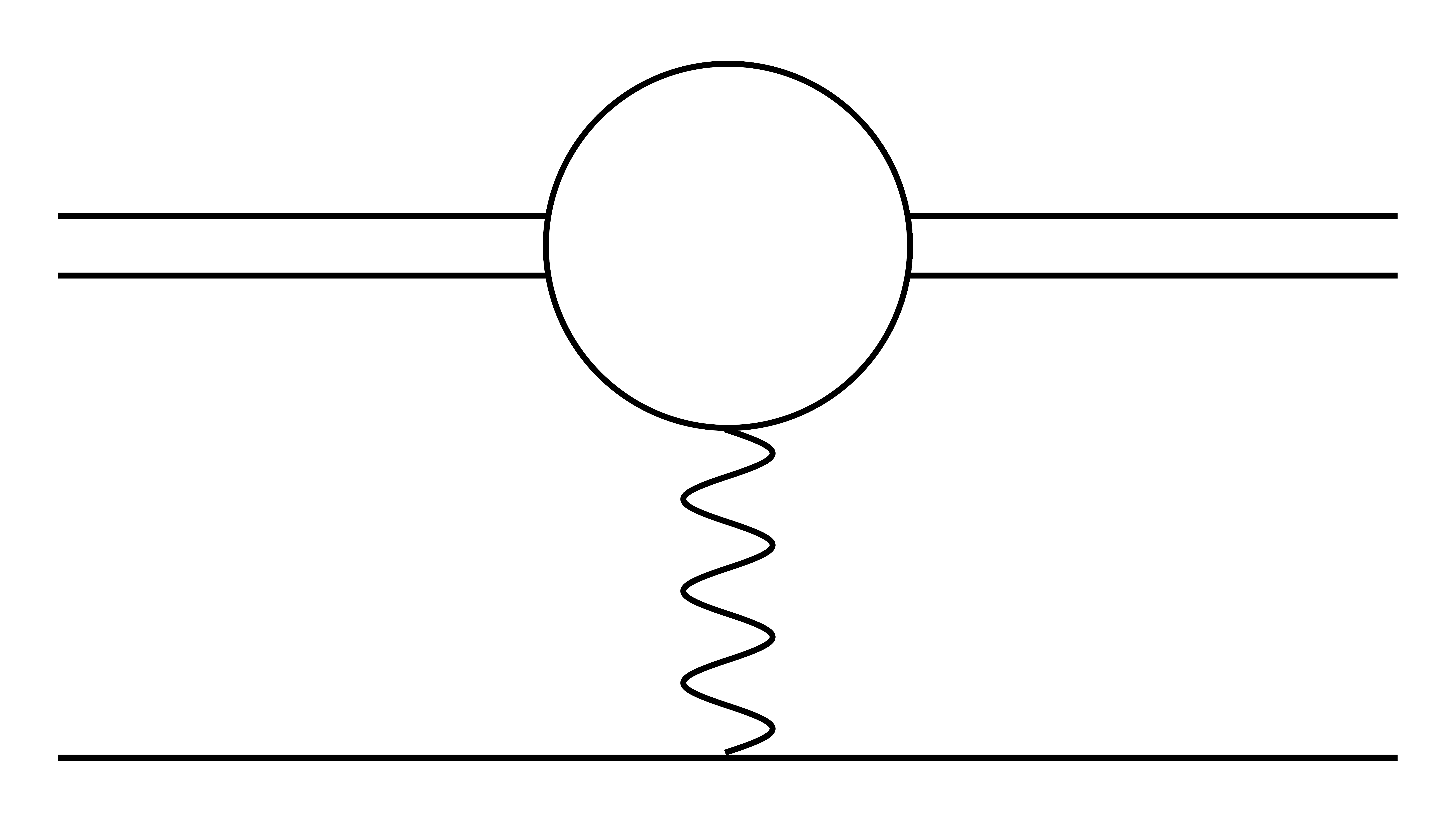}
  \end{center}
  \caption{\label{photonbubble}Magnetic photon exchange on the bubble.}
\end{figure}\end{center}

\subsection{Deuteron to Deuteron Projections Spin}

\hspace{.5cm} $\delta_{kl}$ -term:

\begin{equation}
  =12(-1)^S\delta_{S S'}\delta_{L L'}\begin{Bmatrix}
\frac{1}{2}  & 1 & \frac{1}{2} \\
1 & S & 1 
\end{Bmatrix}
\end{equation}

$k_k p_l$ -term:

\begin{eqnarray}
 && =12(-1)^{1+2S'+L'-L-J+J_1}C^{000}_{L1L''}C^{000}_{L''1L'}\bar{J_1}(\bar{L} \bar{L}''\bar{S}\bar{S}')^{\frac{1}{2}}\nonumber \\
&\times& \begin{Bmatrix}
L' & 1 & L'' \\
\frac{1}{2}  & J_1 & \frac{1}{2} 
\end{Bmatrix}
\begin{Bmatrix}
1 & \frac{1}{2} & S' \\
L' & J & J_1
\end{Bmatrix} 
\begin{Bmatrix}
1 & \frac{1}{2} & S \\
1 & L'' & L\\
1 & J_1 & J
\end{Bmatrix} 
\end{eqnarray}

$k_l p_k$ -term:

\begin{eqnarray}
 && =12(-1)^{\frac{1}{2}+S'+L'-L-J-J_1}C^{000}_{L1L''}C^{000}_{L''1L'}\bar{J_1}(\bar{L} \bar{L}''\bar{S}\bar{S}')^{\frac{1}{2}}\nonumber \\
&\times& \begin{Bmatrix}
L'' & 1 & L' \\
1  & J_1 & 1 
\end{Bmatrix}
\begin{Bmatrix}
 \frac{1}{2} & 1 & S' \\
L' & J & J_1
\end{Bmatrix} 
\begin{Bmatrix}
\frac{1}{2} & 1 & S \\
1 & L'' & L\\
\frac{1}{2} & J_1 & J
\end{Bmatrix} 
\end{eqnarray}

$k_l k_k$ -term:

\begin{eqnarray}
 && =12(-1)^{\frac{1}{2}+2S'+S+L+L'+2J_1}C^{000}_{11L''}C^{000}_{LL''L'}\bar{J_1}(\bar{L}''\bar{L}'\bar{S}\bar{S}')^{\frac{1}{2}}\nonumber \\
&\times&\begin{Bmatrix}
L'' & 1 & 1\\
\frac{1}{2}  & J_1 & \frac{1}{2} 
\end{Bmatrix}
\begin{Bmatrix}
1 & 1 & 1 \\
\frac{1}{2}  & S & J_1
\end{Bmatrix} 
\begin{Bmatrix}
1 & \frac{1}{2}  & S' \\
L'' & S & J_1
\end{Bmatrix} 
\begin{Bmatrix}
S' & L'' & S \\
L & J & L'
\end{Bmatrix} 
\end{eqnarray}

\subsection{Singlet to Deuteron Projections Spin}

$\delta_{kl}$ -term:

\begin{equation}
  =2\sqrt{3}\delta_{S' \frac{1}{2}}\delta_{LL'}
\end{equation}

$k_k p_l$ -term:

\begin{equation}
  =2\sqrt{6}(-1)C^{000}_{L1L''}C^{000}_{L''1L'}(\bar{L} \bar{L}''\bar{S}')^{\frac{1}{2}}
\begin{Bmatrix}
\frac{1}{2} & 1 & S' \\
1 & L''  & L'\\
S & L & J
\end{Bmatrix} 
\end{equation}

$k_l p_k$ -term:

\begin{equation}
  =2\sqrt{6}(-1)^{S'+L+L''-2J}C^{000}_{L1L''}C^{000}_{L''1L'}(\bar{L} \bar{L}''\bar{S}')^{\frac{1}{2}}
\begin{Bmatrix}
L' & 1 & L'' \\
\frac{1}{2} & J & S' 
\end{Bmatrix}
\begin{Bmatrix}
L & 1 & L'' \\
\frac{1}{2} & J & \frac{1}{2}
\end{Bmatrix} 
\end{equation}

$k_l k_k$ -term:

\begin{equation}
  =2\sqrt{6}(-1)^{1+S+2S'+L+L''-J}C^{000}_{11L''}C^{000}_{LL''L'}(\bar{L} \bar{L}''\bar{S}')^{\frac{1}{2}}
\begin{Bmatrix}
L'' & 1 & 1 \\
\frac{1}{2} & \frac{1}{2} & S'
\end{Bmatrix}
\begin{Bmatrix}
S' & L'' & \frac{1}{2} \\
L & J & L'
\end{Bmatrix} 
\end{equation}

\subsection{Deuteron to Singlet Projections Spin}

\hspace{.5cm}$\delta_{kl}$ -term:

\begin{equation}
  =2\sqrt{3}\delta_{S \frac{1}{2}}\delta_{LL'}
\end{equation}

$k_k p_l$ -term:

\begin{equation}
  =2\sqrt{6}(-1)^{\frac{1}{2}+S+L+L''-2J}C^{000}_{L1L''}C^{000}_{L''1L'}(\bar{L}' \bar{L}''\bar{S})^{\frac{1}{2}}
\begin{Bmatrix}
L'' & 1 & L' \\
\frac{1}{2} & J & \frac{1}{2} 
\end{Bmatrix}
\begin{Bmatrix}
L & 1 & L'' \\
\frac{1}{2} & J & S
\end{Bmatrix} 
\end{equation}

$k_l p_k$ -term:

\begin{equation}
  =2\sqrt{6}(-1)^{\frac{1}{2}-S-L+L''}C^{000}_{L1L''}C^{000}_{L''1L'}(\bar{L} \bar{L}''\bar{S})^{\frac{1}{2}}
\begin{Bmatrix}
\frac{1}{2} & 1 & \frac{1}{2} \\
1 & L'' & L' \\
S & L & J 
\end{Bmatrix}
\end{equation}

$k_l k_k$ -term:

\begin{equation}
  =2\sqrt{6}(-1)^{\frac{3}{2}+S}C^{000}_{11L''}C^{000}_{LL''L'}(\bar{L} \bar{L}''\bar{S})^{\frac{1}{2}}
\begin{Bmatrix}
1 & 1 & L'' \\
\frac{1}{2} & S & \frac{1}{2} 
\end{Bmatrix}
\begin{Bmatrix}
L & L'' & L' \\
\frac{1}{2} & J & S
\end{Bmatrix} 
\end{equation}

\subsection{Singlet to Singlet Projections Spin}

0

\subsection{Isospin Projections}

\hspace{.6cm}Deuteron to Deuteron: $2(\kappa_{0}^2-\kappa_{0}\kappa_{1})$
\vspace{0.5cm}

Singlet to Deuteron: $\frac{2}{\sqrt{3}}\kappa_{1}^2-\frac{2}{\sqrt{3}}\kappa_{0}\kappa_{1}$
\vspace{0.5cm}

Deuteron to Singlet: $\frac{2}{\sqrt{3}}\kappa_{1}^2-\frac{2}{\sqrt{3}}\kappa_{0}\kappa_{1}$
\vspace{0.5cm}

Singlet to Singlet: $2\kappa_{0}^2-\frac{4}{3}\kappa_{1}^2-\frac{2}{3}\kappa_{0}\kappa_{1}$

\section{Magnetic photon exchange contributions to $nd$ scattering, one-loop non-planar diagram: Projections}

Here I give the results of projecting all the terms in the diagram in Fig.~\ref{photonnonplanar}. The Lagrangian that describes the photon vertices are given in Appendix \ref{RelevantIntegrals}. 

\begin{center}\begin{figure}[ht]
  \begin{center}
  \includegraphics[scale=0.07]{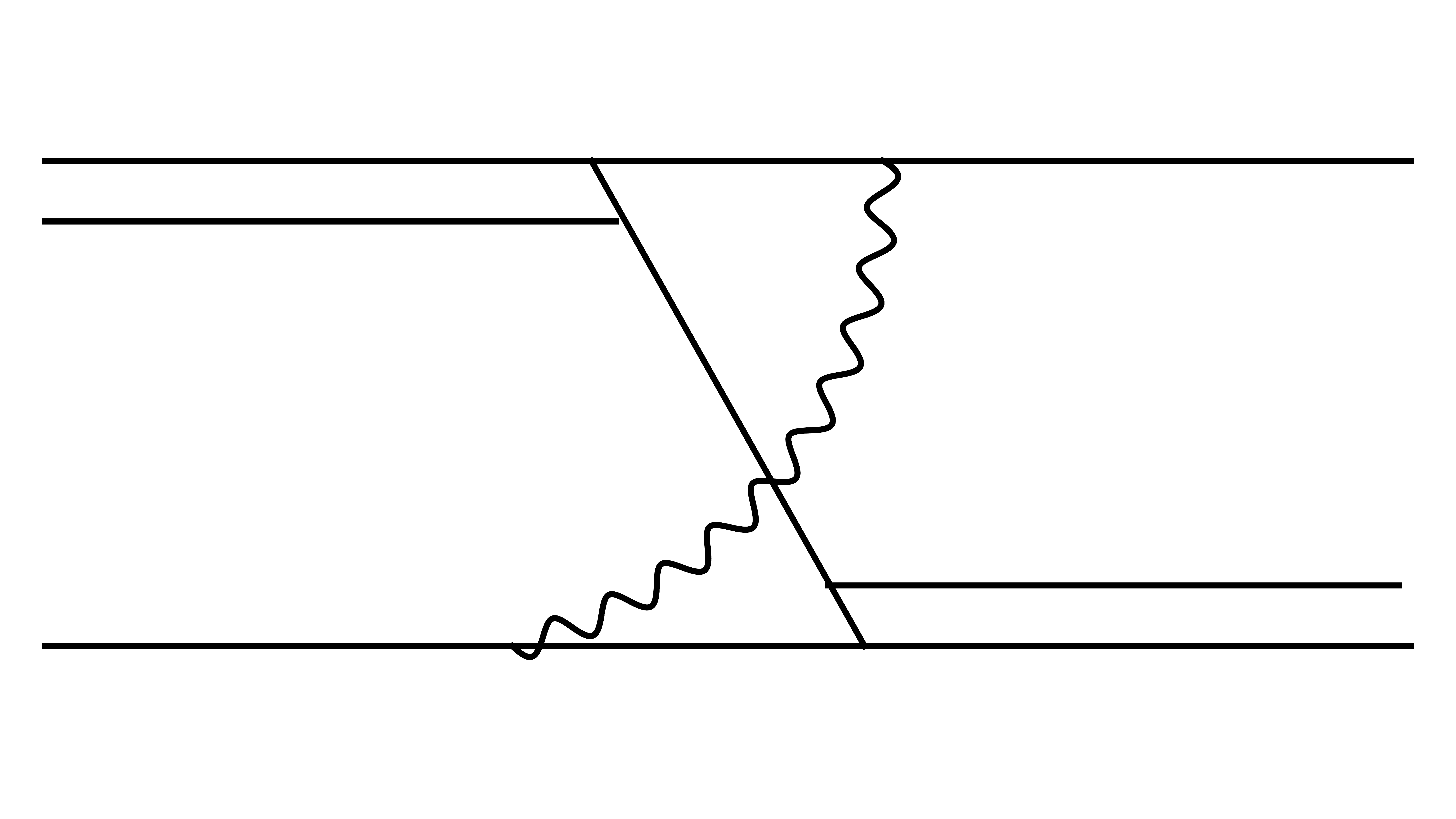}
  \end{center}
  \caption{\label{photonnonplanar}Magnetic photon exchange non-planar diagram.}
\end{figure}\end{center}

\subsection{Deuteron to Deuteron Projections Spin}

\hspace{.5cm}$\delta_{kl}$ -term:

\begin{equation}
  =-36\delta_{S S'}\delta_{L L'}\begin{Bmatrix}
\frac{1}{2}  & \frac{1}{2} & 1 \\
\frac{1}{2} & 1 & \frac{1}{2}\\
1 & \frac{1}{2} & S
\end{Bmatrix}
\end{equation}

$k_k p_l$ -term:

\begin{eqnarray}
 && =36(-1)^{L''-L-S-S'}C^{000}_{L1L''}C^{000}_{L''1L'}\bar{J_1}(\bar{L}' \bar{L}''\bar{S}\bar{S}')^{\frac{1}{2}}\nonumber \\
&\times& \begin{Bmatrix}
1 & \frac{1}{2} & \frac{1}{2} \\
1  & J_1 & \frac{1}{2} 
\end{Bmatrix}
\begin{Bmatrix}
1 & \frac{1}{2} & \frac{1}{2} \\
1 & S & J_1
\end{Bmatrix} 
\begin{Bmatrix}
1 & \frac{1}{2} & \frac{1}{2} \\
1 & S' & J_1
\end{Bmatrix} 
\begin{Bmatrix}
J_1 & 1 & S' \\
1 & L'' & L'\\
S & L & J
\end{Bmatrix} 
\end{eqnarray}

$k_l p_k$ -term:

\begin{eqnarray}
 && =36(-1)^{2J_1+L+L'+L''+S-J}C^{000}_{L1L''}C^{000}_{L''1L'}\bar{J_1}(\bar{L}' \bar{L}''\bar{S}\bar{S}')^{\frac{1}{2}}\nonumber \\
&\times& \begin{Bmatrix}
1 & \frac{1}{2} & \frac{1}{2} \\
1  & J_1 & \frac{1}{2} 
\end{Bmatrix}
\begin{Bmatrix}
1 & \frac{1}{2} & \frac{1}{2} \\
1 & S & J_1
\end{Bmatrix} 
\begin{Bmatrix}
1 & \frac{1}{2} & \frac{1}{2} \\
1 & S' & J_1
\end{Bmatrix} 
\begin{Bmatrix}
L & 1 & L'' \\
J_1 & J & S
\end{Bmatrix} 
\begin{Bmatrix}
L' & 1 & L'' \\
J_1 & J & S'
\end{Bmatrix} 
\end{eqnarray}

$k_l k_k$ -term:

\begin{eqnarray}
 && =36(-1)^{J_1+L+L''+2S'}C^{000}_{11L''}C^{000}_{LL''L'}\bar{J_1}(\bar{L}'\bar{L}''\bar{S}\bar{S}')^{\frac{1}{2}}\nonumber \\
&\times& \begin{Bmatrix}
1 & \frac{1}{2} & \frac{1}{2} \\
1  & J_1 & \frac{1}{2} 
\end{Bmatrix}
\begin{Bmatrix}
1 & \frac{1}{2} & \frac{1}{2} \\
1 & S & J_1
\end{Bmatrix} 
\begin{Bmatrix}
1 & \frac{1}{2} & \frac{1}{2} \\
1 & S' & J_1
\end{Bmatrix} 
\begin{Bmatrix}
J_1 & 1 & S' \\
L'' & S & 1
\end{Bmatrix} 
\begin{Bmatrix}
L & L'' & L' \\
S' & J & S
\end{Bmatrix} 
\end{eqnarray}

\subsection{Singlet to Deuteron Projections Spin}

\hspace{.5cm}$\delta_{kl}$ -term:

\begin{equation}
  =6\sqrt{3}\delta_{S' \frac{1}{2}}\delta_{LL'}\begin{Bmatrix}
1 & \frac{1}{2} & \frac{1}{2} \\
1 & S' & \frac{1}{2} 
\end{Bmatrix}
\end{equation}

$k_k p_l$ -term:

\begin{equation}
  =6\sqrt{3}(-1)^{\frac{1}{2}+S'+L+L''-2J}C^{000}_{L1L''}C^{000}_{L''1L'}(\bar{L} \bar{L}''\bar{S}\bar{S}')^{\frac{1}{2}}
\begin{Bmatrix}
1 & \frac{1}{2} & \frac{1}{2} \\
1 & S' & \frac{1}{2} 
\end{Bmatrix}
\begin{Bmatrix}
L'' & 1& L' \\
1 & \frac{1}{2}  & S'\\
L &\frac{1}{2} & J
\end{Bmatrix} 
\end{equation}

$k_l p_k$ -term:

\begin{equation}
  =6\sqrt{3}(-1)^{\frac{1}{2}+S'+L+L''-2J}C^{000}_{L1L''}C^{000}_{L''1L'}(\bar{L} \bar{L}''\bar{S}\bar{S}')^{\frac{1}{2}}
\begin{Bmatrix}
1 & \frac{1}{2} & \frac{1}{2} \\
1 & S' & \frac{1}{2} 
\end{Bmatrix}
\begin{Bmatrix}
\frac{1}{2} & 1& S' \\
L' & J& L''
\end{Bmatrix} 
\begin{Bmatrix}
L & 1& L'' \\
\frac{1}{2}  & J& \frac{1}{2} 
\end{Bmatrix} 
\end{equation}

$k_l k_k$ -term:

\begin{equation}
  =6\sqrt{3}(-1)^{\frac{3}{2}+2S'+L+L''-J}C^{000}_{11L''}C^{000}_{LL''L'}(\bar{L} \bar{L}''\bar{S}\bar{S}')^{\frac{1}{2}}
\begin{Bmatrix}
1 & \frac{1}{2} & \frac{1}{2} \\
1 & S' & \frac{1}{2} 
\end{Bmatrix}
\begin{Bmatrix}
L'' & 1& 1 \\
\frac{1}{2} & \frac{1}{2} & S'
\end{Bmatrix} 
\begin{Bmatrix}
S' & L''& \frac{1}{2}  \\
L  & J& L'
\end{Bmatrix} 
\end{equation}

\subsection{Deuteron to Singlet Projections Spin}

\hspace{.5cm}$\delta_{kl}$ -term:

\begin{equation}
  =6\sqrt{3}\delta_{S \frac{1}{2}}\delta_{LL'}\begin{Bmatrix}
1 & \frac{1}{2} & \frac{1}{2} \\
1 & S & \frac{1}{2} 
\end{Bmatrix}
\end{equation}

$k_k p_l$ -term:

\begin{equation}
  =6\sqrt{3}(-1)^{\frac{1}{2}+S+L+L''-2J}C^{000}_{L1L''}C^{000}_{L''1L'}(\bar{L} \bar{L}''\bar{S}\bar{S}')^{\frac{1}{2}}
\begin{Bmatrix}
1 & \frac{1}{2} & \frac{1}{2} \\
1 & S & \frac{1}{2} 
\end{Bmatrix}
\begin{Bmatrix}
L'' & 1& L' \\
1 & \frac{1}{2}  & \frac{1}{2}\\
L & S & J
\end{Bmatrix} 
\end{equation}

$k_l p_k$ -term:

\begin{equation}
  =6\sqrt{3}(-1)^{\frac{1}{2}+S+L+L''-2J}C^{000}_{L1L''}C^{000}_{L''1L'}(\bar{L} \bar{L}''\bar{S}\bar{S}')^{\frac{1}{2}}
\begin{Bmatrix}
1 & \frac{1}{2} & \frac{1}{2} \\
1 & S & \frac{1}{2} 
\end{Bmatrix}
\begin{Bmatrix}
\frac{1}{2} & 1& \frac{1}{2} \\
L' & J& L''
\end{Bmatrix} 
\begin{Bmatrix}
L & 1& L'' \\
\frac{1}{2}  & J& S 
\end{Bmatrix} 
\end{equation}

$k_l k_k$ -term:

\begin{equation}
  =6\sqrt{3}(-1)^{\frac{1}{2}+L+L''-J}C^{000}_{11L''}C^{000}_{LL''L'}(\bar{L} \bar{L}''\bar{S}\bar{S}')^{\frac{1}{2}}
\begin{Bmatrix}
1 & \frac{1}{2} & \frac{1}{2} \\
1 & S & \frac{1}{2} 
\end{Bmatrix}
\begin{Bmatrix}
L'' & 1& 1 \\
\frac{1}{2} & S & \frac{1}{2}
\end{Bmatrix} 
\begin{Bmatrix}
S' & L''& S \\
L  & J& L'
\end{Bmatrix} 
\end{equation}

\subsection{Singlet to Singlet Projections Spin}

\hspace{.5cm}$\delta_{kl}$ -term:

\begin{equation}
  =3\delta_{LL'}
\end{equation}

$k_k p_l$ -term:

\begin{equation}
  =3(-1)^{1+L+L''-2J}C^{000}_{L1L''}C^{000}_{L''1L'}(\bar{L} \bar{L}''\bar{S}\bar{S}')^{\frac{1}{2}}
\begin{Bmatrix}
L'' & 1& L' \\
1 & \frac{1}{2}  & \frac{1}{2}\\
L & \frac{1}{2} & J
\end{Bmatrix} 
\end{equation}

$k_l p_k$ -term:

\begin{equation}
  =3(-1)^{1+L+L''-2J}C^{000}_{L1L''}C^{000}_{L''1L'}(\bar{L} \bar{L}''\bar{S}\bar{S}')^{\frac{1}{2}}
\begin{Bmatrix}
\frac{1}{2} & 1& \frac{1}{2} \\
L' & J& L''
\end{Bmatrix} 
\begin{Bmatrix}
L & 1& L'' \\
\frac{1}{2}  & J& \frac{1}{2}  
\end{Bmatrix} 
\end{equation}

$k_l k_k$ -term:

\begin{equation}
  =3(-1)^{\frac{1}{2}+L+L''-J}C^{000}_{11L''}C^{000}_{LL''L'}(\bar{L} \bar{L}''\bar{S}\bar{S}')^{\frac{1}{2}}
\begin{Bmatrix}
L'' & 1& 1 \\
\frac{1}{2} & \frac{1}{2} & \frac{1}{2}
\end{Bmatrix} 
\begin{Bmatrix}
S' & L''& \frac{1}{2} \\
L  & J& L'
\end{Bmatrix} 
\end{equation}

\subsection{Isospin Projections}

\hspace{.6cm}Deuteron to Deuteron: $(\kappa_{0}-\kappa_{1})^2$
\vspace{0.5cm}

Singlet to Deuteron: $\sqrt{3}\kappa_{0}^2-\frac{1}{\sqrt{3}}\kappa_{1}^2-\frac{2}{\sqrt{3}}\kappa_{0}\kappa_{1}$
\vspace{0.5cm}

Deuteron to Singlet: $\sqrt{3}\kappa_{0}^2-\frac{1}{\sqrt{3}}\kappa_{1}^2-\frac{2}{\sqrt{3}}\kappa_{0}\kappa_{1}$
\vspace{0.5cm}

Singlet to Singlet: $-\kappa_{0}^2+\frac{5}{3}\kappa_{1}^2-\frac{2}{3}\kappa_{0}\kappa_{1}$

\chapter{Integral equations}
\label{ch:integral-equations}
\section{Analytical methods}
\label{analyticalmethods}
In this section I will formulate some important facts about a class of integral equations and give some proofs or proof sketches of these facts \cite{petrovskii1996lectures,nla.cat-vn1868446}. I will also give some discussion of the numerical methods used.
 The following form for linear integral equations is called 
Fredholm integral equation of the second kind:
\begin{equation}
y(p)=f(p)+\int_a^b K(p,q)y(q)dq
\end{equation}
where $f(p)$ and $K(p,q)$ are known functions and this is an equation for the unknown function $y(p)$. $K(p,q)$ is called the kernel of the equation.
All the functions and the integral are defined on a finite interval $(a,b)$. 
The theory of this integral equation is very similar to the theory of linear algebraic equations. To understand the analogy and to derive theorems about this equations let's break the interval
$(a,b)$ into $n$ pieces of equal length and approximate the integral by a sum. Denoting:
\begin{equation}
\Delta q=\frac{b-a}{n}
\end{equation}
\begin{equation}
p_i=a+i\frac{b-a}{n}
\end{equation}
\begin{equation}
 q_j=a+j\frac{b-a}{n}
\end{equation}
we find from the original equation the following equations:
\begin{equation}
y(p_i)=f(p_i)+\sum_j K(p_i,q_j)y(q_j) \Delta q, \hspace{0.4cm}  0 \le i\le n
\end{equation}
With further notations: $y(p_i)=y_i$, $f(p_i)=f_i$ and $K(p_i,q_j) \Delta q=K_{ij}$ we arrive at a system of linear algebraic equations:
\begin{equation}
y_i=f_i+\sum_j K_{ij} y_j, \hspace{0.4cm}  0\leq i,j \leq n
\end{equation}
which can be put into the following form:
\begin{equation}
\sum_j (\delta_{ij}-K_{ij}) y_j =f_i
\end{equation}
Now the analogy between the theory of linear integral equations and the theory of linear algebraic equations becomes obvious. The theorems about linear algebraic equations serve as a 
motivation to formulate analogous theorems about integral equations. 
All the theorems about linear algebraic equations involve the concept of matrices and determinants, hence our purpose will be to formulate these theorems without using determinants so that they can be 
generalized to integral equations.

First of all we know that if the determinant of the matrix $(\delta_{ij}-K_{ij})$ is non-zero then this system of equations has a unique solution for arbitrary numbers $f_i$. But if this determinant is zero
then our system of linear equations has at least one non-trivial solution if $f_i=0$ for all $i$. So we can formulate the following ``determinant-free'' statements:

\textbf{Statement 1}:

Either a system of linear algebraic equations has a unique solution for arbitrary inhomogeneous parts or the corresponding homogeneous system of equations has at least one non-trivial solution. 

\textbf{Statement 2}:
If for a given system of algebraic equations the first part of statement 1 is true then the first part of the statement 1 is also true for the transposed system of equations. 

Using the quantum mechanical notation for vectors the system of linear equations can be put in the following form:
\begin{equation} \label{solaeq} 
M \ket{y}=\ket{f}
\end{equation}
where $M$ is the matrix given by $(\delta_{ij}-K_{ij})$.

Now let's assume that the second part of the statement 1 is true for this system of equations, hence it is also true for the transposed system of equations. In this case one can ask the question: for which values
of $f_i$ does the system of equations still have a solution? To answer this question let's denote a solution of the transposed homogeneous equation by $\ket{z}$. Then the following is true:
\begin{equation}
 M^T \ket{z}=\bra{z}M=0
\end{equation}
so we can contract the equation \eqref{solaeq} with the dual vector  $\bra{z}$:
\begin{equation} 
\bra{z}M \ket{y}=\bra{z}\ket{f}
\end{equation}
but the left hand side of this equation is zero by the definition of $\bra{z}$, so we have:
\begin{equation} 
\bra{z}\ket{f}=0
\end{equation}
So we derived that a necessary condition for an inhomogeneous system of linear algebraic equations that satisfies the second part of the statement 1 to have a solution is that the vector $f$ be perpendicular 
to all of the solutions of the transposed homogeneous system of equations. With linear algebra it can be shown that this is also a sufficient condition, so we have the third statement:

\textbf{Statement 3}:

Given a system of inhomogeneous linear algebraic equations, for which the second part of the statement 1 is true, it will have a solution if and only if $\sum_i f_i z_i=0$ for all the solutions
$z_i$ of the transposed homogeneous of the original system of the equations.

Note that in this case the solution to the inhomogeneous system of equations is not unique because adding any solution of the homogeneous system of equations to a solution of the inhomogeneous 
system of equations we can find new solutions.

Having these three statements in hand we can try to formulate corresponding theorems for integral equations:

\textbf{Theorem 1}: Given a Fredholm integral equation of the second kind it will either have a unique solution for any inhomogeneous part, or the corresponding homogeneous equation has at least one solution.

\textbf{Theorem 2}: For a given Fredholm integral equation of the second kind for which the first part of the statement 1 is true, the first part of the statement will also be true for the transposed equation. 

\textbf{Theorem 3}: For a given Fredholm integral equation of the second kind for which the second part of the statement 1 is true, it will have a solution if and only if 
\begin{equation} 
\int f(q) z(q) dq =0
\end{equation}
for any solution of the transposed homogeneous integral equation $z(q)$. 

These theorems are called Fredholm theorems and are true for a large class of functions $f(p)$ and $K(q,p)$. Next I will prove these theorems for certain cases. Before doing that, I want to mention that in practice
integral equations are numerically solved using the machinery described here. Even the Fadeev equations for $nd$ scattering with infinite cutoff are solved by picking a very large cutoff (cutoff is very large if
it is much larger than the characteristic energy scales of the scattering problem), then discretizing the integration interval, arriving at a system of linear algebraic equations, which can be solved numerically.
In this process one usually hopes that increasing the cutoff and the number of mesh points will make the numerical solution converge, approaching to the actual solution. This happens, for example,
 for the quartet channel $nd$ scattering amplitude. Sometimes the numerical solution does not show any sign of convergence though; it keeps oscillating as the cutoff increases. This happens, for
example, for the LO doublet channel $nd$ scattering and the LO three-boson system without three-body forces. This usually means that the renormalization of the theory is not done correctly and one needs to include more conterterms in the 
Lagrangian to make the scattering amplitude converge.

A very important class of integral equations arises when one considers so-called separable kernels which are of the form:
\begin{equation} \label{}
K(p,q)=\sum_{i=1}^m a_i(p)b_i(q).
\end{equation}
Without loss of generality we can assume that all $a_i(p)$ are linearly independent and that all $b_i(q)$ are linearly independent. 
Substituting this into the equation we can see that the dependence of the function $y(p)$ on the variable $p$ can be pulled out of the integral and determined easily:
\begin{equation}
y(p)=f(p)+\sum_{i=1}^ma_i(p)\int b_i(q) y(q)dq.
\end{equation}
Denoting $\int b_i(q) y(q)dq=C_i$ we have:
\begin{equation}
y(p)=f(p)+\sum_{i=1}^m  C_i a_i(p).
\end{equation}
We have determined the dependence of $y(p)$ on $p$ up to some undetermined constants $C_i$. To determine these constants we substitute this solution back into the equation finding:
 \begin{equation}
f(p)+\sum_{i=1}^m  C_i a_i(p)=f(p)+\sum_{i=1}^ma_i(p)\int b_i(q)  [f(q)+\sum_{j=1}^m  C_j a_j(q)] dq.
\end{equation}
Canceling $f(p)$ from both sides and using the fact that $a_i(p)$ are linearly independent we find:
 \begin{equation}
 C_i =\int b_i(q) f(q) dq    +   \sum_{j=1}^m    \int b_i(q)   a_j(q) dq    C_j 
\end{equation}
for any $0 \leq  i \leq m$.
Making the notations: $\int b_i(q) f(q) dq=f_i$ and $\int b_i(q)   a_j(q) dq=K_{ij}$ we can put the last equation in the following form:
 \begin{equation}
  \sum_{j=1}^m (\delta_{ij}-K_{ij}) C_j  = f_i    
   \end{equation}
for any $0 \leq  i \leq m$.
This is a system of linear algebraic equations for the unknown constants $C_i$.
So we have proven that for this class of integral equation the solution can be actually reduced to the solution of an ordinary system of linear algebraic equations.
It is easy to see that to any solution of the integral equation corresponds a unique solution to this system of linear algebraic equations and vise versa. Using this correspondence 
between the solutions of the integral equation and the solutions of the system of linear algebraic equations the Fredholm theorems can be proven easily for this class of 
integral equations. 

Next let's consider integral equations which have ``small'' kernels and prove that for this class of equations the first part of the theorem 1 is always true, with smallness to be defined later. 
Before doing that let's introduce the following notation:
 \begin{equation}
[K_2 \circ K_1] (p,q) =\int K_2(p,s) K_1(s,q) ds.
\end{equation}
It can be shown that if the two functions $K_1(p,q)$ and $K_2(p,q)$ are uniformly continuous then $[K_2 \circ K_1] (p,q)$ is uniformly continuous also:
\begin{eqnarray} 
&&    | [K_2 \circ K_1] (p_1,q_1) -[K_2 \circ K_1] (p_2,q_2) |    \nonumber \\
&=&     |\int K_2(p_1,s) K_1(s,q_1) ds-\int K_2(p_2,s) K_1(s,q_2) ds|               \nonumber \\
&\leq&    |\int K_2(p_1,s) [K_1(s,q_1) -K_1(s,q_2)] ds|+  |\int  [K_2(p_1,s) -K_2(p_2,s)]  K_1(s,q_2) ds. |\nonumber \\ 
\end{eqnarray}
Denoting the upper limit of the functions $K_1(p,q)$ and $K_2(p,q)$ by $M$, and the length of integration interval by $L$, we see that for any $\epsilon$ there is a $\delta$ such that for any 
$|p_1-p_2|<\delta$ and $|q_1-q_2|<\delta$:
 \begin{equation}
|K_1(s,q_1) -K_1(s,q_2)| <\frac{\epsilon}{2ML}
\end{equation}
and
\begin{equation}
|K_2(p_1,s) -K_2(p_2,s)| <\frac{\epsilon}{2ML}.
\end{equation}
Substituting these inequalities into previous inequality we find that $| [K_2 \circ K_1] (p_1,q_1) -[K_2 \circ K_1] (p_2,q_2) | <\epsilon$, which proves the uniform continuity of the function $[K_2 \circ K_1] (p,q)$.

Now let's take an integral equation of the following form:
\begin{equation}
y(p)=f(p)+\lambda \int K(p,q)y(q)dq,
\end{equation}
with $K(p,q)$ and $f(p)$ uniformly continuous and $\lambda$ some small parameter. A natural way of finding an approximate solution to the equation would be to do an interpolation. At the first step we would 
take $y(p)=f(p)$, then we would substitute this back into the same equation to find $y(p)=f(p)+\lambda \int K(p,q)f(q)dq$ etc., and we would hope that this sequence of functions eventually converges. 
Another way of finding the same formal solution would be do write the function $y(p)$ as series in powers of $\lambda$:
\begin{equation} \label{expansion}
y(p)=y_0(p)+\lambda y_1(p)+\lambda^2 y_2(p)+...,
\end{equation}
then substitute this into the equation and match the coefficients of the powers of $\lambda$. After matching we find for the functions $y_k(p)$:
\begin{eqnarray} 
&&   y_0(p)=f(p)  \nonumber \\
&&   y_{k+1}(p)= \int K(p,q) y_k(q) dq, \hspace{0.4cm}  k=0,1,2,...           \nonumber \\
\end{eqnarray}

Doing the iterations it is easy to see:
\begin{equation}
y_k(p)= \int K^{(k)}(p,q)f(q)dq, \hspace{0.4cm}  k=1,2,...  
\end{equation}
where using the notation introduced above:
\begin{equation}
K^{(n)}(p,q)=K \circ K \circ ...\text{n times}...  \circ K
\end{equation}
Using what we have proved earlier it can be shown that all the functions $K^{(k)}(p,q)$ and $y_k(p)$ are uniformly continuous. Denote the upper limit of the function $|K(p,q)|$ by $M$: $|K(p,q)|<M$, 
and the length of the integration interval by $L$. Using the definition of $K^{(k)}(p,q)$ we find:
\begin{equation}
K^{(k)}(p,q)\leq M^k L^{k-1}
\end{equation}
As $f(p)$ is uniformly continuous too, we also have an upper limit for this function: $|f(p)|<F$.
Using the definition of $y_k(p)$ we find an estimate:
\begin{equation}
|y_k(p)| \leq M^kL^k F.
\end{equation}
Substituting this last equation into the equation \eqref{expansion} we find that this series will be uniformly convergent for all $\lambda$ in the circle:
\begin{equation}
|\lambda| < \frac{1}{ML}
\end{equation}
So as far as $\lambda$ is in this circle the function $y(p)$ defined by the series \eqref{expansion} will be uniformly continuous and will satisfy the integral equation. 
To show that this solution is unique let's take two solutions $y_1(p)$ and $y_2(p)$ to our equation. Substituting these functions into the equation and subtracting from each other we find:
\begin{equation}
y_2(p)-y_1(p)=\lambda \int K(p,q) [y_2(q)-y_1(q)] dq
\end{equation}
Denoting the upper limit of $|y_2(p)-y_1(p)|$ by $Y$ we find:
\begin{equation} \label{Y=0}
Y \leq |\lambda| MLY=cY,
\end{equation}
where $c=|\lambda| ML$.
Now using $|\lambda| < \frac{1}{ML}$ we also have that $c \leq 1$. This can be consistent with the inequality \eqref{Y=0} only if $Y=0$. Hence the series solution that we find for the integral equation is unique.

So we have proven the Fredholm theorems for two kinds of kernels. More general case of uniformly continuous kernels is proven by 
noting that any uniformly continuous function $K(p,q)$ defined on a finite square
 can be approximated by a polynomial in powers of $q$ and coefficients that are only functions of $p$ and we already know how to deal with this kind of kernels. Then the difference of $K(p,q)$ and
 this approximating polynomial can be made arbitrarily small, which is also something that we have already covered.

\section{Numerical methods}

\begin{figure}[H]
	\begin{center}
		\hspace{-1cm}\includegraphics[width=140mm]{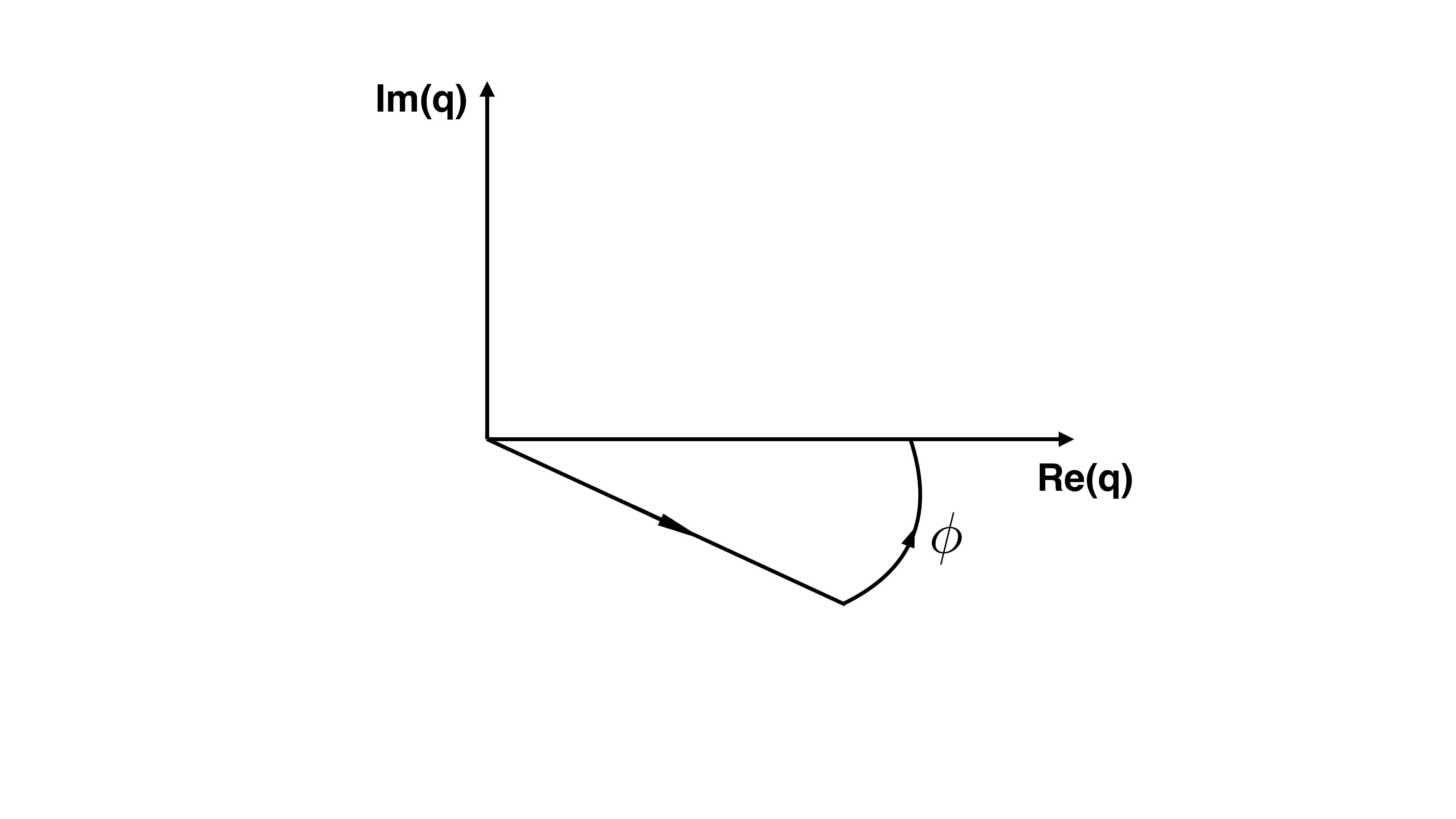} 
	\end{center}
\caption{\label{contour}The integration contour in the complex $q$ plane.}
\end{figure}

In practice the analytical methods introduced in the previous section of this appendix \ref{analyticalmethods} are not very useful for the
equations that we encounter in the \EFT, because the kernels of the integral equations are not separable. Instead we break the integration 
intervals into smaller intervals and numerically solve the system of linear algebraic equations obtained as described in section \ref{analyticalmethods}. 
For definiteness let's consider the LO quartet channel equation again (see Eq.~\eqref{ndQLOinteq}).
 \begin{eqnarray}\label{ndQLOinteq}
&&t ^ {l}_{0} (k,p)= -\frac{y_t^2M_N}{pk} Q_l\left(\frac{p^2+k^2-M_NE}{pk}\right)\nonumber \\ 
&-& \frac{2}{\pi} \int\limits_{0}^{\Lambda} dq q^2 t ^ {l}_{0} (k,q) \frac{1}{\sqrt{\frac{3q^2}{4}-M_NE-i\epsilon}-\gamma_t} \frac{1}{pq} Q_l \left(\frac{p^2+q^2-M_NE-i\epsilon}{pq}\right). \nonumber \\ 
\end{eqnarray}
We can see that the integrand of the integral term in Eq.~\eqref{ndQLOinteq} has singularities along the real axis. Because of these singularities
we need to increase the number of mesh points to make the solution to eventually converge numerically. Alternatively we could change 
the integration contour into the complex $q$ plane. In doing so we need to make sure the new integration contour satisfies some conditions. 
Firstly it has to start and end at $q=0$ and $q=\Lambda$, and secondly, in the region of the complex $q$ plane separated by the 
new contour and the real axis our integrand has to have no singularities. 
Then by applying Cauchy's integral theorem we wouldn't have changed the value of the integral. We can put both of the variables $q$ and $p$ 
on this contour before solving the equation. Then after having the solution function on this contour we can substitute it into the same 
equation, but this time choosing $p$ on the real axis, and we would get the solution on the real axis. 
The advantage of this approach is that if the new contour is chosen such that the integrand has no singularities on it, 
then we would need fewer mesh points to get a converging solution. This method of solving Faddeev equations was introduced by 
J.H. Hetherington and L.H. Schick \cite{PhysRev.137.B935}. The kernel in Eq.~\eqref{ndQLOinteq} has two sources of singularities. 
The first one is the deuteron propagator, which gives the singularity on the real axis. The second one is the function $Q_l \left(\frac{p^2+q^2-M_NE-i\epsilon}{pq}\right)$, which can give branch cuts on the complex plane which are not the real axis. Therefore the new contour has 
to be chosen appropriately, such that it avoids all of the singularities. This is done by investigating the analytical properties of the kernel of the 
integral equation \cite{PhysRev.176.1855}. For the equations we solve we choose the contour given in Fig.~\ref{contour} 
with $\phi=\tan^{-1}(\frac{2k}{\gamma_t})$ and the reference \cite{PhysRev.176.1855} shows that there are no singularities on this contour.

\chapter{Relevant integrals}
\label{RelevantIntegrals}
\section{Two-body $P$-wave contributions to $nd$ scattering, one-loop diagram}

The integrals that come up calculating the one-loop diagram with the two-body $P$-wave interactions are given in Eq.~\eqref{eqn:tensor}, Eq.~\eqref{eqn:vector} and 
Eq.~\eqref{eqn:scalar}. Here I will derive these integrals and give a couple general results.

\begin{eqnarray} \label{eqn:tensor}
&&I_{kl}(\vec{k},\vec{p},a^2,b^2)= \int\frac{d^3q}{(2\pi)^3}  \frac{1}{(\vec{q}-\frac{\vec{k}}{2})^2+a^2} \frac{1}{(\vec{q}-\frac{\vec{p}}{2})^2+b^2}\vec{q}_k \vec{q}_l\nonumber \\
&=&\frac{4\pi}{3(2\pi)^3}\Lambda\delta_{kl}\nonumber \\
&-&\frac{\pi^2}{4(2\pi)^3}\delta_{kl} \left\{ a+b-\frac{4(a^2-b^2)^2}{(a+b)(\vec{k}-\vec{p})^2}+8\frac{\tan^{-1}(\alpha)}{|\vec{k}-\vec{p}|}\frac{(a^2-b^2)^2+(a^2+b^2)\frac{(\vec{k}-\vec{p})^2}{2}+\frac{(\vec{k}-\vec{p})^4}{16}}{(\vec{k}-\vec{p})^2} \right\} \nonumber \\
&-&\frac{\pi^2}{4(2\pi)^3}k_kk_l\left\{ \frac{12(a^2-b^2)^2}{(a+b)(\vec{k}-\vec{p})^4}+\frac{5a-3b}{(\vec{k}-\vec{p})^2}-8\frac{\tan^{-1}(\alpha)}{|\vec{k}-\vec{p}|}\frac{3(a^2-b^2)^2+(3a^2-b^2)\frac{(\vec{k}-\vec{p})^2}{2}+\frac{3(\vec{k}-\vec{p})^4}{16}}{(\vec{k}-\vec{p})^4} \right\} \nonumber \\
&-&\frac{\pi^2}{4(2\pi)^3}p_kp_l\left\{ \frac{12(b^2-a^2)^2}{(a+b)(\vec{k}-\vec{p})^4}+\frac{5b-3a}{(\vec{k}-\vec{p})^2}-8\frac{\tan^{-1}(\alpha)}{|\vec{k}-\vec{p}|}\frac{3(a^2-b^2)^2+(3b^2-a^2)\frac{(\vec{k}-\vec{p})^2}{2}+\frac{3(\vec{k}-\vec{p})^4}{16}}{(\vec{k}-\vec{p})^4} \right\} \nonumber \\ 
&+&  \frac{\pi^2}{4(2\pi)^3} (k_kp_l+k_lp_k) \nonumber \\
&\times&\left\{  \frac{12(a^2-b^2)^2}{(a+b)(\vec{k}-\vec{p})^4}+\frac{a+b}{(\vec{k}-\vec{p})^2}-8\frac{\tan^{-1}(\alpha)}{|\vec{k}-\vec{p}|}\frac{3(a^2-b^2)^2+(a^2+b^2)\frac{(\vec{k}-\vec{p})^2}{2}-\frac{(\vec{k}-\vec{p})^4}{16}}{(\vec{k}-\vec{p})^4} \right \}  \nonumber \\
\end{eqnarray} 
and:

\begin{eqnarray}  \label{eqn:vector}
&&I_{l}(\vec{k},\vec{p},a^2,b^2)= \int\frac{d^3q}{(2\pi)^3}   \frac{1}{(\vec{q}-\frac{\vec{k}}{2})^2+a^2} \frac{1}{(\vec{q}-\frac{\vec{p}}{2})^2+b^2}\vec{q}_l  \nonumber \\
&=&\frac{\pi}{(2\pi)^2}k_l\left\{ \frac{b-a}{(\vec{k}-\vec{p})^2}+2\frac{\tan^{-1}(\alpha)}{|\vec{k}-\vec{p}|}\frac{(a^2-b^2)+\frac{(\vec{k}-\vec{p})^2}{4}}{(\vec{k}-\vec{p})^2} \right\} \nonumber \\
&+&\frac{\pi}{(2\pi)^2}p_l\left\{ \frac{a-b}{(\vec{k}-\vec{p})^2}+2\frac{\tan^{-1}(\alpha)}{|\vec{k}-\vec{p}|}\frac{(b^2-a^2)+\frac{(\vec{k}-\vec{p})^2}{4}}{(\vec{k}-\vec{p})^2} \right\} \nonumber \\ 
\end{eqnarray}
and:
\begin{equation} \label{eqn:scalar}
I(\vec{k},\vec{p},a^2,b^2)= \int\frac{d^3q}{(2\pi)^3}   \frac{1}{(\vec{q}-\frac{\vec{k}}{2})^2+a^2} \frac{1}{(\vec{q}-\frac{\vec{p}}{2})^2+b^2}=\frac{\pi^2}{(2\pi)^3} 4\frac{\tan^{-1}(\alpha)}{|\vec{k}-\vec{p}|}
\end{equation}
where

\begin{equation}
\alpha=\frac{|\vec{k}-\vec{p}|}{2(a+b)}
\end{equation}
The integral given in \eqref{eqn:vector} can be expressed in terms of \eqref{eqn:scalar}, and integral given in \eqref{eqn:tensor} can be expressed in terms of \eqref{eqn:vector} and \eqref{eqn:scalar} \cite{Passarino:1978jh,Ellis:2011cr}. To prove this let's take \eqref{eqn:tensor} and express it in terms of \eqref{eqn:vector} and \eqref{eqn:scalar}, the case of \eqref{eqn:vector} will be analogous and simpler. Notice that in \eqref{eqn:tensor} we can change the integration variable $\vec{q} \rightarrow \vec{q}-\frac{\vec{p}}{2}$ to find it expressed in terms of \eqref{eqn:vector} and \eqref{eqn:scalar} with appropriate parameters and the following integral: (after redefining $\frac{\vec{k}-\vec{p}}{2}$)

\begin{equation}
\int\frac{d^3q}{(2\pi)^3}  \frac{1}{(\vec{q}-\frac{\vec{k}}{2})^2+a^2} \frac{1}{\vec{q}\ ^2+b^2}\vec{q}_k \vec{q}_l
\end{equation}
Now notice that the answer of this integral has to be of the following form:

\begin{equation}
I_{kl}(\vec{k},\vec{p},a^2,b^2)= A\delta_{kl}+B k_kk_l 
\end{equation}
Where $A$ and $B$ are just rotational scalars. 
To find $A$ and $B$ we can contract both sides of the last equation with $\delta_{kl}$ and $k_kk_l$. This will give us the following two equations respectively:
\begin{equation} \label{eqn:system1}
\int\frac{d^3q}{(2\pi)^3}  \frac{1}{(\vec{q}-\frac{\vec{k}}{2})^2+a^2} \frac{1}{\vec{q} \ ^2+b^2}\vec{q} \ ^2= 3A+B \vec{k}^2
\end{equation}
\begin{equation} \label{eqn:system2}
\int\frac{d^3q}{(2\pi)^3}  \frac{1}{(\vec{q}-\frac{\vec{k}}{2})^2+a^2} \frac{1}{\vec{q}\ ^2+b^2}(\vec{q} \cdot \vec{k})^2 = A\vec{k}^2+B\vec{k}^4
\end{equation}
These equations give a system of linear algebraic equations for the constants $A$ and $B$, but we first need to calculate the integrals in the left hand sides of these equations.

Let's first see how the calculation goes for the left hand side of the first equation. Adding and subtracting $b^2$ to the numerator the integral breaks into two:

\begin{equation}
\int\frac{d^3q}{(2\pi)^3}  \frac{1}{(\vec{q}-\frac{\vec{k}}{2})^2+a^2} \frac{1}{\vec{q}\ ^2+b^2}\vec{q}\ ^2=\int\frac{d^3q}{(2\pi)^3}  \frac{1}{(\vec{q}-\frac{\vec{k}}{2})^2+a^2}-b^2\int\frac{d^3q}{(2\pi)^3}  \frac{1}{(\vec{q}-\frac{\vec{k}}{2})^2+a^2} \frac{1}{\vec{q}\ ^2+b^2}
\end{equation}
The second integral here can be recognized as \eqref{eqn:scalar} and the first one can be calculated either directly or by replacing the integration variable to $\vec{q} \rightarrow \vec{q}-\frac{\vec{p}}{2}$ to give:

\begin{equation}
\int\frac{d^3q}{(2\pi)^3}  \frac{1}{(\vec{q}-\frac{\vec{k}}{2})^2+a^2}=\frac{1}{2\pi^2}(\Lambda-\frac{a\pi}{2})
\end{equation}
where $\Lambda$ is the cutoff for $|\vec{q}|$. Here and in what follows after we neglect all the positive powers of $\frac{1}{\Lambda}$. 

Now lets turn to the calculation of the left hand side of the second equation in our system of equations for $A$ and $B$. Here we can do the following substitution:

\begin{equation}
\vec{q} \cdot \vec{k}=-((\vec{q}-\frac{\vec{k}}{2})^2+a^2)+(\vec{q}\ ^2+b^2)+\frac{k^2}{4}+a^2-b^2
\end{equation}
After the substitution we find:

\begin{eqnarray}  \label{eqn:tricky1}
&&\int\frac{d^3q}{(2\pi)^3}  \frac{1}{(\vec{q}-\frac{\vec{k}}{2})^2+a^2} \frac{1}{\vec{q}\ ^2+b^2}(\vec{q} \cdot \vec{k})^2 \nonumber \\
&=&-\int\frac{d^3q}{(2\pi)^3}  \frac{1}{\vec{q}\ ^2+b^2}\vec{q} \cdot \vec{k}+\int\frac{d^3q}{(2\pi)^3}  \frac{1}{(\vec{q}-\frac{\vec{k}}{2})^2+a^2} \vec{q} \cdot \vec{k} \nonumber \\
&+&(\frac{k^2}{4}+a^2-b^2)\int\frac{d^3q}{(2\pi)^3}  \frac{1}{(\vec{q}-\frac{\vec{k}}{2})^2+a^2} \frac{1}{\vec{q}\ ^2+b^2}\vec{q} \cdot \vec{k}\nonumber \\
\end{eqnarray} 
The first integral in this equation is obviously equal to zero, because of symmetry considerations. In the third integral we do the same substitution ($\vec{q} \cdot \vec{k}=-((\vec{q}-\frac{\vec{k}}{2})^2+a^2)+(\vec{q}\ ^2+b^2)+\frac{\vec{k}^2}{4}+a^2-b^2$) to find:

\begin{eqnarray}
&&\int\frac{d^3q}{(2\pi)^3}  \frac{1}{(\vec{q}-\frac{\vec{k}}{2})^2+a^2} \frac{1}{\vec{q}\ ^2+b^2}\vec{q} \cdot \vec{k}      \nonumber \\
&=&-\int\frac{d^3q}{(2\pi)^3} \frac{1}{\vec{q}\ ^2+b^2}  +\int\frac{d^3q}{(2\pi)^3}  \frac{1}{(\vec{q}-\frac{\vec{k}}{2})^2+a^2} \nonumber \\
&+&(\frac{k^2}{4}+a^2-b^2)\int\frac{d^3q}{(2\pi)^3}  \frac{1}{(\vec{q}-\frac{\vec{k}}{2})^2+a^2} \frac{1}{\vec{q}\ ^2+b^2}
\end{eqnarray} 
Here we know the first two integrals and the third one is again given by \eqref{eqn:scalar}. 

The trickiest integral is the middle term in \eqref{eqn:tricky1}. The naive change of integration variable $\vec{q} \rightarrow \vec{q}-\frac{\vec{k}}{2}$ gives the result $\frac{k^2}{(2\pi)^2}(\Lambda-\frac{a\pi}{2})$, which has the correct finite piece but incorrect infinite piece. The problem is that after the shift of the integration variable the integrand looks rotationally symmetric, but as the limits of the integration are changed also, this symmetry is just illusionary. This means that the change of variables is of course allowed, but one has to be careful about changing the limits of integration too. For example after the mentioned change of variables the integral breaks into two parts one of which seem to be the same as the first terms in \eqref{eqn:tricky1} and hence it should be zero, but it is in fact a different integral and is not equal to zero. Instead of changing the integration variables, I calculated this integral directly integrating first over $\phi$ then $\theta$ then $q$, and I got the following result:

\begin{equation} \label{eqn:tricky}
\int\frac{d^3q}{(2\pi)^3}  \frac{1}{(\vec{q}-\frac{\vec{k}}{2})^2+a^2} \vec{q} \cdot \vec{k}=\frac{1}{6\pi^2}k^2\Lambda-\frac{ak^2}{8\pi}
\end{equation}
So putting everything together we are able to solve the system of equations \eqref{eqn:system1} and \eqref{eqn:system2} for the constants $A$ and $B$ and from there it is easy to get to the answer of the integral in \eqref{eqn:tensor}.

There is a simpler way to calculate the integral in \eqref{eqn:tricky}, then the direct integration over the angles and the magnitude. To describe this I will need to use a lemma proven in what follows.

The discussion here is based on the following lemma.

\textbf{Lemma:} Suppose there is a function of one variable $f(x)$ that is analytic near infinity and integrable on any given interval on the real axes. Construct the following two functions of $\Lambda$: 

\begin{equation} \label{eqn:not-shifted}
\int_{-\Lambda}^{\Lambda} f(x) dx
\end{equation}
and

\begin{equation} \label{eqn:shifted}
\int_{-\Lambda+a}^{\Lambda+a} f(x) dx
\end{equation}
where $\Lambda$ and $a<\Lambda$ are arbitrary positive numbers such that $f(x)$ is analytic outside of a circle with radius smaller than $\Lambda-a$. Then the difference of the two functions \eqref{eqn:not-shifted} and \eqref{eqn:shifted} is an analytic function of $\Lambda$ near infinity and its series does not contain the $\Lambda-$free term.

Basically this lemma is just stating that after the shift of the integration limits the constant part of the integral does not change and only the non-zero powers of the infinity change. Note that the integral in \eqref{eqn:shifted} is the same as $\int_{-\Lambda}^{\Lambda} f(x+a) dx$, so we can rephrase the statement in the lemma saying that the $\Lambda-$free term in the series of the function $\int_{-\Lambda}^{\Lambda} (f(x)-f(x-a)) dx$ is zero. The generalization of this lemma will state that the $\Lambda-$free term of the series of the function $\int (f(\vec{q})-f(\vec{q}-\vec{k})) d^3q$ is zero, where the integration is done over a three dimensional ball of radius $\Lambda$. This generalized version can be used to calculate the infinite part of the integral \eqref{eqn:tricky} without any change of variables, then to calculate the finite part of it after changing the variables $\vec{q} \rightarrow \vec{q}-\frac{\vec{k}}{2}$, but not changing the limits of integration, as we know that the finite part is the same before and after the change of variables. 

\textbf{Proof of Lemma:} Let's do the following calculation:

\begin{eqnarray} \label{eqn:calcpart1}
&& \int_{-\Lambda}^{\Lambda} f(x) dx- \int_{-\Lambda+a}^{\Lambda+a} f(x) dx\nonumber \\
&=&   \int_{-\Lambda}^{-\Lambda+a} f(x) dx   +   \int_{-\Lambda+a}^{\Lambda} f(x) dx   -   \int_{-\Lambda+a}^{\Lambda} f(x) dx      -     \int_{\Lambda}^{\Lambda+a} f(x) dx    \nonumber \\
&=&   \int_{-\Lambda}^{-\Lambda+a} f(x) dx  -   \int_{\Lambda}^{\Lambda+a} f(x) dx    \nonumber \\
\end{eqnarray} 
In both of those integrals the integration variable $x$ is such that $|x|>\Lambda-a$, hence $f(x)$ is analytic and can be written as:

\begin{equation} 
f(x)=\sum_{n=-\infty}^{\infty} a_{n}x^n
\end{equation}
At this point we can see that the first statement in the lemma is a statement from complex analysis: any integral of an analytic function in the region of its analyticity is an analytic function.

Substituting this into the equation \eqref{eqn:calcpart1} we get:

\begin{eqnarray} \label{eqn:calcpart2}
&& \int_{-\Lambda}^{\Lambda} f(x) dx- \int_{-\Lambda+a}^{\Lambda+a} f(x) dx\nonumber \\
&=&   \int_{-\Lambda}^{-\Lambda+a} \sum_{n=-\infty}^{\infty} a_{n}x^n dx  -   \int_{\Lambda}^{\Lambda+a} \sum_{n=-\infty}^{\infty} a_{n}x^n dx   \nonumber \\
&=&   a_{-1} \int_{-\Lambda}^{-\Lambda+a}\frac{1}{x} dx  -  a_{-1}\int_{\Lambda}^{\Lambda+a}\frac{1}{x} dx   \nonumber \\
&+&   \sum_{n\neq-1} a_{n} \int_{-\Lambda}^{-\Lambda+a}x^n dx  -   \sum_{n\neq-1}a_{n}\int_{\Lambda}^{\Lambda+a}x^n dx   \nonumber \\
\end{eqnarray} 
The first line of this equation, after doing the integration can be simplified into $\log(1-\frac{2a}{\Lambda+a})$, which goes to zero as $\Lambda \rightarrow \infty$ and hence contains only negative powers of $\Lambda$ and the constant term is zero. The second line can be written as: 

\begin{eqnarray} 
&& \sum_{n\neq-1} a_{n} \int_{-\Lambda}^{-\Lambda+a}x^n dx  -   \sum_{n\neq-1}a_{n}\int_{\Lambda}^{\Lambda+a}x^n dx   \nonumber \\
&=&  \sum_{n<-1} a_{n} \int_{-\Lambda}^{-\Lambda+a}x^n dx  -   \sum_{n<-1}a_{n}\int_{\Lambda}^{\Lambda+a}x^n dx   \nonumber \\
&+&  \sum_{n>-1} a_{n} \int_{-\Lambda}^{-\Lambda+a}x^n dx  -   \sum_{n>-1}a_{n}\int_{\Lambda}^{\Lambda+a}x^n dx   \nonumber \\
\end{eqnarray} 
and after doing the integration:

\begin{eqnarray} 
&& \sum_{n\neq-1} a_{n} \int_{-\Lambda}^{-\Lambda+a}x^n dx  -   \sum_{n\neq-1}a_{n}\int_{\Lambda}^{\Lambda+a}x^n dx   \nonumber \\
&=&  \sum_{n<-1} \frac{a_{n}}{n+1} ((-\Lambda+a)^{n+1}-(-\Lambda)^{n+1}-(\Lambda+a)^{n+1}+\Lambda^{n+1})   \nonumber \\
&+&  \sum_{n>-1} \frac{a_{n}}{n+1} ((-\Lambda+a)^{n+1}-(-\Lambda)^{n+1}-(\Lambda+a)^{n+1}+\Lambda^{n+1})   \nonumber \\
\end{eqnarray} 
All the summands in the sum of the first line of the last equation go to zero as $\Lambda \rightarrow \infty$ and hence, with the same argument as in the first line of the equation  \eqref{eqn:calcpart2}, this line contains only negative powers of $\Lambda$ and the constant term is zero. In the second line we can open the brackets and it becomes obvious that for any given $n$ the constant term in the first term cancels the constant term in the third term.

A much easier way to prove the much stronger version of the second statement of the lemma is to notice that the difference function is an odd function of 
$\Lambda$, hence when put into series in powers of $\Lambda$ it will only have odd powers. This proves that only the odd powers of $\Lambda$ change and all the even powers do not. \qed

The applications of this lemma go beyond the calculation of the integral \eqref{eqn:tricky}, for example many of the steps of the calculation of the integrals given in \eqref{eqn:scalar}, \eqref{eqn:vector} and \eqref{eqn:tensor} are justified by this. More generally in Feynman parameter technique a change of the integration variable always occurs, which in principal may lead to an alteration of the answer and it is not always well explained why it does not. Also in dimensional regularization it is always considered that the same kind of change of variables is always allowed and this lemma can serve as a justification for that, because in dimensional regularization the dimension of the space is rendered such that the integrals always converge, hence they only have the finite part, which according to the lemma does not change. 

The generalization of the lemma to the higher dimensional integrals needs to be verified in the future.

The integral \eqref{eqn:scalar} can be calculated using position space techniques \cite{Savage:1999cm,Binger:1999rq}. The idea is to use Fourier transformations. First note that:

\begin{equation}
\int d^3x \frac{e^{-a |\vec{x}|}}{|\vec{x}|} e^{i(\vec{q}+\frac{\vec{k}}{2}) \cdot \vec{x}}=\frac{4\pi i}{a^2+(\vec{q}+\frac{\vec{k}}{2})^2}
\end{equation}
using this the integral \eqref{eqn:scalar} can be written as:

\begin{eqnarray}
&&  I(\vec{k},\vec{p},a^2,b^2)= \int\frac{d^3q}{(2\pi)^3}   \frac{1}{(\vec{q}-\frac{\vec{k}}{2})^2+a^2} \frac{1}{(\vec{q}-\frac{\vec{p}}{2})^2+b^2} \nonumber \\
&=& -\frac{1}{(4\pi)^2} \int\frac{d^3q}{(2\pi)^3} \int d^3x \frac{e^{-a |\vec{x}|}}{|\vec{x}|} e^{i(\vec{q}-\frac{\vec{k}}{2}) \cdot \vec{x}}  \int d^3y \frac{e^{-b |\vec{y}|}}{|\vec{y}|} e^{i(\vec{q}-\frac{\vec{p}}{2}) \cdot \vec{y}}
\end{eqnarray}
Where $\vec{x}$ and $\vec{y}$ are three dimensional spatial vectors. Doing the $\vec{q}$ integration first we find a delta function of the vector $\vec{x}+\vec{y}$, which then can be used to do the $\vec{y}$ 
integration. After that the $\vec{x}$ integration becomes trivial giving the result in \eqref{eqn:scalar}.

\section{Magnetic photon exchange diagrams}

\begin{center}\begin{figure}[ht]
  \begin{center}
  \includegraphics[scale=0.2]{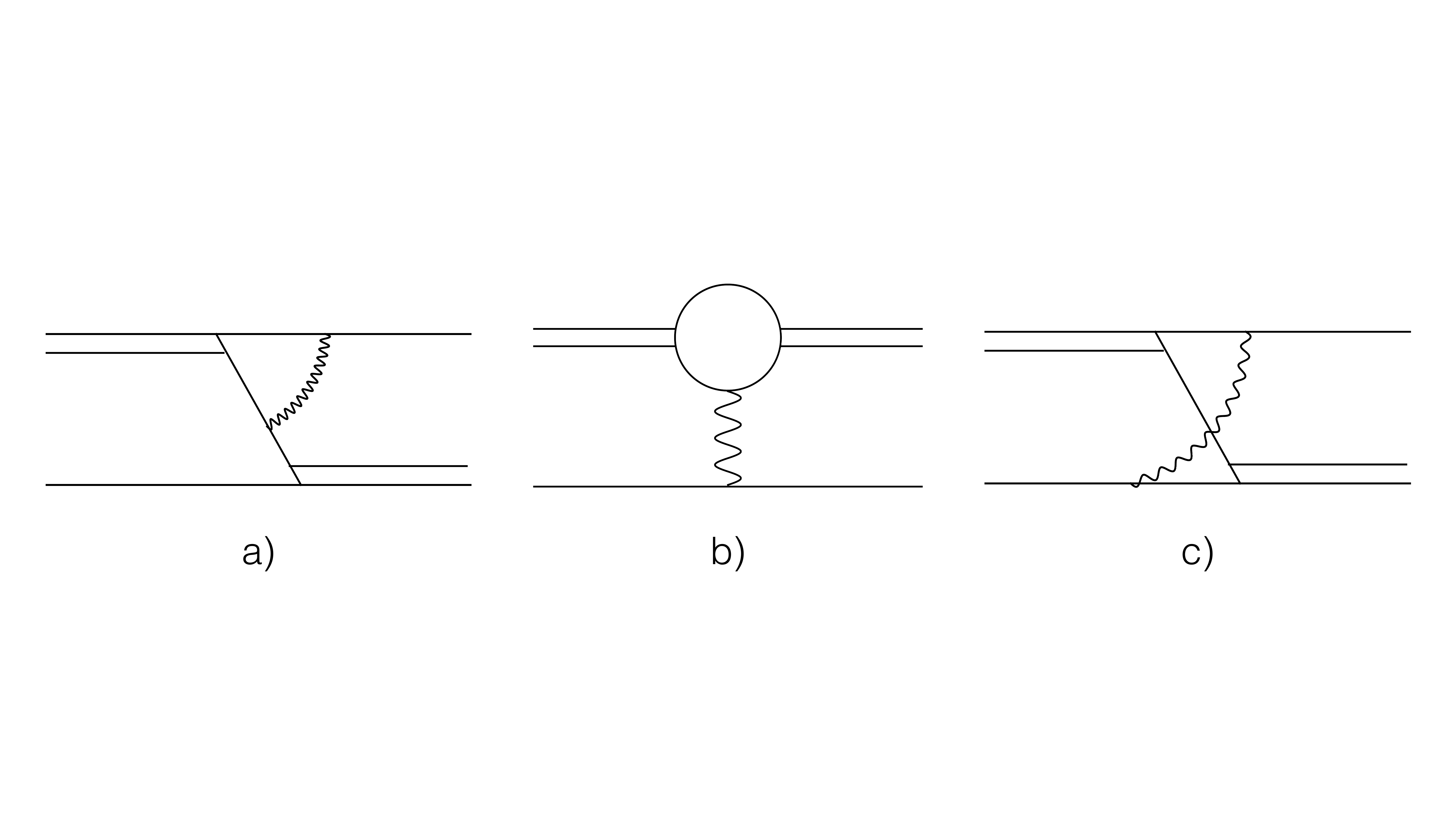}
  \end{center}
  \caption{\label{magneticphoton}Magnetic photon exchange contributions.}
\end{figure}\end{center}

In the $nd$ elastic scattering process there is only one electrically charged particle, the proton, so we don't have Coulomb interaction diagrams, but the neutrons and the proton can interact through their magnetic moments. 
These interactions are given by the Lagrangian in Eq.~\eqref{magnetic} \cite{Chen:1999tn}:

 \begin{equation} \label{magnetic}
\mathcal{L}_B=\frac{e}{2M_N}N^{\dagger}(\kappa_0+\kappa_1\tau_3)\vec{\sigma} \cdot \vec{B}N
\end{equation}
where $\kappa_0$ and $\kappa_1$ are the magnetic moments of the neutron and the proton, and $\vec{B}$ is given by the vector potential 
$\vec{B}=\vec{\nabla}\cross \vec{A}$. The diagrams that contribute to $nd$ scattering and involve this interaction vertex are of higher order than $\ntlo$, 
so they are not included in our calculations. Some of the lowest order diagrams that involve this interaction contribute at N$^5$LO and are given in Fig.~\ref{magneticphoton}.

To see that these diagrams contribute at N$^5$LO let's use the rules described in Section \ref{powercountingsection}: 
the diagrams have two $dNN$ vertices together scaling as $\sim y_t^2\sim\frac{1}{M}$, three nucleon propagators $\sim\frac{M^3}{Q^6}$, one photon propagator $\sim\frac{1}{Q^2}$, 
a non-relativistic loop integration $\sim\frac{Q^5}{M}$, and two photon vertices $\sim\alpha(\frac{ Q}{M})^2$, 
where $\alpha$  is the fine structure constant and it can be approximated as $\alpha\sim0.01$. Putting all of these together we get for the magnetic-photon diagrams' scaling: 
$\sim \alpha \frac{1}{MQ}$. As we work at energy scales where the power counting parameter is about $\frac{Q}{M}\sim0.3$ we can estimate the scaling of the fine structure constant as 
$\alpha\sim(\frac{Q}{M})^4$. Substituting this we find for the scaling of the diagrams: $\sim (\frac{Q}{M})^5 \frac{1}{Q^2}$. Recalling that the LO contributions scale as $\sim\frac{1}{Q^2}$, 
we see that the magnetic-moment diagrams contribute at N$^5$LO.

As I already mentioned we have not included the contributions from these diagrams in our calculation of the $\ntlo$ $nd$ scattering amplitude, 
but we have calculated them and they can be useful in the future, for the higher order calculations of this amplitude. 
Here I will just mention that the spin-isospin projections of these 
diagrams is done by using all the techniques described in Appendix \ref{Projections} and they pose no extra complications. The integration part of the first two 
diagrams a) and b) is done using the integrals Eqs.\eqref{eqn:scalar}, \eqref{eqn:vector} and \eqref{eqn:tensor} given in the first section of this appendix. 
The only new complication that we encounter comes from the loop-integration part of the diagram c). A new type of integral needs to be calculated which is given by:

\begin{equation} \label{newscalar}
I_c(\vec{k},\vec{p},a^2,b^2)= \int\frac{d^3q}{(2\pi)^3}  \frac{1}{q^2} \frac{1}{(\vec{q}-\frac{\vec{k}}{2})^2+a^2} \frac{1}{(\vec{q}-\frac{\vec{p}}{2})^2+b^2}.
\end{equation}
Note that the only difference between this integral and the one in Eq.~\eqref{eqn:scalar} is the extra denominator $\frac{1}{q^2}$ that come from the photon propagator. 
The integral in Eq.~\eqref{eqn:scalar} can be calculated using position space techniques, as described earlier, or it can also be done using Feynman parametrization. 
Neither of these approaches works for Eq.~\eqref{newscalar}. 
Using the position space techniques, after using the delta function that comes from integrating over $\vec{q}$ 
we get two more integrals left as opposed to one when doing Eq.~\eqref{eqn:scalar} and it doesn't seem to show any way to proceed. Using Feynman parametrization 
we get two parameters as opposed to one when doing Eq.~\eqref{eqn:scalar}, the integration over the first parameter is fairly simple, but it gives expressions involving 
square roots and $\arctan$ functions. After that we have one more parameter left in the integral, and it shows no hope of being integrable in a closed form. So a new approach 
is necessary if we want to find a closed form for this integral. I found a simple change of integration variable that simplifies it considerably and expresses it in terms 
of Eq.~\eqref{eqn:scalar}. Defining a new vector:

\begin{equation}
\vec{l}=\frac{\hat{q}}{|\vec{q}|}
\end{equation}
we can change the integration over $\vec{q}$ to an integration over $\vec{l}$ finding:

\begin{eqnarray}\label{IctoI}
&& I_c(\vec{k},\vec{p},a^2,b^2)=\int\frac{d^3q}{(2\pi)^3} \frac{1}{q^2}  \frac{1}{(\vec{q}-\frac{\vec{k}}{2})^2+a^2} \frac{1}{(\vec{q}-\frac{\vec{p}}{2})^2+b^2} \nonumber \\
&=&\frac{1}{\kappa_1^2 \kappa_2^2}\int\frac{d^3l}{(2\pi)^3}  \frac{1}{(\vec{l}-\frac{\vec{k}}{2 \kappa_{1}^2})^2+\frac{a^2}{\kappa_{1}^4}} \frac{1}{(\vec{l}-\frac{\vec{p}}{2 \kappa_{2}^2})^2+\frac{b^2}{\kappa_{2}^4}},
\end{eqnarray}
where

\begin{equation}
\kappa_1^2=a^2+\frac{k^2}{2}
\end{equation}
and 

\begin{equation}
\kappa_2^2=b^2+\frac{p^2}{2}.
\end{equation}
The last line of Eq.~\eqref{IctoI} is just the integral Eq.~\eqref{eqn:scalar}, so we get the final answer:

\begin{equation}
 I_c(\vec{k},\vec{p},a^2,b^2)=\frac{1}{\kappa_1^2 \kappa_2^2} I\left(\frac{\vec{k}}{2 \kappa_{1}^2},\frac{\vec{p}}{2 \kappa_{2}^2},\frac{a^2}{\kappa_{1}^4},\frac{b^2}{\kappa_{2}^4}\right)=\frac{1}{2\pi \kappa_1^2\kappa_2^2} \frac{1}{|\frac{\vec{k}}{\kappa_1^2}-\frac{\vec{p}}{\kappa_2^2}|}\tan^{-1}\left(\frac{|\frac{\vec{k}}{\kappa_1^2}-\frac{\vec{p}}{\kappa_2^2}|}{2(\frac{a}{\kappa_1^2}+\frac{b}{\kappa_2^2})}\right).
\end{equation}


\bibliographystyle{unsrtnat}
\cleardoublepage
\normalbaselines 
\addcontentsline{toc}{chapter}{Bibliography} 
\bibliography{./References}

\end{document}